\documentclass[a4paper,11pt,nofootinbib]{article}
\pdfoutput=1

\usepackage{jheppub,fancyhdr,graphicx,epstopdf,color,amsmath,cases,slashed,hyperref}
\usepackage{makecell}
\usepackage{float}
\usepackage{caption,subcaption}  
\usepackage{dcolumn}   
\usepackage{amsmath, amssymb}        
\usepackage{latexsym,array,multirow,verbatim,enumerate,cancel}
\usepackage{slashed}
\usepackage{bigcenter}
\usepackage[utf8]{inputenc}
\usepackage{footnote}
\makesavenoteenv{tabular}
\makesavenoteenv{table}

\def\issue(#1,#2,#3){{\bf #1}, #2 (#3)}

\catcode`\@=11
\def\lsim{\mathrel{\mathpalette\@versim<}}
\def\gsim{\mathrel{\mathpalette\@versim>}}
\def\@versim#1#2{\vcenter{\offinterlineskip
\ialign{$\m@th#1\hfil##\hfil$\crcr#2\crcr\sim\crcr } }}
\catcode`\@=12

\catcode`@=12

\newcommand{\met}{$\cancel E_T$}

\newcommand{\newc}{\newcommand}
\newc{\wt}{\widetilde}
\newc{\ra}{\rightarrow}

\def\beq {\begin{equation}}
\def\eeq {\end{equation}}
\def\bi {\begin{itemize}}
\def\ei {\end{itemize}}
\def\bea {\begin{eqnarray}}
\def\eea {\end{eqnarray}}

\def \met{\slashed{E}_T}

\newcommand{\br}{\begin{eqnarray}}
\newcommand{\er}{\end{eqnarray}}
\newcommand{\be}{\begin{equation}}
\newcommand{\ee}{\end{equation}}

\newcommand{\ch}{\widetilde \chi^\pm}

\newcommand{\ifb} {\rm {fb}^{-1}}

\def \ch2p {{\wt\chi_2^+}}

\def \ch2m {{\wt\chi_2^-}}

\newc{\dmchi}{\Delta m_{\wt\chi}}

\def\issue(#1,#2,#3){{\bf #1}, #2 (#3)}

\usepackage{array}
\newcolumntype{L}[1]{>{\raggedright\let\newline\\\arraybackslash\hspace{0pt}}m{#1}}
\newcolumntype{C}[1]{>{\centering\let\newline\\\arraybackslash\hspace{0pt}}m{#1}}
\newcolumntype{R}[1]{>{\raggedleft\let\newline\\\arraybackslash\hspace{0pt}}m{#1}}

\hypersetup{
   colorlinks=true,       
   linkcolor=blue,        
   citecolor=blue,         
   filecolor=magenta,      
   }

\title{Resonant heavy Higgs searches at the HL-LHC}

\preprint{IPPP/18/107}

\author[a]{Amit Adhikary,}
\author[b]{Shankha Banerjee,}
\author[a]{Rahool Kumar Barman,}
\author[a]{Biplob Bhattacherjee}
\affiliation[a]{Centre for High Energy Physics, Indian Institute of Science, Bangalore 560012, India}
\affiliation[b]{Institute for Particle Physics Phenomenology, Department of Physics, Durham University,
Durham DH1 3LE, United Kingdom}
\emailAdd{amitadhikary@iisc.ac.in}
\emailAdd{shankha.banerjee@durham.ac.uk}
\emailAdd{rahoolbarman@iisc.ac.in}
\emailAdd{biplob@iisc.ac.in}
\date{\today}

\abstract
{In this work, we show the importance of searches for heavy resonant scalars ($H$) and pseudoscalars ($A$). Taking cue from the present searches, we make projections for searches in an extended scalar sector at the high luminosity run of the Large Hadron Collider. We study the three most relevant search channels, \textit{i.e.}, $H \to hh, \; H/A \to t\bar{t}$ and $b\bar{b}H/A$. Upon studying multifarious final states for the resonant double Higgs production, we find that the $b\bar{b}\gamma\gamma$ ($\sigma(pp \to H \to hh) \in [81.27,14.45]$ fb for $m_H \in [300,600]$ GeV at 95\% C.L.) and $b\bar{b}b\bar{b}$ ($[5.4,2.5]$ fb for $m_H \in [800,1000]$ GeV at 95\% C.L.) channels are the most constraining. For the $b\bar{b}H$ channel, we can exclude $\sigma(pp \to b\bar{b}H) \in [22.2,3.7]$ fb for $m_H \in [300,500]$ GeV. Finally, we consider the phenomenological Minimal Supersymmetric Standard Model as an example and impose various present constraints and our future direct search-limits and obtain strong constraints on the $m_A-\tan{\beta}$ parameter space, where $m_A$ and $\tan{\beta}$ are respectively the mass of the pseudoscalar and the ratio of the vacuum expectation values of the two Higgs doublets. Assuming that the heavy Higgs boson decays only to Standard Model (SM) states, we find that the $H \to hh \to b\bar{b}\gamma\gamma$ ($H \to t\bar{t}$) channel excludes $\tan{\beta}$ as low as 4 ($m_A \in [400,800]$ GeV) at 95\% CL. This weakens up to $\sim 5.5$ when the $b\bar{b}H$ channel dominates. Upon allowing for non-SM decay modes, the limits weaken.}

\begin{document}
\maketitle

\section{Introduction}
\label{intro}
The Higgs boson discovered in 2012, was the last missing piece in the Standard Model of particle physics (SM). The SM, however, is inadequate to explain the nature and existence of dark matter, the small but non-negligible masses of the neutrinos, the asymmetry between baryons and anti-baryons, to name a few. Besides, SM can not explain the hierarchy problem which is inherent in the theory. Well motivated theories including supersymmetry have the potential to solve some of these limitations. There are additional fundamental theoretical requirements that the SM can not satisfy. The aforementioned experimental observations and theoretical requirements compel us to look for physics beyond the Standard Model (BSM). However, the possibilities being innumerable, it is extremely difficult to ascertain the nature of such new physics. Since the discovery of the Higgs boson, and a growing convergence of its properties with the SM expectations~\cite{Khachatryan:2016vau, CMS-PAS-HIG-16-020, ATLAS-CONF-2016-067, ATLAS-CONF-2016-079,ATLAS-CONF-2016-081,  CMS:2017jkd, ATLAS-CONF-2016-112, CMS-PAS-HIG-15-003, Sirunyan:2017khh, ATLAS-CONF-2018-021, Sirunyan:2018hoz, Aaboud:2018urx, Aaboud:2018zhk, Sirunyan:2018kst, CMS-PAS-HIG-16-006, ATLAS-CONF-2017-014}, the new physics possibilities are gradually getting strongly constrained. Searches for BSM are being performed at the Large Hadron Collider (LHC) by the CMS, ATLAS, ALICE and LHCb collaborations. Except for some excitement in the flavour physics sector, there have not been any strong hints for new physics in the form of new particles or significant deviations in couplings with respect to the SM. Even though supersymmetry is perhaps one of the most elegant theories of our time, it comes with additional new particles, which need to be discovered at some stage. Even though searches are being conducted for a considerable region of parameter space for the Minimal Supersymmetric Standard Model (MSSM), there are more non-traditional searches which need to be performed. The MSSM parameter space has been extensively studied in light of the constraints from cosmology, flavour physics, and Run-I plus Run-II data from LHC~\cite{Carena:2011aa, Arbey:2011ab, Baer:2012mv, Arbey:2012dq, Altmannshofer:2012ks,  Cheung:2013kla, Chowdhury:2015yja, Bhattacherjee:2015sga, Bechtle:2016kui, Bechtle:2012jw, Djouadi:2013lra, Bechtle:2015pma, Buchmueller:2013rsa, Scopel:2013bba, deVries:2015hva, Barr:2016sho, Kowalska:2016ent, Han:2016xet, Buckley:2016kvr, Zhao:2017qpe,Bagnaschi:2016afc,Bagnaschi:2016xfg,Barman:2016jov,Bagnaschi:2017tru,Costa2018}. However, there are simple extensions of the MSSM that can weaken the present bounds considerably. On the positive side, the LHC can potentially pin down the Higgs couplings to weak bosons and most of the fermions at the level of $\mathcal{O}(5-10\%)$~\cite{CMS-PAS-FTR-18-011, ATL-PHYS-PUB-2018-054, Peskin:2012we, Peskin:2013xra}. However, as has been shown in numerous experimental~\cite{Aad:2015xja, Aad:2015uka, ATLAS-CONF-2016-049, ATLAS-CONF-2016-071, ATLAS-CONF-2016-004, CMS-PAS-HIG-16-028, CMS-PAS-HIG-16-002, CMS-PAS-B2G-16-008, CMS-PAS-HIG-16-032, CMS-PAS-HIG-16-011, CMS-PAS-HIG-17-002, CMS-PAS-HIG-17-006, Aaboud:2016xco, Aaboud:2018sfw} (including future extrapolations~\cite{ATL-PHYS-PUB-2014-019, ATL-PHYS-PUB-2017-001, CMS-DP-2016-064}) and phenomenological studies~\cite{Osland:1998hv, Baur:2003gp,  Liu:2004pv, Dib:2005re, Wang:2007zx, Pierce:2006dh, Kanemura:2008ub, Contino:2010mh, Grober:2010yv, Dolan:2012ac, Dolan:2012rv, Kribs:2012kz, Contino:2012xk, Dawson:2012mk, Dolan:2012ac, Ellwanger:2013ova, Nishiwaki:2013cma, Barr:2013tda, Chen:2013emb, deLima:2014dta, Chen:2014xra, Baglio:2014nea, Hespel:2014sla,  Slawinska:2014vpa, Englert:2014uqa, Goertz:2014qta, Maltoni:2014eza, Baglio:2014nea, Hespel:2014sla, Azatov:2015oxa, Lu:2015jza, Lu:2015qqa, Carvalho:2015ttv, Bian:2016awe, Banerjee:2016nzb, Gorbahn:2016uoy, Carvalho:2016rys, Cao:2016zob, Crivellin:2016ihg, Grober:2016wmf, Huang:2017jws, Kribs:2017znd, DiLuzio:2017tfn, Nakamura:2017irk, Alves:2017ued, Adhikary:2017jtu, Chala:2018ari, Basler:2018dac}, the measurement of the elusive triple Higgs coupling ($\lambda_{hhh}$) is a difficult feat at the LHC. Studies show that future colliders are expected to be more adept in constraining or even measuring this coupling to a great precision~\cite{Plehn:2005nk, Binoth:2006ym, Yao:2013ika, Liu:2014rba, Barr:2014sga, He:2015spf, Papaefstathiou:2015iba, Azatov:2015oxa, Kotwal:2016tex, Fuks:2017zkg, Banerjee:2018yxy}. In order to be completely sure whether or not there is any extended Higgs sector, it is of utmost importance to measure the Higgs quartic coupling, $\lambda_{hhhh}$ and the Higgs trilinear coupling, $\lambda_{hhh} = \lambda_{hhhh} v$, where $v$ is the vacuum expectation value of the SM Higgs boson. Now, independent measurements of the Higgs couplings to the gauge bosons and fermions will constrain $v$ and we already have a precise Higgs mass measurement. To confirm this sector of the SM, one needs to measure $\lambda_{hhh}$ or $\lambda_{hhhh}$~\cite{Glover:1987nx,Boudjema:1995cb, Plehn:1996wb, Djouadi:1999rca,Baur:2002qd,Baur:2003gp}.

In the following sections, we focus on the various production and decay processes of a resonant scalar, \textit{viz.}, a resonant decay to a pair of SM-like Higgs bosons, to a pair of top quarks, a heavy pseudoscalar, $A$, decaying to an SM-like Higgs boson and a $Z$-boson and the production of a heavy scalar in association with a pair of bottom quarks. The final theme of this work is in the context of the phenomenological MSSM (pMSSM). However, our results are presented in such a way that they can be mapped onto most models with an extended scalar sector. Table~\ref{tab:tab1} summarises the various bounds set on the double-Higgs production cross-section by CMS and ATLAS in the non-resonant and resonant categories. The resonant production results are mostly interpreted in terms of spin-0 and spin-2 hypotheses. Besides, there are many supersymmetric interpretations for the resonant scalar searches. As an example, for the $b\bar{b}\tau^+\tau^-$ resonant search performed by CMS~\cite{Sirunyan:2017djm}, the MSSM parameters $m_A$ (mass of the $CP$-odd scalar, $A$) and $\tan{\beta}$ (ratio of the vacuum expectation values of the two Higgs doublets in the model, \textit{viz.}, $H_u$ and $H_d$) are excluded in the range 230 GeV $< m_A <$ 360 GeV and $\tan{\beta} \lesssim 2$, at 95\% CL. Thus, besides measuring deviations to the Higgs self-coupling, there are other possible channels to look for in order to establish an extended scalar sector. Some of these new channels include the production of the SM-like Higgs in association with a $Z$-boson reconstructing a resonance. Another possible channel is the production of a pair of top-quarks. Now, the first of these channels can be via a heavy pseudoscalar resonance~\cite{Khachatryan:2015tha, Aaboud:2017cxo}, whereas the $t\bar{t}$ production can be either through a heavy scalar or pseudoscalar~\cite{Carena:2016npr, Aaboud:2017hnm}. However, both these channels can also come about from a heavy $Z^{\prime}$~\cite{CMS:2016zte, Aaboud:2017ahz, Aaboud:2017cxo, Aaboud:2018mjh, Aaboud:2018bun}. Now, in order to be sure whether the $Zh$ or $t\bar{t}$ production is via a spin-0 or spin-1 resonance, one needs to delve deeper into the angular observables. Lastly, we also study the effects of the high $\tan{\beta}$ regime for a heavy scalar produced in association with a pair of $b$-quarks and decaying to a pair of $\tau$-leptons~\cite{Aaboud:2017sjh}.

\begin{table}
\begin{center}
 \begin{tabular}{||c | c | c | c | c||} 
 \hline
 Channel                                                          & CMS (NR)     & CMS (R)         & ATLAS (NR)   & ATLAS (R)        \\ [0.5ex] 
                                                                  & ($\times$SM) & [fb, (GeV)]     & ($\times$SM) & [fb, (GeV)]      \\
 \hline\hline
 $b\bar{b}b\bar{b}$                                               & $75$           & $1500-45$         & $13$           & $2000-2$      \\
 ~\cite{Sirunyan:2018tki,CMS-PAS-HIG-17-009,Aaboud:2018knk}       &              & $(260-1200)$      &              & $(260-3000)$  \\
 \hline
 $b\bar{b}\gamma\gamma$                                           & $24$           & $240-290$         & $22$           & $1100-120$    \\
 ~\cite{CMS-PAS-HIG-17-008,Sirunyan:2018iwt,Aaboud:2018ftw}       &              & $(250-900)$       &              & $(260-1000)$  \\
 \hline
 $b\bar{b}\tau^+\tau^-$                                           & $30$           & $3110-70$         & $12.7$         & $1780-100$   \\
 ~\cite{Sirunyan:2017djm,Aaboud:2018sfw}                          &              & $(250-900)$       &              & $(260-1000)$  \\
 \hline
 $\gamma\gamma W W^*$~\cite{Aaboud:2018ewm}                       &              &                 & $200$          & $40000-6100$ \\
 ($\gamma\gamma\ell\nu jj$)                                       &              &                 &              & $(260-500)$   \\

 \hline
 $b\bar{b}\ell\nu\ell\nu$                                         & $79$           & $20500-800$       & $300$          & $6000-170$     \\
 ~\cite{Sirunyan:2017guj,Aaboud:2018zhh}                          &              & $(300-900)$       &              & $(500-3000)$   \\ [1ex] 
 \hline
 \hline
 $WW^*WW^*$                                                       &              &                 & $160$          & $9300-2800$    \\
 ~\cite{Aaboud:2018ksn}                                           &              &                 &              & $(260-500)$    \\ [1ex] 
 \hline
\end{tabular}
\caption{Bounds obtained on the di-Higgs cross-sections (in fb) from CMS and ATLAS studies dedicated to the search for non-resonant (NR) and resonant (R) double Higgs production in various channels. The numbers in brackets show the range of the heavy scalar mass considered in that particular study.}
\label{tab:tab1}
\end{center}
\end{table}

Our paper is organised as follows. We study the reach of the HL-LHC for the $H \to h h$ channel in various final states, in section~\ref{sec:Htohh}, by showing the 95\% CL bounds on $\sigma(p p \to H \to h h)$ as functions of the heavy Higgs mass, $m_H$. Following the Higgs pair production, we address the couplings of the $CP$-even heavy Higgs to a pair of top quarks and to a pair of bottom quarks in sections~\ref{sec:Htott} and~\ref{sec:bbH} respectively. In section~\ref{sec:pMSSM}, we use the previous results to recast our limits in the purview of the phenomenological MSSM (pMSSM) and show the future reach of these searches in the $m_A-\tan{\beta}$ parameter space. We finally summarise our results and present our future outlook in section~\ref{sec:summary}.

\section{The $pp\to H\to hh$ Channel}
\label{sec:Htohh}

As discussed in the introduction, the objective of this work is to scrutinise the viable scalar extensions of the Standard Model (SM). In this section, we focus on a heavy $CP$-even scalar produced via gluon fusion and subsequently decaying to a pair of SM-like Higgs bosons. The decay width of the heavy Higgs boson is chosen to be $\sim 1~{\rm GeV}$ which is within the resolution of the detector. We would like to mention that the heavy Higgs search limits derived in the course of this analysis would stand valid only if the detector resolution is greater than the chosen heavy Higgs decay width.
In the next five subsections, we study the reach of the HL-LHC in constraining the resonant Higgs pair production cross-section, $\sigma(p p \to H \to h h)$, upon studying multifarious channels, \textit{viz.}, $b\bar{b}\gamma\gamma, b\bar{b}b\bar{b}, b\bar{b}\tau^+\tau^-, b\bar{b}W^+W^-$ and $\gamma\gamma W^+W^-$. Many of these channels with $\tau$-leptons and $W$-bosons give different signatures upon considering leptonic or hadronic modes. We study all possible final states giving importance to the total rate as well the cleanliness. Unless otherwise stated, we generate the signal samples with \texttt{Pythia 6} and for the background samples, we use \texttt{MG5\_aMC@NLO}~\cite{Alwall:2014hca}. The showering and hadronisation is performed within the \texttt{Pythia 6}~\cite{Sjostrand:2001yu} framework. The $b$-tagging efficiency and mistag efficiencies of a $c$-jet or a light jet posing as a $b$-tagged jet are employed as functions of the transverse momentum of the jet~\cite{Sirunyan:2017ezt}. Unless explicitly mentioned, the \texttt{CTEQ6l} PDF set has been used throughout this work. Also, to take into account the detector effects, we use the fast-detector simulation package, \texttt{Delphes-3.4.1}~\cite{deFavereau:2013fsa}.

\begin{figure}[htb!]
\centering
\includegraphics[trim=0 550 0 70,clip,width=12 cm]{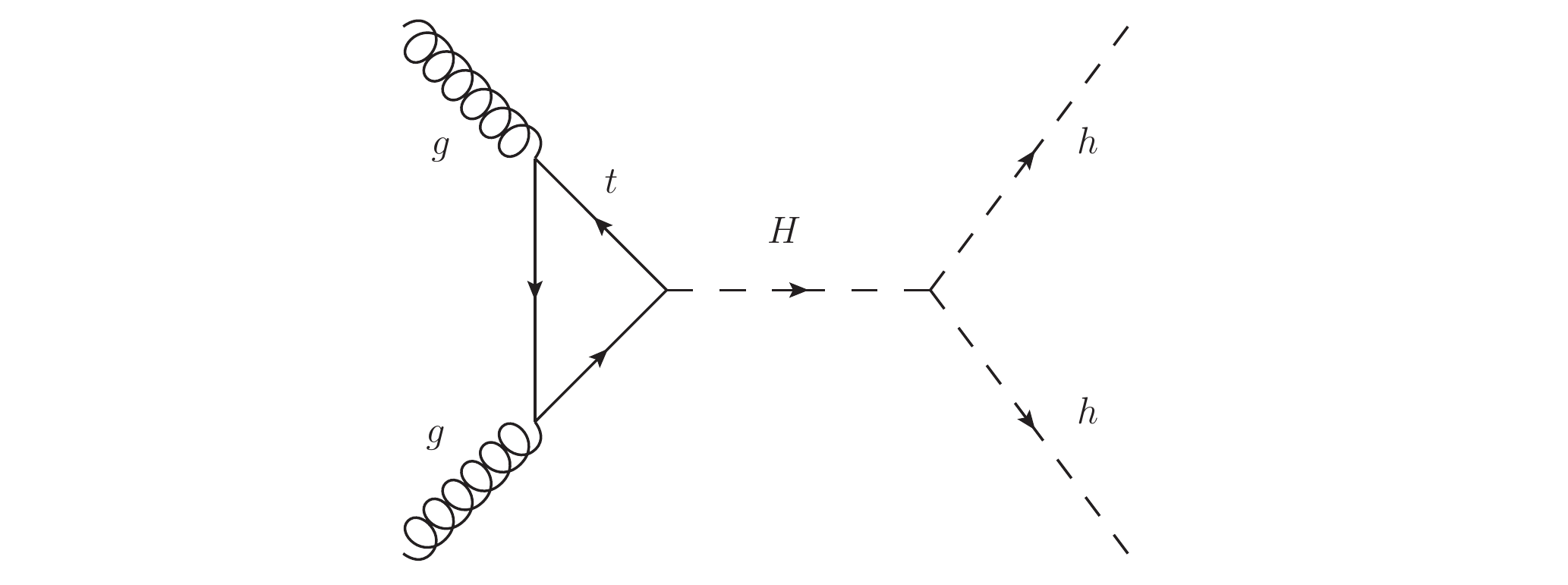}
\caption{Feynman diagram for the signal production from $pp\to H\to hh$. The SM Higgses then decay to the corresponding final states \textit{viz.} $h\to b\bar{b}$ and $h\to \gamma\gamma$ give rise to $b\bar{b}\gamma\gamma$ final state.}
\label{FD:bbaa}
\end{figure}

\subsection{The $b\bar{b}\gamma\gamma$ Channel}
\label{sec:Htohhtobbgaga}

The $b\bar{b}\gamma\gamma$ final state is the golden channel when it comes to studying the non-resonant double Higgs production. The cleanliness of this channel, owing to smaller backgrounds, triumphs over the reduced rate ($Br(h\to \gamma\gamma) \sim 0.2\%$). Here however, we turn to a resonant scalar production which decays to a pair of SM-like Higgs bosons. Our goal is to ascertain the reach of the HL-LHC in measuring $\sigma(p p \to H \to h h)$ for a range of heavy Higgs masses ($m_H$). One of the reasons for this final state being a favourite is that the reconstruction and identification precision of photons at the LHC is very high.

\begin{figure}[htb!]
\centering
\includegraphics[trim=0 560 0 80,clip,width=\textwidth]{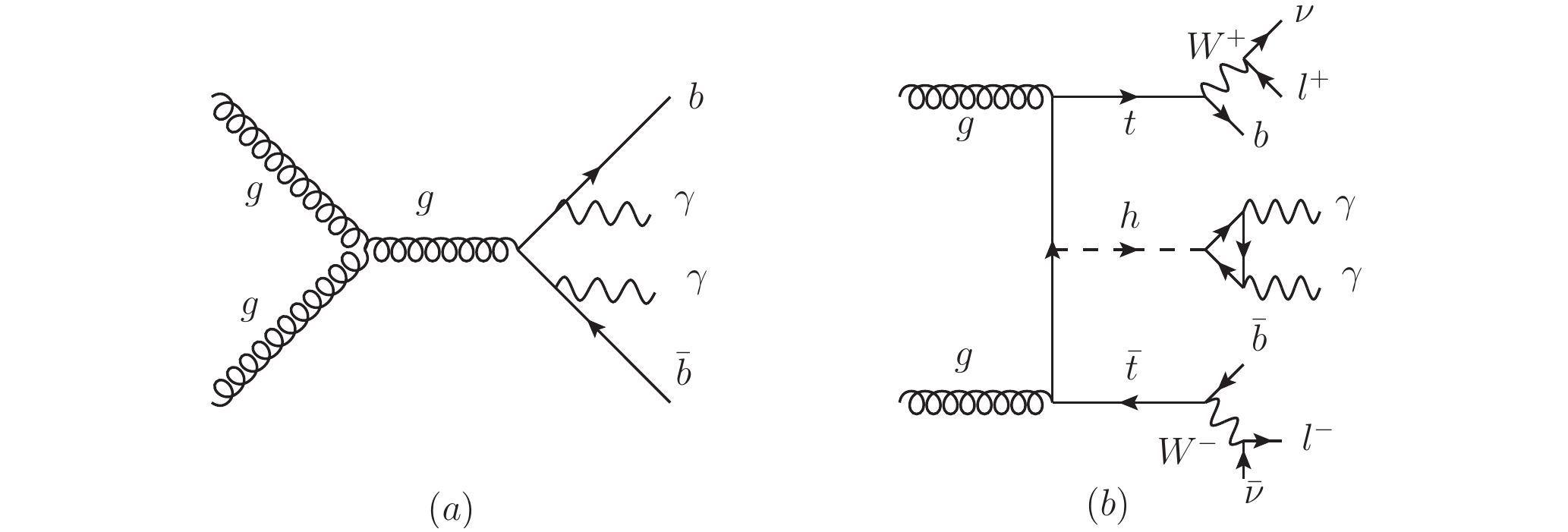}
\caption{Feynman diagrams illustrating the (a) $b\bar{b}\gamma\gamma$ and (b) $t\bar{t}h$ background processes, corresponding to the $b\bar{b}\gamma\gamma$ search channel.}
\label{FD:bbaa}
\end{figure}

Even though the signal seems to have a clean final state, there are several backgrounds at play which need to be dealt with carefully. The major backgrounds (Fig.~\ref{FD:bbaa}) typically have the form of $hh+X$ which includes the SM double Higgs production, $h+X$ which includes $Zh$, $hb\bar{b}$ and $t\bar{t}h$, and null-Higgs processes like $t\bar{t}+t\bar{t}\gamma$ where leptons may fake as photons, $b\bar{b}\gamma\gamma+c\bar{c}\gamma\gamma+jj\gamma\gamma$ (henceforth termed as $b\bar{b}\gamma\gamma^*$) where for the latter two, the light-jets may fake $b$-jets. Other fake backgrounds include $b\bar{b} j \gamma + c\bar{c} j \gamma$ (we will refer to it as Fake 1 category), $b\bar{b} j j$ (referred to as the Fake 2 category), where the $c$-jets may pose as $b$-jets and the light jet may mimic a photon, and the single Higgs processes, \textit{viz.}, $hjj + hc\bar{c}$ (classified henceforth as the $hjj^*$ category), where the light-jets and $c$-jets may mimic $b$-jets. One of the major differences between most of the backgrounds and the signal lies in the invariant mass distribution of the $b$-jets, $m_{b\bar{b}}$. However, even when the $m_{b\bar{b}}$ distribution of the signal (as well as the non-resonant SM di-Higgs background) peaks around the SM-like Higgs mass, $m_h$, it is broad and can have considerable overlap with the $m_{b\bar{b}}$ distribution either ensuing from a $Z$-boson or from a continuum. It should be noted that the most dominant backgrounds come from the QCD-QED $b\bar{b}\gamma\gamma^*$, $t\bar{t}h$ and SM-like di-Higgs processes. The former being a continuum, covers a large part of the kinematic variable space with the signal. The SM-like di-Higgs background also has very large overlap with the signal. One of the easiest way to break this degeneracy is to utilise the $m_{b\bar{b}\gamma \gamma}$ or reconstructed $m_{hh}$ distribution which has a clear peak around the heavy scalar mass, $m_H$, for the signal.
 
We generate the QCD-QED $b\bar{b}\gamma\gamma$ and $Z\gamma\gamma \to b\bar{b} \gamma\gamma$ backgrounds upon merging with one additional jet. We employ the MLM merging scheme~\cite{Mangano:2006rw} where the extra jet contains gluon, light quarks, $c$- as well as $b$-quarks. Among the $h+X$ category, the $Zh$ is generated with the Higgs boson decaying to a pair of photons and the $Z$-boson decaying to a pair of bottom quarks. Furthermore, the $t\bar{t}h$ and $b\bar{b}h$ backgrounds are generated with $h\to \gamma\gamma$. The major fake backgrounds with jets in the final state are generated with the aforementioned jet definition with one exception. We define both of the jets in the $jj\gamma\gamma$ channel in a way as to have no overlap with the $b\bar{b}\gamma\gamma$ background.  In case of the $t\bar{t}+X$ backgrounds, we generate the $t\bar{t}$ events with both of the top quarks decaying leptonically which ultimately fakes as photons. However, for the $t\bar{t}\gamma$ background, we require one of the tops to decay leptonically and the other hadronically. Finally, we generate separate single Higgs backgrounds via gluon fusion in association with $c$-quarks and also with light jets. The separation between the $hc\bar{c}$ and $hjj$ backgrounds are necessary in order to appropriately take into account the different fake rates for $c \to b$ and $j \to b$. All of these backgrounds are generated with specific cuts at the generation level which we summarise in Appendix~\ref{sec:appendixA}.

\begin{figure}
\centering
\includegraphics[scale=0.37]{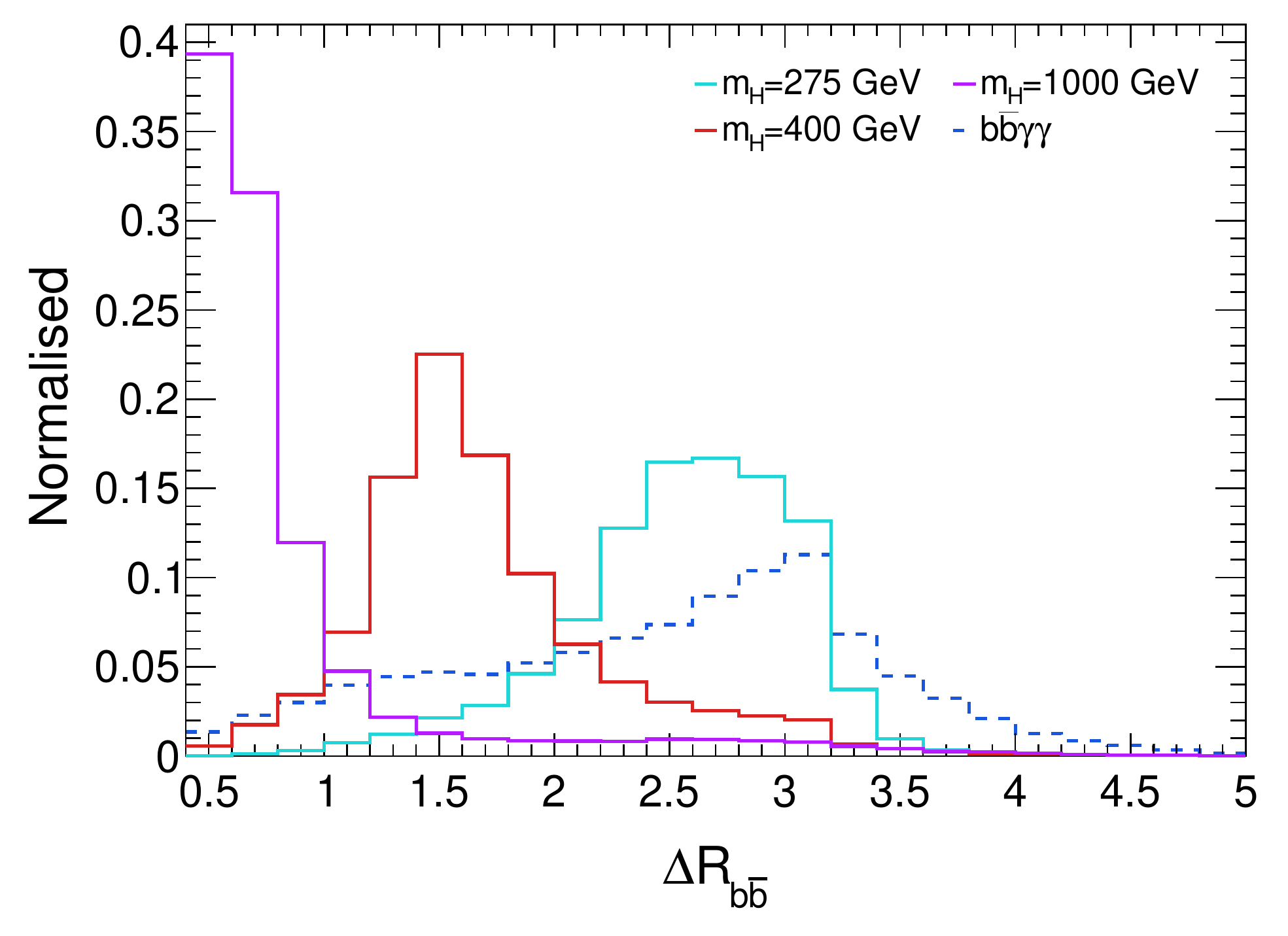}\includegraphics[scale=0.37]{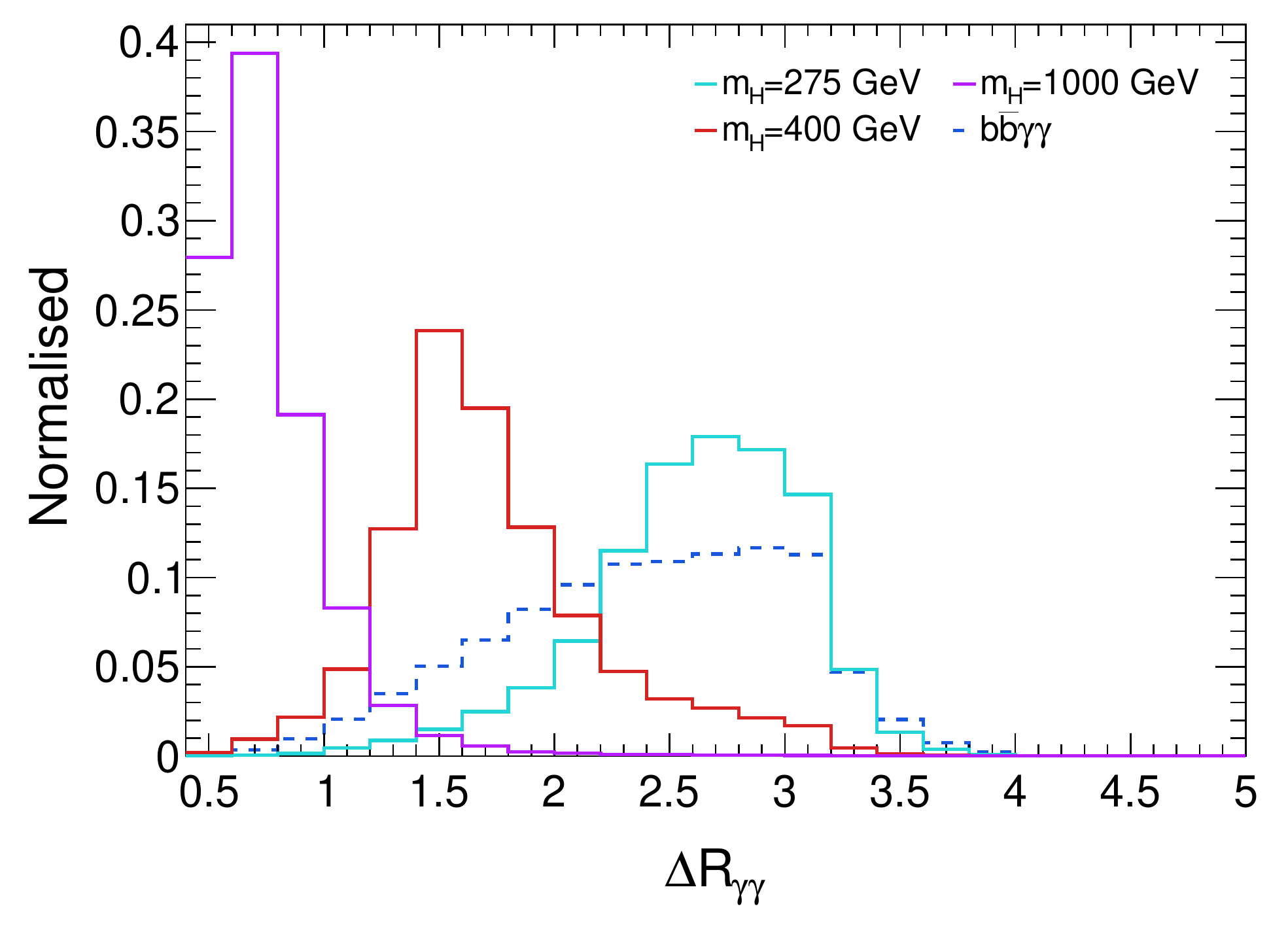}
\caption{Normalised distributions of $\Delta R_{bb}$ and $\Delta R_{\gamma\gamma}$ for heavy Higgs masses, $m_H$ $=275,~400~\text{and}~1000$ GeV with dominant $b\bar{b}\gamma\gamma$ background.}
\label{dr}
\end{figure}

\begin{figure}
\centering
\includegraphics[scale=0.40]{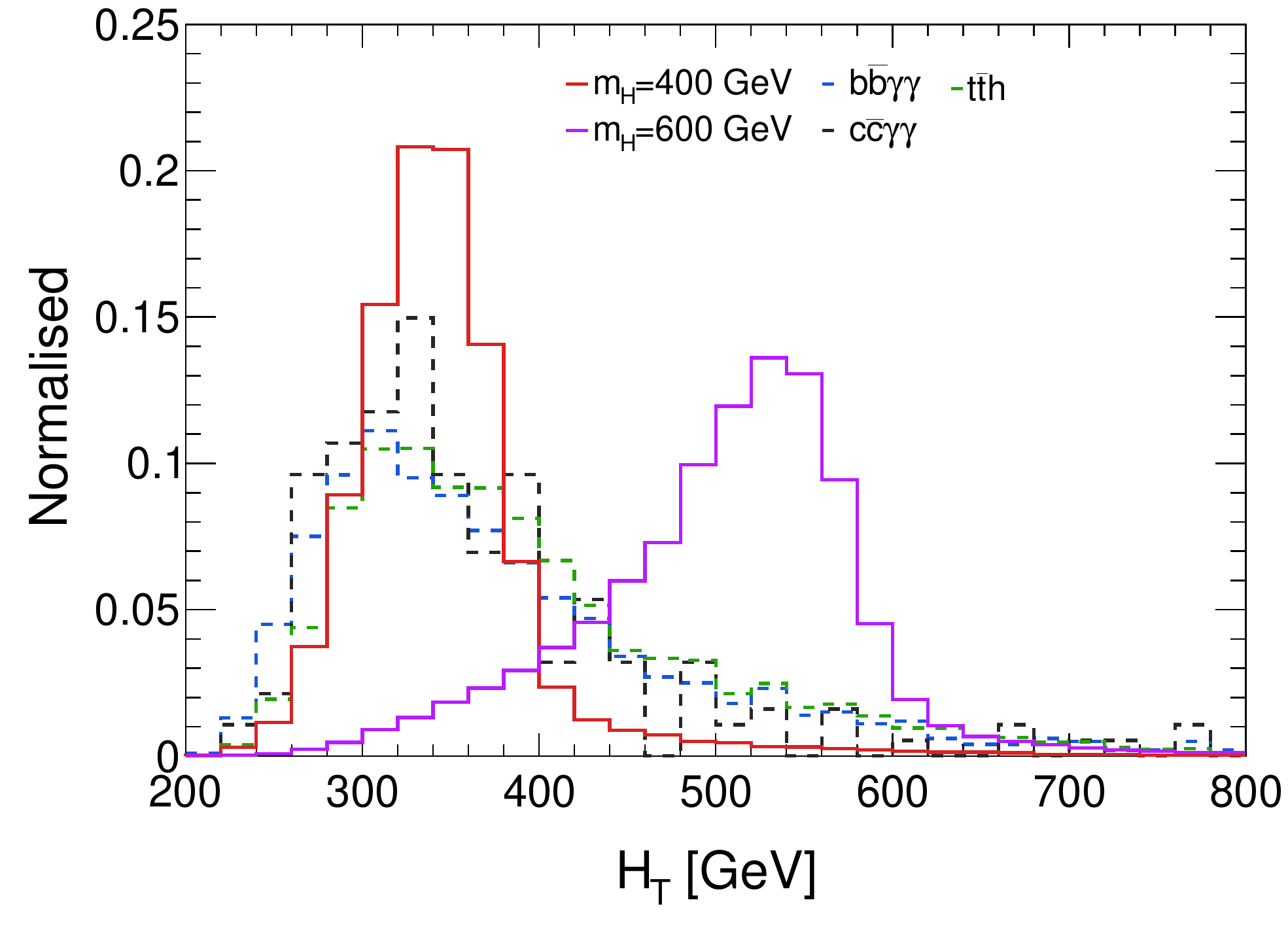}
\caption{Normalised distribution of $H_T$ for heavy Higgs masses of $m_H$ $=400,~600$ GeV with dominant backgrounds.}
\label{ht}
\end{figure}

\begin{figure}
\centering
\includegraphics[scale=0.40]{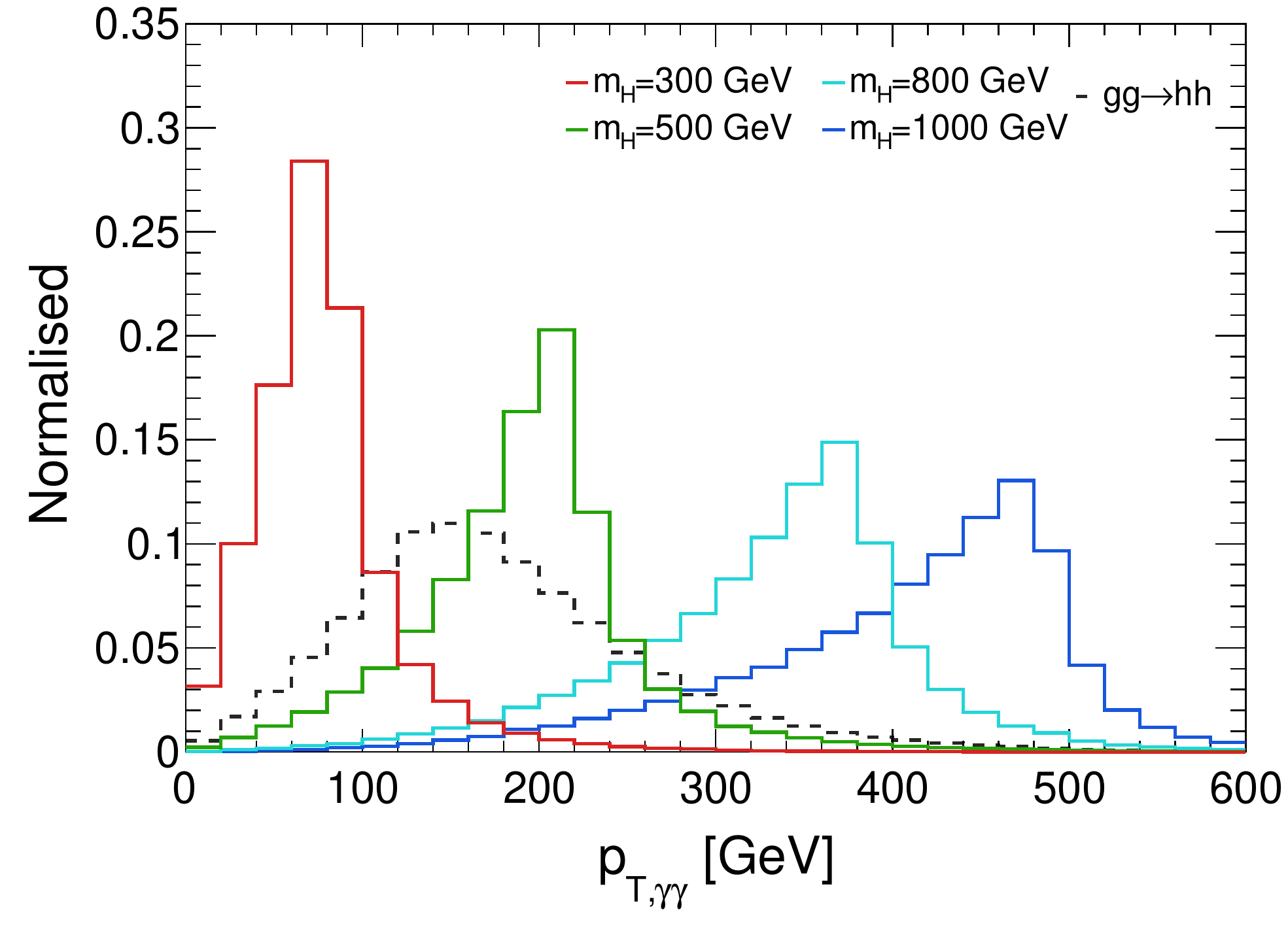}
\caption{Normalised distribution of $p_{T,\gamma\gamma}$ for heavy Higgs masses of $m_H=300,~500,~800~\text{and}~1000$ GeV along with the SM di-Higgs production from gluon-gluon fusion process.}
\label{PtH}
\end{figure}

\begin{figure}
\centering
\includegraphics[scale=0.37]{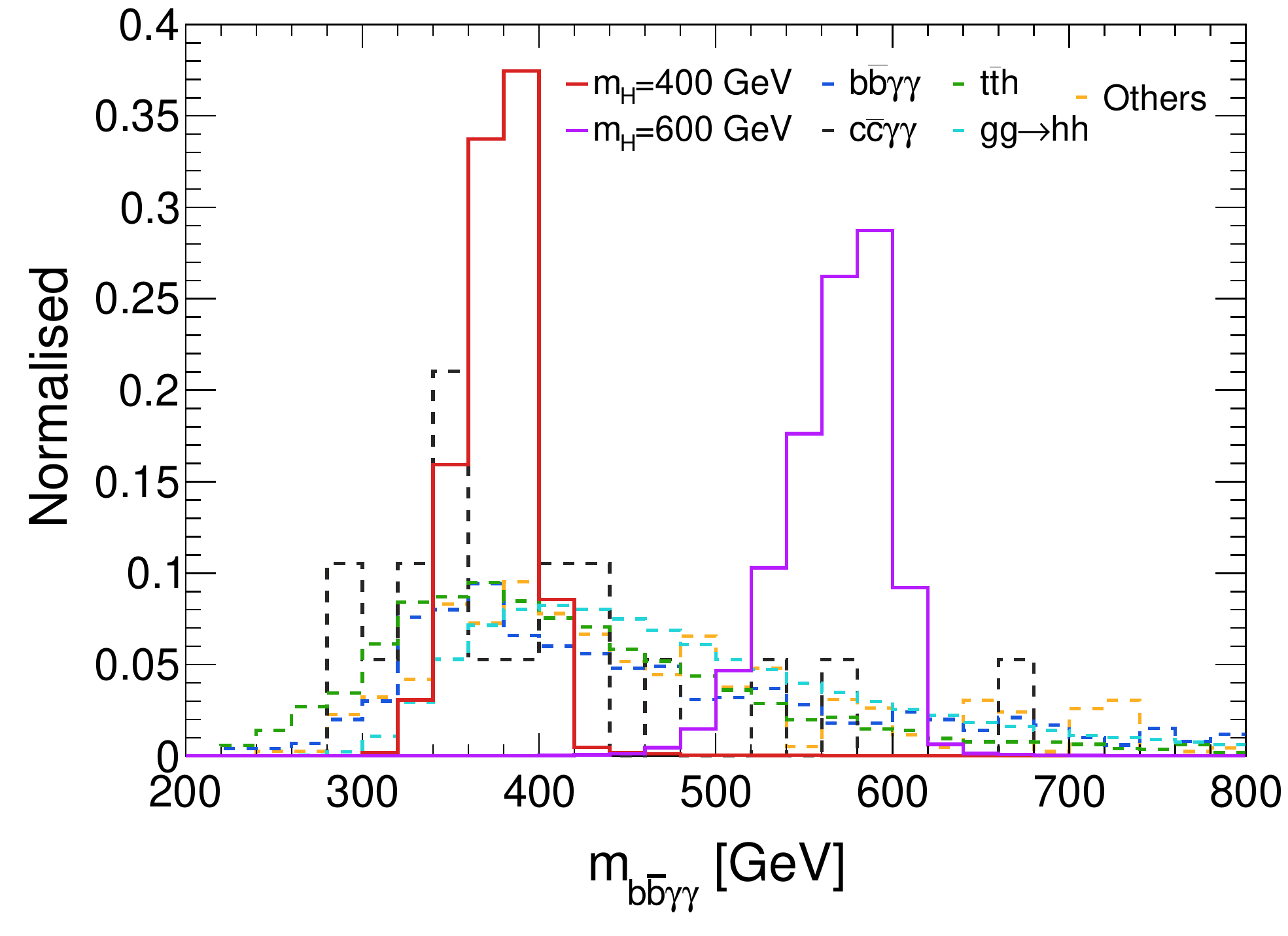}
\includegraphics[scale=0.37]{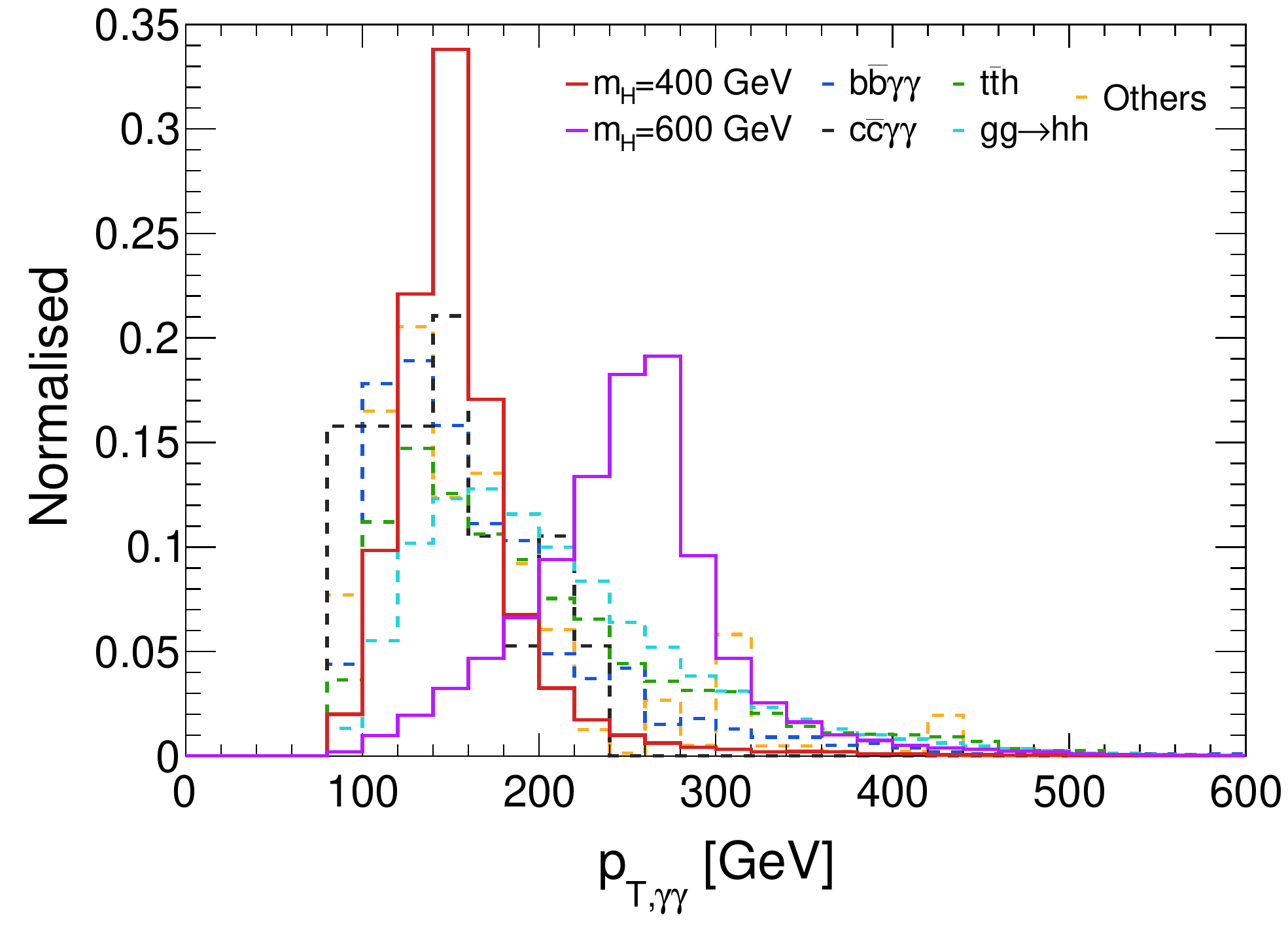}
\caption{The $m_{b\bar{b}\gamma\gamma}$ and $p_{T,\gamma\gamma}$ distributions for heavy Higgs masses of $m_H = 400$ and $600$ GeV with backgrounds. Here the heavy Higgs boson is searched for in the $bb\gamma\gamma$ final state. The distributions are shown after imposing the fixed cuts.}
\label{Hfig:invhh_ptaaa}
\end{figure}

\begin{figure}
\centering
\includegraphics[scale=0.45]{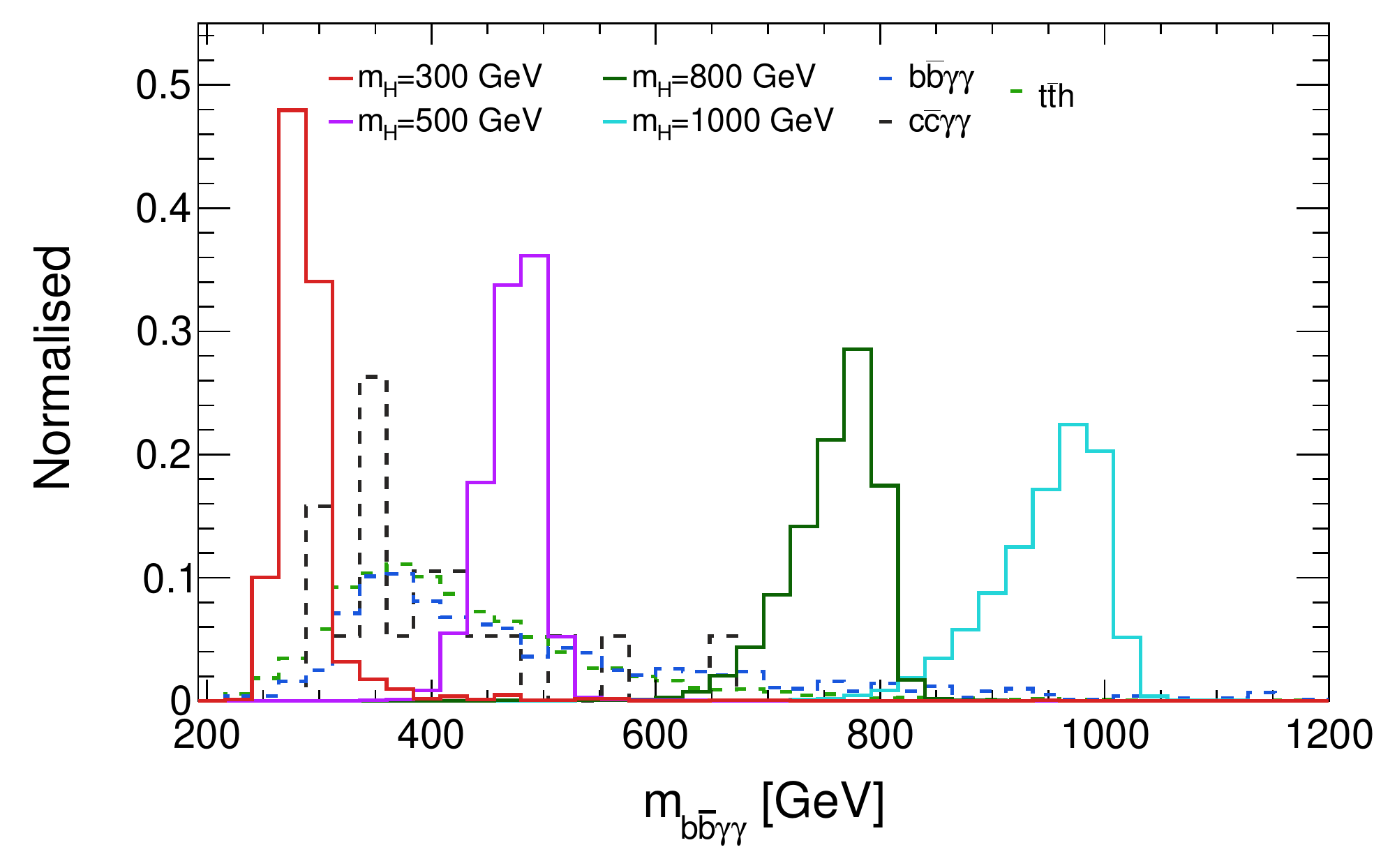}
\caption{Normalised distribution of $m_{b\bar{b}\gamma\gamma}$ for heavy Higgs masses of $m_H=300,~500,~800~\text{and}~1000$ GeV with dominant backgrounds.}
\label{mH}
\end{figure}

The idea of this section is to understand the reach of the HL-LHC in constraining models with extended scalar sectors. We thus employ optimised search strategies for a varied range of scalar masses. We vary $m_H$ in the mass range 275 GeV and 1 TeV. Specifically, we consider the following benchmark points, \textit{viz.}, $m_H=275, 300, 350, 400, 450, 500, 550, 600, 800$ GeV and 1 TeV. In line with our previous work~\cite{Adhikary:2017jtu}, we first perform a classical cut and count analysis to gauge the sensitivity of various benchmark points. We closely follow various cuts from the ATLAS projection study~\cite{ATL-PHYS-PUB-2017-001}. Namely, we require exactly two $b$-tagged jets and two photons in the final state. The photons are required to have transverse momenta, $p_T > 10$ GeV and a pseudorapidity coverage of $|\eta| < 2.5$. Moreover, the two photons are also required to lie within the pseudorapidity range, $|\eta_{\gamma}| < 1.37$ (barrel region) or $1.52 < |\eta_{\gamma}|< 2.37$ (endcap region). After imposing these basic requirements, we apply some stronger selection cuts in order to enhance the signal to background ratio, $S/B$. We require the invariant mass of the pair of photons, $m_{\gamma\gamma}$, to reconstruct sharply about the SM-like Higgs mass in the range (122,128) GeV. Furthermore, we veto events containing lepton(s) in the final state in order to reduce the impact of the $t\bar{t}h$ background when it decays semi-leptonically or leptonically. We also impose lower bounds on the transverse momenta of the leading and sub-leading $b$-jets and photons. Moreover, upon inspecting the distribution of $\Delta R_{\gamma\gamma}$ and $\Delta R_{b\bar{b}}$ (Fig.~\ref{dr}), we find that with larger values of $m_H$, the SM-like Higgs bosons are more boosted yielding more collimated final states.
\begin{table}
\begin{center}
\begin{tabular}{||c||}\hline 
Fixed cuts \\ \hline
$122~{\rm GeV} < m_{\gamma \gamma} < 128~{\rm GeV}$ \\
$N_\ell =0$\\
$p_{T,b} > 40 \; (30)~{\rm GeV},~ p_{T,\gamma} > 30 \; (30)~{\rm GeV}$\\ 
$0.4 < \Delta R_{\gamma \gamma} < (3.0/2.0/1.5)$, $0.4 < \Delta R_{bb} < (3.0/2.0/1.5)$, $\Delta R_{\gamma b} > 0.4$\\
$90~{\rm GeV} < m_{bb} < 130~{\rm GeV}$ \\\hline\hline
\end{tabular}
\caption{Applied fixed cuts for the cut-based analysis in the $b\bar{b}\gamma\gamma$ channel.}
\label{tab:bbgamgam_fix_cut}
\end{center}
\end{table}
We thus require $\Delta R_{\gamma \gamma}$ and $\Delta R_{b\bar{b}}$ to lie in the range (0.4,3.0), (0.4,2.0) and (0.4,1.5) for $m_H=275, 300$ and 350 GeV, $m_H = 400, 450, 500, 550$ and 600 GeV, and $m_H =800$ GeV and 1 TeV respectively. Besides, we require the invariant mass of the $b$-jets, $m_{b\bar{b}}$ to lie in the range (90,130) GeV. This choice is related to account for the jet-energy correction and has been described in Ref.~\cite{Adhikary:2017jtu}. We summarise these cuts in Table~\ref{tab:bbgamgam_fix_cut}. As the next logical step, we delve deeper into the kinematics. We reconstruct the invariant mass of the $b\bar{b}\gamma\gamma$ system, $m_{b\bar{b}\gamma\gamma}$ and its total visible energy, $H_T$. These two variables are intrinsically correlated. Also, from Fig.~\ref{ht}, it is evident that the $H_T$ distribution is broader leading to more background contamination as compared to the $m_{b\bar{b}\gamma\gamma}$ distribution which we show in Fig.~\ref{Hfig:invhh_ptaaa} (left). Thus, we proceed with $m_{b\bar{b}\gamma\gamma}$ in order to further optimise our analysis.
We also reconstruct the transverse momentum of the SM-like Higgs decaying to a pair of photons, $p_{T,\gamma\gamma}$. As can be seen from the $p_{T,\gamma\gamma}$ distribution in Fig.~\ref{Hfig:invhh_ptaaa}, the spectrum is harder for heavier values of $m_H$. We choose $p_{T,\gamma\gamma} > 50$ GeV for $m_H = 275$ and 300 GeV and for all other masses, we choose the transverse momentum of this reconstructed Higgs to be larger than 100 GeV. Thus, after these fixed cuts, we perform a simplified optimisation with the $m_{b\bar{b}\gamma\gamma}$ variable in order to enhance $S/B$. These cuts are finally tabulated in Table~\ref{tab:bbgamgam_cut} where we also present the signal efficiency and the background yield at an integrated luminosity, $\mathcal{L}= 3$ ab$^{-1}$ with $m_H$ being varied. The signal efficiency, $\epsilon$, here points to the ratio of the total number of signal events remaining after all the cuts applied in sequence to the generated number of events. The second column refers to the range of $m_{b\bar{b}\gamma\gamma}$ that optimises the signal and the third column denotes the minimum $p_T$ for the diphoton system. This optimisation is different as compared to the SM di-Higgs production scenario as shown in Ref.~\cite{Adhikary:2017jtu}. This difference is related to the kinematics of the event topology. As an example, the $p_{T,\gamma\gamma}$ distribution changes with the heavy Higgs mass as shown in Fig.~\ref{PtH}. From this figure, it is evident that different optimisation is required for each mass point and also for the SM scenario. We must note that in Table~\ref{tab:bbgamgam_cut}, the choices of the upper and lower cuts on $m_{b\bar{b}\gamma\gamma}$ for the different heavy Higgs masses can be understood from Fig.~\ref{mH}. These ranges are obtained after optimising for each value of $m_H$. Finally, we provide a detailed cut-flow table for $m_H=400$ GeV in Table~\ref{tab:bbgamgam_H400} with a complete information for the backgrounds including the perturbative order at which the rates are considered as well as the number of events remaining at the HL-LHC.

\begin{table}[htb!]
\begin{bigcenter}
\scalebox{0.7}{%
\begin{tabular}{||c|c|c|c|c||}
\hline
 Heavy Higgs mass, & \multicolumn{2}{c|}{Optimised cuts}           & \multicolumn{2}{c||}{After all cuts}   \\\cline{2-5}

 $m_H$ (GeV)      & $m_{b\bar{b}\gamma\gamma}$ (GeV) & $p_{T,\gamma\gamma}>$ (GeV)  & Signal Efficiency ($\epsilon$) & Background yield at $3000 \; \textrm{fb}^{-1}$   \\\hline

 $275$  & [$235$ , $275$] & $50$   & $0.012$ & $30.01$ \\\hline
 $300$  & [$255$ , $305$] & $50$   & $0.024$ & $55.62$ \\\hline
 $350$  & [$300$ , $355$] & $100$  & $0.024$ & $23.33$ \\\hline
 $400$  & [$345$ , $405$] & $100$  & $0.032$ & $15.80$ \\\hline
 $450$  & [$395$ , $455$] & $100$  & $0.042$ & $13.75$ \\\hline
 $500$  & [$445$ , $510$] & $100$  & $0.051$ & $10.87$ \\\hline
 $550$  & [$460$ , $570$] & $100$  & $0.068$ & $14.39$ \\\hline
 $600$  & [$460$ , $615$] & $100$  & $0.076$ & $18.11$ \\\hline
 $800$  & [$560$ , $830$] & $100$  & $0.091$ & $9.54$ \\\hline
 $1000$ & [$780$ , $1030$]& $100$  & $0.090$ & $2.31$  \\\hline
\end{tabular}}
\end{bigcenter}
\caption{Details of the final optimised cuts with signal efficiency and background yields after applying all cuts.}
\label{tab:bbgamgam_cut}
\end{table}

\begin{table}[htb!]
\begin{bigcenter}
\scalebox{0.7}{%
\begin{tabular}{||c||c|c|c|c|c|c|c|c|c|c||}
\hline
 & Signal Efficiency & \multicolumn{9}{c||}{Event rates with $\mathcal{L} = 3$ ab$^{-1}$} \\ \cline{3-11}
 Cut flow   & $pp\to H$ & \multicolumn{9}{c||}{SM Backgrounds} \\
\cline{3-11} 
 & $\to hh\to 2b2\gamma$) & $hh \to 2b2\gamma$ & $hb\bar{b}$ & $t\bar{t}h$ & $Zh$ & $Z\gamma\gamma$ & $b\bar{b}\gamma\gamma *$~\footnote{$b \bar{b} \gamma \gamma + c \bar{c} \gamma \gamma + j j \gamma \gamma $.}  & Fake 1~\footnote{$b \bar{b} j \gamma + c \bar{c} j \gamma$.} & Fake 2~\footnote{$b \bar{b} j j$.} & $hjj^*$~\footnote{$(gg\to hjj) + (gg\to hc\bar{c})$}  \\
\cline{1-11}  
Order         &                       & NNLO~\cite{hhtwiki}         & NNLO (5FS) +                   & NLO~\cite{bkg_twiki_cs}       & NNLO (QCD) +                &  LO  & LO  & LO & LO & LO\\ 
                                     &             &                & NLO (4FS)~\cite{bkg_twiki_cs}  &                               & NLO EW~\cite{bkg_twiki_cs}  &      &     &    &    & \\
\cline{1-11}

$m_{\gamma\gamma}$   & $0.123$   & $39.71$  &  $36.68$   & $397.97$   & $62.21$   & $32.86$  & $1071.38$  & $837.45$   & $403.98$  &  $9.60$  \\\hline

$N_\ell$             & $0.122$   & $39.70$  &  $36.68$   & $290.10$   & $62.21$   & $32.86$  & $1071.34$  & $837.40$   & $403.98$  &  $9.60$  \\\hline

$p_{T,b/\gamma}$     & $0.081$   & $27.65$  &  $16.34$   & $197.83$   & $35.87$   & $14.00$  & $510.73$   & $361.01$   & $183.70$  &  $4.91$  \\\hline

$\Delta R$ cuts      & $0.052$   & $20.56$  &  $5.09$    & $36.73$    & $22.32$   & $4.86$   & $56.24$    & $35.60$    & $27.05$   &  $1.53$  \\\hline

$m_{bb}$             & $0.036$   & $14.19$  &  $1.41$    & $12.74$    & $4.43$    & $1.02$   & $16.44$    & $11.47$    & $7.47$    &  $0.41$  \\\hline

$p_{T,\gamma\gamma}$ & $0.035$   & $14.01$  &  $1.36$    & $12.29$    & $4.28$    & $0.98$   & $15.53$    & $10.90$    & $6.70$    &  $0.40$  \\\hline

$m_{b\bar{b}\gamma\gamma}$   & $0.032$  & $2.96$  &  $0.29$  & $3.31$ & $0.87$    & $0.21$   & $3.84$     & $3.18$     & $1.03$    &  $0.08$  \\\hline\hline

\end{tabular}}
\end{bigcenter}
\caption{The cut-flow table for heavy Higgs of mass $400$ GeV. The table also shows the various perturbative orders at which the cross-sections have been used.}
\label{tab:bbgamgam_H400}
\end{table}

\begin{figure}
\centering
\includegraphics[scale=0.5]{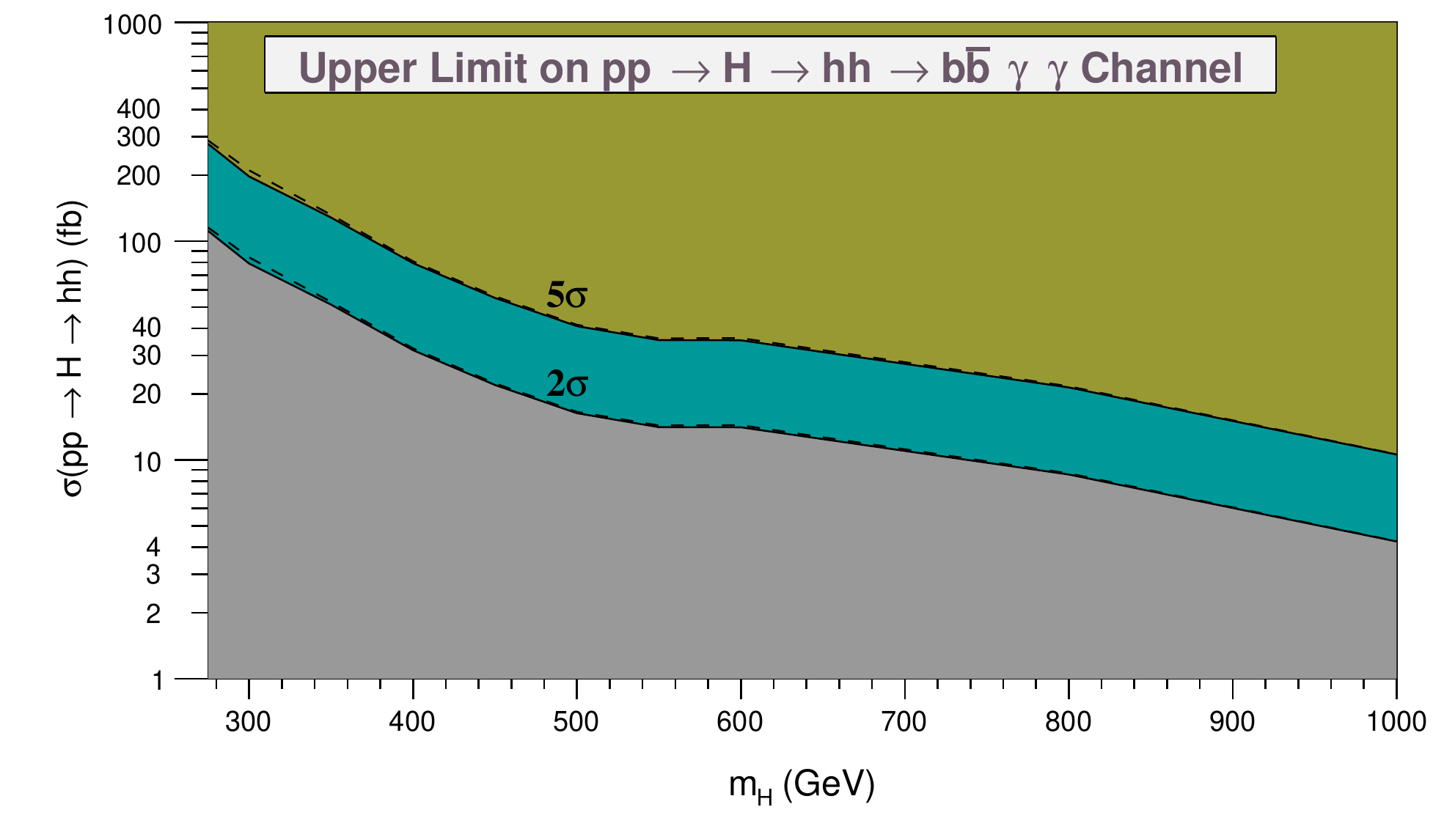}
\caption{Upper limit on $\sigma(pp\to H\to hh)$ (fb) as a function of $m_{H}$ (GeV) for the $b\bar{b}\gamma\gamma$ channel with cut-based analysis. The solid (dashed) lines show the 2$\sigma$-5$\sigma$ band on taking 0\% (5\%) systematic uncertainties.}
\label{Hfig:2b2ga_cut}
\end{figure}

Utilising these results, we derive the projected upper limits on the production cross section of the heavy Higgs in a \textit{model independent} manner~\footnote{We consider the cut-based optimisations as final as we did not obtain any observable improvement with a multivariate analysis.}. We calculate the cross-section reach by using the significance formula: $S/\sqrt{B}=N$, where $N$ denotes the number of confidence intervals. Here, the signal yield, $S$, is defined as $\sigma(pp\to H\to hh \to b\bar{b} \gamma \gamma) \times \mathcal{L} \times \epsilon$ and $B$ represents the total background yield after the cut-based analysis. With this, we derive $\sigma(pp\to H\to hh)$ at the $N\sigma$ level, with $N=2~\textrm{and}~5$, respectively, corresponding to a $95\%$ and $99.7\%$ confidence level (CL) upper limit, also referred to as the exclusion and discovery limits, respectively.
We show the final results in Fig.~\ref{Hfig:2b2ga_cut} with the upper limit on $\sigma(pp\to H\to hh)$ as a function of $m_H$ and we display the $2\sigma$ and $5\sigma$ lines. The $2\sigma$ upper limit is strong between $400$ GeV and 1 TeV, varying between $31.74$ fb and $4.24$ fb. Upon adding $5\%$ systematic uncertainty~\footnote{The significance formula for a systematic uncertainty of $N\%$ has the form: $S/\sqrt{B + \left(N*B/100\right)^{2}}$}, the upper limit becomes $32.35$ fb and $4.25$ fb respectively within the previously mentioned mass range. It must be noted that the upper limit is mildly affected by incorporating a systematic uncertainty of $\sim$5\%. The reason can be attributed to the fact that the signal over background ratio ($S/B$) is high.

Next, we perform a multivariate analysis in order to improve upon the cut-based analysis. We use the following variables:
\begin{equation}
\begin{split}
m_{bb},~\Delta R_{bb},~p_{T, \gamma\gamma},~\Delta R_{\gamma\gamma},~m_{b\bar{b}\gamma\gamma},~\Delta R_{b_i \gamma_j},\\\Delta R_{bb,\gamma\gamma},~p_{T, b_1},~p_{T, b_2},~p_{T, \gamma_1},~p_{T, \gamma_2} \nonumber
\end{split}
\end{equation}
Here, the variable names have their usual meaning. The $\Delta R_{b_i \gamma_j}$ is the distance in the $\eta-\phi$ plane between the $b$-jets and photons with $i~{\rm and}~j=1,2$. Also, $\Delta R_{bb,\gamma\gamma}$ is the $\Delta R$ separation between the system of $b$-jets and the two photon system. However, after performing this analysis we obtain comparable results. Thus, we do not show the results of the multivariate analysis in this section.



\subsubsection{The $pp\to A\to Zh$ Channel}
\label{sec:A2Zh}

With the accumulation of more data, we are on the brink of accepting the fact that in MSSM or in generic two Higgs doublet models, the SM-like Higgs is in the decoupling regime with its coupling to the SM gauge bosons being proportional to $\sin(\beta - \alpha)$, with $\alpha$ and $\tan{\beta}$ being the mixing angle in the neutral $CP$-even sector and the ratio of the two vacuum expectation values of the two doublets, respectively. For the up and down type fermions, the Yukawa couplings for the SM-like Higgs boson are proportional to $\cos{\alpha}/\sin{\beta}$ and $\sin{\alpha}/\cos{\beta}$ respectively. In the decoupling regime, $\sin(\beta - \alpha) \sim 1$ and hence the decay width of $A \to Z h$ which is proportional to the coupling $\cos(\beta - \alpha)$, is small. In a non-decoupling regime, $pp \to A \to Zh$ can give us deep insight into two scalars simultaneously. 

The CMS~\cite{CMS-PAS-FTR-13-024} collaboration has derived projected upper limits on $\sigma(pp \to A \to Zh)$ from searches in the $\ell\ell b\bar{b}$~($Z\to \ell\ell,~h\to b\bar{b}$) final state for HL-LHC. In the present study, keeping in continuation to the analysis prescribed in the previous section (Section~\ref{sec:Htohhtobbgaga}), we explore the prospects of directly probing $A$ in the $A \to Zh \to b\bar{b}\gamma\gamma$ channel in the context of HL-LHC.

\begin{figure}
\centering
\includegraphics[scale=0.37]{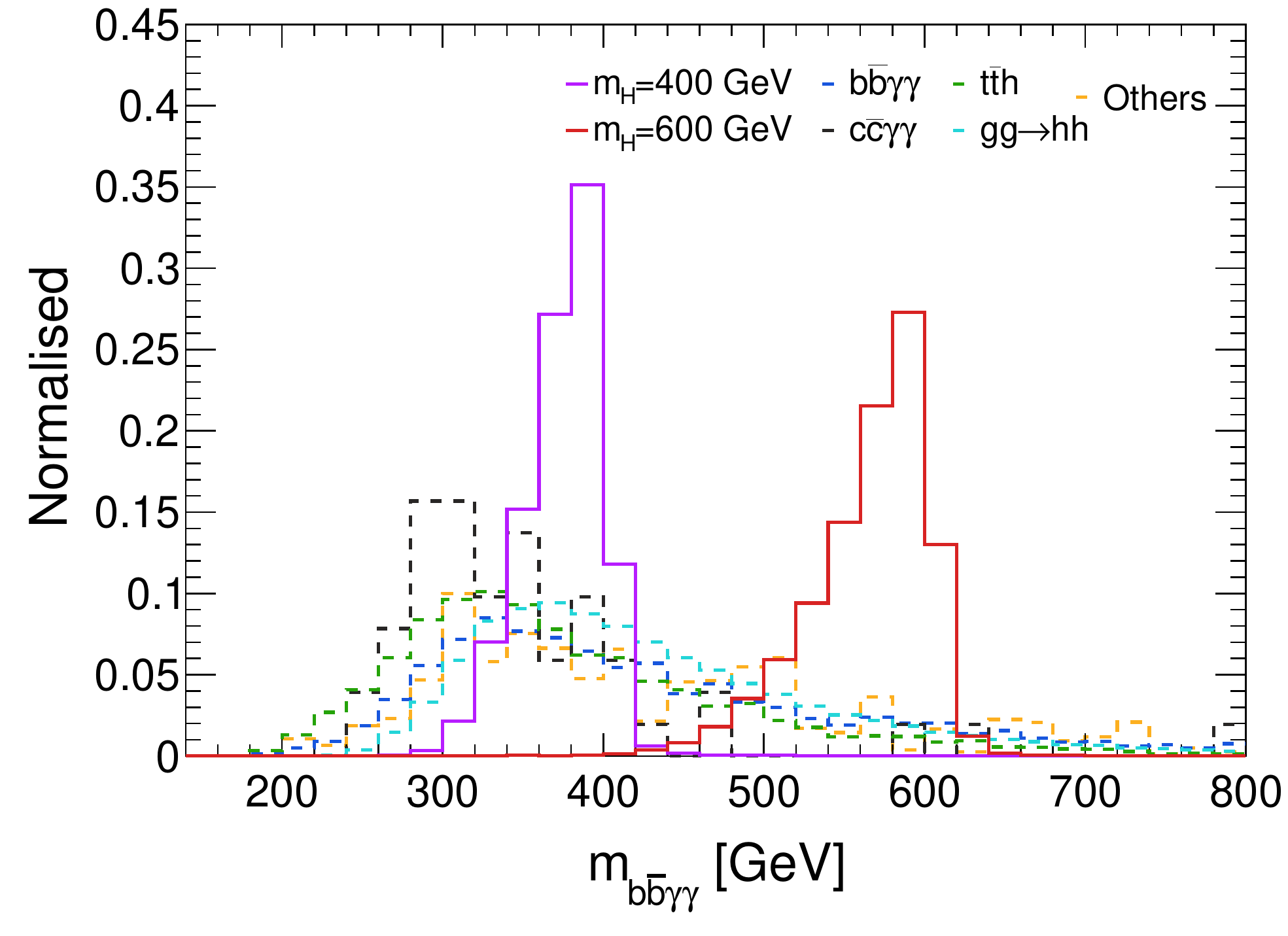}
\includegraphics[scale=0.37]{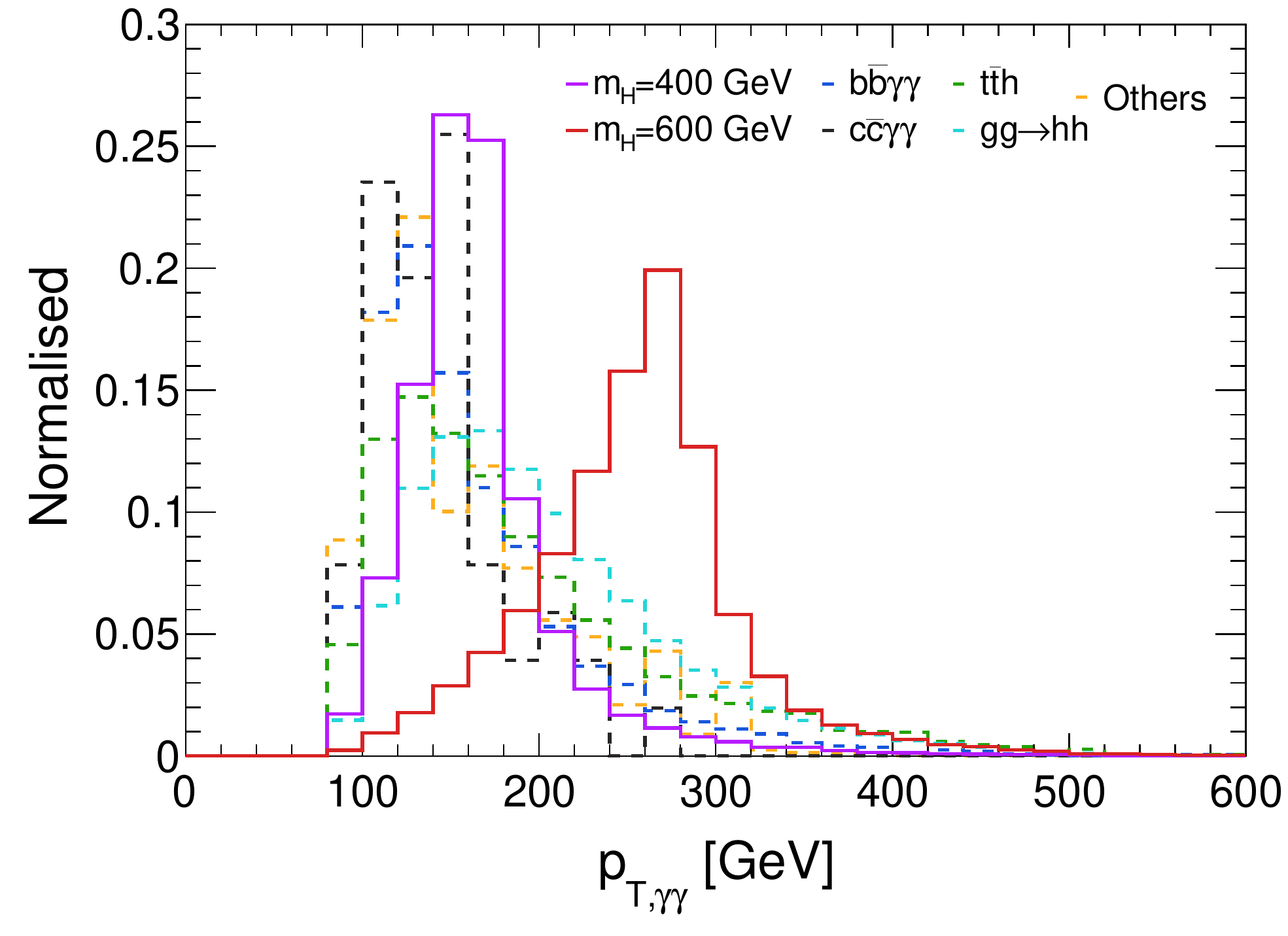}
\caption{The $m_{b\bar{b}\gamma\gamma}$ and $p_{T,\gamma\gamma}$ distributions for heavy Higgs masses of $m_A = 400$ and $600$ GeV with backgrounds. Here the heavy pseudoscalar is searched for in the $bb\gamma\gamma$ final state. The distributions are shown after imposing the fixed cuts.}
\label{A2zh:invhh_invbb}
\end{figure}

\begin{figure}
\centering
\includegraphics[scale=0.5]{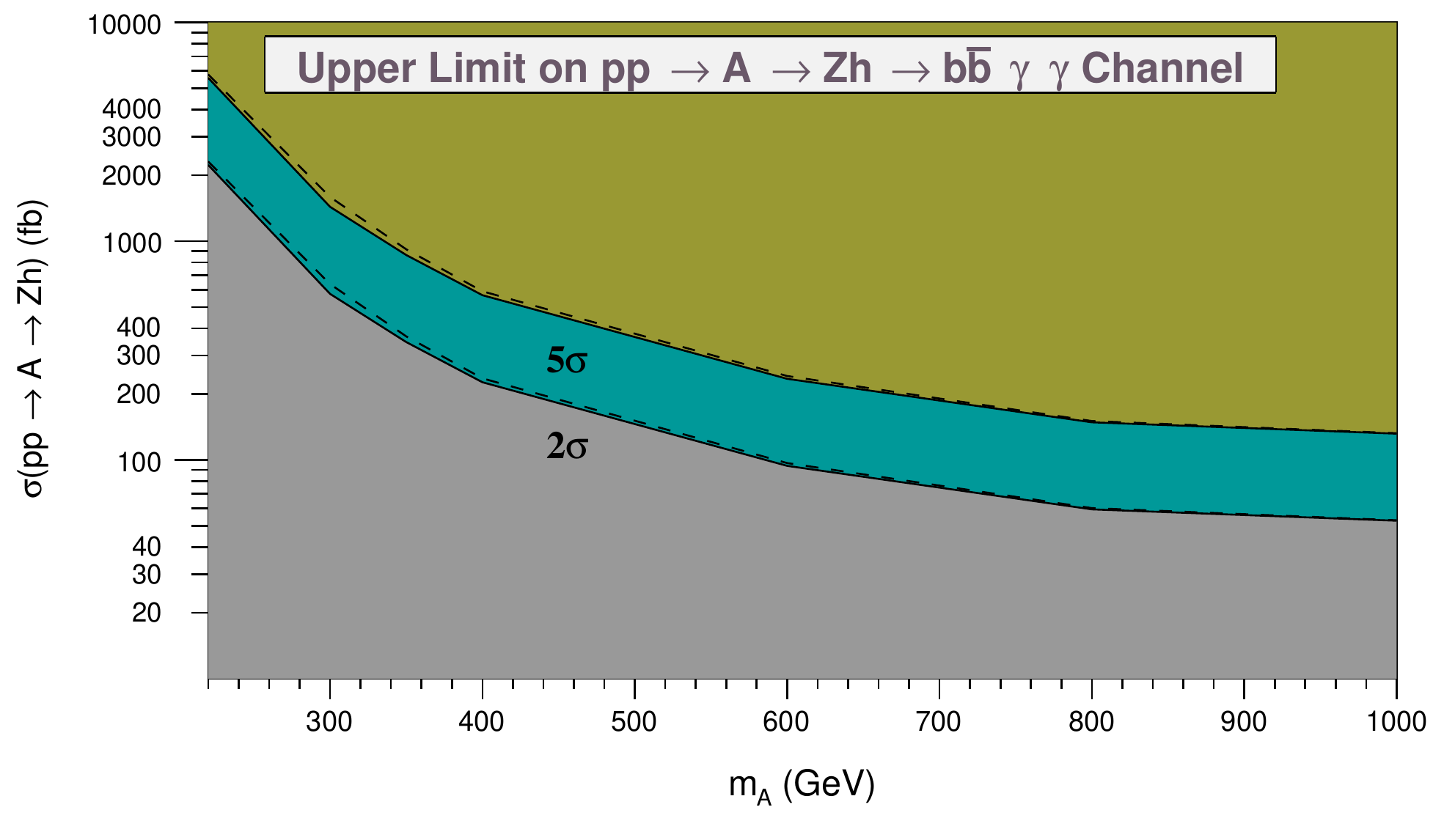}
\caption{Upper limit on $\sigma(pp\to A \to Zh)$ (fb) as a function of $m_{A}$ (GeV) for the $b\bar{b}\gamma\gamma$ channel with cut-based analysis. The solid (dashed) lines show the 2$\sigma$-5$\sigma$ band on taking 0\% (5\%) systematic uncertainties.}
\label{A2zh:cutbased}
\end{figure}

We will remain agnostic to the fact that the prospects of observing $A \to Z h$ in the decoupling regime may be extremely small. The main difference here is the fact that in the previous analysis both the diphoton and the $b\bar{b}$ pairs are required to peak around the SM-like Higgs boson mass, whereas in the present case, the $b$-jets are required to peak around the $Z$-boson mass. We follow a cut-based analysis as before and optimise the $m_{b\bar{b}\gamma\gamma}$ and $p_{T,\gamma\gamma}$ cuts for different values of $m_A$. These variables are shown to have substantial discriminatory power and are shown in Fig.~\ref{A2zh:invhh_invbb} for $m_A = 400$ GeV and $600$ GeV. Details of the fixed and optimised cuts are presented in Table~\ref{tab1:zhbbaa} and Table~\ref{tab2:zhbbaa} respectively. In Table~\ref{tab:zhbbaa_A400}, we show the cut-flow table for $m_A = 400$ GeV. After a full optimisation, we show the 95\% and 99.7\% CL exclusion for $\sigma(p p \to A \to Zh)$ in Fig.~\ref{A2zh:cutbased}. The bounds are weaker than their $H \to hh$ counterpart mainly because of a larger overlap with the $Zh$ background. It is to be noted that the projected upper limits on $\sigma(pp \to A \to Zh)$ derived from searches in the $b\bar{b}\gamma\gamma$ channel (see Figure~\ref{A2zh:cutbased}) are weaker than the projected limit obtained by CMS from searches in the $\ell\ell b\bar{b}$~\cite{CMS-PAS-FTR-13-024} final state.

\begin{table}
\begin{center}
\begin{tabular}{||c||}\hline 
Fixed cuts \\ \hline
$122~{\rm GeV} < m_{\gamma \gamma} < 128~{\rm GeV}$ \\
$p_{T,b} > 35 \; (25)~{\rm GeV},~ p_{T,\gamma} > 30 \; (30)~{\rm GeV}$\\ 
$0.4 < \Delta R_{\gamma \gamma} < (3.0/2.0/1.5)$, $0.4 < \Delta R_{bb} < (3.0/2.0/1.5)$, $\Delta R_{\gamma b} > 0.4$\\
$55~{\rm GeV} < m_{bb} < 100~{\rm GeV}$ \\\hline\hline
\end{tabular}
\caption{Applied fixed cuts for the cut-based analysis.}
\label{tab1:zhbbaa}
\end{center}
\end{table}

\begin{table}[htb!]
\begin{bigcenter}
\scalebox{0.7}{%
\begin{tabular}{||c|c|c|c|c||}
\hline
 Heavy Pseudoscalar mass, & \multicolumn{2}{c|}{Optimised cuts}           & \multicolumn{2}{c||}{After all cuts}   \\\cline{2-5}

 $m_A$ (GeV)      & $m_{b\bar{b}\gamma\gamma}$ (GeV) & $p_{T,\gamma\gamma}>$ (GeV)  & Signal Efficiency ($\times 10^{-2}$) & Background yield at $3000 \; \textrm{fb}^{-1}$   \\\hline

 $220$  & [$170$ , $235$]  & $50$   & $0.48$  & $30.40$ \\\hline
 $300$  & [$255$ , $305$]  & $50$   & $3.24$  & $91.87$ \\\hline
 $350$  & [$290$ , $360$]  & $100$  & $4.01$  & $51.08$ \\\hline
 $400$  & [$345$ , $420$]  & $100$  & $5.00$  & $34.10$ \\\hline
 $600$  & [$470$ , $625$]  & $100$  & $10.19$ & $24.28$ \\\hline
 $800$  & [$590$ , $830$]  & $100$  & $10.68$ & $10.69$ \\\hline
 $1000$ & [$780$ , $1040$] & $100$  & $7.50$  & $4.15$ \\\hline\hline
\end{tabular}}
\end{bigcenter}
\caption{Details of the final optimised cuts with signal efficiency and background yields after applying all cuts.}
\label{tab2:zhbbaa}
\end{table}

\begin{table}[htb!]
\begin{bigcenter}
\scalebox{0.7}{%
\begin{tabular}{||c||c|c|c|c|c|c|c|c|c|c||}
\hline
 & Signal Efficiency & \multicolumn{9}{c||}{Event rates with $3000 \; \textrm{fb}^{-1}$ of integrated luminosity} \\ \cline{3-11}
 Cut flow   & for $pp\to A\to Zh$ & \multicolumn{9}{c||}{SM Backgrounds} \\
\cline{3-11} 
 & $\to 2b2\gamma$ & $hh \to 2b2\gamma$ & $hb\bar{b}$ & $t\bar{t}h$ & $Zh$ & $Z\gamma\gamma$ & $b\bar{b}\gamma\gamma *$~\footnote{$b \bar{b} \gamma \gamma + c \bar{c} \gamma \gamma + j j \gamma \gamma $.}  & Fake 1~\footnote{$b \bar{b} j \gamma + c \bar{c} j \gamma$.} & Fake 2~\footnote{$b \bar{b} j j$.} & $hjj^*$~\footnote{$(gg\to hjj) + (gg\to hc\bar{c})$}  \\
\cline{1-11}  

$m_{\gamma\gamma}$           & $0.115$   & $39.71$  &  $36.68$   & $397.97$   & $62.21$   & $32.86$  & $1071.38$  & $837.45$   & $403.98$  &  $9.60$  \\\hline

$p_{T,b/\gamma}$             & $0.091$   & $32.19$  &  $26.29$   & $314.39$   & $47.00$   & $19.82$  & $670.56$   & $483.95$   & $241.41$  &  $7.12$  \\\hline

$\Delta R$ cuts              & $0.071$   & $23.08$  &  $8.28$    & $62.03$    & $25.88$   & $5.86$   & $72.77$    & $48.40$    & $34.27$   &  $2.32$  \\\hline

$m_{bb}$                     & $0.059$   & $9.02$   &  $3.03$    & $23.85$    & $22.56$   & $5.03$   & $33.45$    & $21.94$    & $15.72$   &  $0.88$  \\\hline

$p_{T,\gamma\gamma}$         & $0.058$   & $8.89$   &  $2.86$    & $22.77$    & $21.50$   & $4.80$   & $31.30$    & $20.48$    & $13.91$   &  $0.84$  \\\hline

$m_{b\bar{b}\gamma\gamma}$   & $0.050$   & $2.94$   &  $0.64$    & $6.39$     & $5.97$    & $1.28$   & $8.16$     & $5.68$     & $2.83$    &  $0.21$  \\\hline\hline

\end{tabular}}
\end{bigcenter}
\caption{The cut-flow table for a pseudoscalar mass of 400 GeV. The various perturbative orders used in the calculations are the same as in Table~\ref{tab:bbgamgam_H400}.}
\label{tab:zhbbaa_A400}
\end{table}

\subsection{The $b\bar{b}b\bar{b}$ Channel}
\label{sec:Htohhtobbbb}

After having studied the cleanest possible di-Higgs final state, we turn our attention to the one with the largest rate, \textit{viz.}, $p p \to H \to h h \to 4b$. Several searches have already been conducted in this channel~\cite{Aaboud:2016xco, Aaboud:2018knk, CMS-PAS-HIG-17-009, CMS-PAS-HIG-16-026} and provide some of the strongest bounds in both the non-resonant and resonant sectors. ATLAS~\cite{Aaboud:2018knk} has computed the observed (expected) upper bound on $\sigma(p p \to h h \to 4b)$ to be 13 (20.7) times that of the SM expectation with an integrated luminosity of 27.5 fb$^{-1}$. This channel has been further combined in ATLAS' HL-LHC projection~\cite{ATL-PHYS-PUB-2018-053} alongside $p p \to h h \to b\bar{b}\gamma\gamma/b\bar{b}\tau\tau$. The above channel will gain an improvement in sensitivity according to the Ref.~\cite{ATL-PHYS-PUB-2018-053} because of a projected 8\% improvement in $b$-tagging efficiency, besides having larger data sets. In this work, we closely follow the search strategy used by the ATLAS collaboration in Ref.~\cite{Aaboud:2016xco}. Even though this channel has the highest signal rate, the enormous multijet and $t\bar{t}$ backgrounds may considerably overwhelm the signal yield.

\begin{figure}[htb!]
\centering
\includegraphics[trim=0 640 0 70,clip,width=\textwidth]{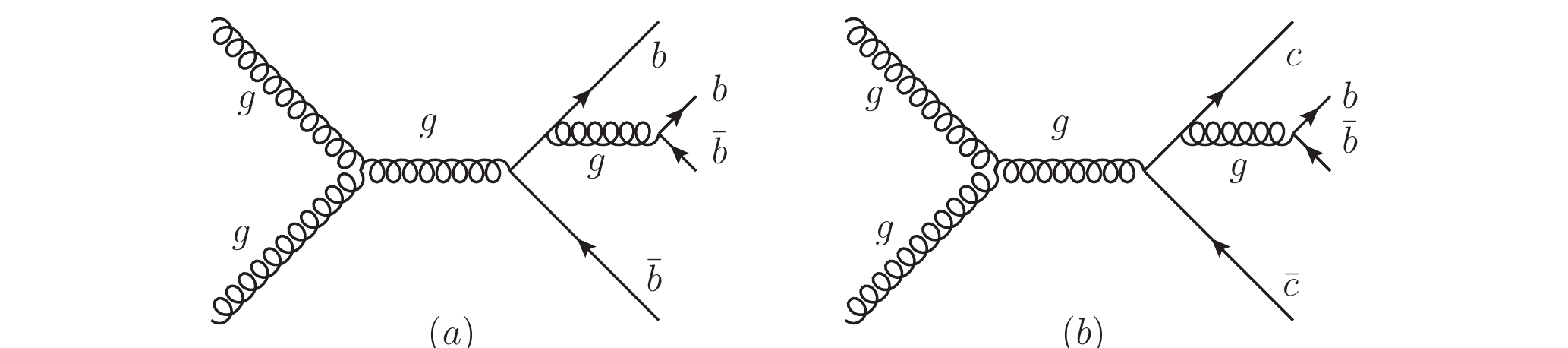}
\caption{Feynman diagrams for dominant (a) $b\bar{b}b\bar{b}$ and (b) $b\bar{b}c\bar{c}$ background for the $b\bar{b}b\bar{b}$ channel.}
\label{FD:bbbb}
\end{figure}

The dominant backgrounds (Fig.~\ref{FD:bbbb}) include the multijet production from QCD processes and the top pair production. For the multijet backgrounds, we dissect our sample generation into three different categories each having at least two $b$-quarks, \textit{viz.}, $b\bar{b}b\bar{b}$, $b\bar{b}c\bar{c}$ and $b\bar{b}jj$, in order to have sufficient statistics to take into account the different tagging efficiencies and fake rates. We do not generate the $h$ + jets and $Z$ + jets backgrounds separately but we include their tree-level diagrams while generating the multijet backgrounds as they have negligible contributions~\cite{Aaboud:2016xco}. We must mention here that we do not consider other possible sources of multijet production \textit{viz.}, $c\bar{c}c\bar{c}$, $c\bar{c}jj$ etc. as these processes will be highly suppressed (with respect to $b\bar{b}b\bar{b}$) upon multiplying by the fake efficiency factors, in succession. We generate the $t\bar{t}$ background with the top quark decaying to a $b$-quark and a $W$-boson. The $W$-bosons are then further decayed to $c\bar{s}$ or $\bar{c}s$. We avoid the $W \to u\bar{d}$ mode as the probability of a light jet faking a $b$-jet is $\sim 10$ times smaller than that of a $c$-jet. Lastly, we also consider the subdominant background coming from the non-resonant di-Higgs production ($gg\to hh$) and also from $t\bar{t}b\bar{b}$ (including $t\bar{t}Z/t\bar{t}h$). 

We select events containing exactly 4 $b$-tagged jets with the requirement of $p_{T,b}>60~\text{GeV}$ and $|\eta_{b}|<2.5$. The scalar sum, $H_{T}$, of the transverse momenta of all the visible particles in an event must fulfil, $H_{T}>300$ GeV. Finally, we form two di-jet systems from these four $b$-jets. The two jets within a dijet system must satisfy $0.4<\Delta R_{bb}<1.5$. We choose the leading (sub-leading) di-jet system to have $p_T>200 \; (150)$ GeV~\footnote{These are preliminary cuts before performing the multivariate analysis. The rationale behind these cuts are (a) some of these cuts have been applied at the generation level on some of the backgrounds in order to have better control over event statistics owing to large production cross-sections (see Appendix~\ref{sec:appendixA}) and (b) some other cuts are applied by observing the kinematic distributions of these observables. However, we apply stronger cuts on these variables in the following where we optimise them alongside other correlated variables through a more sophisticated multivariate analysis.}. Furthermore, to reduce the contamination from the $t\bar{t}$ background, we reconstruct the top by combining extra jets in an event with the di-jet systems. These jets must be within $\Delta R <1.5$ in the $\eta-\phi$ plane with the di-jet system. If an event contains exactly one extra jet, then we choose the di-jet system which is closest to it and combine to form a top quark system, $m_{t_1}$. However, when there are two such jets, we compute the minimum of all possible $\Delta R$ combinations between these two jets and the two di-jets before reconstructing two other top masses, $m_{t_2}$ and $m_{t_3}$. Because for our signal, we do not expect any proper top quark reconstruction, we thus veto events if the reconstructed mass of any of these possible choices for the top quark exceeds 120 GeV. After imposing this cut the $t\bar{t}$ background reduces to half with more than $80\%$ of the signal events still to spare. We detail these cuts one by one alongside the signal efficiency and cross-sections for the background processes in Table~\ref{tab:bbbb_precut}.

\begin{table}[htb!]
\begin{bigcenter}
\scalebox{0.7}{%
\begin{tabular}{||c||c|c|c|c||c|c|c|c|c|c||}
\hline
 & \multicolumn{4}{c||}{Signal Efficiency ($\times 10^{-3}$)} & \multicolumn{6}{c||}{Cross section [fb]} \\ \cline{6-11}
 Cut flow   & \multicolumn{4}{c||}{($pp\to H\to hh\to b\bar{b}b\bar{b}$)} & \multicolumn{6}{c||}{SM Backgrounds} \\
\cline{2-11} 
 & \multicolumn{4}{c||}{For $m_H$ of}                 & $b\bar{b}b\bar{b}$ & $b\bar{b}c\bar{c}$ & $t\bar{t}$ & $b\bar{b}jj$  & $hh \to 4b$ & $t\bar{t}b\bar{b}$ \\ \cline{2-5}
 & $400$ GeV  & $600$ GeV & $800$ GeV & $1000$ GeV  &     &      &    &     &  &  \\
\cline{1-11}  
 
\cline{1-11}  
Order    & -   & - & - & -  & LO  &  LO  & NNLO + NNLL~\cite{ttbarNNLO} & LO  & NNLO~\cite{hhtwiki} & LO \\ \hline

$4$ $b$-jets            & $5.40$   & $30.79$  &  $56.36$   & $67.29$   & $278.15$   & $26.85$  & $2.83$   & $66.74$  & $0.16$  & $8.29$   \\\hline

$H_{T}$                 & $5.05$   & $30.69$  &  $56.32$   & $67.26$   & $263.80$   & $25.48$  & $2.78$   & $64.15$  & $0.16$  & $8.27$   \\\hline
 
$\Delta R_{bb}$         & $1.67$   & $24.80$  &  $48.00$   & $57.34$   & $93.67$    & $7.94$   & $1.72$   & $11.90$  & $0.12$  & $1.87$   \\\hline

$p_{T,\textrm{di-jet}}$ & $0.41$   & $21.19$  &  $46.44$   & $56.52$   & $54.25$    & $4.69$   & $1.28$   & $7.54$   & $0.09$  & $1.61$   \\\hline

$m_{t}$                 & $0.33$   & $18.25$  &  $38.34$   & $45.83$   & $46.43$    & $3.80$   & $0.20$   & $6.36$   & $0.08$  & $0.66$   \\\hline\hline

\end{tabular}}
\end{bigcenter}
\caption{Cut-flow table before performing the multivariate analysis, in the $b\bar{b}b\bar{b}$ channel. 
}
\label{tab:bbbb_precut}
\end{table}

Finally, after all the aforementioned cuts are applied in succession, we check for any possible improvement upon performing a multivariate analysis. We utilise the BDT algorithm for our purposes and choose the following nine kinematic variables with maximal potency,

\begin{eqnarray}
 p_{T,\textrm{di-jet},k},~m_{\textrm{di-jet},k},~\Delta \phi_{bb,\textrm{di-jet},k},~\Delta\eta_{\textrm{di-jets}},~\Delta R_{\textrm{di-jets}},~m_{b\bar{b}b\bar{b}}
 \nonumber.
\end{eqnarray}

\begin{figure}
\centering
\includegraphics[scale=0.37]{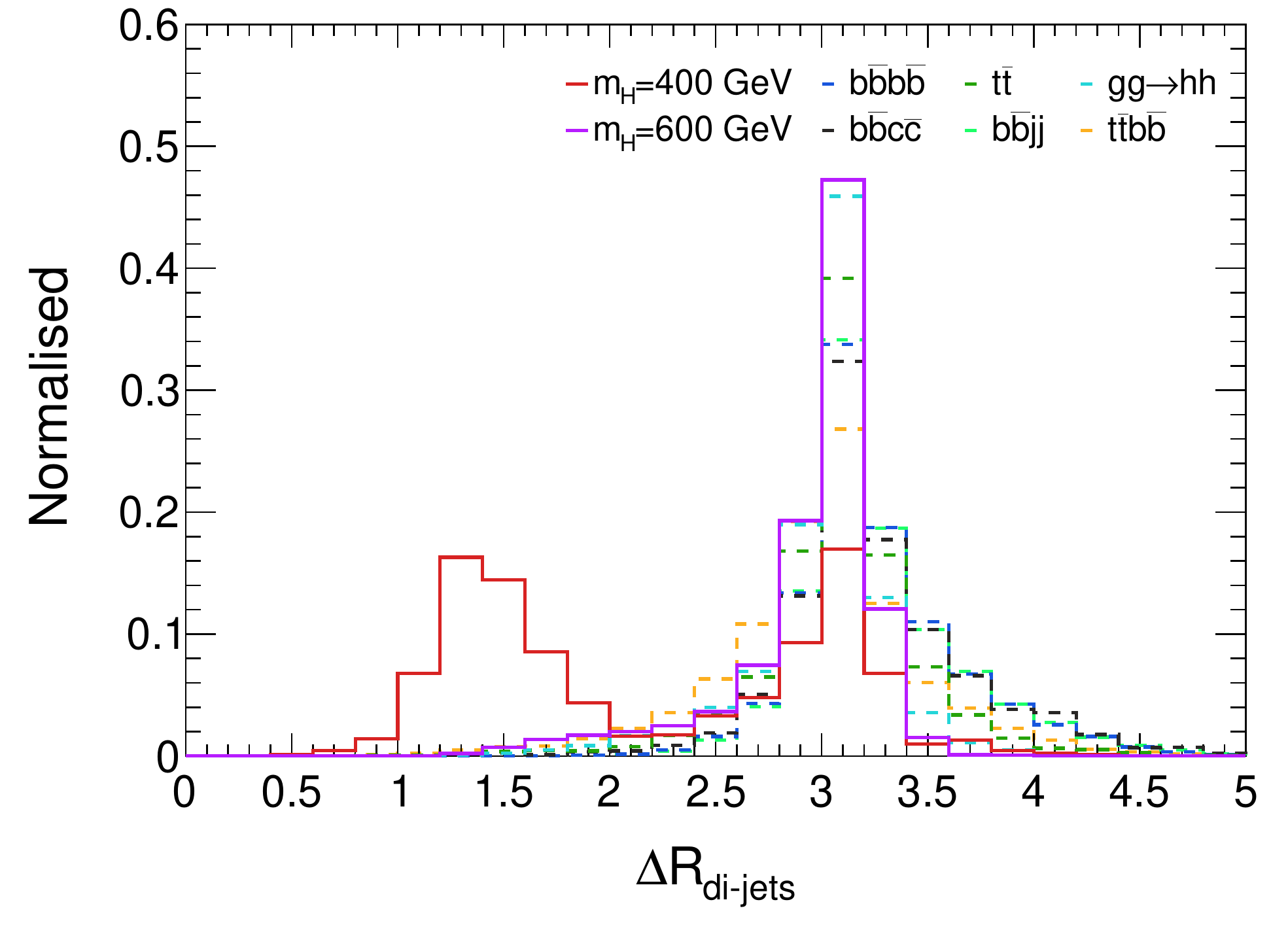}
\includegraphics[scale=0.37]{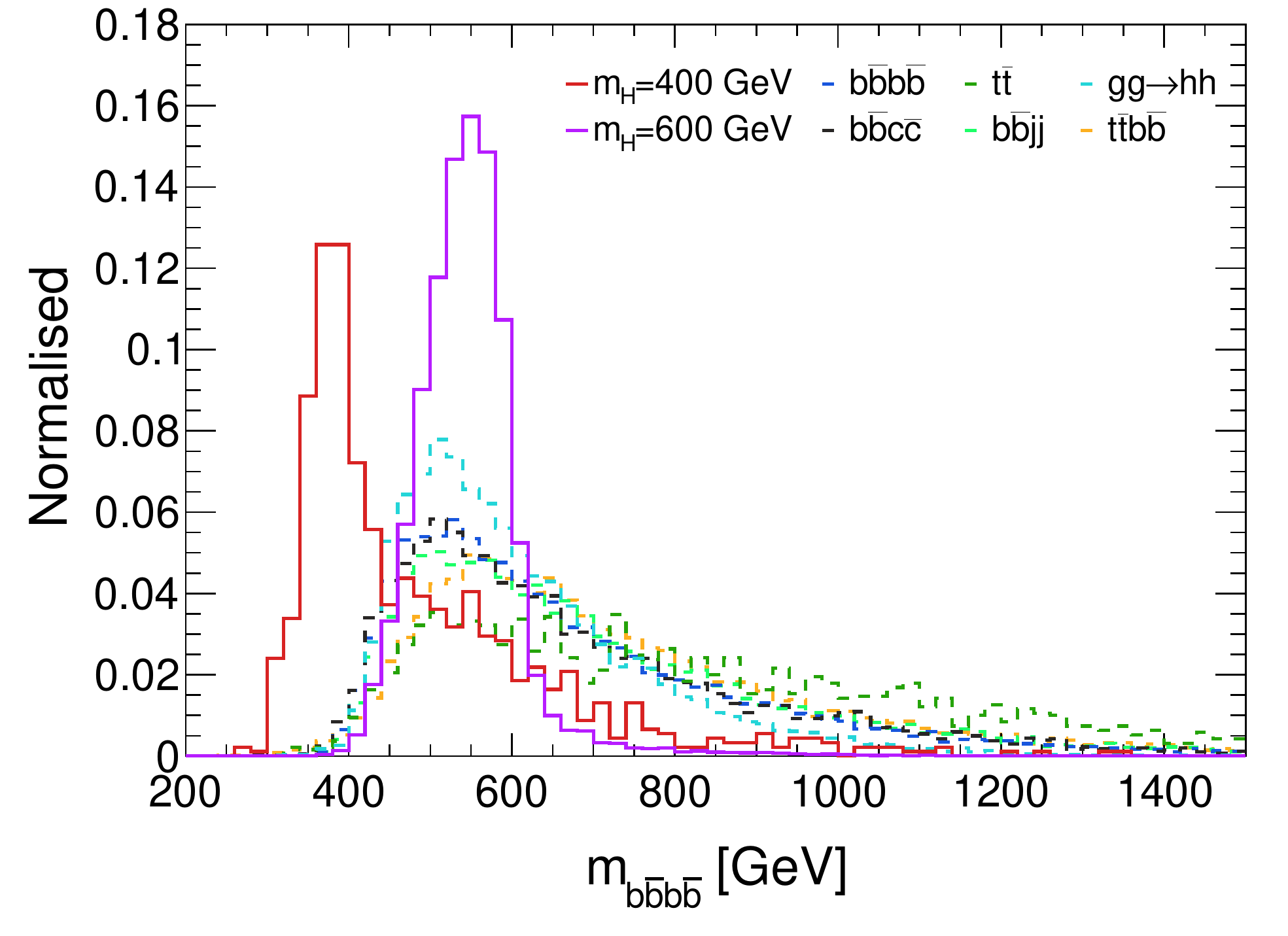} \\
\includegraphics[scale=0.37]{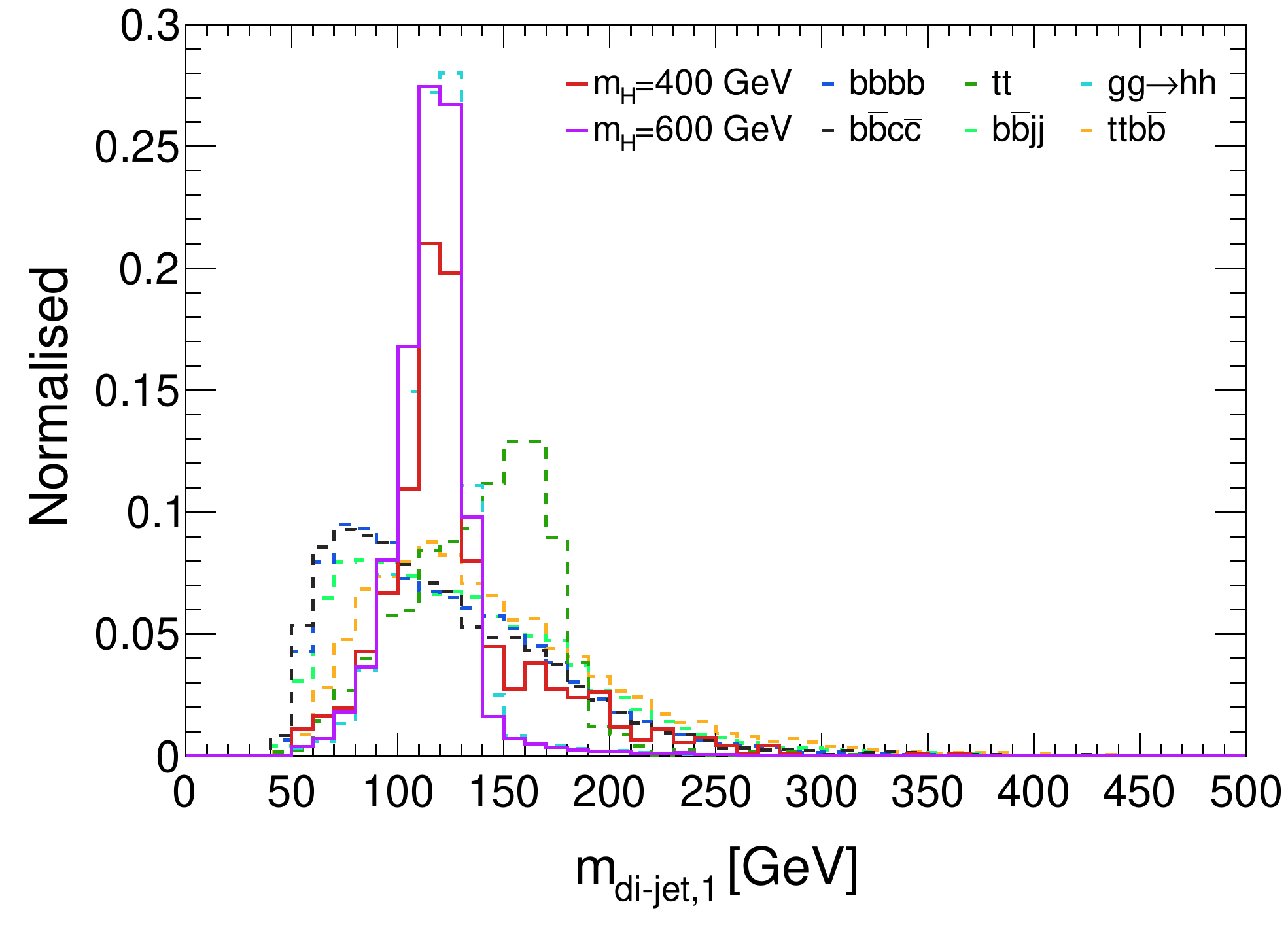}
\includegraphics[scale=0.37]{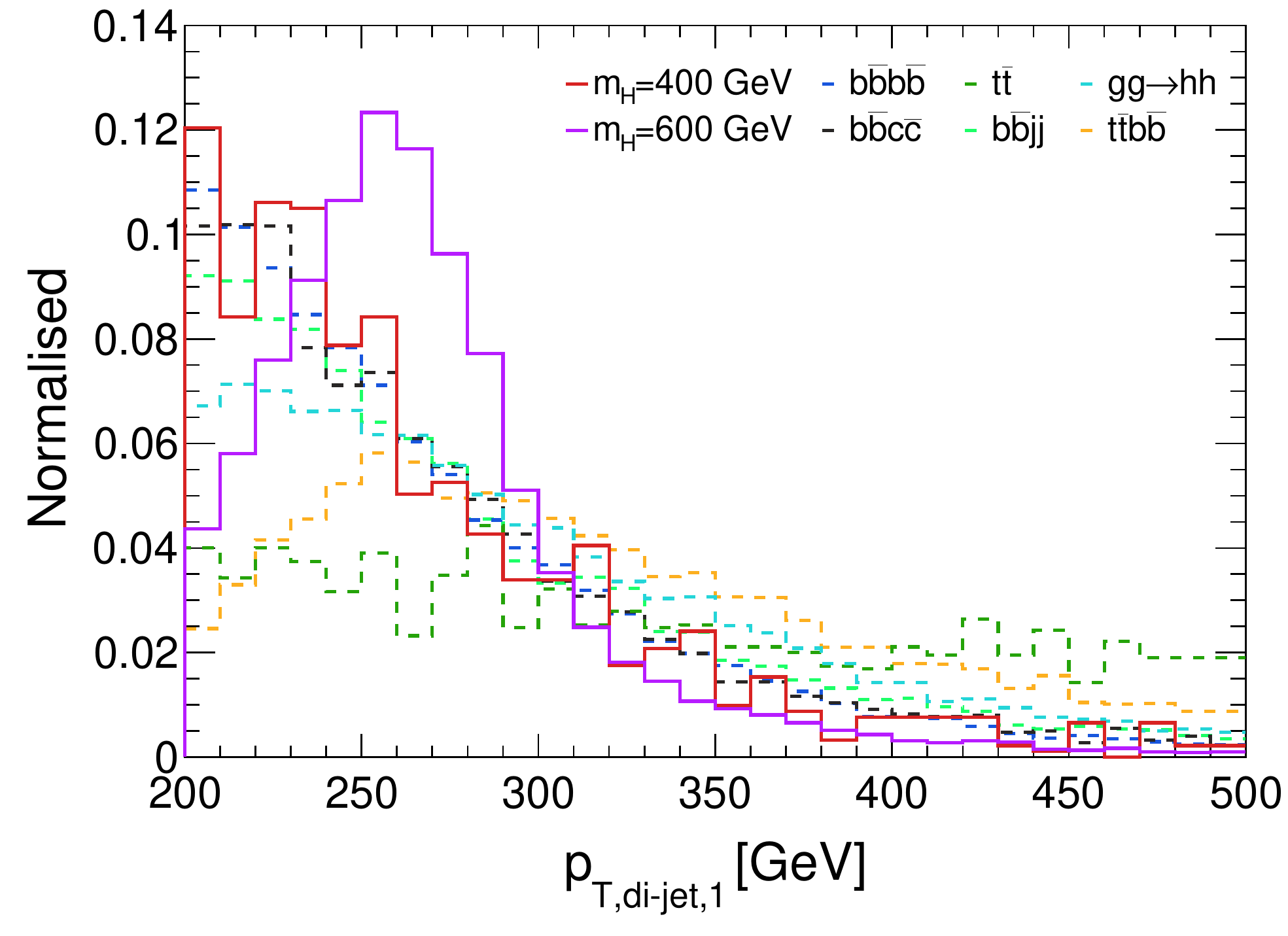}
\caption{The $\Delta R_{\textrm{di-jets}}$, $m_{4b}$, $m_{\textrm{di-jet,1}}$ and $p_{T,\textrm{di-jet,1}}$ distributions for heavy Higgs masses of $m_H = 400$ and $600$ GeV with backgrounds. Here the heavy Higgs boson is searched for in the $4b$ final state. The distributions are shown after imposing the cuts mentioned in Table~\ref{tab:bbbb_precut}.}
\label{Hfig:bbbb1}
\end{figure}

Here we use the kinematic variables reconstructed from the two di-jet systems \textit{viz.}, invariant mass ($m_{\textrm{di-jet}}$), transverse momentum ($p_{T,\textrm{di-jet}}$) and azimuthal angle separation between the $b$-jets forming the dijet systems ($\Delta\phi_{bb,\textrm{di-jet}}$). The subscript $k=1, 2$ refers to the $p_T$ ordering of the di-jets. We also take the separation in the $\eta$ and $\eta$-$\phi$ plane between the two di-jets, \textit{viz.}, $\Delta\eta_{\textrm{di-jets}}$ and $\Delta R_{\textrm{di-jets}}$ respectively. $m_{b\bar{b}b\bar{b}}$ is the invariant mass of the four $b$-jet system. The top four variables with the highest discriminatory power are shown in Fig.~\ref{Hfig:bbbb1}. We can see that the lower masses have significantly longer tails while performing the mass reconstructions.

\begin{figure}
\centering
\includegraphics[scale=0.5]{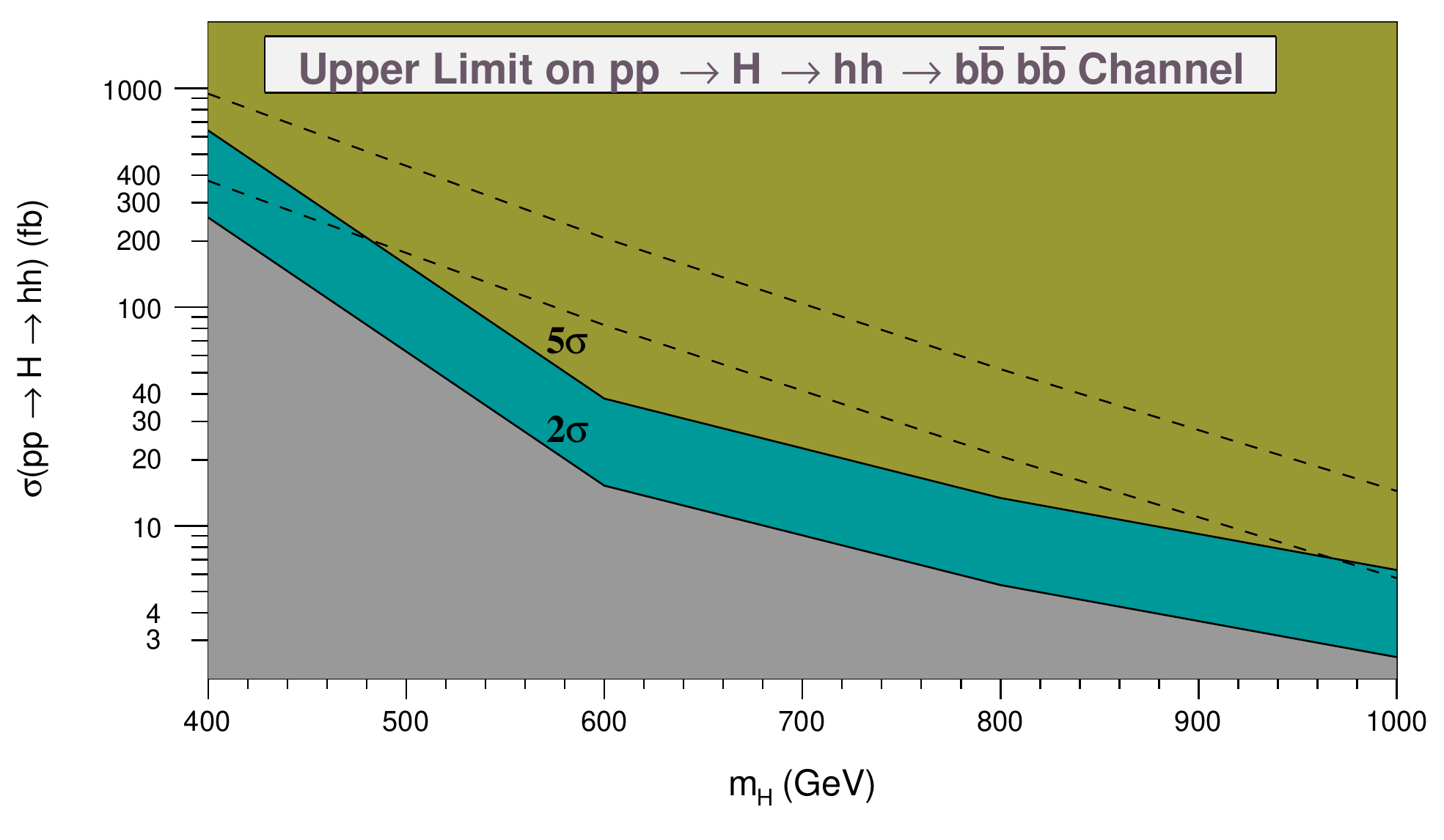}
\caption{Upper limit on $\sigma(pp\to H\to hh)$ (fb) as a function of $m_{H}$ (GeV) for the $b\bar{b}b\bar{b}$ channel. The solid (dashed) lines show the 2$\sigma$-5$\sigma$ band on taking 0\% (5\%) systematic uncertainties.}
\label{Hfig:bbbb2}
\end{figure}

Finally, in Table~\ref{tab1:bbbb}, we present the background yields after the BDT optimisation has been completed. Like in the previous section, we translate these results into an exclusion diagram showing the upper limits on $\sigma(pp\to H\to hh)$ as a function of the heavy Higgs mass. We show these in Fig.~\ref{Hfig:bbbb2}. The limit is very strong between $600$ GeV and $1$ TeV with the 95\% CL upper limit varying between $15.26$ fb and $2.51$ fb. The upper limit becomes between $[82.70,~5.77]$ fb within the aforementioned range by adding $5\%$ systematic uncertainty,.

\begin{center}
\begin{table}[htb!]
\centering
\scalebox{0.6}{%
\begin{tabular}{|c|c|c|c|}\hline
(a) & Process & Order & Events \\ \hline \hline
\multirow{7}{*}{Background}
 & $b\bar{b}b\bar{b}$               & LO                          & $203.60$ \\  
 & $b\bar{b}c\bar{c}$               & LO                          & $121.51$ \\  
 & $b\bar{b}jj$       		        & LO                          & $46.11$ \\  
 & $t\bar{t}$       		        & NNLO~\cite{ttbarNNLO}       & $10.65$ \\ 
 & $pp\to hh$    		            & NNLO~\cite{hhtwiki}         & $6.77$ \\
 & $t\bar{t}b\bar{b}$       		& LO                          & $77.28$ \\  
  \cline{2-4}  
 & \multicolumn{2}{c|}{Total}       & $465.92$ \\ \hline
\end{tabular}}
\bigskip

\scalebox{0.6}{%
\begin{tabular}{|c|c|c|}\hline
(b) & Process & Events \\ \hline \hline
\multirow{7}{*}{Background}
 & $b\bar{b}b\bar{b}$               & $8950.94$ \\  
 & $b\bar{b}c\bar{c}$               & $731.91$ \\  
 & $b\bar{b}jj$       	            & $1263.16$ \\  
 & $t\bar{t}$       	            & $74.56$ \\ 
 & $pp\to hh$                       & $103.65$ \\
 & $t\bar{t}b\bar{b}$       		& $230.05$ \\  
 \cline{2-3}  
 & \multicolumn{1}{c|}{Total}       & $11354.27$ \\ \hline
\end{tabular}}
\quad
\scalebox{0.6}{%
\begin{tabular}{|c|c|c|}\hline
(c) & Process & Events \\ \hline \hline
\multirow{7}{*}{Background}
 & $b\bar{b}b\bar{b}$               & $4304.59$ \\  
 & $b\bar{b}c\bar{c}$               & $358.89$ \\  
 & $b\bar{b}jj$       	            & $624.17$ \\  
 & $t\bar{t}$       		        & $130.07$ \\ 
 & $pp\to hh$    	                & $50.96$ \\
 & $t\bar{t}b\bar{b}$       		& $152.77$ \\
 \cline{2-3}  
 & \multicolumn{1}{c|}{Total}       & $5621.45$ \\ \hline
\end{tabular}}
\quad
\scalebox{0.6}{%
\begin{tabular}{|c|c|c|}\hline
(d) & Process & Events \\ \hline \hline
\multirow{7}{*}{Background}
 & $b\bar{b}b\bar{b}$               & $1228.85$ \\  
 & $b\bar{b}c\bar{c}$               & $127.17$ \\  
 & $b\bar{b}jj$       	            & $219.03$ \\  
 & $t\bar{t}$       	            & $63.26$ \\ 
 & $pp\to hh$    		            & $18.44$ \\
 & $t\bar{t}b\bar{b}$       		& $52.97$ \\ 
 \cline{2-3}  
 & \multicolumn{1}{c|}{Total}       & $1709.72$ \\ \hline
\end{tabular}}
\caption{Respective background yields for the $b\bar{b}b\bar{b}$ channel after the BDT analyses optimised for a heavy Higgs mass of $(a)~400$ GeV, $(b)~600$ GeV, $(c)~800$ GeV and $(d)~1$ TeV. The tables also list the perturbative order at which the cross-sections are considered.}
\label{tab1:bbbb}
\end{table}
\end{center}

\section*{Discussion about $m_H=400$ GeV }
\label{sec:4b_400}

The $95\%$ and $99.7\%$ CL upper limits on $\sigma(pp\to H\to hh)$ for the heavy Higgs with a mass around $400$ GeV is very large (256.66 fb and 641.65 fb respectively) as compared to the other mass points, even after the BDT optimisation. The reason for this is the following. The signal efficiency for $m_H = 400$ GeV reduces by $\sim$ 67\% after imposing the $\Delta R_{bb}$ selection as can be seen from Table~\ref{tab:bbbb_precut}. Since the heavy Higgs mass ($400$ GeV) is near the threshold of the non-resonant di-Higgs production, the SM-like Higgs bosons for the resonant case are produced with low $p_T$. This further leads to the Higgs decay products being widely separated in the $\eta-\phi$ plane and thus obviously does not satisfy our di-jet selection criteria of $\Delta R_{bb} <1.5$ within each di-jet system. With the sole intention of improving the sensitivity, we adopt a $\chi^2$ minimisation technique as described below. We define a new kinematic variable $\chi^2_{hh}$ for the events which do not satisfy the $\Delta R_{bb} < 1.5$ selection criteria as follows

\begin{eqnarray}
\chi^2_{hh} &\equiv& \min_{\Delta R_{bb}}  \left [ 
\frac{\left ( m^2_{\textrm{di-jet},1} - m^2_h \right )^2}{\sigma_{h1}^4} \,   +
\frac{\left ( m^2_{\textrm{di-jet},2} - m^2_h \right )^2}{\sigma_{h2}^4}  \, \right ], 
\end{eqnarray}

where $m_h=125$ GeV and $\sigma_{hj}=0.1\times m_{\textrm{di-jet},j}$ with $j=1,2$ marks the $p_T$ ordering. Thus, in addition to the events satisfying $\Delta R_{bb} < 1.5$, we also consider those events which contain di-jet pairs with $\Delta R$ separation between the $b$-jets to be more than $1.5$. Following this, we construct the aforementioned $\chi^2_{hh}$ variable for each possible pair of reconstructed di-jet. The event is finally selected if the non-zero minimum value of the $\chi^2_{hh}$ variable is less than 50~\footnote{We checked our results upon choosing $\chi^2_{hh}$ both higher and lower than 50. For higher values, the signal yield increases but the background increases at a higher rate, generating a weaker limit. On the other hand, upon lowering the $\chi^2_{hh}$ value below 50, the signal yield decreases substantially. This makes the value 50 an ideal and optimal choice.}. Upon using this modification, the signal efficiency increases by $\sim 26\%$ at the di-jet selection level while simultaneously increasing the dominant backgrounds like $b\bar{b}b\bar{b}$ by $\sim 5\%$ and $t\bar{t}$ by $\sim 7\%$. However, the limit on the upper limit of the cross-section improves to $245.75$ fb and $614.37$ fb at 95\% and 99.7\% CL respectively.

\subsection{The $b\bar{b}\tau\tau$ Channel}
\label{sec:Htohhtobbtautau}

Next, we turn our attention to one of the best probes for the di-Higgs searches, \textit{viz.}, the $b\bar{b}\tau\tau$ channel. The intricacy and potential of this channel lies in our ability to reconstruct the $\tau$-leptons as these come with neutrinos which show up as missing transverse energy in the detector. This channel gives rise to three phenomenologically different final states, \textit{viz.}, $b\bar{b}\ell^+\ell^-+\slashed{E}_T$, $b\bar{b}\ell \tau_h +\slashed{E}_T$ and $b\bar{b}\tau_h \tau_h +\slashed{E}_T$. In this work, we will only consider the last category, \textit{i.e.}, the one with the fully hadronic $\tau$ decays. The hadronically decayed $\tau$-leptons are termed as $\tau$-hadrons or $\tau$-jets which may either contain one (one-pronged) or three (three-pronged) charged particle(s) inside the jet cone. Thus, it is essential to tag these $\tau$-jets in order to segregate them from regular QCD jets ensuing from the various backgrounds that we will discuss below. We will not discuss the fully leptonic case here as from our previous analysis~\cite{Adhikary:2017jtu} we know that the sensitivity is extremely low even at the HL-LHC.

\begin{figure}[htb!]
\centering
\includegraphics[trim=0 450 0 75,clip,width=\textwidth]{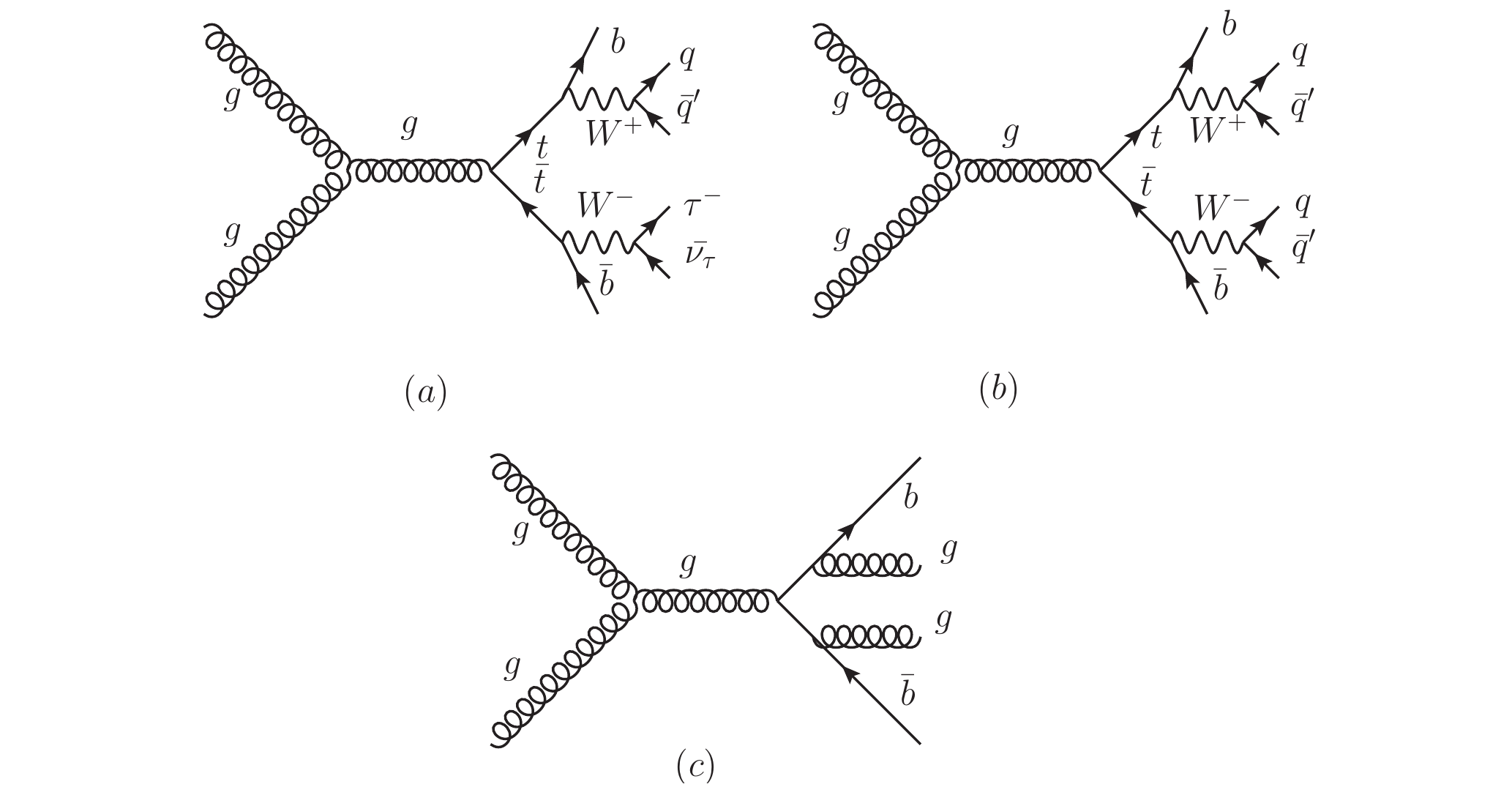}
\caption{Feynman diagrams for the dominant (a) semi-leptonic $t\bar{t}$, (b) fully hadronic $t\bar{t}$ and (d) $b\bar{b}jj$ (where jet (j) can fake as $\tau$ jet) background for the $b\bar{b}\tau\tau$ channel.}
\label{FD:bb2tau}
\end{figure}

We generate two different samples for the dominant $t \bar{t}$ background (Fig.~\ref{FD:bb2tau}), where either both the $W$-bosons decay to jets or where one decays into a lepton ($viz.~e^\pm,\mu^\pm~or~\tau^\pm$) and the other to a pair of jets. The QCD-QED background $\tau\tau b\bar{b}$ also contributes significantly. Besides, we also generate the subdominant backgrounds which include $t\bar{t}h$, $t\bar{t}W$, $t\bar{t}Z$, $b\bar{b} h$, $Zh$ and the non-resonant Higgs pair production \textit{i.e.}, $gg\to hh$. We simulate the $Zh$ background upon considering two processes where in one case the $Z$-boson decays to a pair of bottom quarks and the Higgs boson decays to a pair of $\tau$-leptons and in the other the decays are reversed. Moreover, we also generate the dominant fake background for the hadronic channel in the form of $b\bar{b} j j$ (Fig.~\ref{FD:bb2tau}), where the light jets can be fake $\tau$-tagged jets\footnote{The $\tau$-leptons decay hadronically (each with a branching fraction of $\sim 65\%$) leading to jets in the final state. In our analyses, we use the $\tau$-tagging method as discussed in Ref.~\cite{Bagliesi:2007qx}.}. We detail the generation level cuts for these various backgrounds in Appendix~\ref{sec:appendixA}. Following the generation level cuts, we further apply some basic cuts on the signal and background samples in order to ensure a common kinematic phase space. The $b/\tau$-jets and the leptons ($e, \mu$) are required to lie within $|\eta| < 2.5$ and have $p_{T,b/\tau (\ell)} > 20 \; (10)$ GeV. The light jets must satisfy $p_{T,j} > 20$ GeV and $|\eta_j| < 4.5$. The minimum distance in the $\eta-\phi$ plane between the $b$-jets and the leptons, and also among themselves is required to be $\Delta R > 0.2$. The reconstructed invariant mass of the bottom pair and the visible $\tau$ pair must obey $m_{bb/\tau\tau}>50$ GeV.

\subsubsection{The $b\bar{b}\tau_h\tau_h$ Channel}
\label{sec:Htohhtobbtautau:hh}

\begin{figure}
\centering
\includegraphics[scale=0.37]{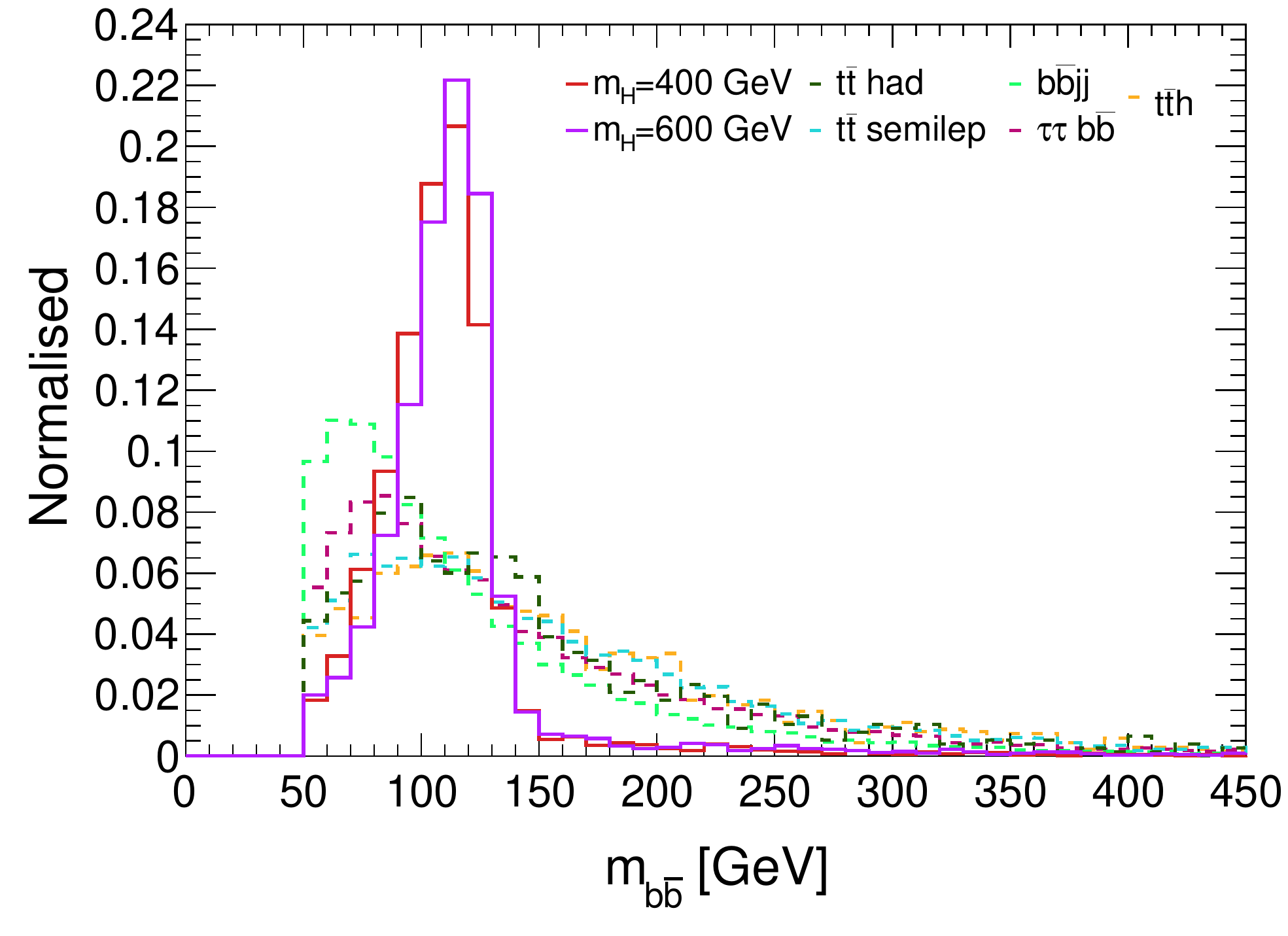}
\includegraphics[scale=0.37]{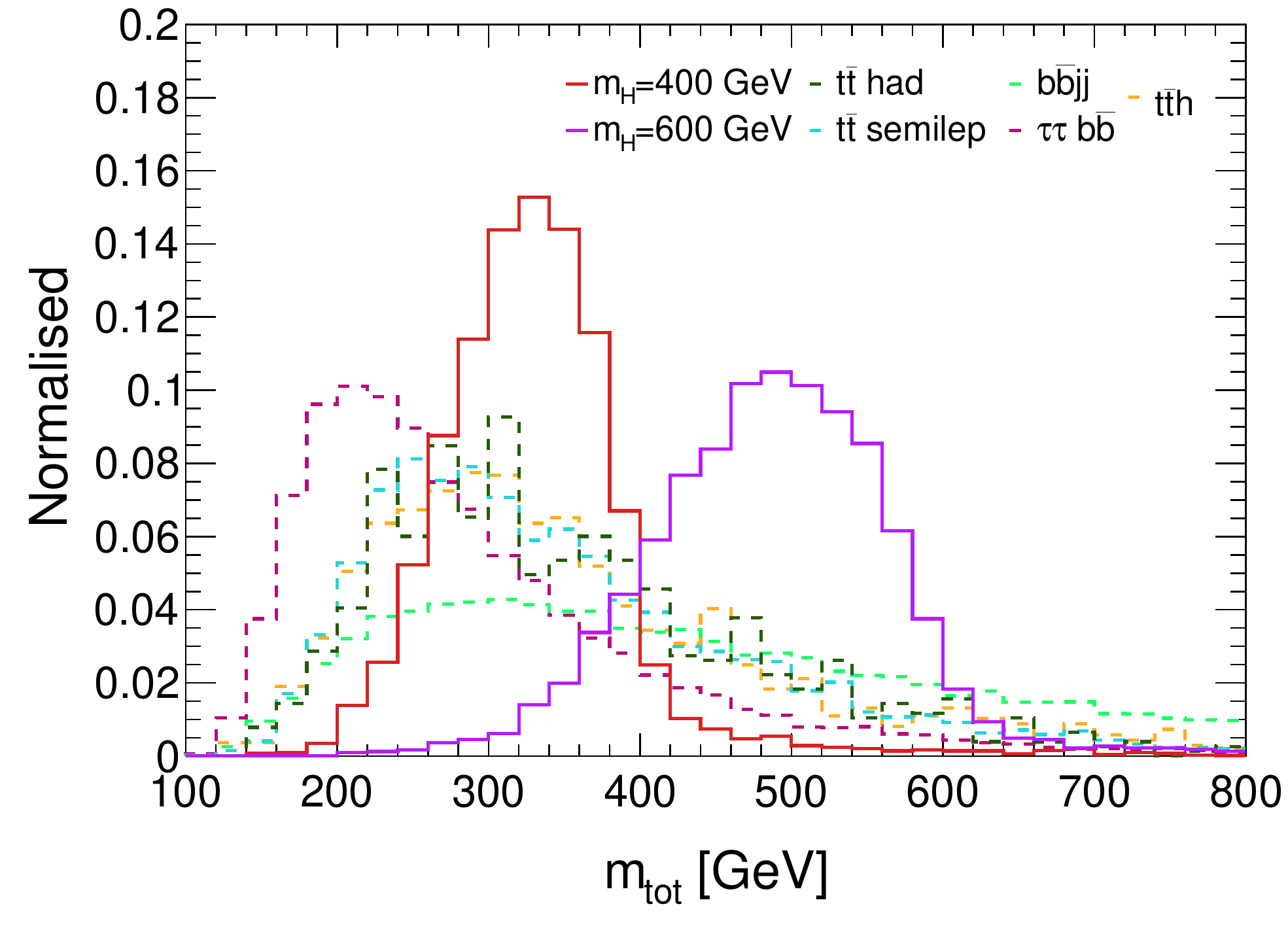} \\
\includegraphics[scale=0.37]{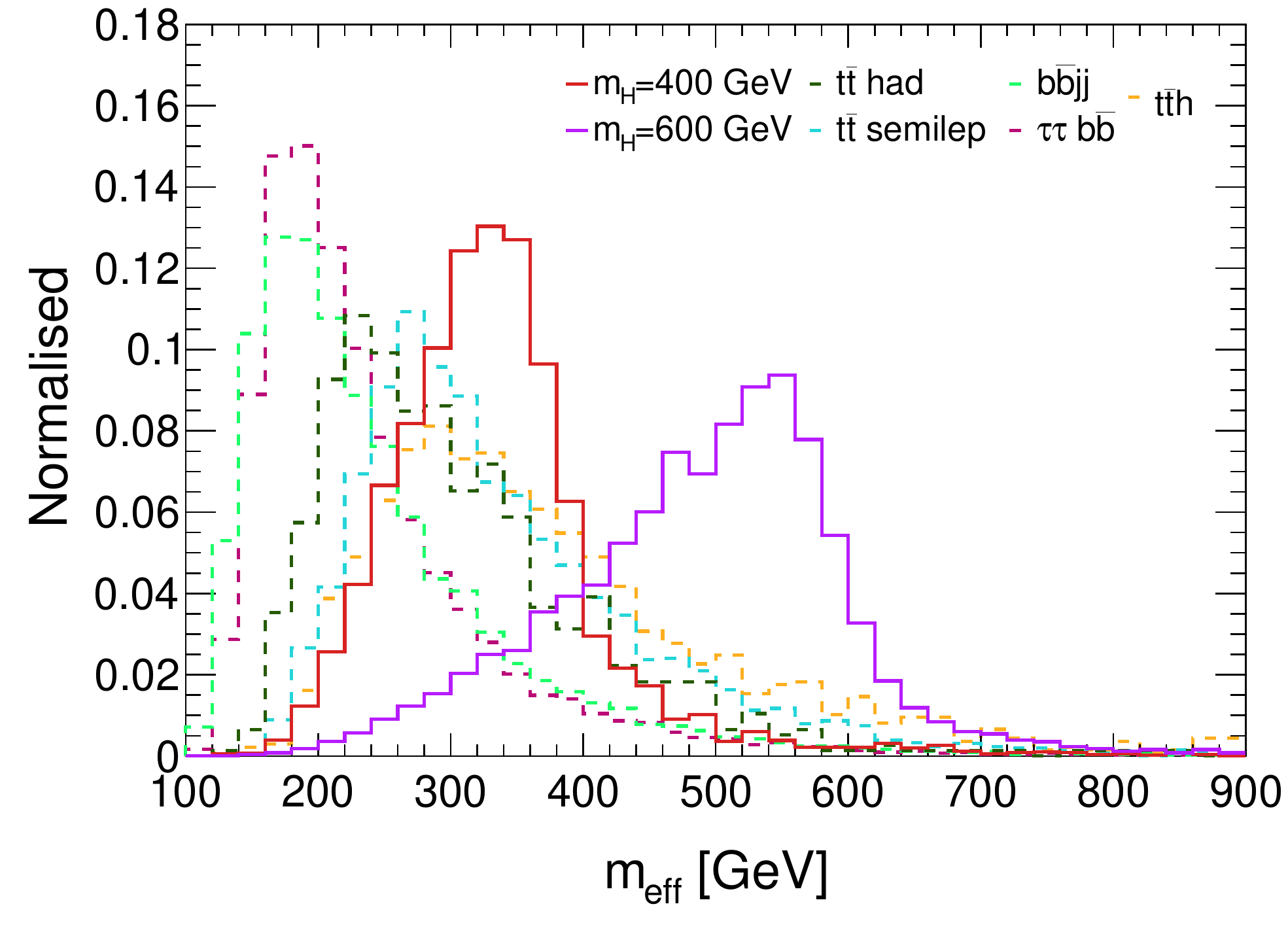}
\includegraphics[scale=0.37]{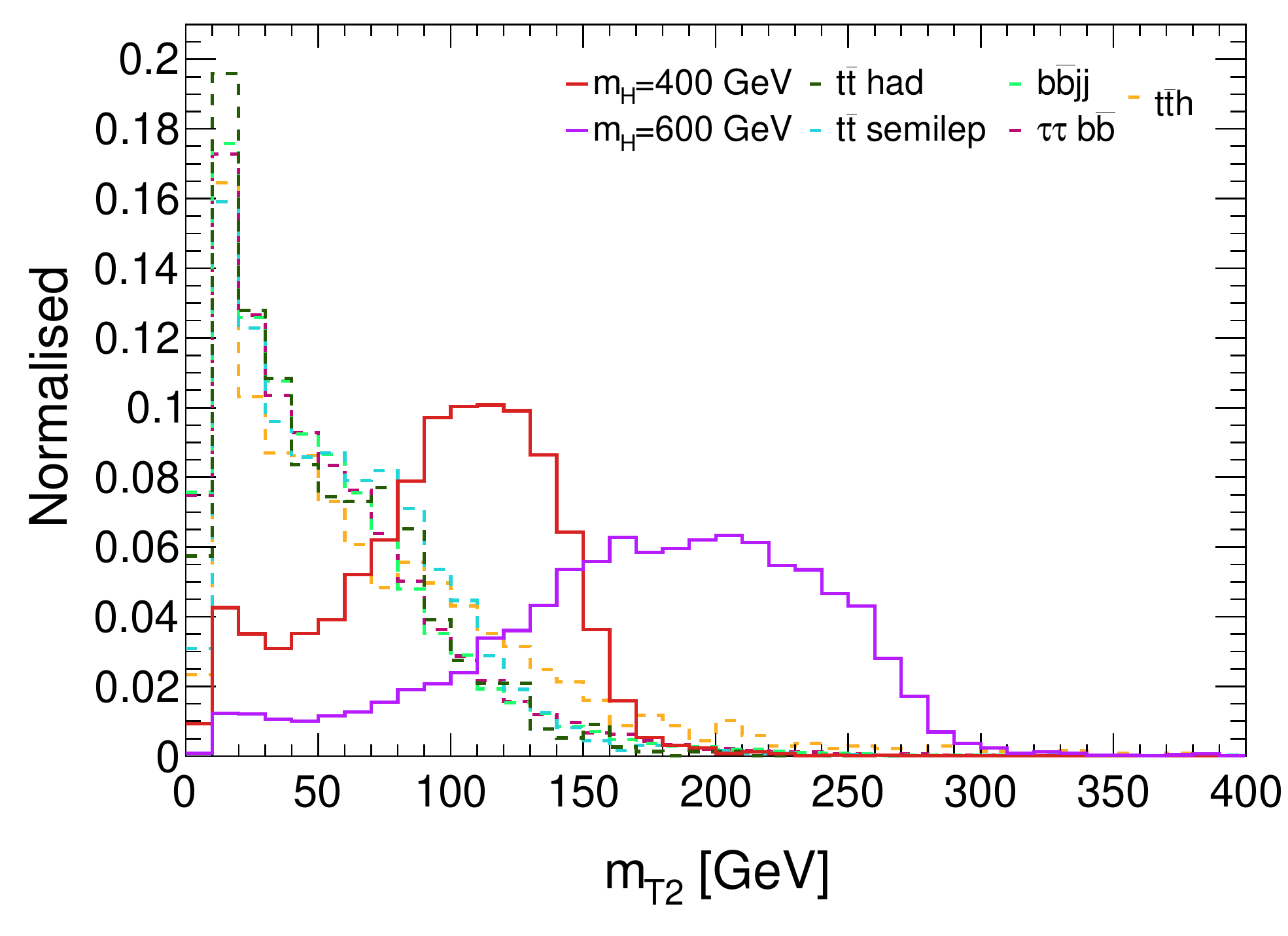} \\
\includegraphics[scale=0.37]{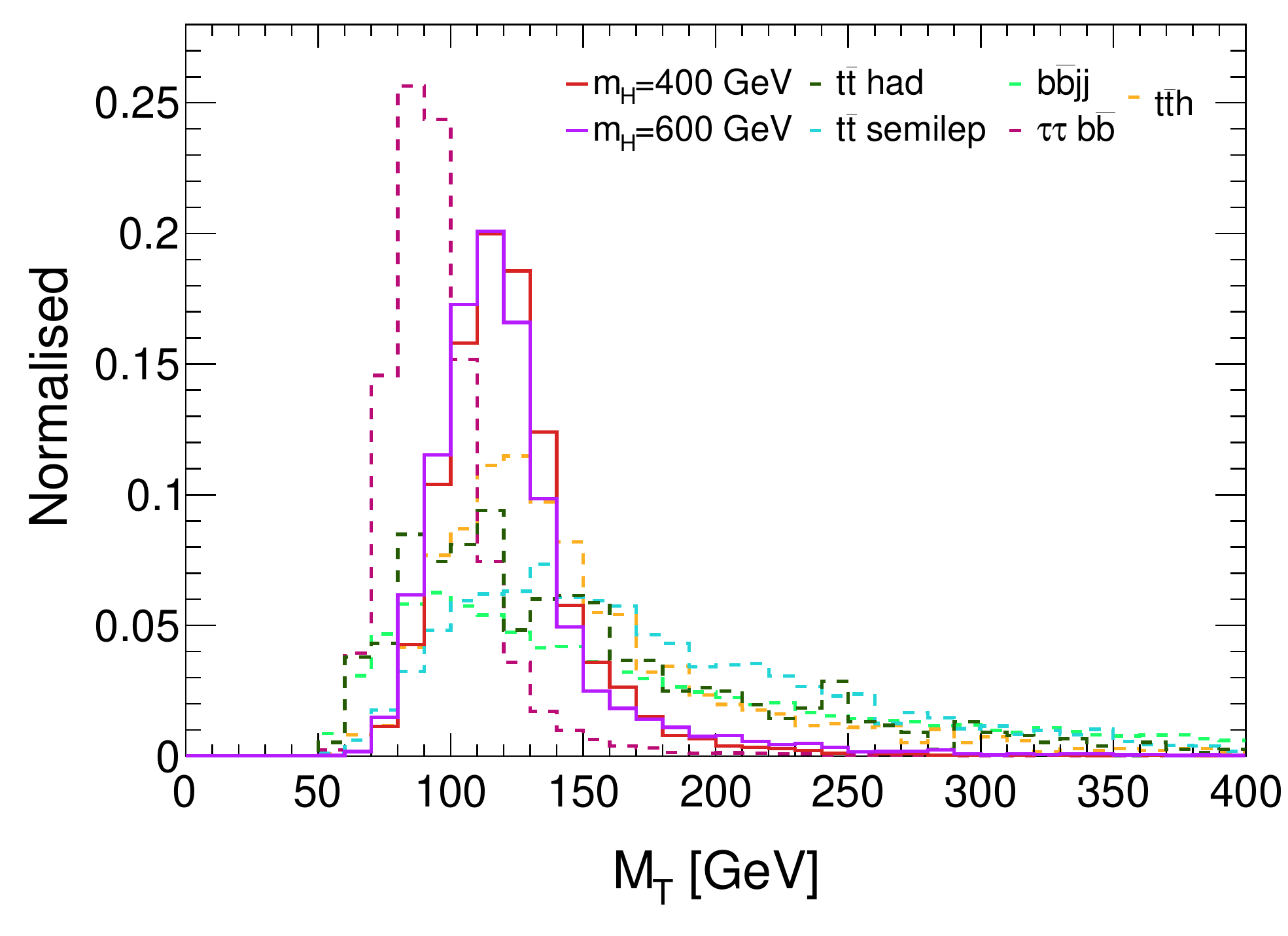}
\caption{The $m_{bb}$, $m_{\textrm{tot}}$, $m_{\textrm{eff}}$, $m_{T2}$ and $M_T$ distributions for heavy Higgs masses of $m_H = 400$ and $600$ GeV with dominant backgrounds. Here the heavy Higgs boson is searched for in the $b\bar{b}\tau_h\tau_h$ final state. The distributions are shown after imposing the basic trigger cuts.}
\label{Hfig:bbthth1}
\end{figure}

\begin{figure}
\centering
\includegraphics[scale=0.5]{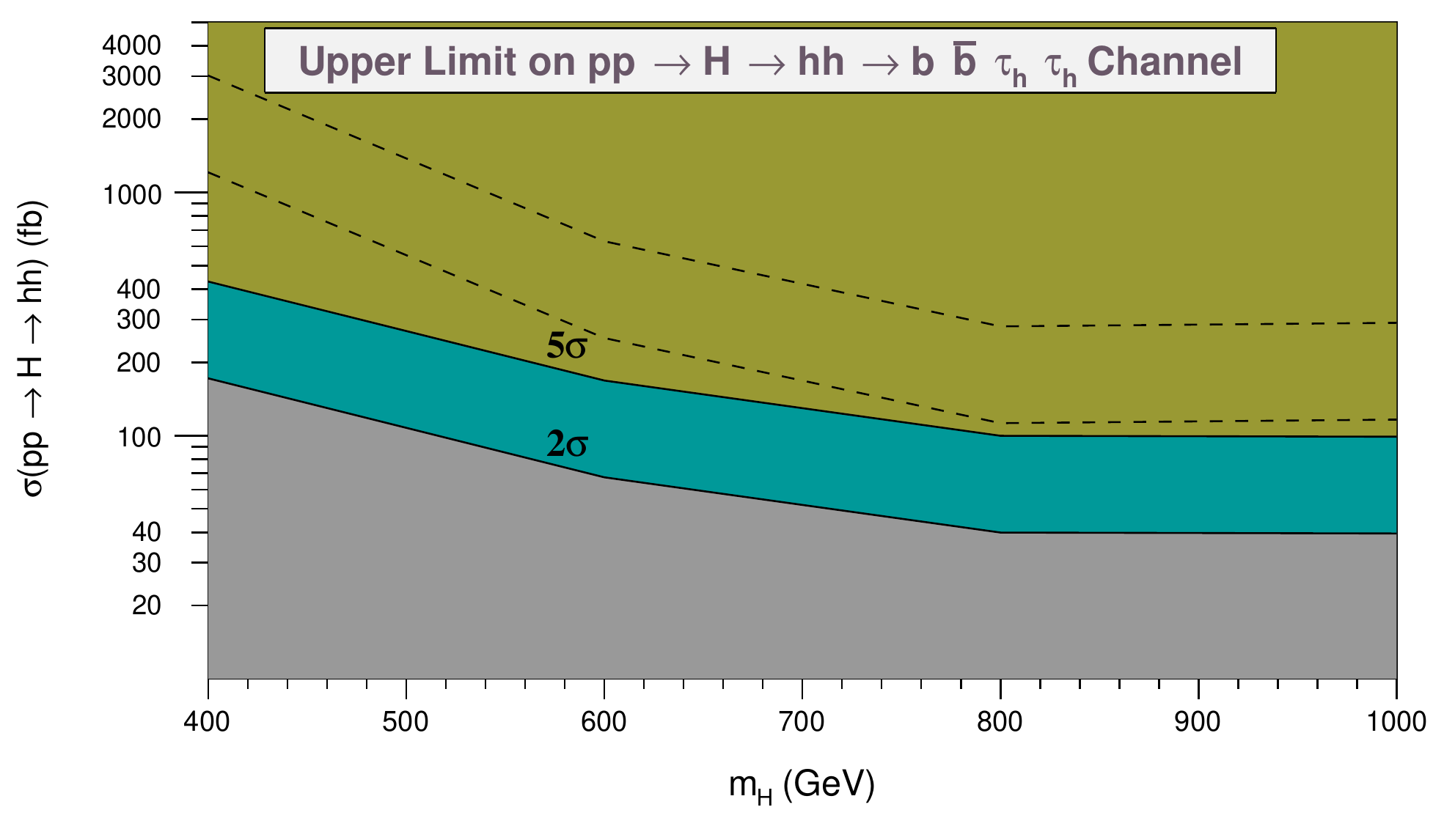}
\caption{Upper limit on $\sigma(pp\to H\to hh)$ (fb) as a function of $m_{H}$ (GeV) for the $b\bar{b}\tau_h\tau_h$ channel. The solid (dashed) lines show the 2$\sigma$-5$\sigma$ band on taking 0\% (5\%) systematic uncertainties.}
\label{Hfig:bbthth2}
\end{figure}

In this sub-section we briefly outline the prospects of searching for the heavy Higgs in the $b\bar{b}\tau_h\tau_h$ final state. In doing so, we select events containing exactly two $b$-tagged jets and two $\tau$-tagged jets alongside the cuts described above. Having seen the strength of the multivariate analyses for this channel, in Ref.~\cite{Adhikary:2017jtu}, rather than opting for the classical cut-based analysis, we perform a BDT analysis with the following 13 variables with the maximal discerning capability:

\begin{equation}
\begin{split}
p_{T,bb},~m_{bb},~p_{T,\tau_h\tau_h},~\Delta R_{\tau_h\tau_h},~\Delta\phi_{\tau_{h1}\met},~\Delta\phi_{\tau_{h2}\met},\\~M_T,~m_{T2},~p_{T,\textrm{tot}},~m_{\textrm{tot}},~m_{\textrm{eff}},~\Delta R_{b_1\tau_{h1}},~\Delta R_{bb,\tau_h\tau_h} \nonumber
\end{split}
\end{equation}

\noindent where, $M_T$ is the transverse mass of the $h \to \tau \tau$ system~\footnote{For this whole section, we use the conventional definition~\cite{Han:2005mu} of $M_T^2 = (\sum\limits_i E_{T,i})^2-(\sum\limits_i \vec{p}_{T,i})^2$, where $i$ runs over the relevant objects. In Section~\ref{sec:bbH}, we use a modified definition of $M_T$.}, $p_{T,\textrm{tot}}$ and $m_{\textrm{tot}}$ are respectively the transverse momenta and mass of the full visible system and $m_{\textrm{eff}}$ is the scalar sum of the transverse mass of all the visible products plus $\slashed{E}_T$. The rest of the variables have usual definitions. The top five variables are shown in fig.~\ref{Hfig:bbthth1}. As can be seen, the $m_{T2}$ variable is particularly useful for heavier Higgs masses as it can be used to completely eradicate the $t\bar{t}$ background. We train the signal and background samples and they are optimised for each benchmark signal point. We list the background events after optimising the BDT and imposing the cut for four values of $m_H$, in Table~\ref{Htab:bbthth}. Finally, we show the upper limit on the heavy Higgs production cross-section (assuming BR($H \to h h$) $=100\%$) in Fig.~\ref{Hfig:bbthth2}. The $95\%$ CL upper limit on the cross-section between $m_H=$ 600 GeV and 1 TeV varies between $67.34$ fb and $39.56$ fb. With $5\%$ systematic uncertainty, the limits become $251.76$ fb and $116.30$ fb respectively.

\begin{center}
\begin{table}[htb!]
\centering
\scalebox{0.7}{%
\begin{tabular}{|c|c|c|c|}\hline
(a) & Process & Order & Events \\ \hline \hline
\multirow{10}{*}{Background}
 & $t\bar{t}$ had               & NNLO~\cite{ttbarNNLO}                       & $662.20$ \\ 
 & $t\bar{t}$ semi-lep          & NNLO~\cite{ttbarNNLO}                       & $5366.58$ \\  
 & $\tau\tau b\bar{b}$      	& LO                                          & $3143.59$ \\ 
 & $t\bar{t}h$              	& NLO~\cite{bkg_twiki_cs}                     & $296.20$ \\  
 & $t\bar{t}Z$             		& NLO~\cite{Lazopoulos:2008de}                & $141.56$\\  
 & $t\bar{t}W$       		    & NLO~\cite{Campbell:2012dh}                  & $33.50$ \\ 
 & $pp\to hh$           		& NNLO~\cite{hhtwiki}                         & $50.46$ \\
 & $b\bar{b}h$                  & NNLO                                        & $2.36$ \\
 & $Zh$                         & NNLO                                        & $132.88$ \\  
 & $b\bar{b}jj$                 & LO                                          & $9558.83$ \\ \cline{2-4}  
 & \multicolumn{2}{c|}{Total}			                                      & $19388.16$ \\ \hline
\end{tabular}}
\bigskip
\scalebox{0.7}{%
\begin{tabular}{|c|c|c|}\hline
(b) & Process & Events \\ \hline \hline
\multirow{10}{*}{Background}
 & $t\bar{t}$ had             & $126.60$ \\ 
 & $t\bar{t}$ semi-lep        & $884.05$ \\  
 & $\tau\tau b\bar{b}$        & $633.83$ \\ 
 & $t\bar{t}h$                & $90.79$ \\  
 & $t\bar{t}Z$                & $57.06$\\  
 & $t\bar{t}W$                & $3.28$ \\ 
 & $pp\to hh$                 & $29.85$ \\
 & $b\bar{b}h$                & $0.36$ \\
 & $Zh$                       & $60.94$ \\  
 & $b\bar{b}jj$               & $3303.97$ \\ \cline{2-3}  
 & \multicolumn{1}{c|}{Total} & $5190.73$ \\ \hline
\end{tabular}}
\quad
\scalebox{0.7}{%
\begin{tabular}{|c|c|c|}\hline
(c) & Process & Events \\ \hline \hline
\multirow{10}{*}{Background}
& $t\bar{t}$ had             & $97.38$ \\ 
& $t\bar{t}$ semi-lep        & $498.06$ \\  
& $\tau\tau b\bar{b}$        & $379.98$ \\ 
& $t\bar{t}h$                & $60.52$ \\  
& $t\bar{t}Z$                & $57.06$\\  
& $t\bar{t}W$                & $6.57$ \\ 
& $pp\to hh$                 & $12.87$ \\
& $b\bar{b}h$                & $0.29$ \\
& $Zh$                       & $24.50$ \\  
& $b\bar{b}jj$               & $1639.38$ \\ \cline{2-3}  
& \multicolumn{1}{c|}{Total} & $2776.61$ \\ \hline
\end{tabular}}
\quad
\scalebox{0.7}{%
\begin{tabular}{|c|c|c|}\hline
(d) & Process & Events \\ \hline \hline
\multirow{10}{*}{Background}
& $t\bar{t}$ had             & $48.69$ \\ 
& $t\bar{t}$ semi-lep        & $460.70$ \\  
& $\tau\tau b\bar{b}$        & $319.31$ \\ 
& $t\bar{t}h$                & $70.61$ \\  
& $t\bar{t}Z$                & $48.29$\\  
& $t\bar{t}W$                & $11.82$ \\ 
& $pp\to hh$                 & $9.92$ \\
& $b\bar{b}h$                & $0.24$ \\
& $Zh$                       & $19.34$ \\  
& $b\bar{b}jj$               & $2068.14$ \\ \cline{2-3}  
& \multicolumn{1}{c|}{Total} & $3057.06$ \\ \hline
\end{tabular}}
\caption{Background yields after the BDT analysis for heavy Higgs mass of $(a)~400$ GeV, $(b)~600$ GeV, $(c)~800$ GeV and $(d)~1000$ GeV for the $b\bar{b}\tau_h\tau_h$ channel.}
\label{Htab:bbthth}
\end{table}
\end{center}

\subsection{The $b\bar{b} WW^*$ Channel}
\label{sec:Htohhtobbww}

In this section, we consider the situation where a heavy scalar decays to a pair of SM-like Higgs bosons with one of them decaying to a pair of $b$-quarks and the other to $WW^*$, leading to three possible final states depending on the decays of the $W$-bosons. We perform our analyses for the fully leptonic (leptons at this stage include $e, \mu, \tau$) and the semi-leptonic channels. We avoid studying the fully hadronic mode as the signal will be overwhelmed by the huge QCD background. 

\begin{figure}[htb!]
\centering
\includegraphics[trim=0 420 0 75,clip,width=\textwidth]{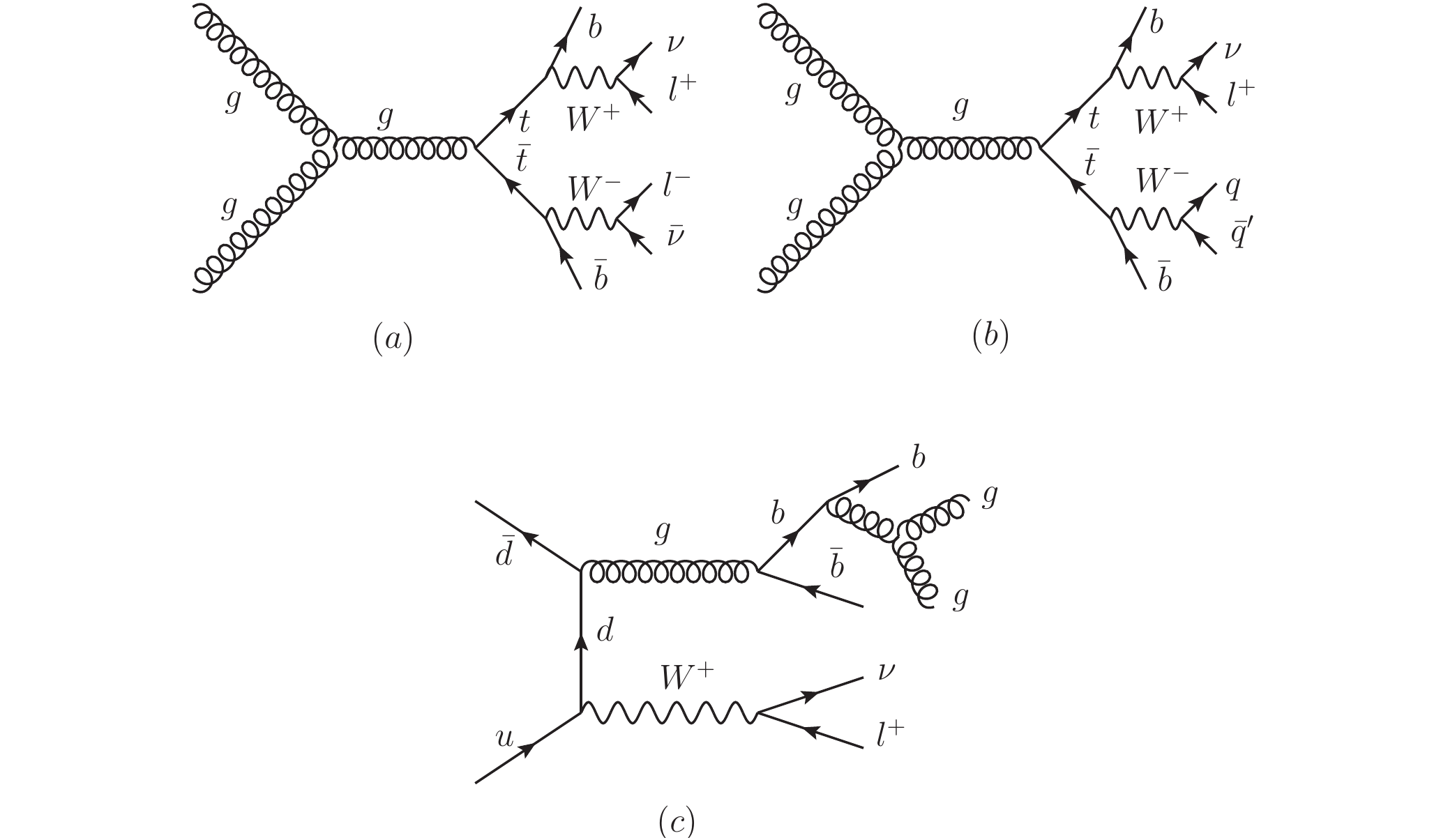}
\caption{Feynman diagrams for (a) leptonic $t\bar{t}$, (b) semi-leptonic $t\bar{t}$ and  (c) $Wb\bar{b}~+$ jets background for the $b\bar{b}WW^*$ channel. Here, $l$ refers to $e,~\mu$ and $\tau$ lepton.}
\label{FD:bbww}
\end{figure}

The dominant contribution to the background (Fig.~\ref{FD:bbww}) for both the channels mentioned above comes from top pair production because of its large production cross-section. We generate this background where either or both the $W$-bosons decay leptonically. The fully hadronic $t\bar{t}$ mode is not considered as a potential background as the fake rate for $j \to \ell$ is negligible for all practical purposes. The fully leptonic $t\bar{t}$ background contributes to the fully leptonic channel final state whereas for the semi-leptonic scenario, the contribution comes from both the fully-leptonic as well as the semi-leptonic $t\bar{t}$. The second most dominant background for the semi-leptonic channel is $Wb\bar{b}~+$ jets, where the $W$-boson decays leptonically ($e, \mu, \tau$). We generate this background upon merging with two additional jets by exploiting the MLM merging scheme~\cite{Mangano:2006rw}. While generating the $Wb\bar{b}~+$ jets background we ensure that there is no double counting ensuing from the semi-leptonic $t\bar{t}$ background. Besides the aforementioned backgrounds, a significant contribution also comes from the $\ell^+\ell^- b\bar{b}$ production where $\ell$ refers to $e,\mu$ and $\tau$. Finally, we also consider the subdominant backgrounds \textit{viz.}, $t\bar{t}h$, $t\bar{t}Z$, $t\bar{t}W$ and the non-resonant $gg\to hh$. 

In this subsection and the following section (section~\ref{sec:Htott}), the top-pair production is the dominant background. Thus, the reconstruction of the top quarks is a very powerful tool in order to reduce the contribution from this background. For the semi-leptonic case, the only source of missing transverse energy, $\met$~\footnote{To incorporate the $\slashed{E}_T$ smearing, we use the standard module of the Delphes ATLAS card.}, arises from the neutrino of the leptonically decaying $W$-boson from the top decay. We reconstruct the top from its decay products~\footnote{First the $W$-boson mass is reconstructed in order to attain the $p_Z$ component of the neutrino.}. The quadratic equation gives two possible solutions for the neutrino $p_z$. Besides, because there are two $b$-jets in the final state, we get four possible choices for the mass of the leptonically decaying top. We use these variables during our analysis. After reconstructing both the tops, we reconstruct the total system from all the final state particles. We also use this later in section~\ref{sec:Htott} which exhibits the same final state. These variables help us greatly in reducing the semi-leptonic $t\bar{t}$ background for high values of $m_H$. 

\begin{figure}
\centering
\includegraphics[scale=0.37]{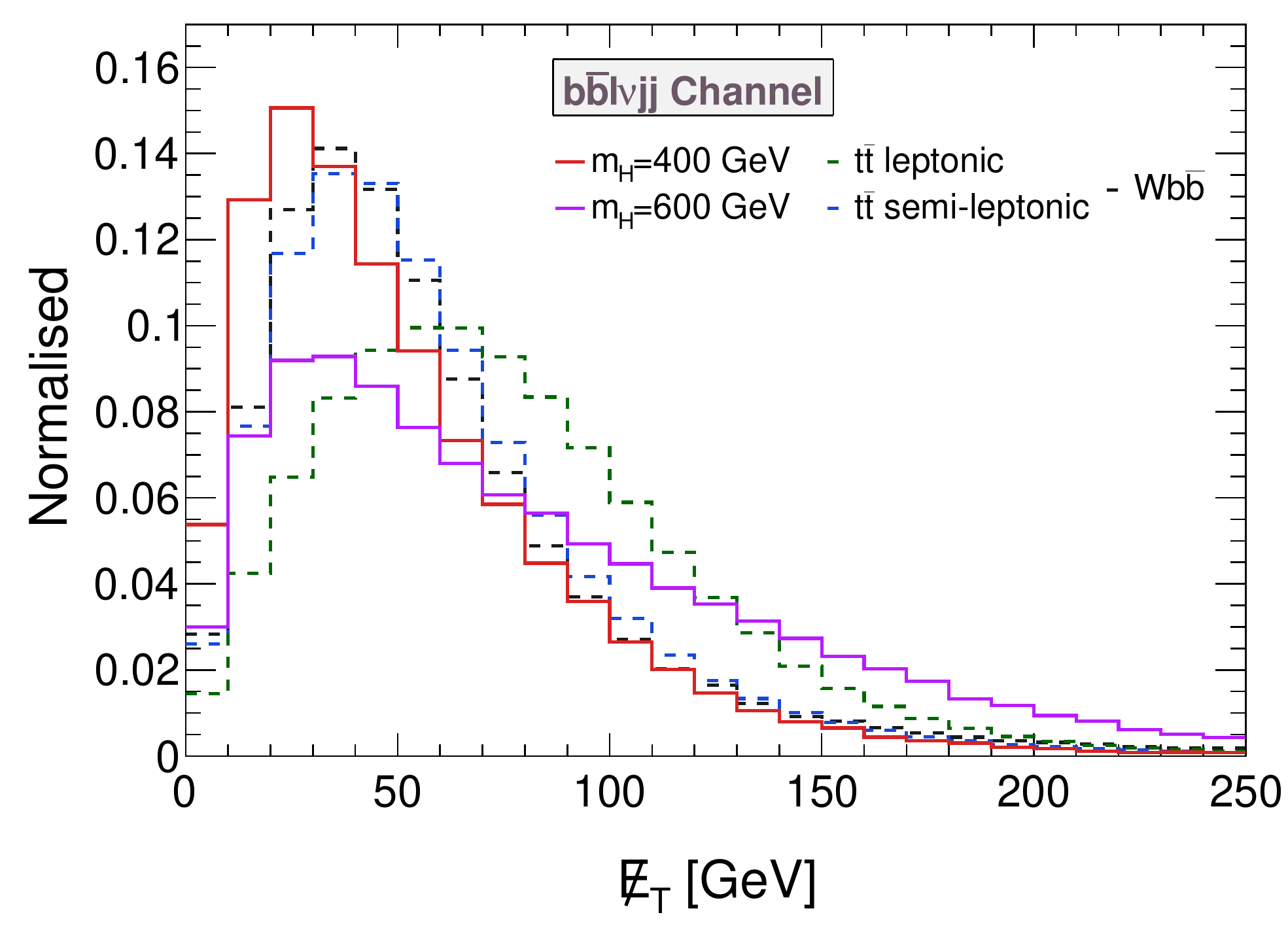}
\includegraphics[scale=0.37]{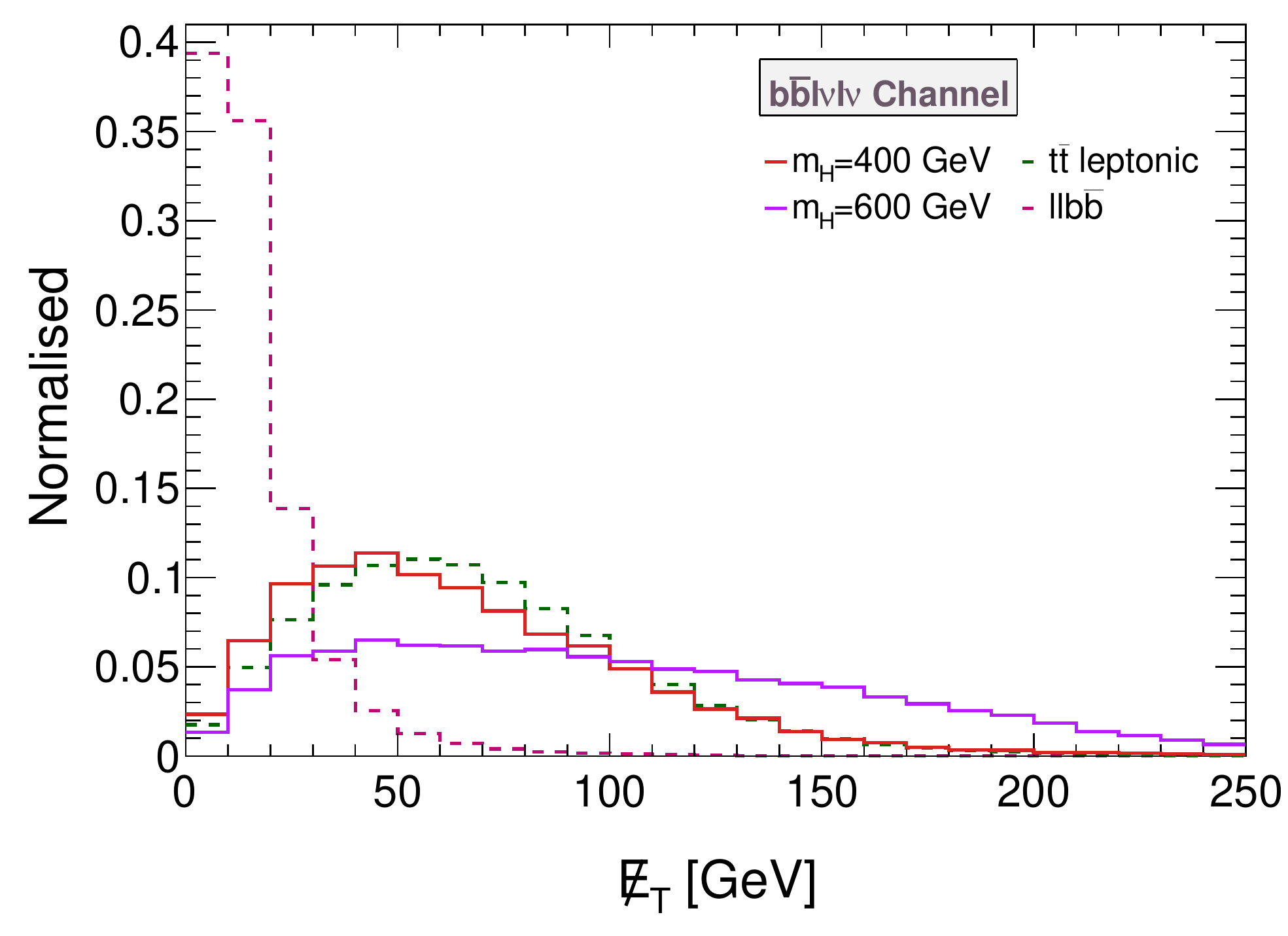}
\caption{The $\met$ distributions for the 1$\ell$ and 2$\ell$ categories for $m_H = 400$ and $600$ GeV with dominant backgrounds. Here the heavy Higgs boson is searched for in the $bbWW^*$ channel. The distributions are shown after imposing the basic trigger cuts.}
\label{Hfig:bbWW1}
\end{figure}

Before embarking on the final analysis, we impose a common set of trigger cuts for both the leptonic and the semi-leptonic channels. The $p_T$, $|\eta|$ and $\Delta R$ cuts for the various objects are discussed in subsection~\ref{sec:Htohhtobbtautau} and also in Appendix~\ref{sec:appendixA}.  Furthermore, we require generation-level cuts on the invariant mass of b-jets, \textit{viz.}, $m_{bb}>50$ GeV. The selected events are also require to have $\met>40$ GeV upon scrutinising the distribution. The $\met$ distribution for the 1$\ell$ and 2$\ell$ cases are shown in Fig.~\ref{Hfig:bbWW1}. Finally, we perform separate multivariate analyses for the two final states upon using the BDTD algorithm. While training samples for both the leptonic and semi-leptonic analyses, we only consider the $t\bar{t}$ background since it constitutes the bulk of the total background. This training is used for testing all other backgrounds which are subdominant in front of $t\bar{t}$.

\subsubsection{The $2\ell 2b+\met$ Channel}
\label{sec:Htohhtobbww:2l}

\begin{figure}
\centering
\includegraphics[scale=0.37]{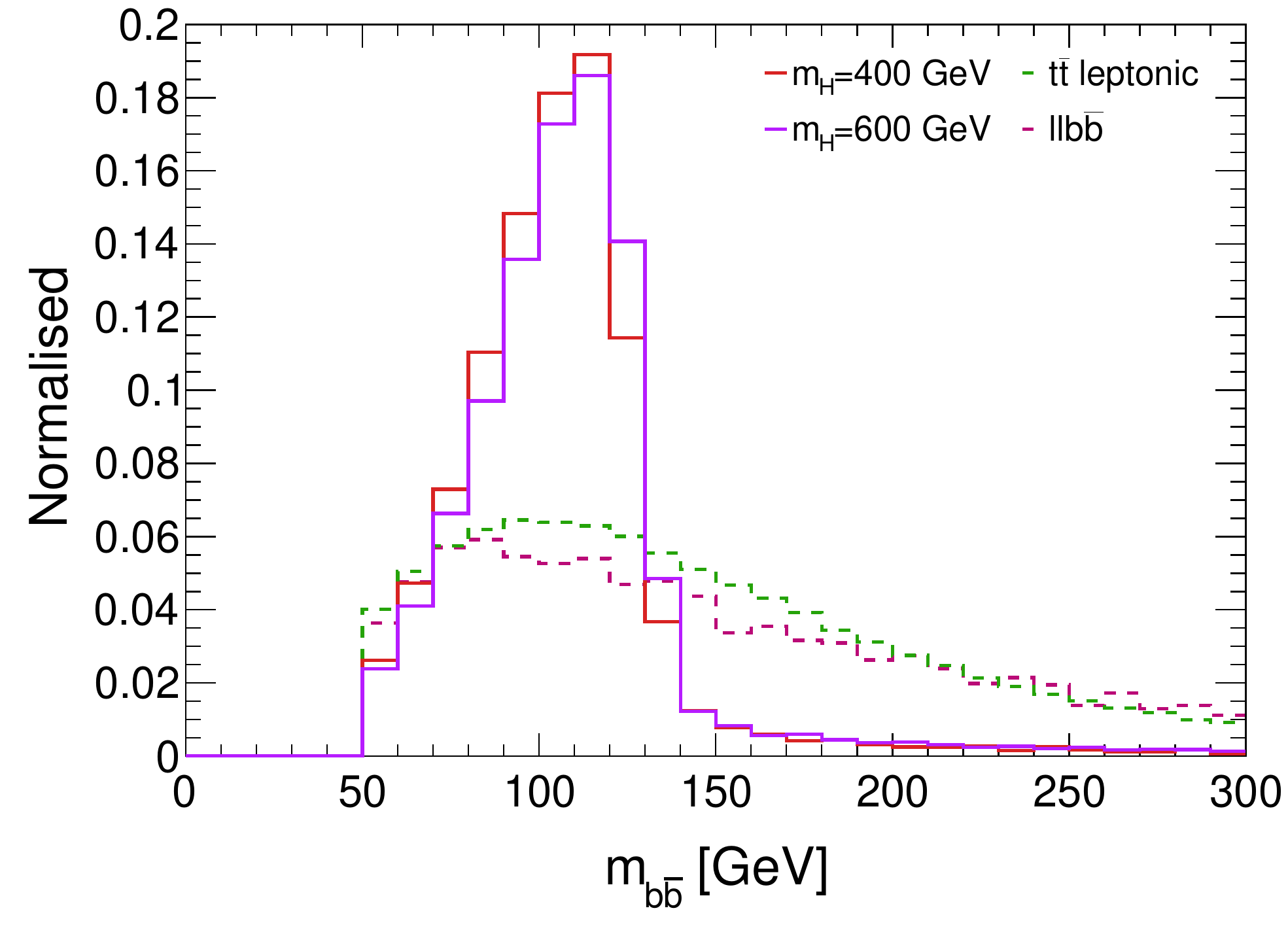}
\includegraphics[scale=0.37]{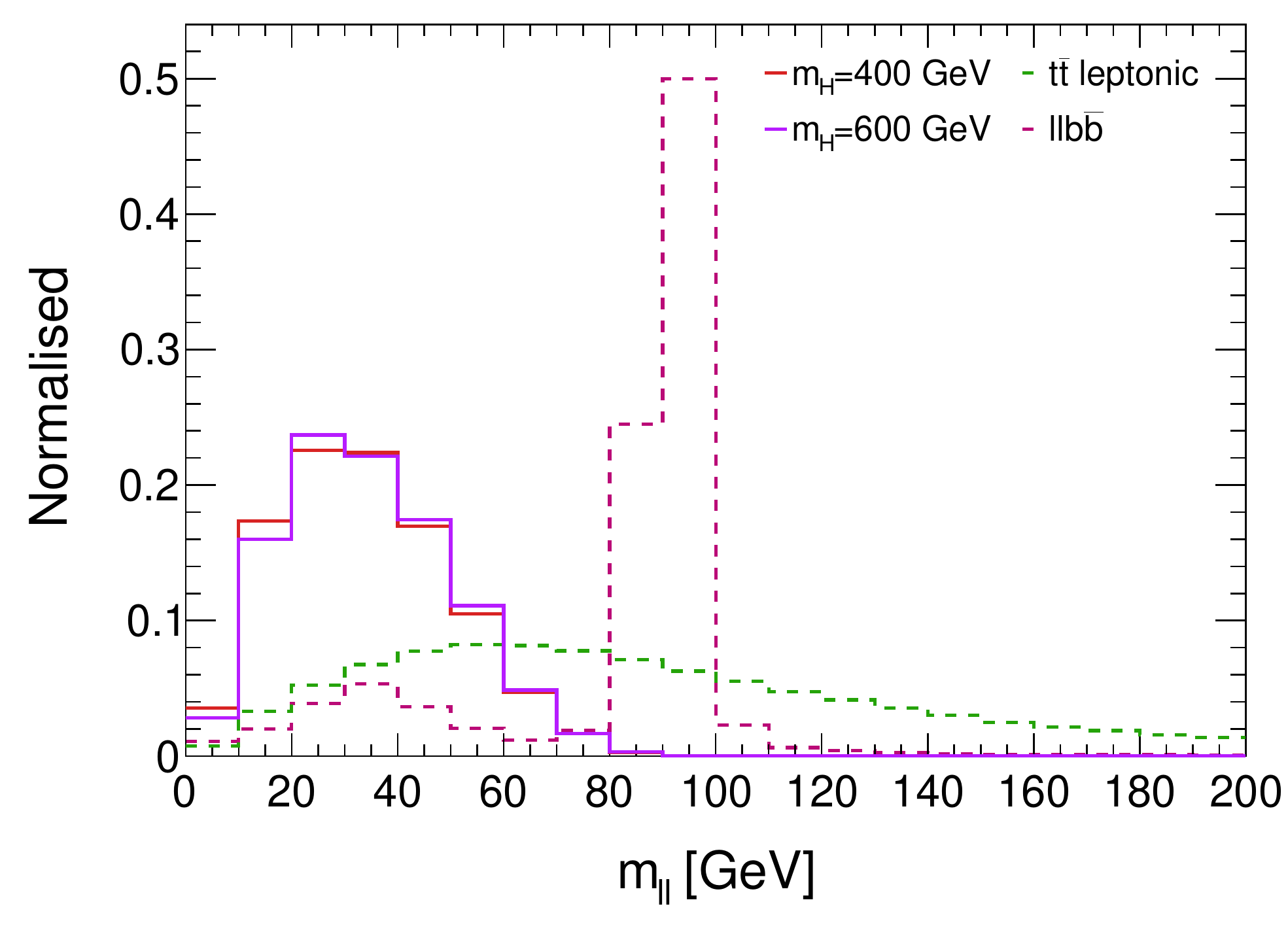} \\
\includegraphics[scale=0.37]{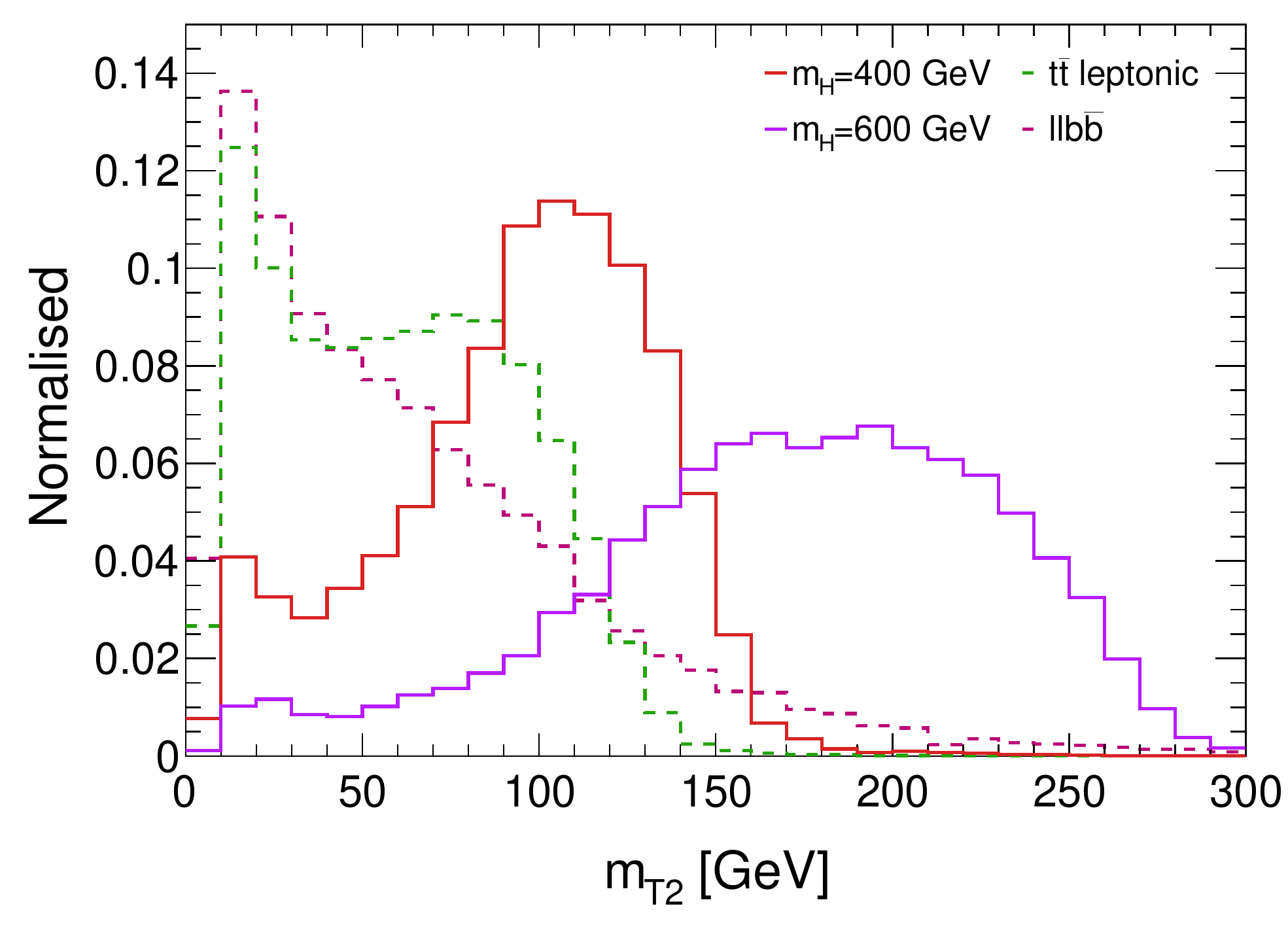}
\includegraphics[scale=0.37]{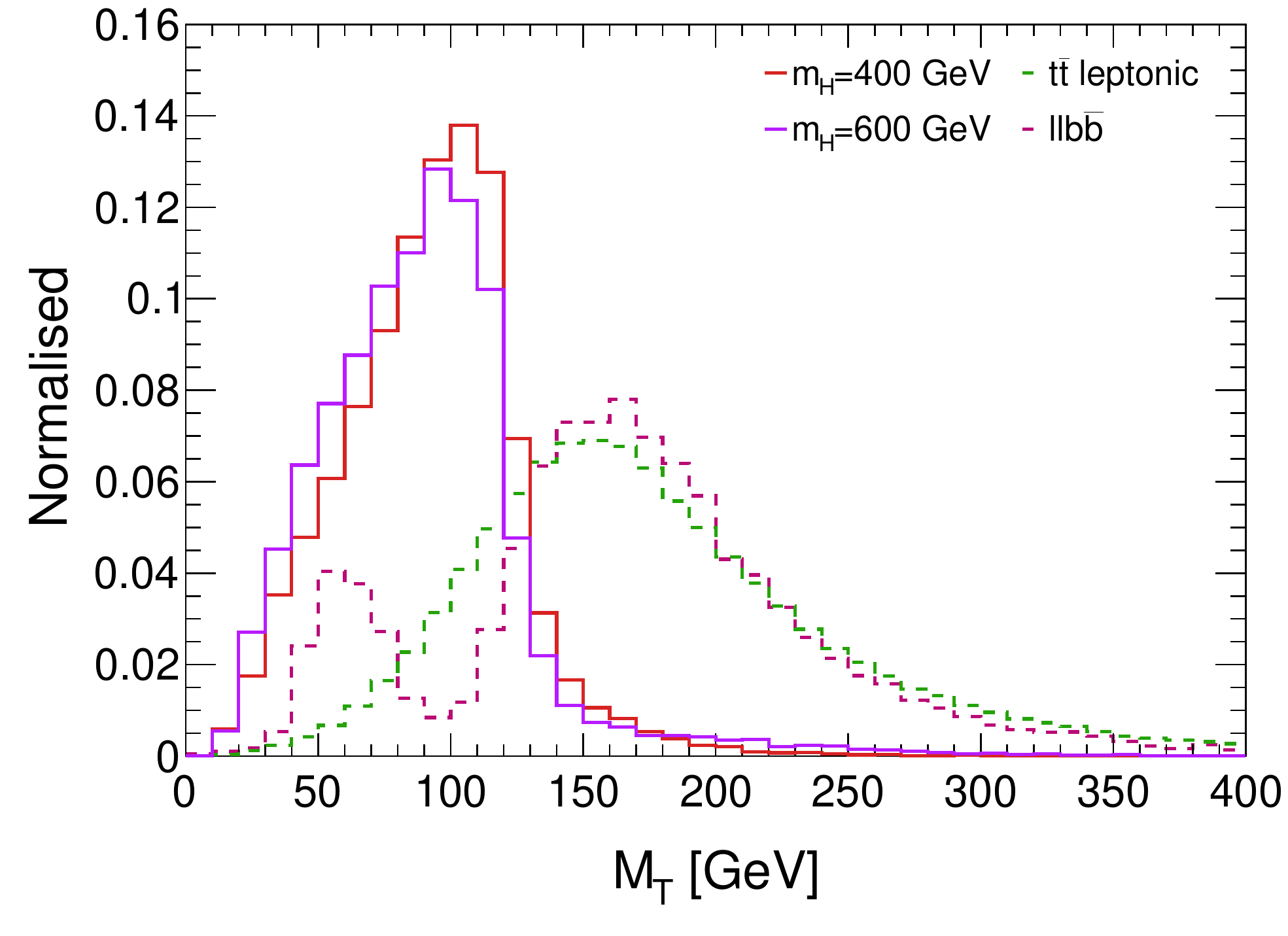}
\caption{The $m_{bb}$, $m_{\ell\ell}$, $m_{T2}$ and $M_T$ distributions for the 2$\ell$ category for $m_H = 400$ and $600$ GeV with dominant backgrounds. Here the heavy Higgs boson is searched for in the $b\bar{b}WW^*$ channel. The distributions are shown before the multivariate analysis.}
\label{fig1:2l2bMET}
\end{figure}

For the fully leptonic final state, we select events with exactly two $b$-tagged jets, and two isolated leptons having opposite charge meeting the trigger criteria as mentioned above. We choose the following set of kinematic variables in order to perform the multivariate analysis:

\begin{equation}
\begin{split}
p_{T, bb},~\eta_{bb},~\phi_{bb},~m_{bb},~\Delta R_{bb},~\Delta\phi_{bb},~p_{T, \ell\ell},~\eta_{\ell\ell},~\phi_{\ell\ell},~m_{\ell\ell},~\Delta R_{\ell\ell},\\~M_T,~m_{T2}, ~m_{\textrm{tot}},~p_{T, \textrm{tot}},~\phi_{\textrm{tot}},~m_{\textrm{eff}},~\Delta R_{b_1\ell_2},~\Delta R_{bb,\ell\ell},~p_{T, \ell_2} \nonumber,
\end{split}
\end{equation}

\noindent where, $M_T$ is the transverse mass of the SM-like Higgs decaying to $W$-bosons. The rest of the variables have either been defined before or have usual meaning. The top four variables are shown in Fig.~\ref{fig1:2l2bMET}. The signal distributions are significantly different from the various backgrounds.

\begin{figure}
\centering
\includegraphics[scale=0.5]{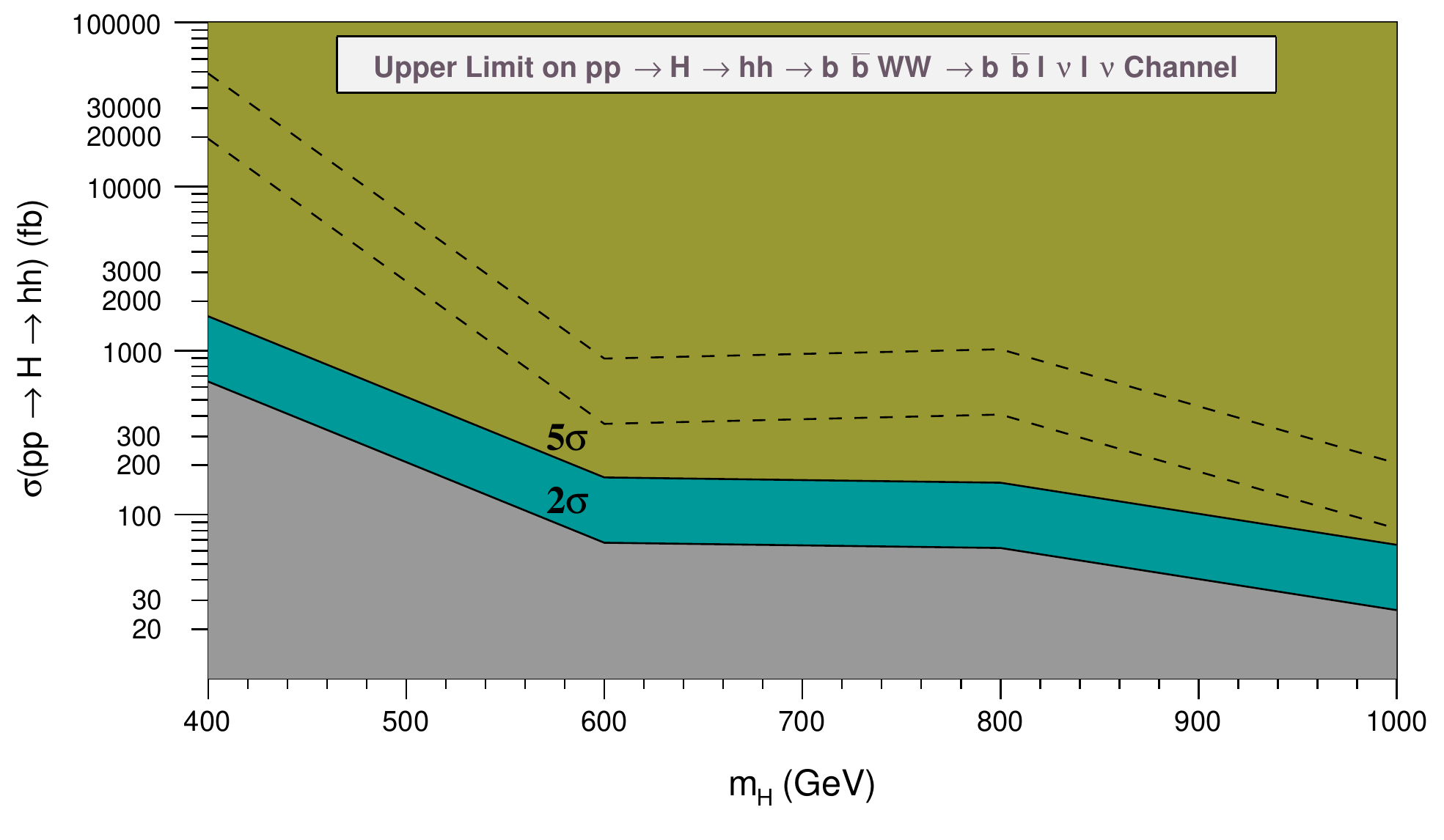}
\caption{Upper limit on $\sigma(pp\to H\to hh)$ (fb) as a function of $m_{H}$ (GeV) for the $2\ell + 2b + \met$ channel. The solid (dashed) lines show the 2$\sigma$-5$\sigma$ band on taking 0\% (5\%) systematic uncertainties.}
\label{fig2:2l2bMET}
\end{figure}

Finally, in Table~\ref{tab1:2l2bMET}, we summarise the number of background events after imposing the optimised cut on the BDT variable. Like in the other channels, we impose an upper limit on $\sigma(pp\to H\to hh)$ as a function of the heavy Higgs mass. This is shown in Fig.~\ref{fig2:2l2bMET}. The $95\%$ CL upper limit varies between $67.41$ fb and $26.18$ fb ($357.51$ fb and $82.21$ fb with $5\%$ systematic uncertainty) within 600 GeV $< m_H <$ 1 TeV and is somewhat weaker than the channels discussed earlier owing to smaller $S/B$.

\begin{center}
\begin{table}[htb!]
\centering
\scalebox{0.7}{%
\begin{tabular}{|c|c|c|c|}\hline
(a) & Process & Order & Events \\ \hline \hline

\multirow{8}{*}{Background}
 & $t\bar{t}$ lep           	    & NNLO~\cite{ttbarNNLO}                       & $356309.30$ \\  
 & $t\bar{t}h$              		& NLO~\cite{bkg_twiki_cs}                     & $1310.44$ \\  
 & $t\bar{t}Z$             		    & NLO~\cite{Lazopoulos:2008de}                & $1264.20$\\  
 & $t\bar{t}W$       		        & NLO~\cite{Campbell:2012dh}                  & $627.97$ \\ 
 & $pp\to hh$           		    & NNLO~\cite{hhtwiki}                         & $90.72$ \\
 & $\ell\ell b\bar{b}$      		& LO                                          & $5013.31$ \\ \cline{2-4}  
 & \multicolumn{2}{c|}{Total}	                                                  & $364615.94$ \\ \hline

\end{tabular}}
\bigskip
\scalebox{0.7}{%
\begin{tabular}{|c|c|c|}\hline
(b) & Process & Events \\ \hline \hline
\multirow{8}{*}{Background}
 & $t\bar{t}$ lep                     & $7056.45$ \\  
 & $t\bar{t}h$                        & $322.80$ \\  
 & $t\bar{t}Z$                        & $640.88$\\  
 & $t\bar{t}W$                        & $114.29$ \\ 
 & $pp\to hh$                         & $37.75$ \\  
 & $\ell\ell b\bar{b}$                & $2678.57$ \\ \cline{2-3}  
 & \multicolumn{1}{c|}{Total}         & $10850.74$ \\ \hline
\end{tabular}}
\quad
\scalebox{0.7}{%
\begin{tabular}{|c|c|c|}\hline
(c) & Process & Events \\ \hline \hline
\multirow{8}{*}{Background}
 & $t\bar{t}$ lep                     & $11954.46$ \\  
 & $t\bar{t}h$                        & $328.30$ \\  
 & $t\bar{t}Z$                        & $812.07$\\  
 & $t\bar{t}W$                        & $185.24$ \\ 
 & $pp\to hh$                         & $20.38$ \\  
 & $\ell\ell b\bar{b}$                & $3233.14$ \\ \cline{2-3}  
 & \multicolumn{1}{c|}{Total}         & $16533.59$ \\ \hline
\end{tabular}}
\quad
\scalebox{0.7}{%
\begin{tabular}{|c|c|c|}\hline
(d) & Process & Events \\ \hline \hline
\multirow{8}{*}{Background}
 & $t\bar{t}$ lep                     & $1286.76$ \\  
 & $t\bar{t}h$                        & $135.72$ \\  
 & $t\bar{t}Z$                        & $386.28$\\  
 & $t\bar{t}W$                        & $49.27$ \\ 
 & $pp\to hh$                         & $10.68$ \\  
 & $\ell\ell b\bar{b}$                & $1674.80$ \\ \cline{2-3}  
 & \multicolumn{1}{c|}{Total}         & $3543.51$ \\ \hline
\end{tabular}}
\caption{Respective background yields for the $2\ell + 2b + \met$ channel after the BDT analyses optimised for $m_H = (a)~400$ GeV, $(b)~600$ GeV, $(c)~800$ GeV and $(d)~1$ TeV. The tables also list the perturbative order at which the cross-sections are considered.}
\label{tab1:2l2bMET}
\end{table}
\end{center}

\subsubsection{The $1\ell 2b 2j+\met$ Channel}
\label{sec:Htohhtobbww:1l}

\begin{figure}
\centering
\includegraphics[scale=0.37]{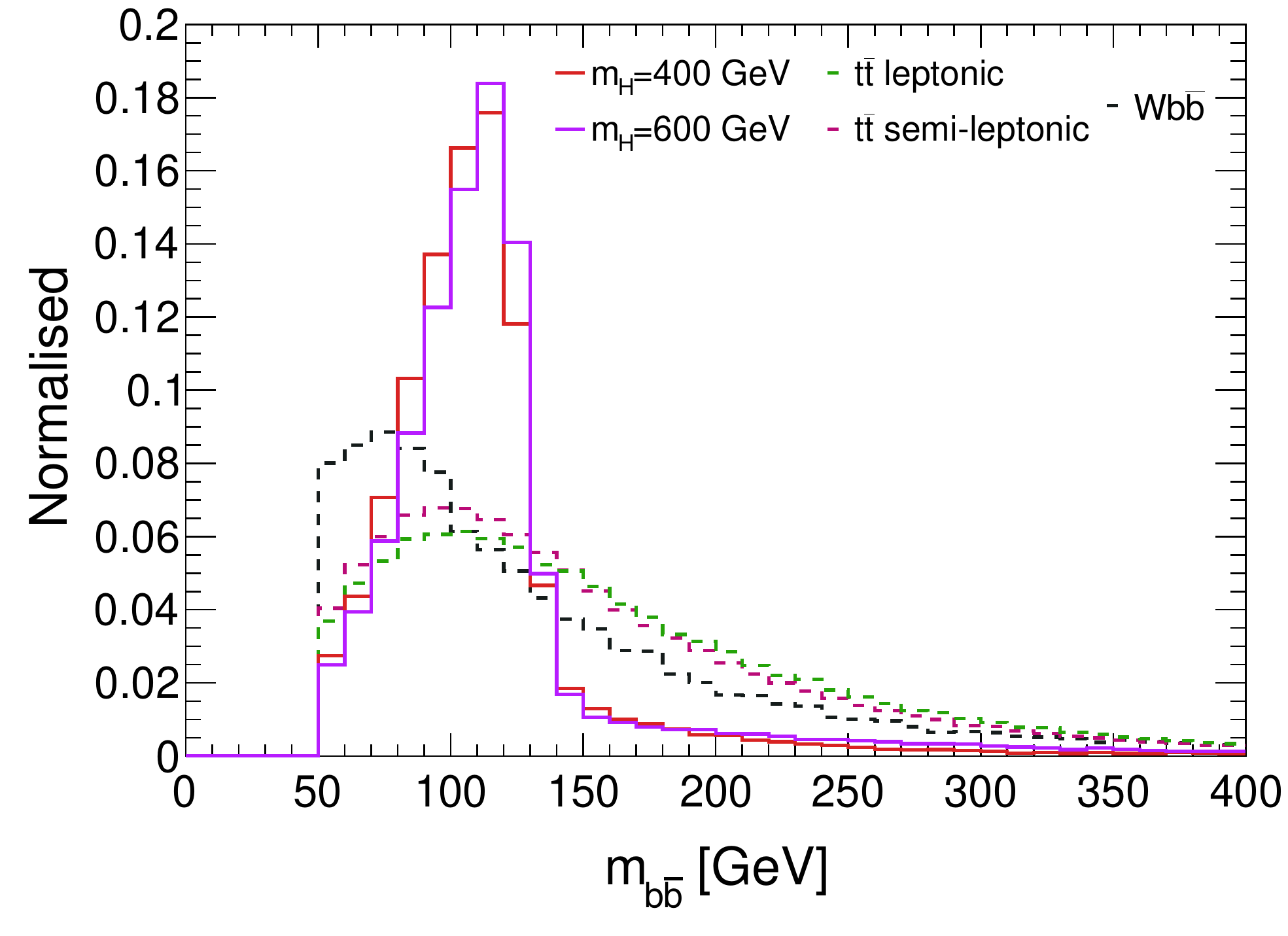}
\includegraphics[scale=0.37]{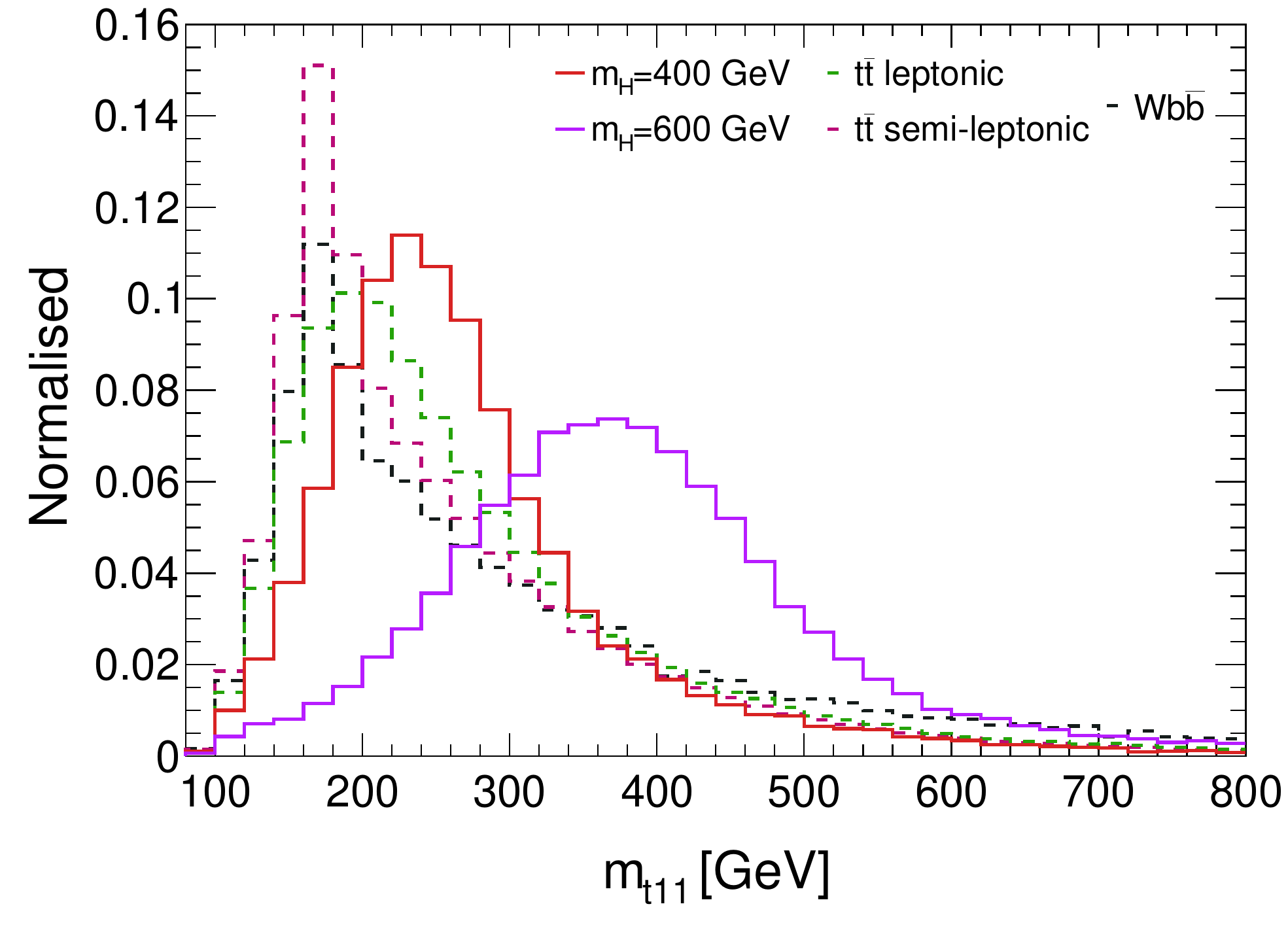} \\
\includegraphics[scale=0.37]{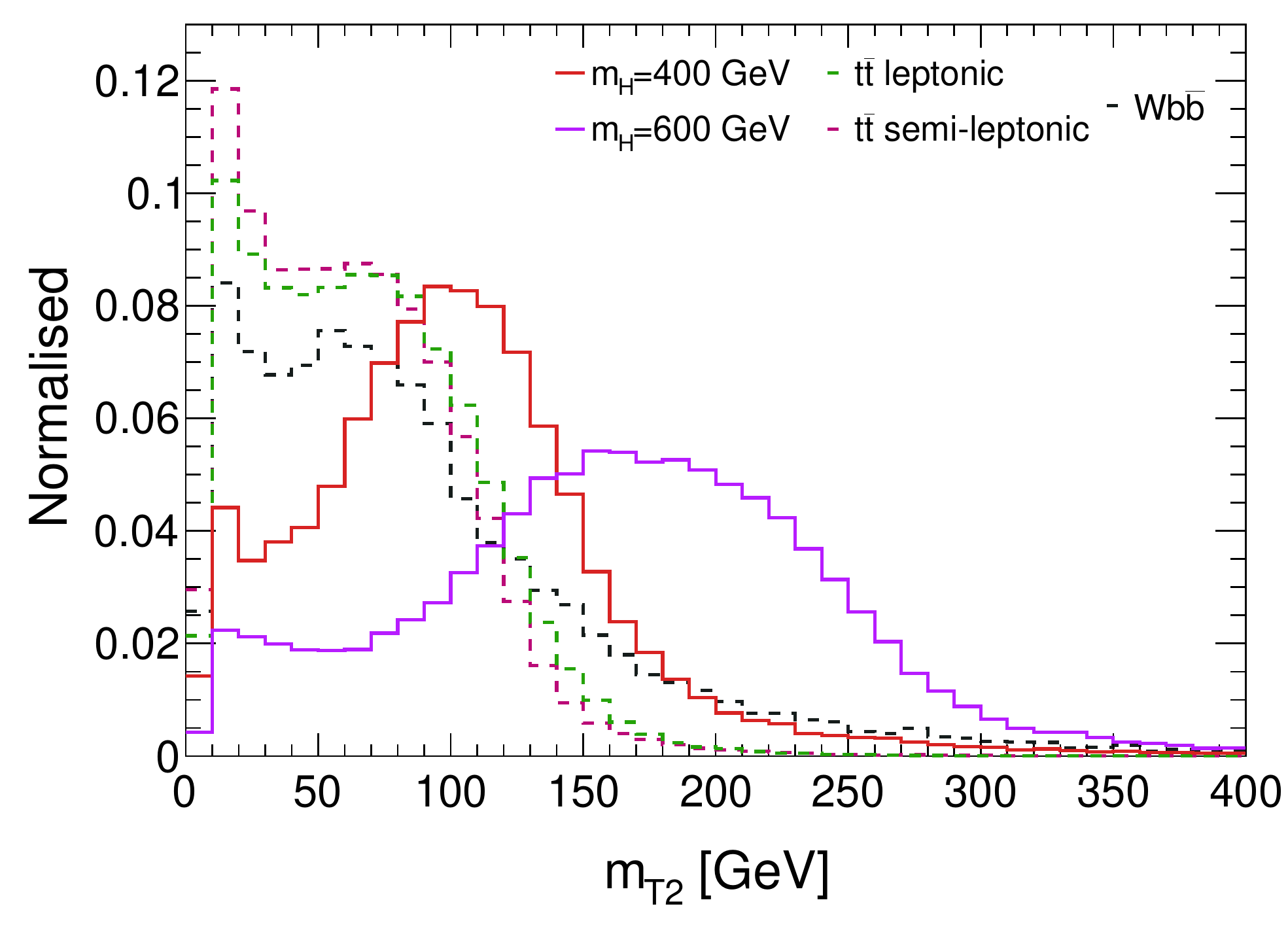}
\includegraphics[scale=0.37]{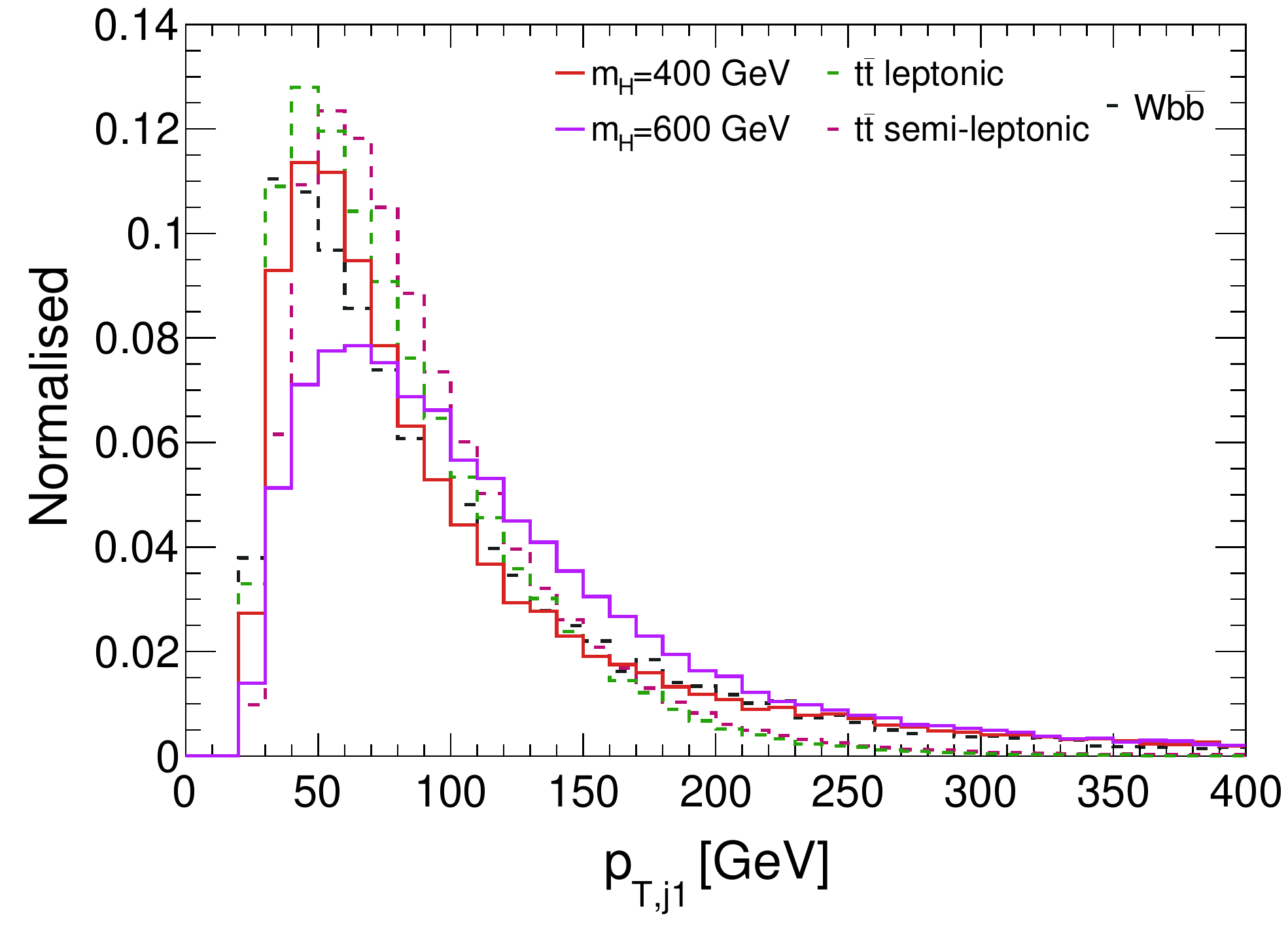}
\caption{The $m_{bb}$, $m_{t11}$, $m_{T2}$ and $p_{T,j_1}$ distributions for the 1$\ell$ category for $m_H = 400$ and $600$ GeV with dominant backgrounds. Here the heavy Higgs boson is searched for in the $b\bar{b}WW^*$ channel. The distributions are shown before the multivariate analysis.}
\label{fig1:1l2jMET}
\end{figure}

Finally, we discuss the potential of the semi-leptonic final state as well. We require events with exactly two $b$-tagged jets, one isolated lepton and at least two light jets satisfying the trigger criteria discussed earlier. Besides, we consider the same set of cuts as for the dileptonic channel before performing the multivariate analysis. We find the following kinematic variables to have the best discriminatory power and use them for our multivariate analysis:

\begin{equation}
\begin{split}
\eta_{bb},~m_{bb},~m_t,~m_{jj},~\Delta R_{jj},~\Delta R_{\ell, jj},~M_T,~m_{T2},\\~m_{bbj_1}, 
~m_{t11},~m_{t12},~p_{T, \ell\nu},~p_{T, b_1},~p_{T, \ell_1},~p_{T, j_1}\nonumber,
\end{split}
\end{equation}

\noindent where, $m_t$ is the transverse mass of the leptonically decaying $W$-boson. $\Delta R_{\ell, jj}$ is the distance in the $\eta-\phi$ plane between the system comprising of the two hardest jets and the lepton. $m_{bbj_{1}}$ refers to the invariant mass of the two $b$-tagged jets and the hardest $p_T$ jet. The reconstructed transverse momentum of the leptonically decaying $W$-boson is denoted as $p_{T, \ell\nu}$. $m_{tij}$ is the mass of the leptonically decaying top quark with the reconstruction procedure outlined before. The first index $i=1,2$ indicates the $p_T$ ordering of the $b$-jet. The second index $j=1,2$ refers to the choice of the $z$-component of the neutrino momentum. The other variables have usual definitions. The best discriminatory variables are listed in Fig.~\ref{fig1:1l2jMET}. However, we can see that the separation power for the 1$\ell$ category is significantly less compared to its 2$\ell$ counterpart.

\begin{figure}
\centering
\includegraphics[scale=0.5]{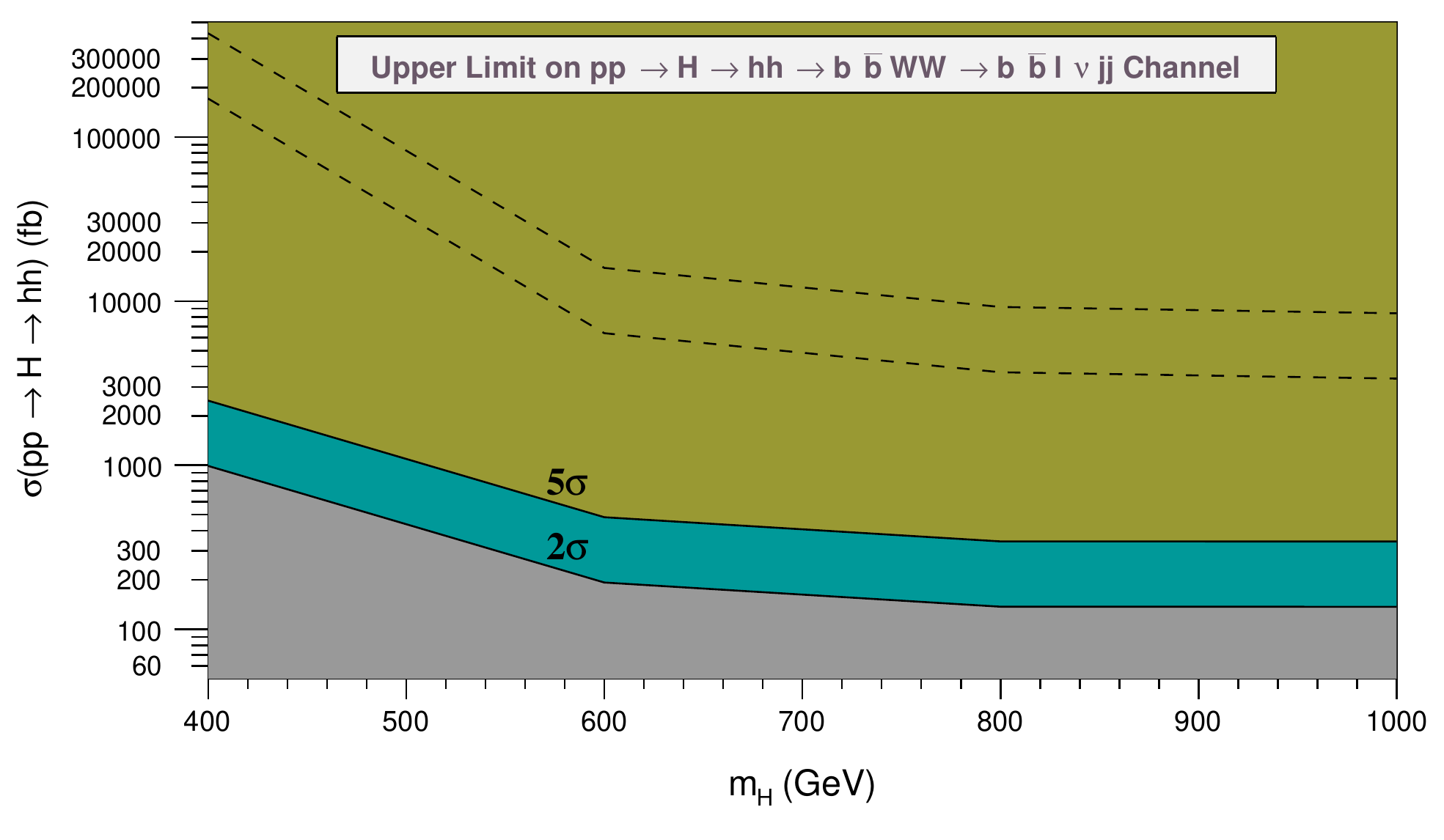}
\caption{Upper limit on $\sigma(pp\to H\to hh)$ (fb) as a function of $m_{H}$ (GeV) for the $1\ell +2j+ 2b + \met$ channel. The solid (dashed) lines show the 2$\sigma$-5$\sigma$ band on taking 0\% (5\%) systematic uncertainties.}
\label{fig2:1l2jMET}
\end{figure}

Coming to the results, table~\ref{tab1:1l2jMET} summarises the background yields after the BDT cut. The upper limit on $\sigma(pp\to H\to hh)$ as a function of $m_H$ are shown in Fig.~\ref{fig2:1l2jMET}. The limits are considerably weak in this channel.

\begin{center}
\begin{table}[htb!]
\centering
\scalebox{0.7}{%
\begin{tabular}{|c|c|c|}\hline
(a) & Process & Events \\ \hline \hline

\multirow{9}{*}{Background}
 & $t\bar{t}$ semi-lep           	 & $9740640.28$ \\ 
 & $t\bar{t}$ lep                	 & $1614225.57$ \\  
 & $Wb\bar{b}+\textrm{jets}$ [LO]	 & $569181.53$ \\   
 & $t\bar{t}h$                  	 & $28364.75$ \\  
 & $t\bar{t}Z$                  	 & $24846.06$\\  
 & $t\bar{t}W$                  	 & $16935.36$ \\ 
 & $pp\to hh$                   	 & $318.61$ \\ 
 & $\ell\ell b\bar{b}$          	 & $20252.87$ \\ \cline{2-3}  
 & \multicolumn{1}{c|}{Total}   	 & $12014765.03$ \\ \hline
\end{tabular}}
\bigskip
\scalebox{0.7}{%
\begin{tabular}{|c|c|c|}\hline
(b) & Process & Events \\ \hline \hline
\multirow{9}{*}{Background}
 & $t\bar{t}$ semi-lep            & $280140.51$ \\ 
 & $t\bar{t}$ lep                 & $40221.77$ \\  
 & $Wb\bar{b}+\textrm{jets}$      & $106228.20$ \\  
 & $t\bar{t}h$                    & $3804.77$ \\  
 & $t\bar{t}Z$                    & $2952.00$\\  
 & $t\bar{t}W$                    & $1958.78$ \\ 
 & $pp\to hh$                     & $87.93$ \\ 
 & $\ell\ell b\bar{b}$            & $1985.36$ \\ \cline{2-3}   
 & \multicolumn{1}{c|}{Total}     & $437379.32$ \\ \hline
\end{tabular}}
\quad
\scalebox{0.7}{%
\begin{tabular}{|c|c|c|}\hline
(c) & Process & Events \\ \hline \hline
\multirow{9}{*}{Background}
 & $t\bar{t}$ semi-lep            & $169281.37$ \\ 
 & $t\bar{t}$ lep                 & $19965.61$ \\
 & $Wb\bar{b}+\textrm{jets}$      & $91249.74$ \\   
 & $t\bar{t}h$                    & $2519.09$ \\  
 & $t\bar{t}Z$                    & $2261.73$\\  
 & $t\bar{t}W$                    & $1630.34$ \\ 
 & $pp\to hh$                     & $42.35$ \\ 
 & $\ell\ell b\bar{b}$            & $848.49$ \\ \cline{2-3}   
 & \multicolumn{1}{c|}{Total}     & $287798.72$ \\ \hline
\end{tabular}}
\quad
\scalebox{0.7}{%
\begin{tabular}{|c|c|c|}\hline
(d) & Process & Events \\ \hline \hline
\multirow{9}{*}{Background}
 & $t\bar{t}$ semi-lep            & $138443.80$ \\ 
 & $t\bar{t}$ lep                 & $19342.98$ \\  
 & $Wb\bar{b}+\textrm{jets}$      & $77935.55$ \\ 
 & $t\bar{t}h$                    & $1842.32$ \\  
 & $t\bar{t}Z$                    & $1919.34$\\  
 & $t\bar{t}W$                    & $1399.78$ \\ 
 & $pp\to hh$                     & $21.38$ \\ 
 & $\ell\ell b\bar{b}$            & $571.21$ \\ \cline{2-3}    
 & \multicolumn{1}{|c|}{Total}    & $241476.36$ \\ \hline
\end{tabular}}
\caption{Respective background yields for the $1\ell+ 2j+ 2b + \met$ channel after the BDT analyses optimised for $m_H = (a)~400$ GeV, $(b)~600$ GeV, $(c)~800$ GeV and $(d)~1$ TeV. The various orders of the signal and backgrounds are same as in Table~\ref{tab1:2l2bMET}.}
\label{tab1:1l2jMET}
\end{table}
\end{center}

\subsection{The $\gamma\gamma WW^*$ Channel}
\label{sec:Htohhtowwgammagamma}

\begin{figure}[htb!]
\centering
\includegraphics[trim=0 270 0 80,clip,width=\textwidth]{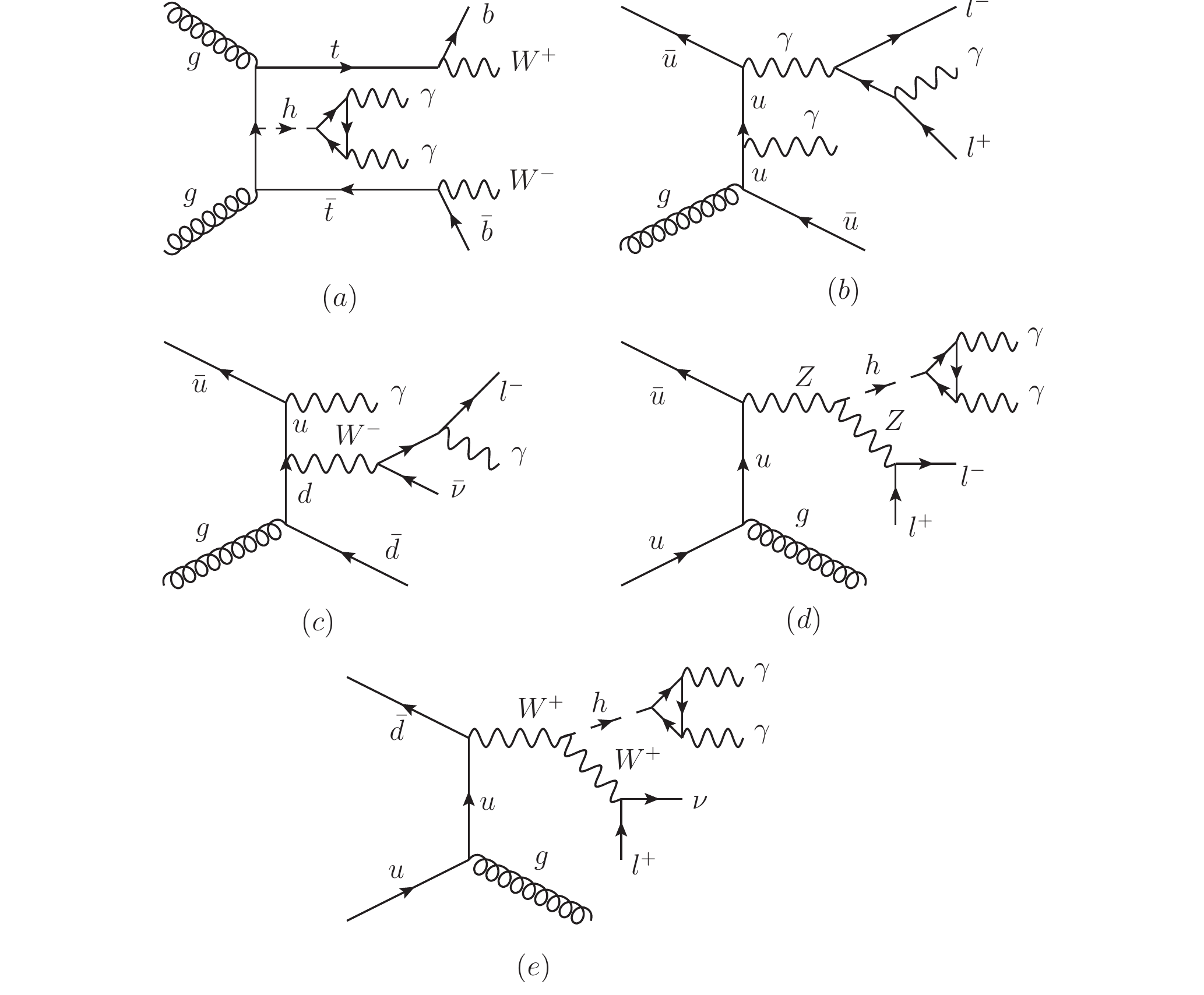}
\caption{Feynman diagrams for (a) $t\bar{t}h$, (b) $\ell \ell \gamma \gamma~+$ jets, (c) $\ell \nu \gamma \gamma~+$ jets, (d) $Zh~+$ jets and (e) $Wh~+$ jets background for the $\gamma\gamma WW^*$ channel.}
\label{FD:wwaa}
\end{figure}

After the $b\bar{b}\gamma\gamma$ channel this is the second most cleanest channel in terms of the final state particles but with the pitfall of having very low event rate. In this channel, one of the SM-like Higgs decays to a pair of photons and the other to lepton(s) through $h\to WW^*$. Similar to the $b\bar{b}WW^*$ analysis in subsection~\ref{sec:Htohhtobbww}, here also we divide the channel into the leptonic and semi-leptonic category. Because of the relatively clean final states, these channels have low contaminations due to backgrounds. We simulate the $Zh$ and $Wh$ backgrounds upon merging with two additional jets (the definition of jet is given in subsection~\ref{sec:Htohhtobbgaga}). Here we decay the $Z$- and the $W$-bosons leptonically ($e, \mu, \tau$). The $Wh$ channel contributes only to the semi-leptonic category. Besides, there are $\ell \nu \gamma \gamma$ and $\ell \ell \gamma \gamma$ productions coming from $\gamma^*$, diagrams containing triple and quartic gauge boson interactions and various other diagrams not involving the Higgs. We generate these two backgrounds upon merging with an additional jet and using the same scheme as before. Next, we also consider the $t\bar{t}h$ background with Higgs-boson decayed to a pair of photons. Finally, we also consider the SM Higgs pair production which is subdominant. We show the Feynman diagram of the dominant backgrounds in Fig~\ref{FD:wwaa}.

Before performing the multivariate analyses, we impose the generic trigger cuts. The $p_T$, $|\eta|$ and $\Delta R$~\footnote{$\Delta R_{\gamma \gamma/\gamma \ell} > 0.4$ and $\Delta R_{\ell\ell > 0.2}$.} cuts are the same as has been defined in subsection~\ref{sec:Htohhtobbtautau}. The above cuts for the photons are the same as those on the leptons. Owing to an excellent resolution for the diphoton invariant mass, we require $122~\textrm{GeV}<m_{\gamma\gamma}<128~\textrm{GeV}$. Finally, we also require $m_{\ell\ell}>20$ GeV because we generate the $\ell \ell \gamma \gamma$ background with this invariant mass cut at the generation level (the details of these cuts are mentioned in Appendix~\ref{sec:appendixA}). We now describe the results of the multivariate analyses for these two channels in the following two subsections. 

\subsubsection{The $\gamma\gamma 1\ell 2j+\met$ Channel}
\label{sec:Htohhtowwgammagamma:1l}

\begin{figure}
\centering
\includegraphics[scale=0.37]{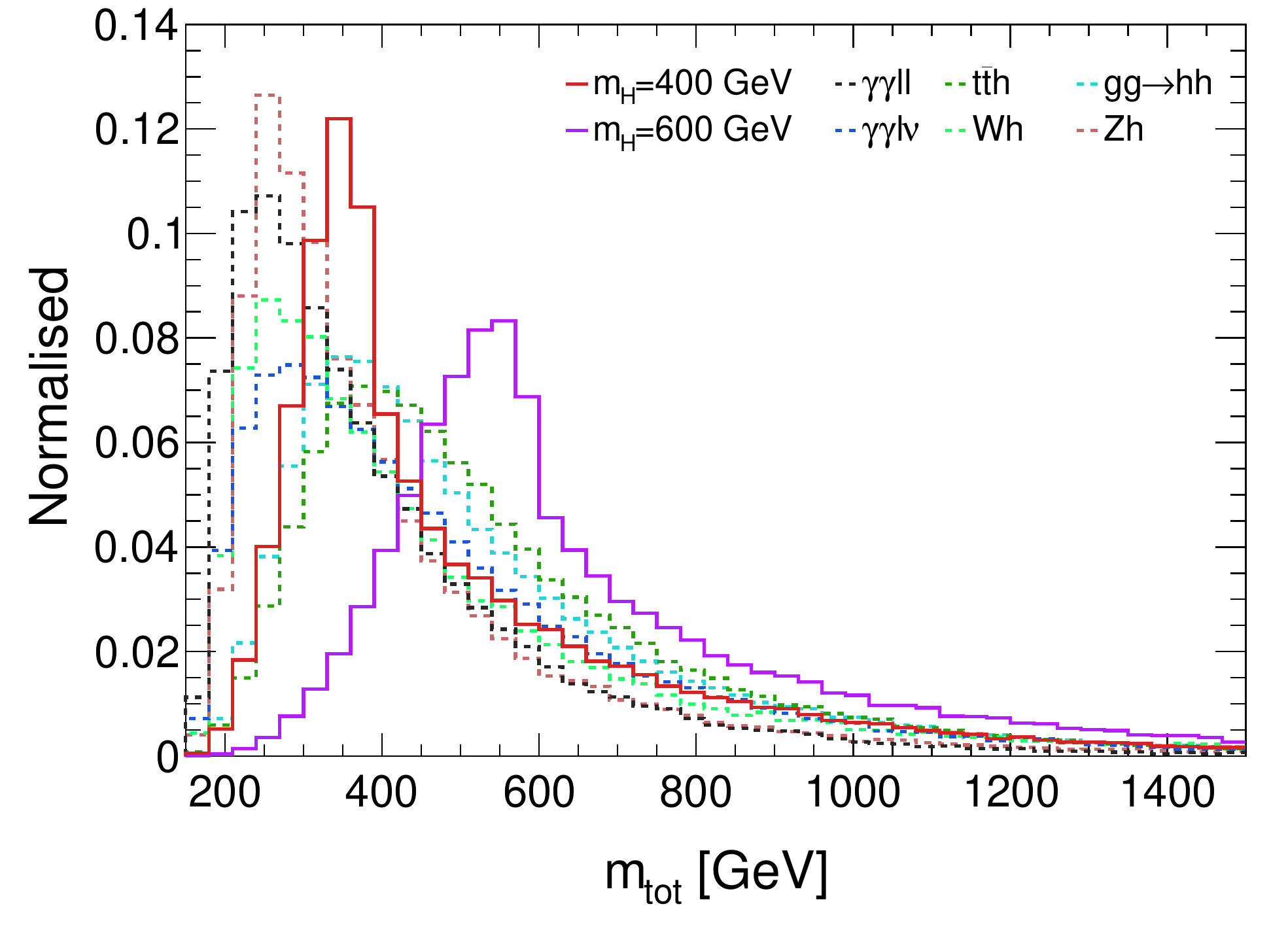}
\includegraphics[scale=0.37]{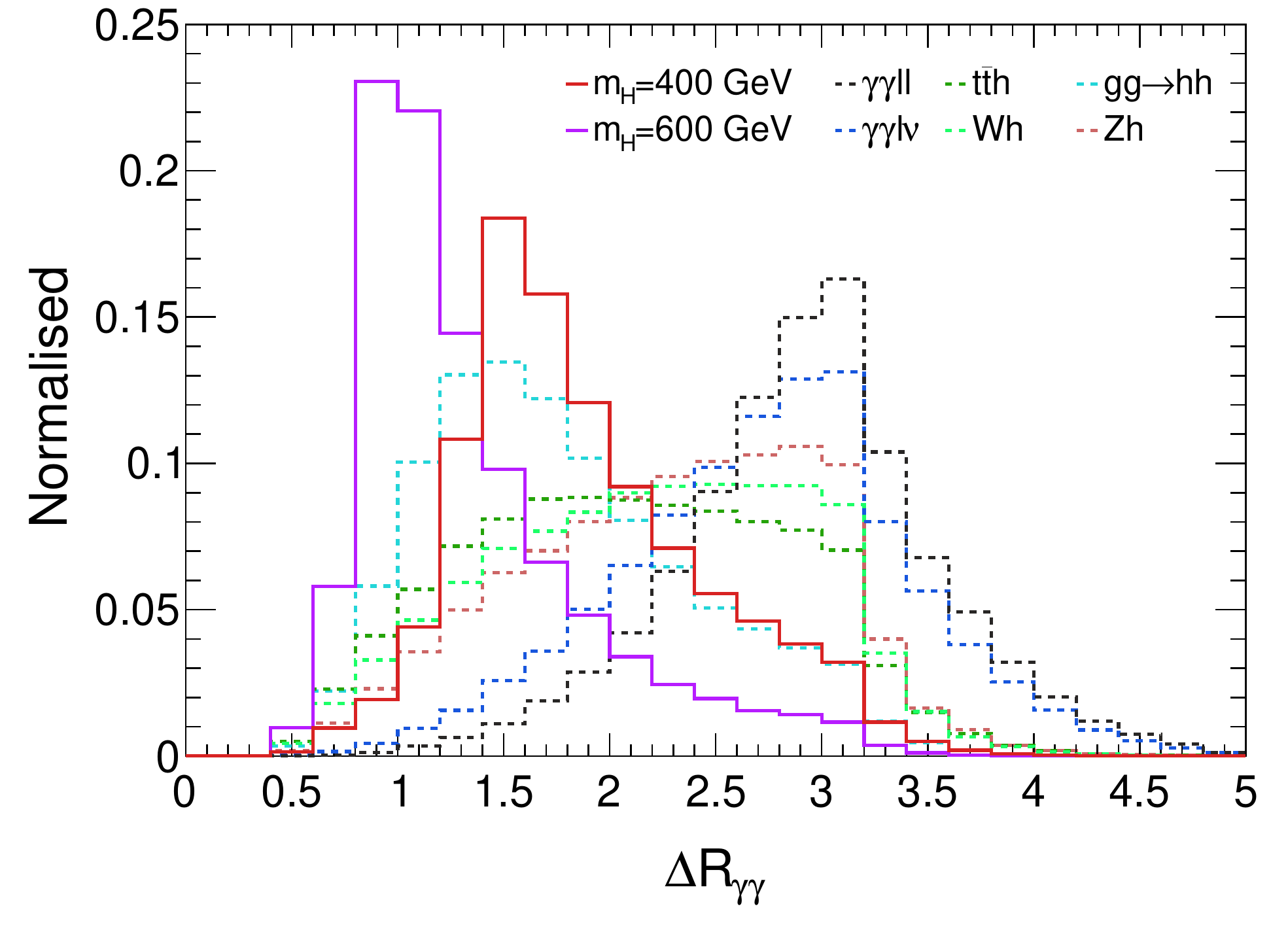} \\
\includegraphics[scale=0.37]{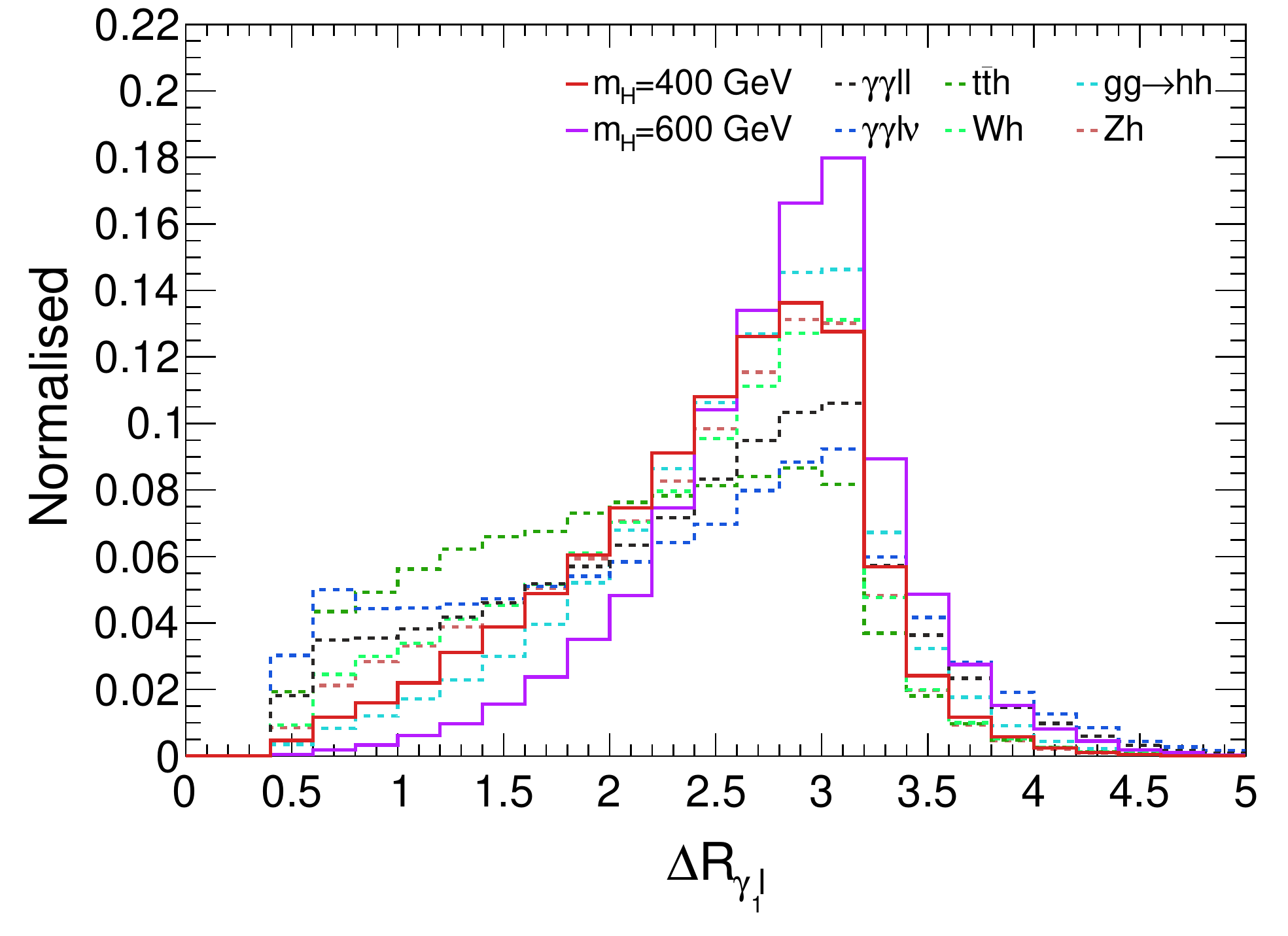}
\includegraphics[scale=0.37]{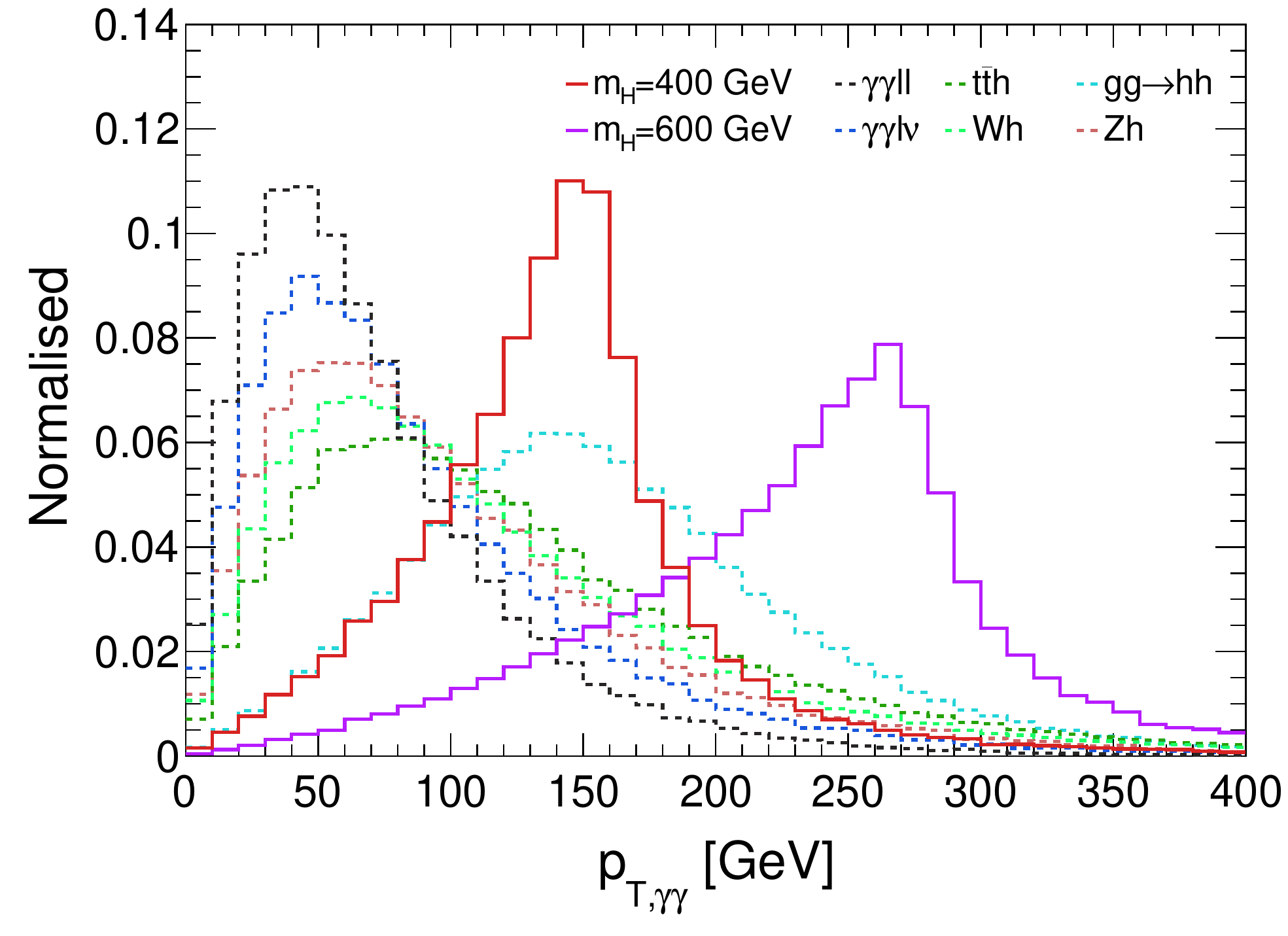} \\
\includegraphics[scale=0.37]{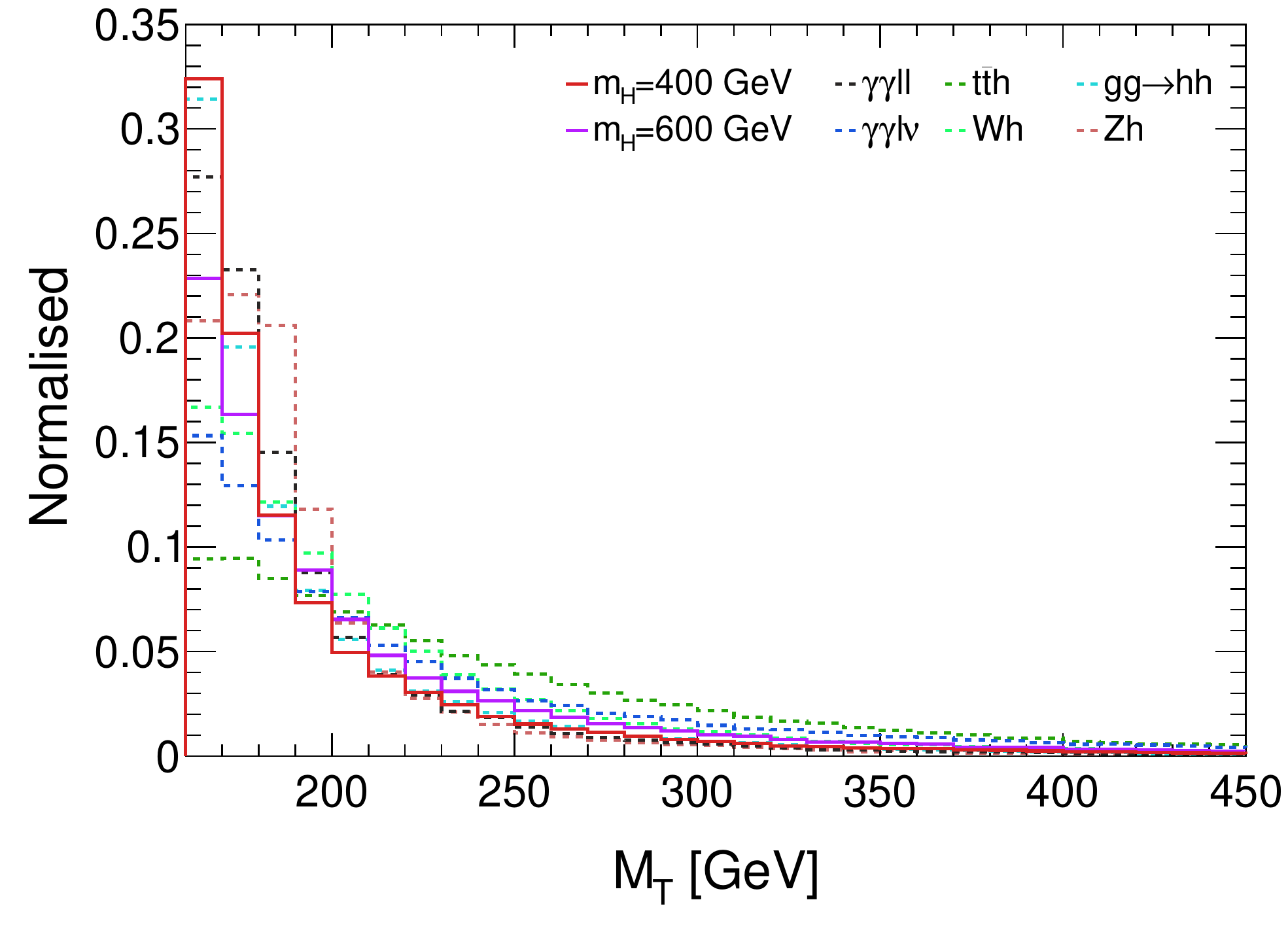}
\caption{The $m_{\textrm{tot}}$, $\Delta R_{\gamma\gamma}$, $\Delta R_{\gamma_1\ell}$, $p_{T,\gamma\gamma}$ and $M_T$ distributions for the 1$\ell$ category for $m_H = 400$ and $600$ GeV with backgrounds. Here the heavy Higgs boson is searched for in the $\gamma\gamma WW^*$ channel. The distributions are shown after imposing the basic trigger cuts.}
\label{2w2ga:1l2gajjMET1}
\end{figure}

\begin{figure}
\centering
\includegraphics[scale=0.5]{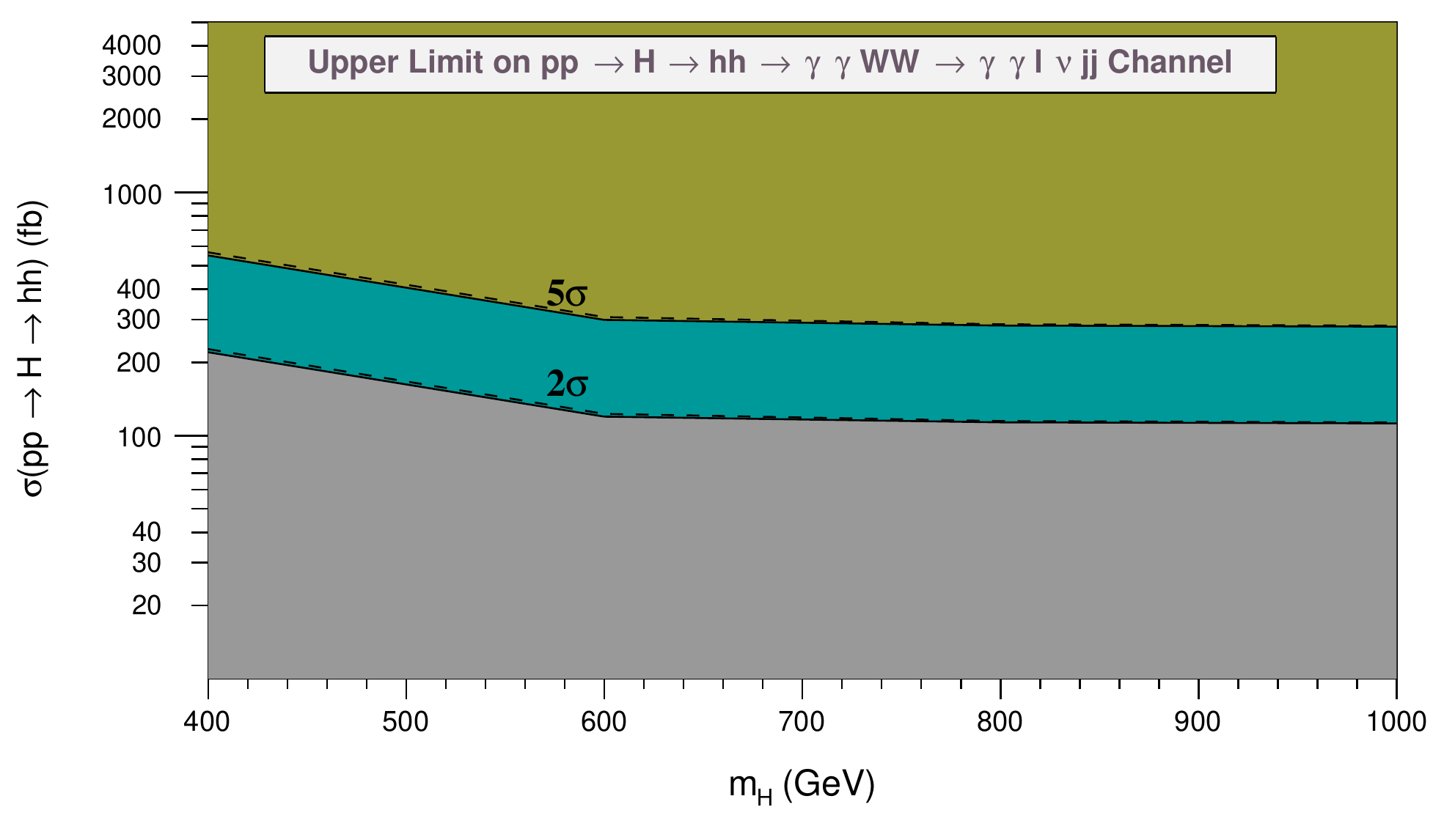}
\caption{Upper limit on $\sigma(pp\to H\to hh)$ (fb) as a function of $m_{H}$ (GeV) for the $\gamma\gamma 1\ell 2j+\met$ channel. The solid (dashed) lines show the 2$\sigma$-5$\sigma$ band on taking 0\% (5\%) systematic uncertainties.}
\label{2w2ga:1l2gajjMET2}
\end{figure}

Before performing the BDT analysis, we select events with exactly two isolated photons, one isolated lepton and at least two jets in the final state, which fulfils all the aforementioned trigger requirements. Like all the other channels, we consider the following variables to train our signal and background samples for the multivariate analysis:

\begin{equation}
p_{T, \gamma\gamma},~\Delta R_{\gamma\gamma},~\Delta\phi_{\ell\met},~mt,~\Delta R_{\ell jj},~m_{tot}, ~m_{eff},~M_T,~\Delta R_{\gamma_1\ell},~\Delta R_{\gamma_2\ell},~p_{T,\gamma_2},~p_{T,\ell},~\met,\nonumber
\end{equation}

\noindent where the variables carry their usual meaning. The five best variables are shown in Fig.~\ref{2w2ga:1l2gajjMET1}. The background yields after the BDT optimisation are shown in Table~\ref{tab1:1l2gajjMET}. In Fig.~\ref{2w2ga:1l2gajjMET2}, we show the upper limit on $\sigma(pp\to H\to hh)$ as a function of $m_H$. The $95\%$ CL upper limit changes from $220.11$ fb ($226.74$ fb with $5\%$ systematic) for $m_H=400$ GeV to $112.20$ fb ($113.24$ fb with $5\%$ systematic) for $m_H=1$ TeV.

\begin{center}
\begin{table}[htb!]
\centering
\scalebox{0.7}{%
\begin{tabular}{|c|c|c|c|}\hline
(a) & Process & Order & Events \\ \hline \hline
\multirow{7}{*}{Background} 
        & $t\bar{t}h$                          & NLO~\cite{bkg_twiki_cs}                   & $6.16$ \\
        & $Zh \; + $ jets                      & NNLO (QCD) + NLO (EW)~\cite{bkg_twiki_cs} & $1.28$ \\
        & $Wh \; +$ jets                       & NNLO (QCD) + NLO (EW)~\cite{bkg_twiki_cs} & $11.27$ \\      
        & $pp\to hh$                           & NNLO~\cite{hhtwiki}                       & $1.35$ \\          
        & $\ell \nu \gamma \gamma \; +$ jets   & LO                                        & $3.33$ \\        
        & $\ell \ell \gamma \gamma \; +$ jets  & LO                                        & $\sim 1.00$ \\ \cline{2-4} 
        & \multicolumn{2}{c|}{Total}                                                       & $24.39$ \\ \hline
\end{tabular}}
\bigskip

\scalebox{0.7}{%
\begin{tabular}{|c|c|c|}\hline
(b) & Process & Events \\ \hline \hline
\multirow{7}{*}{Background} 
        & $t\bar{t}h$                          & $6.94$ \\
        & $Zh \; + $ jets                      & $1.21$ \\
        & $Wh \; +$ jets                       & $9.67$ \\      
        & $pp\to hh$                           & $1.10$ \\          
        & $\ell \nu \gamma \gamma \; +$ jets   & $2.25$ \\        
        & $\ell \ell \gamma \gamma \; +$ jets  & $0.42$ \\ \cline{2-3} 
        & Total                                & $21.59$ \\ \hline
\end{tabular}}
\quad
\scalebox{0.7}{%
\begin{tabular}{|c|c|c|}\hline
(c) & Process & Events \\ \hline \hline
\multirow{7}{*}{Background} 
        & $t\bar{t}h$                          & $3.43$ \\
        & $Zh \; + $ jets                      & $0.37$ \\
        & $Wh \; +$ jets                       & $4.50$ \\      
        & $pp\to hh$                           & $0.25$ \\          
        & $\ell \nu \gamma \gamma \; +$ jets   & $1.10$ \\        
        & $\ell \ell \gamma \gamma \; +$ jets  & $0.13$ \\ \cline{2-3} 
        & Total                                & $9.78$ \\ \hline
\end{tabular}}
\quad
\scalebox{0.7}{%
\begin{tabular}{|c|c|c|}\hline
(d) & Process & Events \\ \hline \hline
\multirow{7}{*}{Background} 
        & $t\bar{t}h$                          & $2.68$ \\
        & $Zh \; + $ jets                      & $0.20$ \\
        & $Wh \; +$ jets                       & $3.38$ \\      
        & $pp\to hh$                           & $0.13$ \\          
        & $\ell \nu \gamma \gamma \; +$ jets   & $1.09$ \\        
        & $\ell \ell \gamma \gamma \; +$ jets  & $0.08$ \\ \cline{2-3} 
        & Total                                & $7.56$ \\ \hline
\end{tabular}}
\caption{Respective background yields for the $\gamma\gamma 1\ell 2j+\met$ channel after the BDT analyses optimised for $m_H = (a)~400$ GeV, $(b)~600$ GeV, $(c)~800$ GeV and $(d)~1$ TeV. The tables also list the perturbative order at which the cross-sections are considered.}
\label{tab1:1l2gajjMET}
\end{table}
\end{center}

\subsubsection{The $\gamma\gamma 2\ell+\met$ Channel}
\label{sec:Htohhtowwgammagamma:2l}

\begin{figure}
\centering
\includegraphics[scale=0.37]{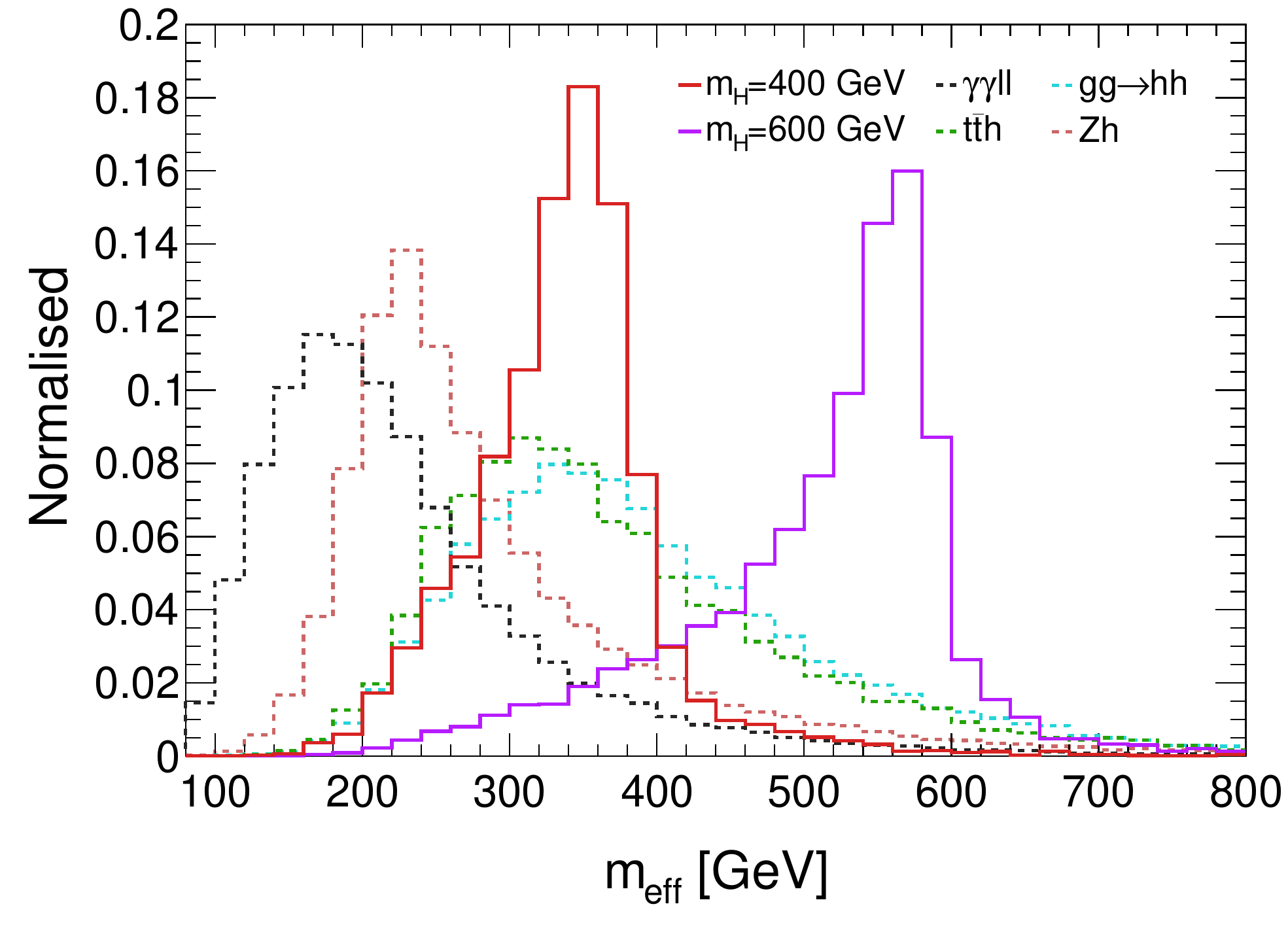}
\includegraphics[scale=0.37]{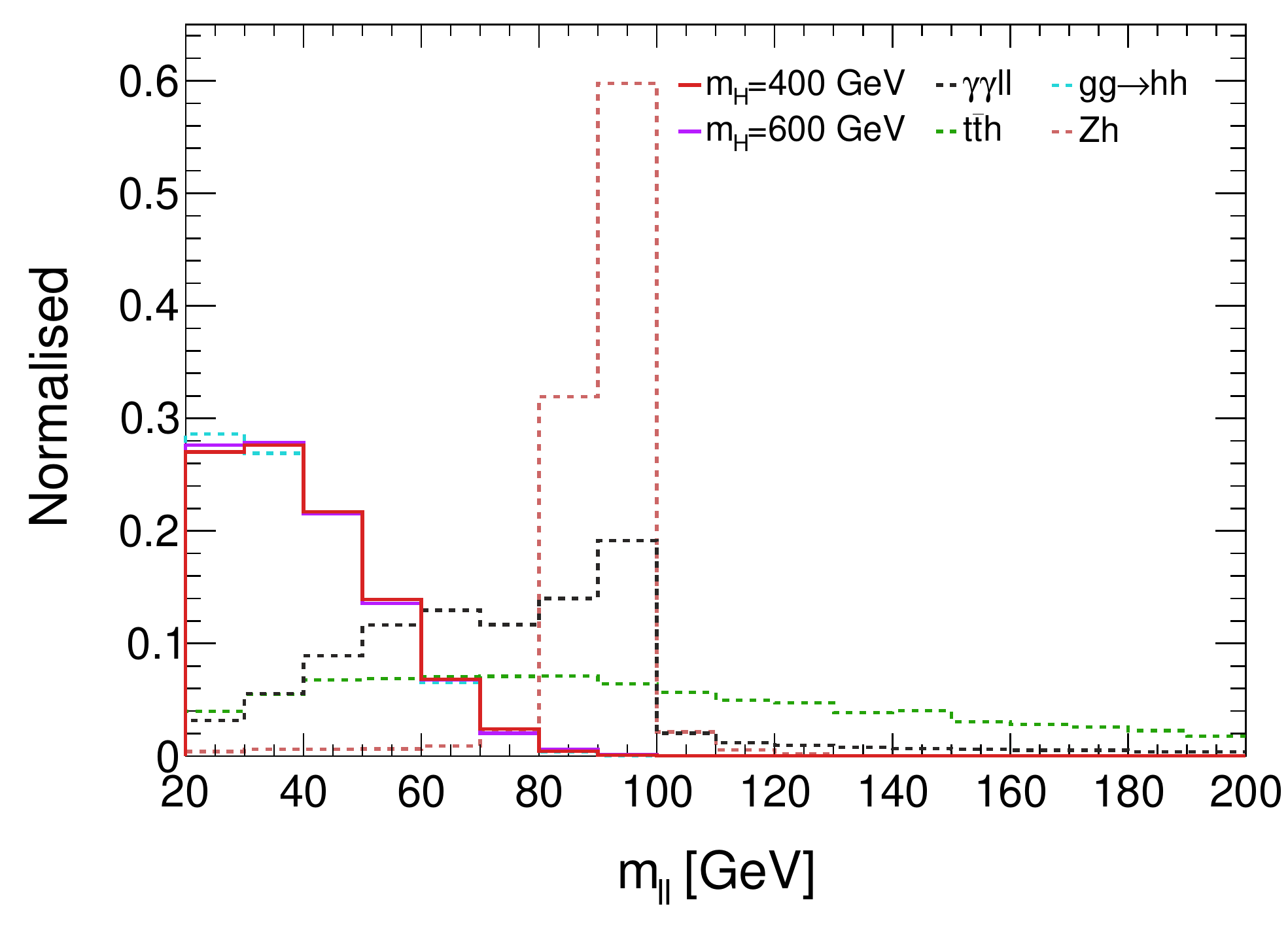} \\
\includegraphics[scale=0.37]{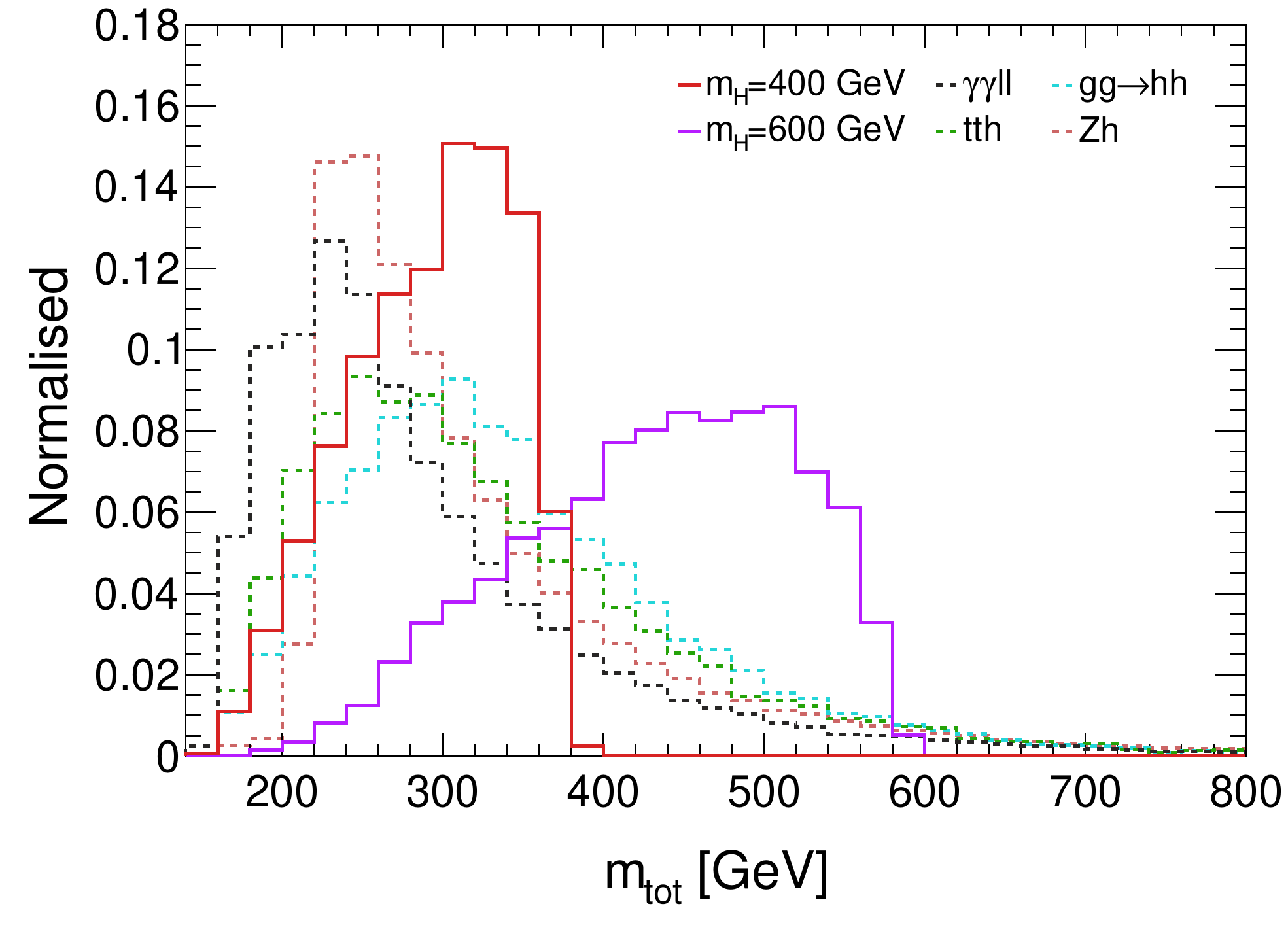}
\includegraphics[scale=0.37]{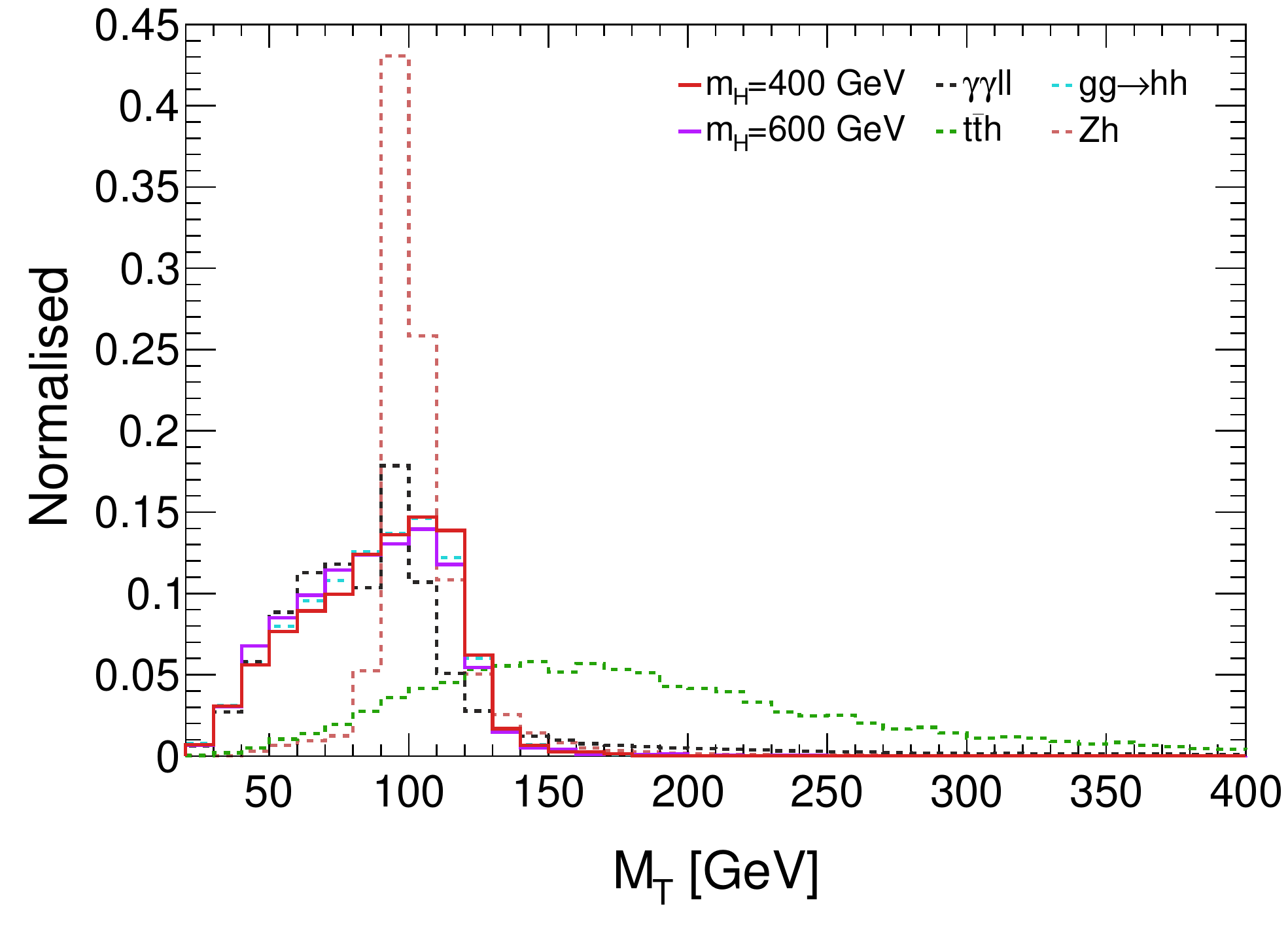} \\
\includegraphics[scale=0.37]{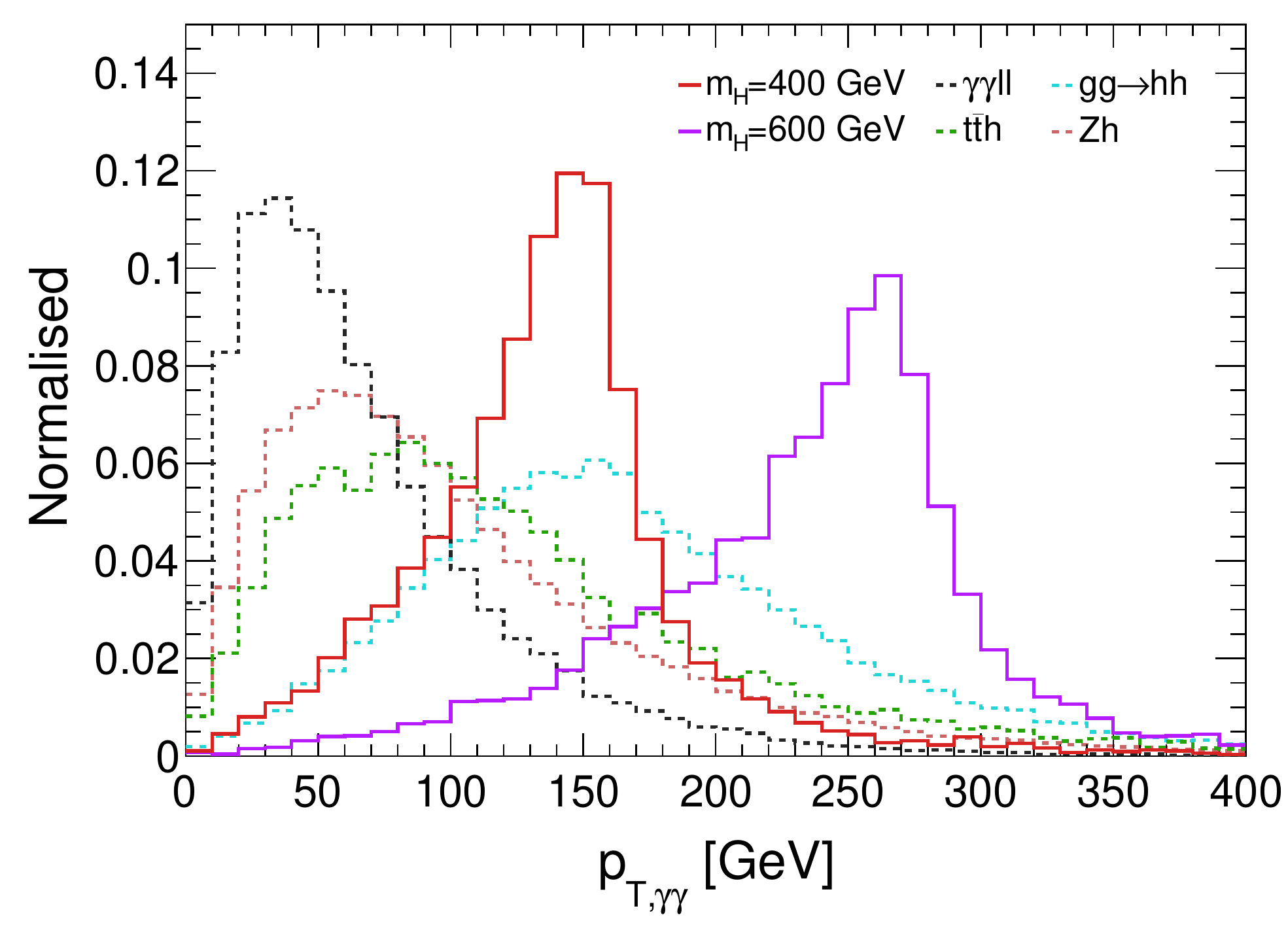}
\caption{The $m_{\textrm{eff}}$, $m_{\ell\ell}$, $m_{\textrm{tot}}$, $M_T$ and $p_{T,\gamma\gamma}$ distributions for the 2$\ell$ category for $m_H = 400$ and $600$ GeV with backgrounds. Here the heavy Higgs boson is searched for in the $\gamma\gamma WW^*$ channel. The distributions are shown after imposing the basic trigger cuts.}
\label{2w2ga:2l2gaMET1}
\end{figure}

\begin{figure}
\centering
\includegraphics[scale=0.5]{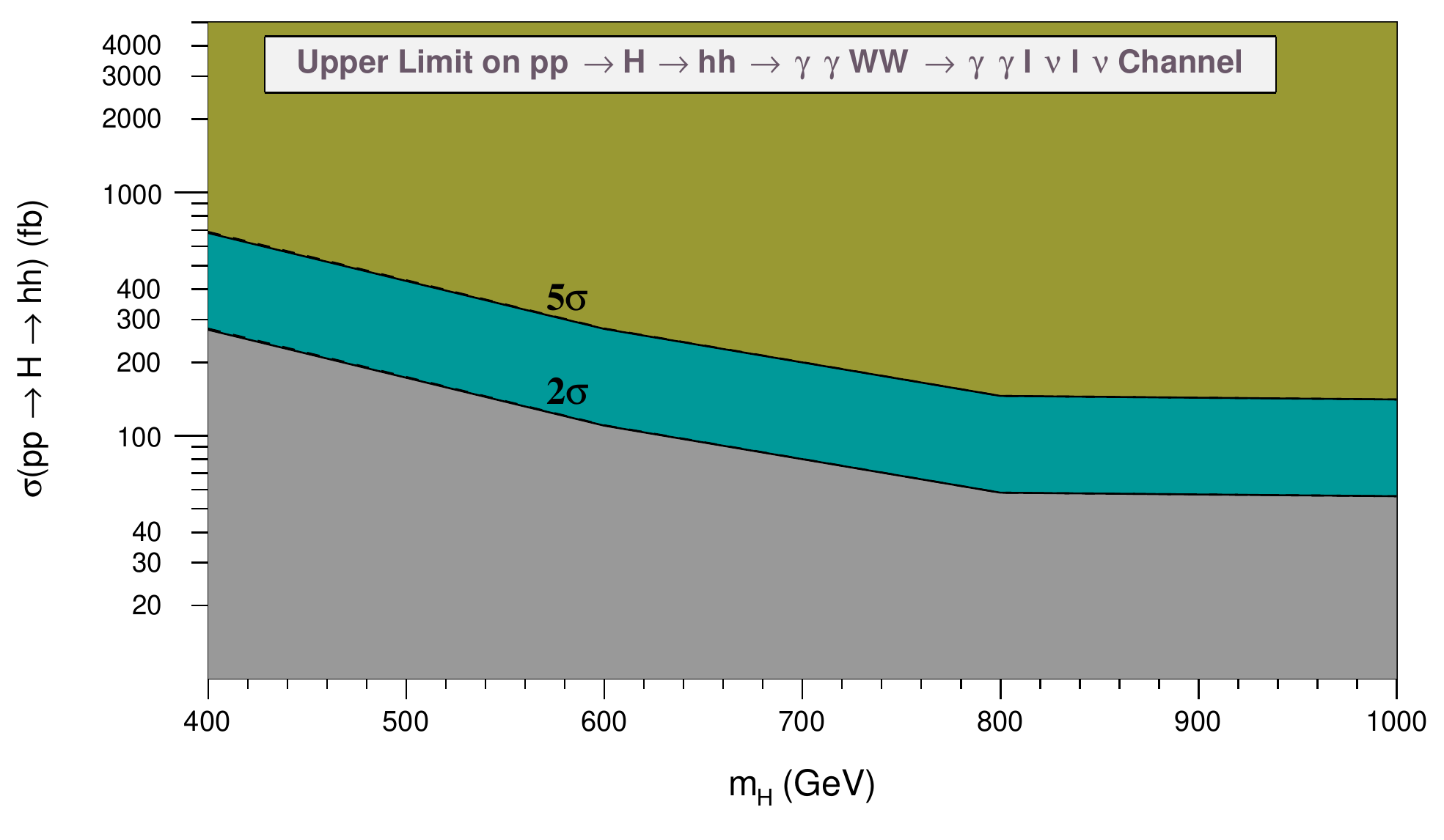}
\caption{Upper limit on $\sigma(pp\to H\to hh)$ (fb) as a function of $m_{H}$ (GeV) for the $\gamma\gamma 2\ell+\met$ channel. The solid (dashed) lines show the 2$\sigma$-5$\sigma$ band on taking 0\% (5\%) systematic uncertainties.}
\label{2w2ga:2l2gaMET2}
\end{figure}

This is the final channel that we study for the $p p \to H \to h h$ case. We choose events with exactly two isolated photons, and two isolated leptons with opposite charge, following the trigger cuts mentioned above. Finally, we choose the following kinematic variables for the multivariate analysis:

\begin{equation}
p_{T, \gamma\gamma},~\Delta R_{\gamma\gamma},~\Delta\phi_{\gamma\gamma},~m_{\ell\ell},~\Delta R_{\ell\ell},~M_T,~m_{\textrm{tot}},~m_{\textrm{eff}},~\Delta R_{\gamma\gamma,\ell\ell},~p_{T,\ell 1}, \nonumber
\end{equation}

\noindent with the usual definitions for the variables. Some of the variables of interest are plotted in Fig.~\ref{2w2ga:2l2gaMET1}. The background yields after the BDT cut are tabulated in Table~\ref{tab2:2l2gaMET} whereas the upper limit on $\sigma(pp\to H\to hh)$ as a function of heavy Higgs mass is shown in Fig.~\ref{2w2ga:2l2gaMET2}. The 95\% CL upper limit for the leptonic scenario is stronger than its semi-leptonic counterpart in the heavy Higgs mass range of 600 GeV and 1 TeV. The upper limit varies in between $109.80$ fb and $56.30$ fb ($110.24$ fb and $56.30$ fb with $5\%$ systematic uncertainty) at 95\% CL, in the aforementioned range.

\begin{center}
\begin{table}[htb!]
\centering
\scalebox{0.7}{%
\begin{tabular}{|c|c|c|}\hline
(a) & Process & Events \\ \hline \hline
\multirow{7}{*}{Background} 
        & $t\bar{t}h$                          & $4.78$ \\
        & $Zh \; + $ jets                      & $1.03$ \\
        & $pp\to hh$                           & $0.74$ \\          
        & $\ell \ell \gamma \gamma \; +$ jets  & $2.44$ \\ \cline{2-3} 
        & Total                                & $8.99$ \\ \hline
\end{tabular}}
\bigskip
\quad
\scalebox{0.7}{%
\begin{tabular}{|c|c|c|}\hline
(b) & Process & Events \\ \hline \hline
\multirow{7}{*}{Background} 
        & $t\bar{t}h$                          & $0.89$ \\
        & $Zh \; + $ jets                      & $0.56$ \\
        & $pp\to hh$                           & $0.37$ \\          
        & $\ell \ell \gamma \gamma \; +$ jets  & $0.64$ \\ \cline{2-3} 
        & Total                                & $2.46$ \\ \hline
\end{tabular}}
\quad
\scalebox{0.7}{%
\begin{tabular}{|c|c|c|}\hline
(c) & Process & Events \\ \hline \hline
\multirow{7}{*}{Background} 
        & $t\bar{t}h$                          & $0.26$ \\
        & $Zh \; + $ jets                      & $0.21$ \\
        & $pp\to hh$                           & $0.12$ \\          
        & $\ell \ell \gamma \gamma \; +$ jets  & $0.24$ \\ \cline{2-3} 
        & Total                                & $0.83$ \\ \hline
\end{tabular}}
\quad
\scalebox{0.7}{%
\begin{tabular}{|c|c|c|}\hline
(d) & Process & Events \\ \hline \hline
\multirow{7}{*}{Background} 
        & $t\bar{t}h$                          & $0.14$ \\
        & $Zh \; + $ jets                      & $0.37$ \\
        & $pp\to hh$                           & $0.05$ \\          
        & $\ell \ell \gamma \gamma \; +$ jets  & $0.14$ \\ \cline{2-3} 
        & Total                                & $0.70$ \\ \hline
\end{tabular}}
\caption{Respective background yields for the $\gamma\gamma 2\ell+\met$ channel after the BDT analyses optimised for $m_H = (a)~400$ GeV, $(b)~600$ GeV, $(c)~800$ GeV and $(d)~1$ TeV. The various perturbative orders for the backgrounds are the same as in Table~\ref{tab1:1l2gajjMET}.}
\label{tab2:2l2gaMET}
\end{table}
\end{center}

\subsection{Summarising the $H \to hh$ channel}
\label{sec:HtohhtoSummarise}

\begin{figure}
\centering
\includegraphics[scale=0.4]{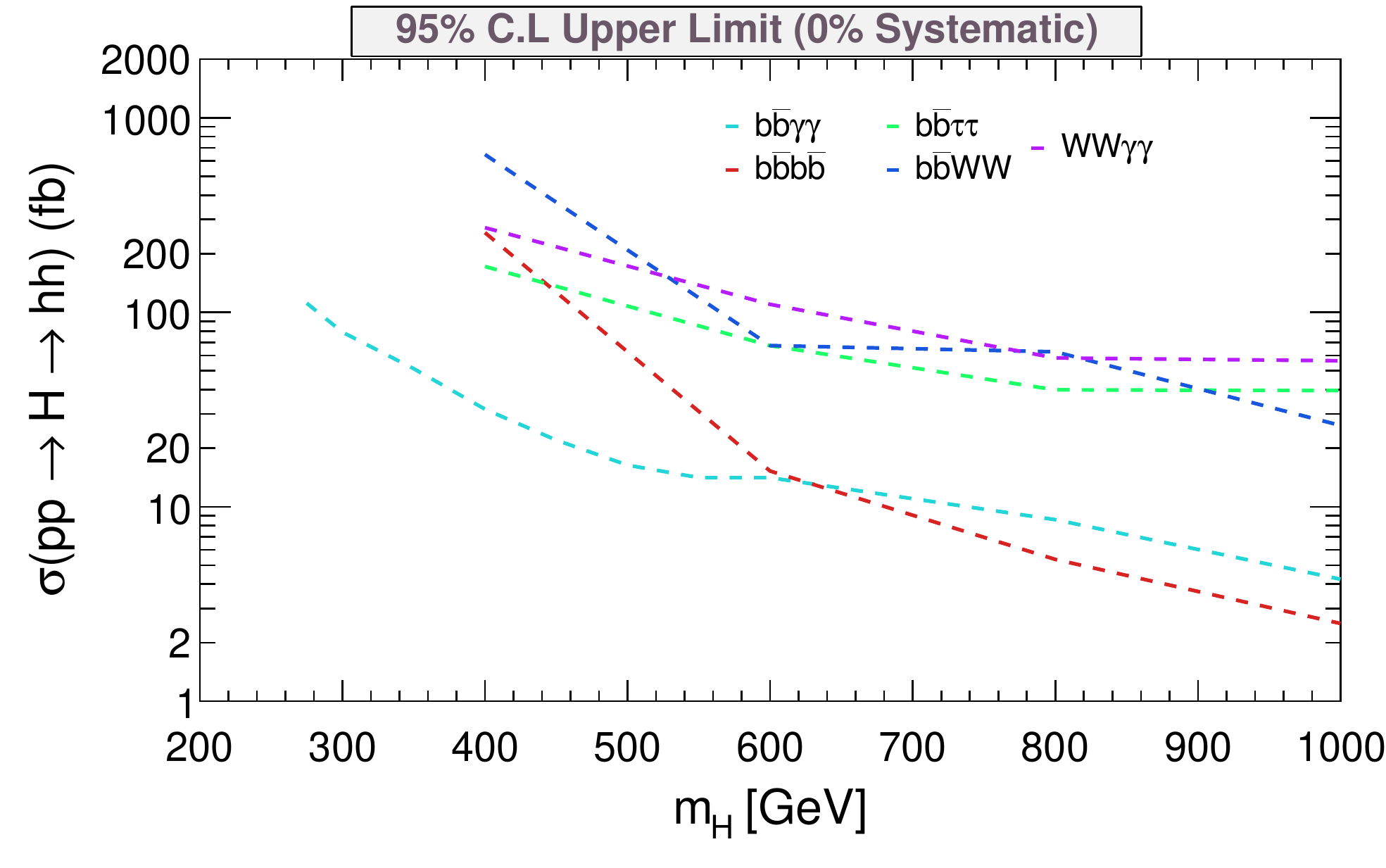}\includegraphics[scale=0.4]{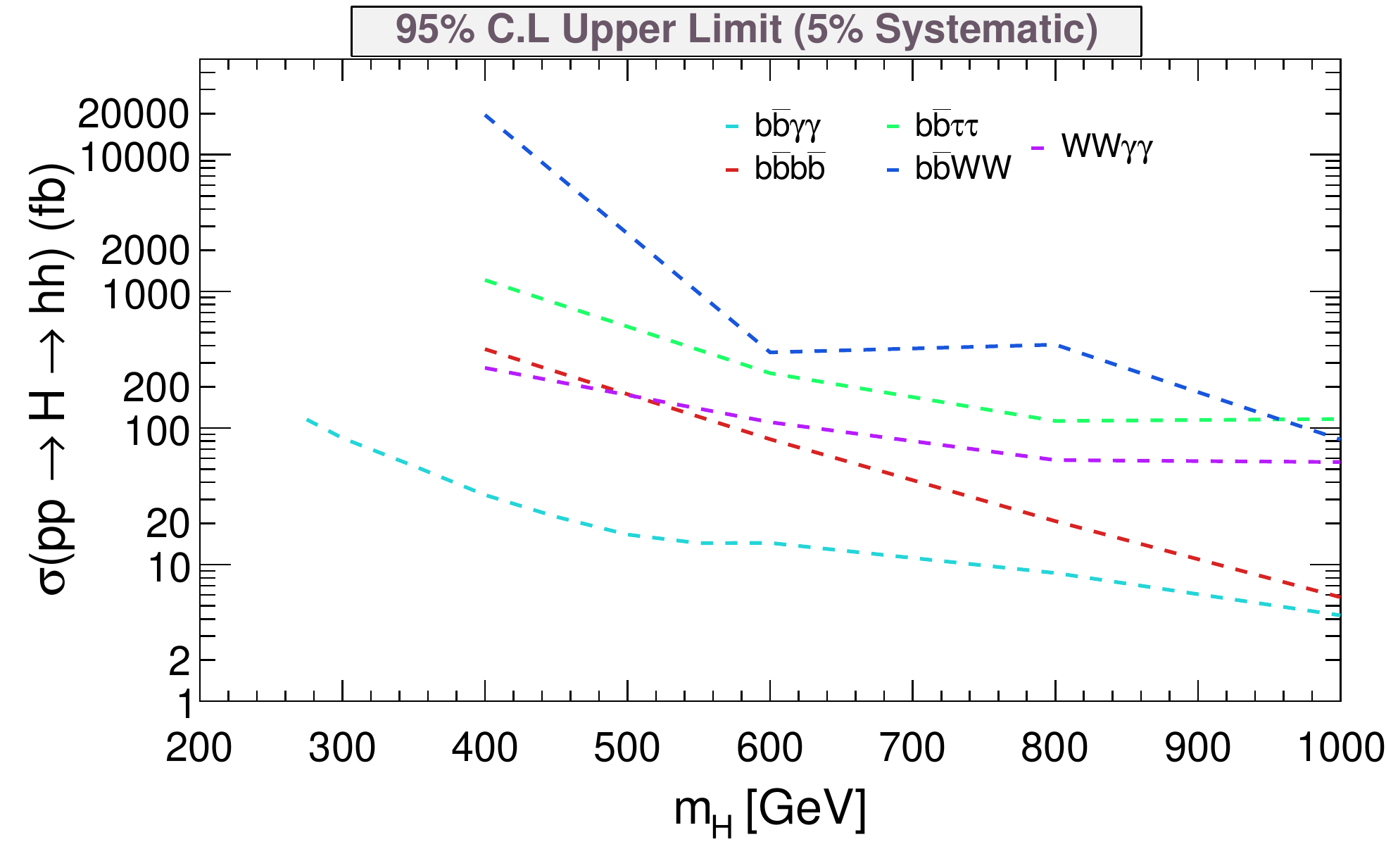}
\caption{95\% CL upper limit on $\sigma(pp\to H \to hh)$ (fb) as a function of $m_{H}$ (GeV) for the $b\bar{b}\gamma\gamma, \; b\bar{b}b\bar{b}, \; b\bar{b}\tau^+\tau^-, \; b\bar{b}WW^*$ ($2\ell$) and $WW^*\gamma\gamma$ ($2\ell$) channels without systematic uncertainty (left) and with $5\%$ systematic uncertainty (right).}
\label{H2hh:summary}
\end{figure}

Having studied five different channels with more than one sub-processes in three instances, we summarise the results in this subsection. The 95\% CL upper limits on $\sigma(p p \to H \to h h)$ for all these channels is shown in Fig.~\ref{H2hh:summary}. We find that the strongest limits come from the $b\bar{b}\gamma\gamma$ and $4b$ channels. The $b\bar{b}\gamma\gamma$ is strongest up to $m_H \sim$ 600 GeV. From 600 GeV onward, the $4b$ channel is more constraining owing to its larger cross-section. The present limits on $\sigma(p p \to H \to h h)$ from the 13 TeV analysis vary between $\sim$ 970 fb (300 GeV) and $\sim$ 225 fb (600 GeV) from the $b\bar{b}\gamma\gamma$ channel~\cite{Aaboud:2018ftw} and between $\sim$ 70 fb (800 GeV) and $\sim$ 25 fb (1 TeV) from the $4b$ analysis~\cite{Aaboud:2018knk}. Our projected limits vary between $79.03$ fb (300 GeV) and $14.10$ fb (600 GeV) from the $b\bar{b}\gamma\gamma$ channel and between $5.36$ fb (800 GeV) and $2.51$ fb (1 TeV) from the $4b$ channel. We find an order of magnitude improvement in the sensitivity. We must note in passing that from the results obtained in Ref.~\cite{Aaboud:2018ksn}, the $H \to hh \to 4W$ category does not show much promise even at the HL-LHC, owing to very small signal yields in all categories. The maximum sensitivity is expected in the $e\mu$ 2-lepton category with 5 expected events.

\section{The $pp\to H\to t\bar{t}$ Channel}
\label{sec:Htott}

\begin{figure}[htb!]
\centering
\includegraphics[trim=0 530 0 70,clip,width=12 cm]{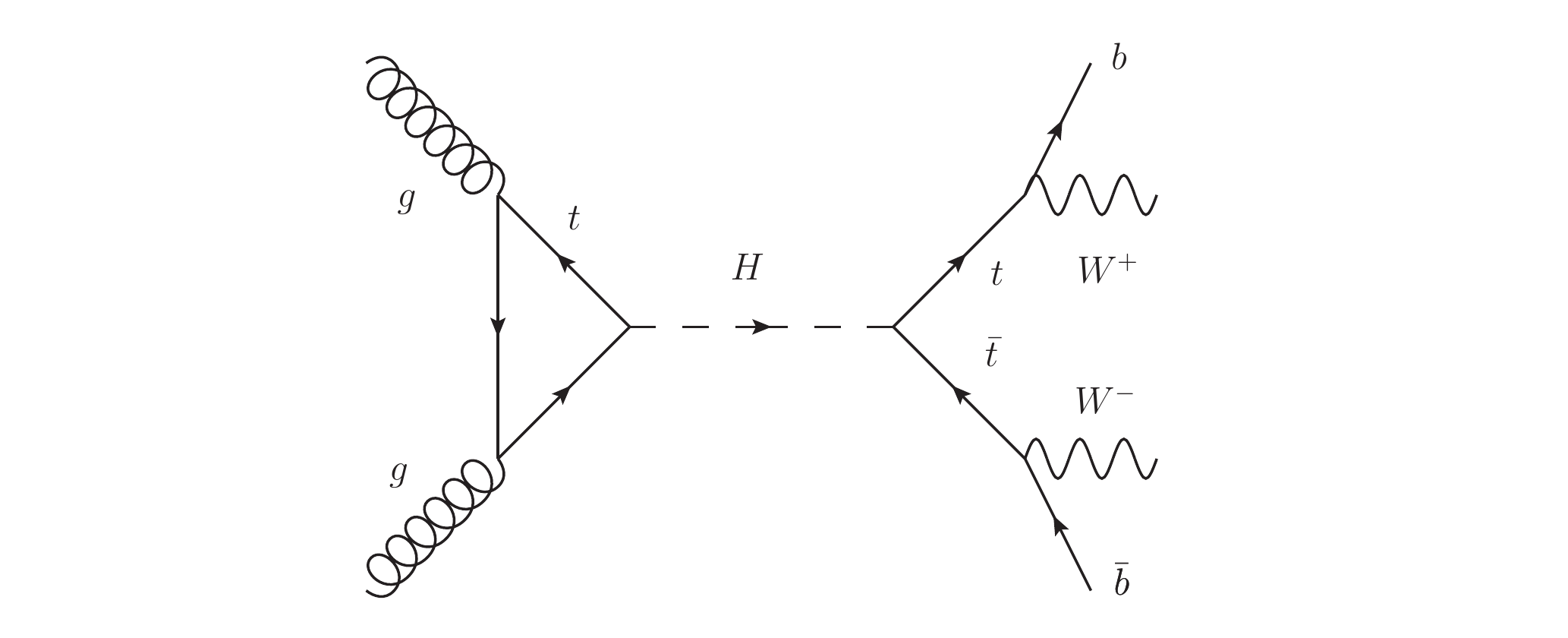}
\caption{Feynman diagram for the $pp\to H\to t\bar{t}$ signal process.}
\label{FD:H2tt}
\end{figure}

After having studied the $H \to hh$ in multifarious channels in detail, we now turn our attention to a heavy scalar (or pseudoscalar) resonance being produced predominantly by gluon fusion and decaying to a pair of top quarks (Fig~\ref{FD:H2tt}). This channel has already gained some attention in the experimental community~\cite{Aaboud:2017hnm,Aaboud:2018mjh}. Searches for resonant scalars, pseudoscalars, $Z^{\prime}$-bosons, Kaluza-Klein gluons and Kaluza-Klein gravitons have been performed. The aim of this section is to try and improve upon these existing searches and provide potential reach of the $\sigma(H \to t \bar{t})$ by studying the fully leptonic and the semi-leptonic final states. The branching ratio of $t \to b W$ being close to 100\% makes the channel essentially become a search for $H \to b\bar{b}W^+W^-$. However, unlike the $H \to h h \to b\bar{b}WW^*$ channel studied in subsection~\ref{sec:Htohhtobbww}, where one of the $W$-bosons is off-shell, here both of them are on-shell. This is the first essential difference between the two channels and the reason why one requires a completely different search strategy for the two cases. In the previous section~\ref{sec:Htohh}, we required BR$(H \to h h)=100\%$. However, in realistic scenarios, if the heavy scalar is produced predominantly via gluon fusion (top/bottom loops), it should also decay to a pair of top quarks (and also bottom quarks) if it is above the $t\bar{t}$ threshold. Similar to the $H \to h h \to b\bar{b} W W^*$ channel, here also we divide the analysis into two parts, \textit{viz.}, the leptonic and the semi-leptonic channels. We apply the same trigger-level cuts to the various objects as sketched in subsection~\ref{sec:Htohhtobbww}. The backgrounds are the same as before. As before, we implement the production and decay of the heavy scalar in the \texttt{Pythia 6} framework.

\subsection{The leptonic Channel}
\label{sec:Htott:2l}

\begin{figure}
\centering
\includegraphics[scale=0.37]{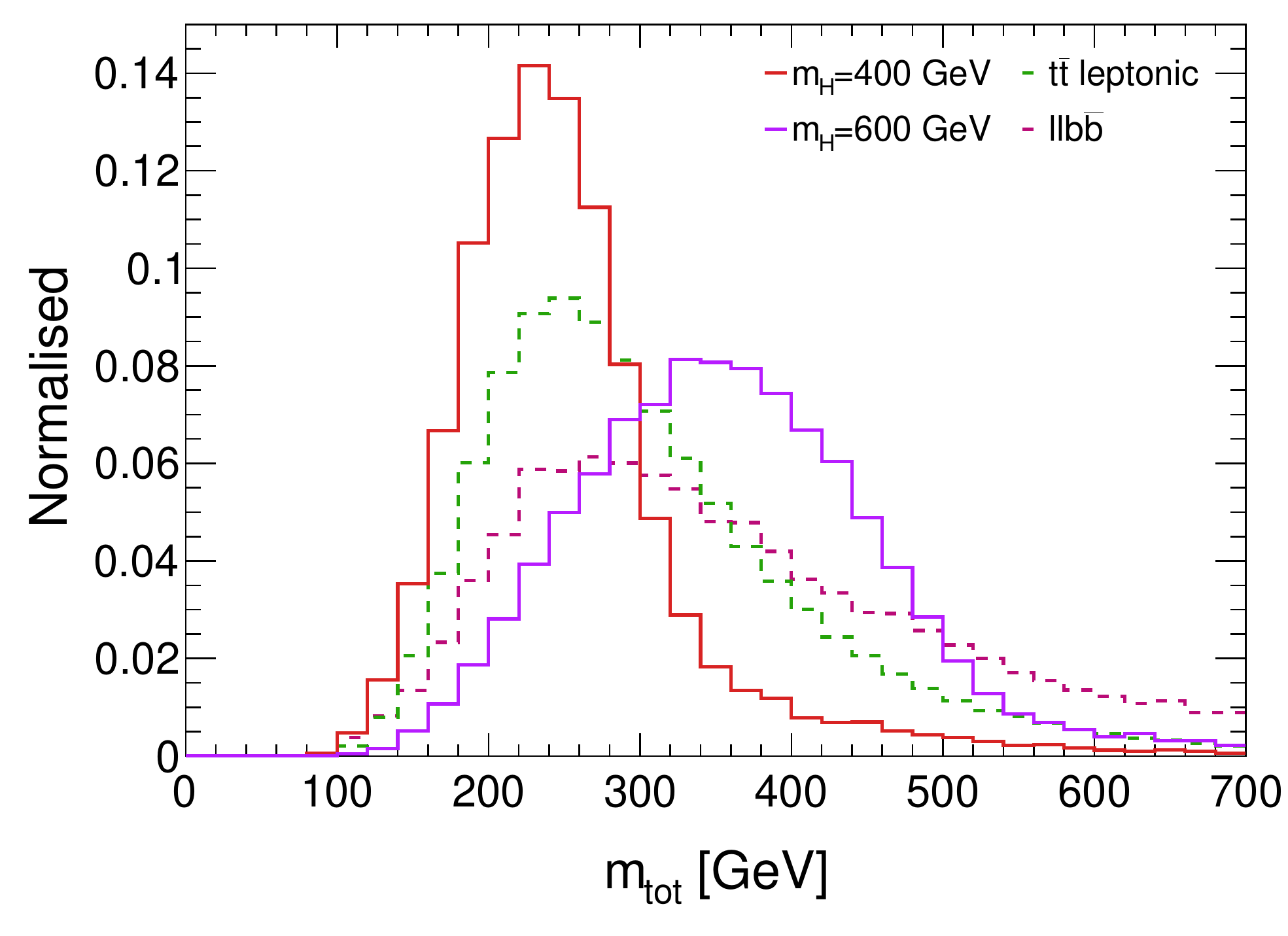}
\includegraphics[scale=0.37]{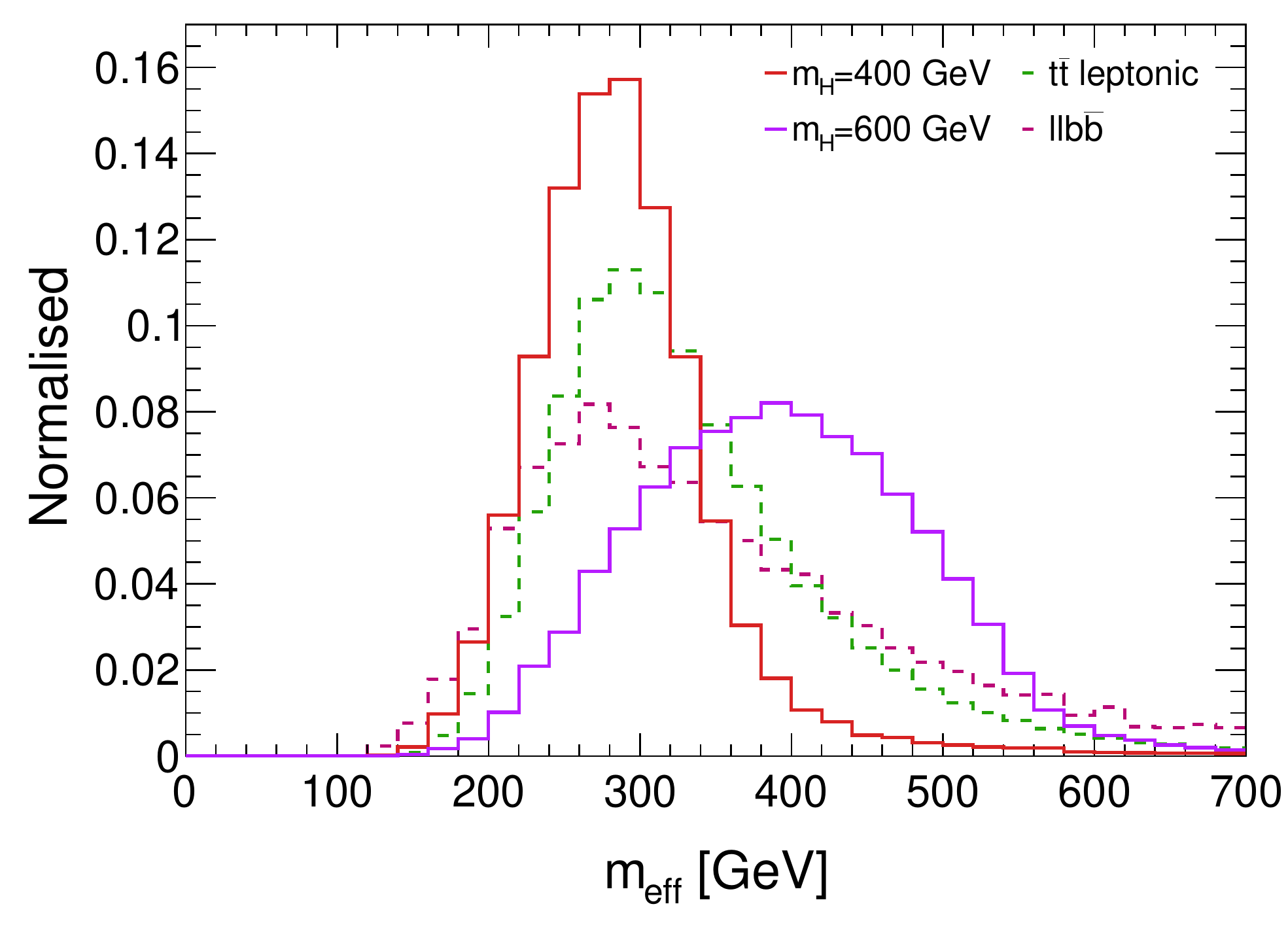} \\
\includegraphics[scale=0.37]{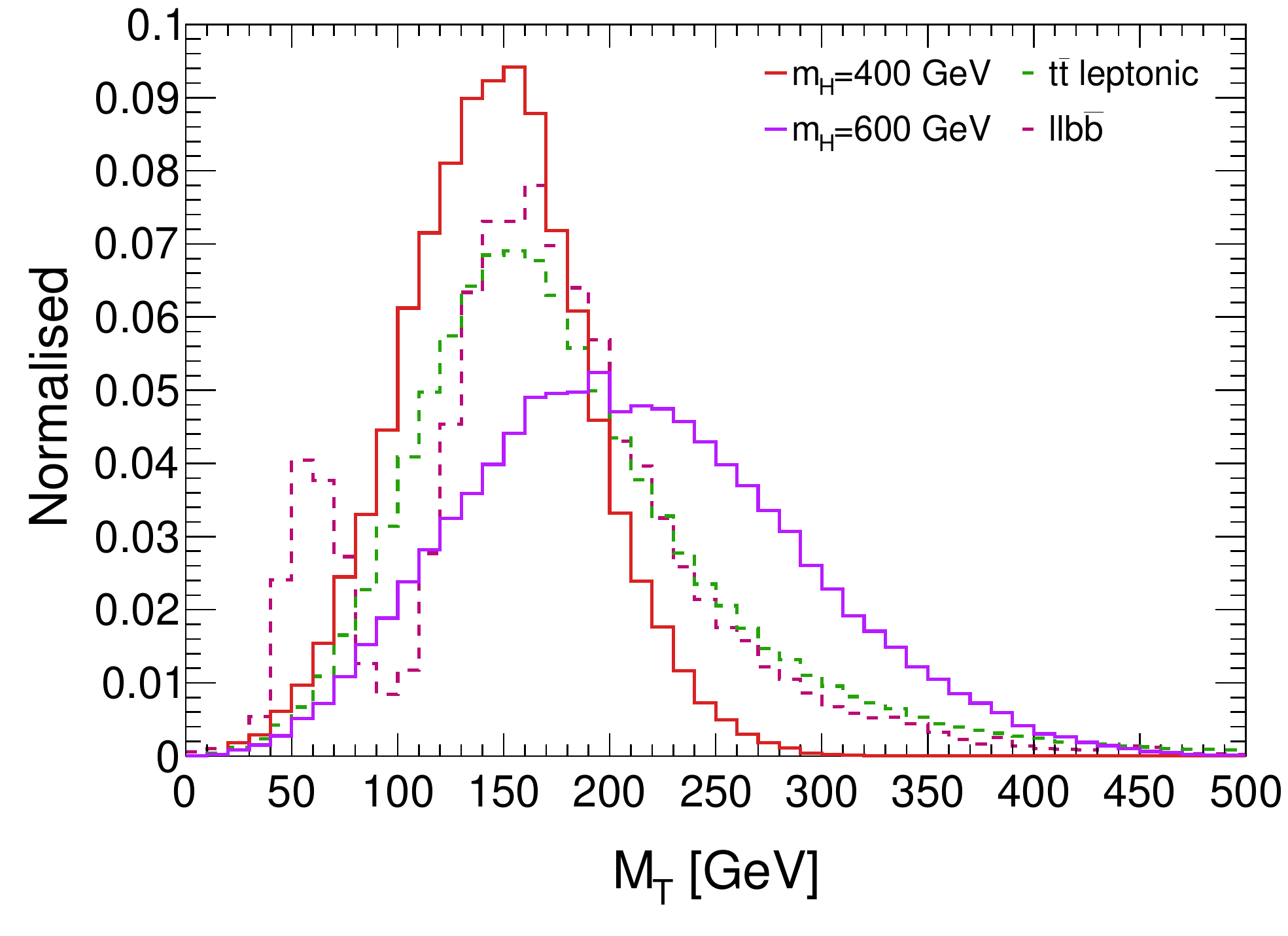}
\includegraphics[scale=0.37]{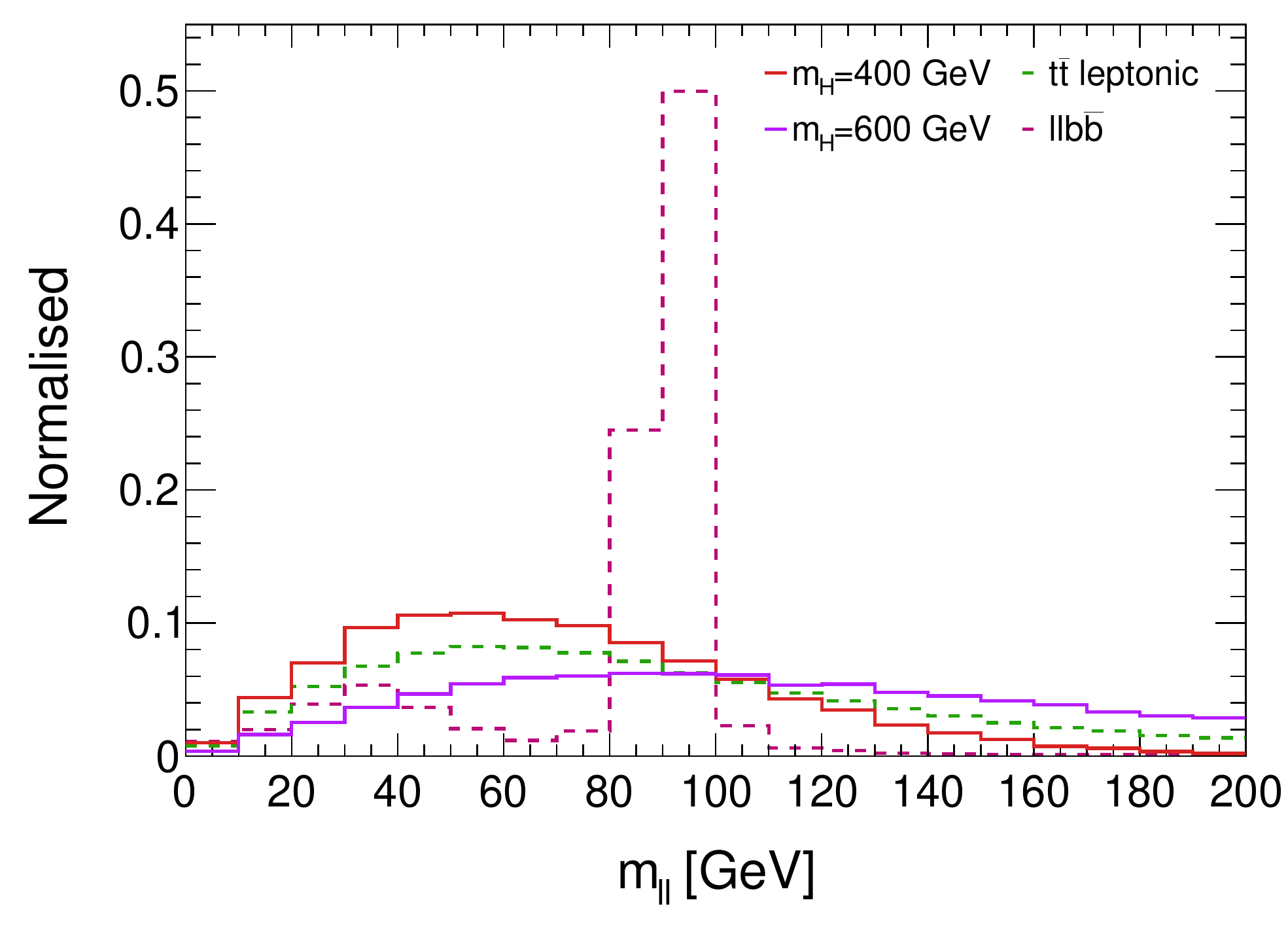}
\caption{The $m_{\textrm{tot}}$, $m_{\textrm{eff}}$, $M_T$ and $m_{\ell\ell}$ distributions for the 2$\ell$ category for $m_H = 400$ and $600$ GeV with dominant backgrounds. Here the heavy Higgs boson is searched for in the $t\bar{t}$ channel. The distributions are shown before doing the multivariate analysis.}
\label{Htt:2l}
\end{figure}

\begin{figure}
\centering
\includegraphics[scale=0.5]{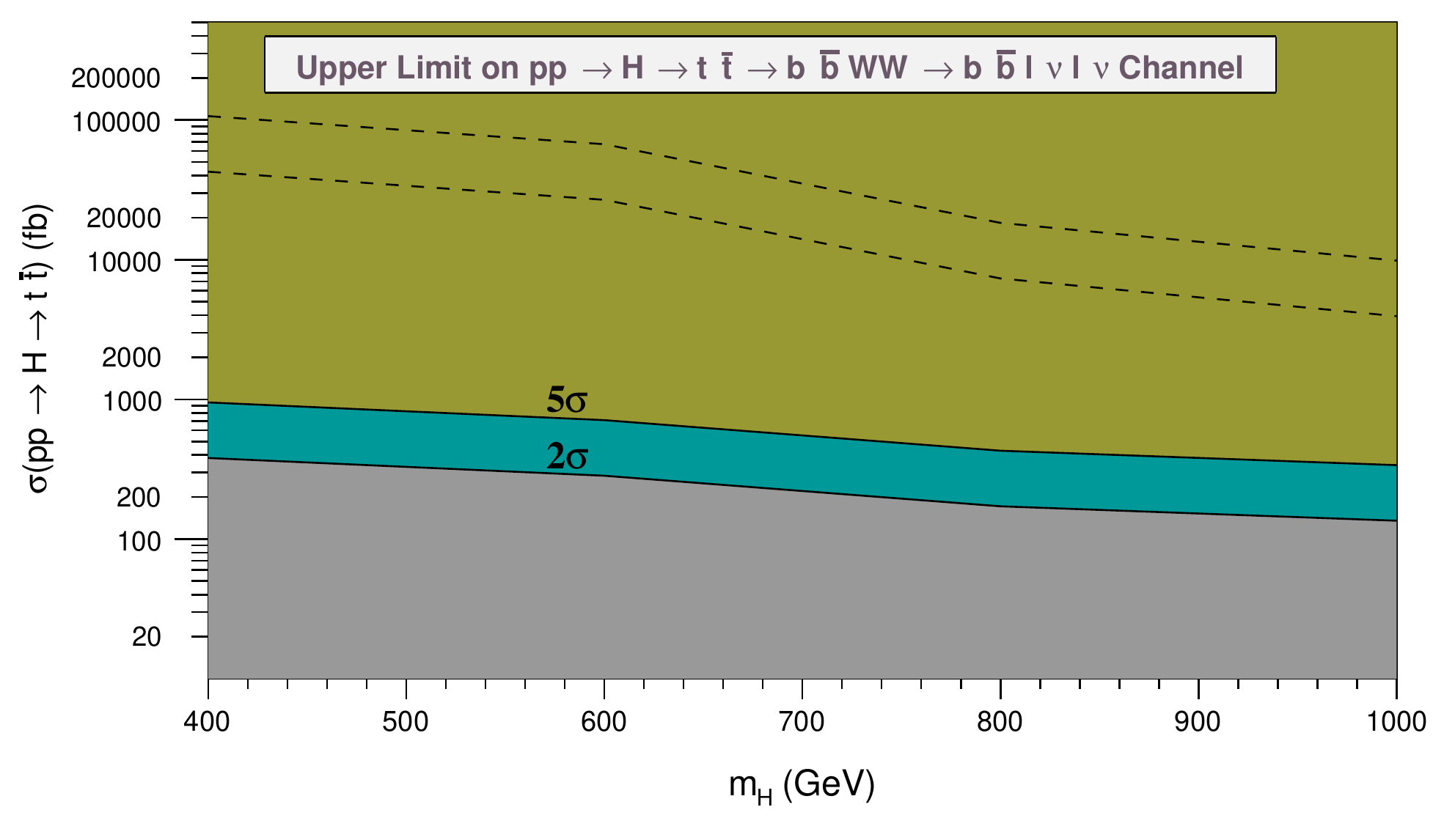}
\caption{Upper limit on $\sigma(pp\to H\to t\bar{t})$ (fb) as a function of $m_{H}$ (GeV) for the $2\ell + 2b + \met$ channel. The solid (dashed) lines show the 2$\sigma$-5$\sigma$ band on taking 0\% (5\%) systematic uncertainties.}
\label{h2ttul:2l2bMET}
\end{figure}

Like in section~\ref{sec:Htohhtobbww:2l}, here also we select events with two oppositely charged isolated leptons and two $b$-tagged jets. Without performing a classical cut-based analysis, we optimise our results to obtain the best-possible sensitivity by employing a boosted decision tree analysis. The set of variables which discriminate the signal from the backgrounds are as follows:

\begin{equation}
p_{T, bb},~\eta_{bb},~\phi_{bb},~m_{bb},~M_T,~m_{\textrm{tot}},~m_{\textrm{eff}},~\Delta R_{b_1\ell_1},~p_{T, b_1},~p_{T, b_2},~p_{T, \ell_1},~p_{T, \ell_2}, \nonumber
\end{equation}

\noindent where all the variables have their usual meaning as mentioned earlier. We would like to mention here that we also consider the $m_{T2}$ variable during our analysis. However, this variable is $\sim 80\%$ correlated with $p_{T, bb}$. Moreover, $m_{T2}$ has a lower BDT ranking as compared to $p_{T, bb}$. We explicitly checked that adding this correlated variable does not improve our BDT sensitivity. Thus for this analysis, we do not use $m_{T2}$ ($m_{T2}$ was used in the $b\bar{b}\tau\tau$ (section~\ref{sec:Htohhtobbtautau}) and $b\bar{b}WW^*$ (section~\ref{sec:Htohhtobbww}) analyses.). The top four discriminatory variables are shown in Fig.~\ref{Htt:2l}. In Table~\ref{H2ttbar:2l2bMET}, the number of background events at an integrated luminosity of 3000 fb$^{-1}$, optimised to maximise the sensitivity for various values of $m_H$ and after imposing cuts on the BDT observable, are presented. Like in all the other channels, we present the 95\% and 99.7\% upper limit on $\sigma(pp\to H\to t\bar{t})$ as a function of $m_H$, in Fig.~\ref{h2ttul:2l2bMET}. We find that the 95\% upper limit on the cross-section lies between $380.43$ fb and $135.25$ fb ($42683.56$ fb and $3940.56$ fb with $5\%$ systematic uncertainty) for $m_H$ varying between 400 GeV and 1 TeV.

\begin{center}
\begin{table}[htb!]
\centering
\scalebox{0.7}{%
\begin{tabular}{|c|c|c|}\hline
(a) & Process & Events \\ \hline \hline

\multirow{8}{*}{Background}
 & $t\bar{t}$ lep             	      & $4979032.27$ \\  
 & $t\bar{t}h$                        & $6211.98$ \\  
 & $t\bar{t}Z$                        & $6769.81$\\  
 & $t\bar{t}W$       		          & $4018.71$ \\ 
 & $pp\to hh$           	          & $111.55$ \\
 & $\ell\ell b\bar{b}$                & $38875.30$ \\ \cline{2-3}  
 & \multicolumn{1}{c|}{Total}         & $5035019.62$ \\ \hline
\end{tabular}}
\bigskip
\quad
\scalebox{0.7}{%
\begin{tabular}{|c|c|c|}\hline
(b) & Process & Events \\ \hline \hline
\multirow{8}{*}{Background}
 & $t\bar{t}$ lep                     & $3520173.16$ \\  
 & $t\bar{t}h$                        & $6832.81$ \\  
 & $t\bar{t}Z$                        & $10547.04$\\  
 & $t\bar{t}W$                        & $5398.14$ \\ 
 & $pp\to hh$                         & $73.68$ \\  
 & $\ell\ell b\bar{b}$                & $29580.72$ \\ \cline{2-3}  
 & \multicolumn{1}{c|}{Total}         & $3572605.55$ \\ \hline
\end{tabular}}
\quad
\scalebox{0.7}{%
\begin{tabular}{|c|c|c|}\hline
(c) & Process & Events \\ \hline \hline
\multirow{8}{*}{Background}
 & $t\bar{t}$ lep                     & $712411.06$ \\  
 & $t\bar{t}h$                        & $2289.83$ \\  
 & $t\bar{t}Z$                        & $4211.79$\\  
 & $t\bar{t}W$                        & $1998.19$ \\ 
 & $pp\to hh$                         & $32.69$ \\  
 & $\ell\ell b\bar{b}$                & $10697.64$ \\ \cline{2-3}  
 & \multicolumn{1}{c|}{Total}         & $731641.20$ \\ \hline
\end{tabular}}
\scalebox{0.7}{%
\begin{tabular}{|c|c|c|}\hline
(d) & Process & Events \\ \hline \hline
\multirow{8}{*}{Background}
 & $t\bar{t}$ lep                     & $326174.10$ \\  
 & $t\bar{t}h$                        & $1349.87$ \\  
 & $t\bar{t}Z$                        & $2866.39$\\  
 & $t\bar{t}W$                        & $1229.00$ \\ 
 & $pp\to hh$                         & $23.20$ \\  
 & $\ell\ell b\bar{b}$                & $7492.23$ \\ \cline{2-3}  
 & \multicolumn{1}{c|}{Total}         & $339134.79$ \\ \hline
\end{tabular}}
\caption{Respective background yields for the $2\ell + 2b + \met$ channel after the BDT analyses optimised for $m_H = (a)~400$ GeV, $(b)~600$ GeV, $(c)~800$ GeV and $(d)~1$ TeV. The various orders of the signal and backgrounds are same as in Table~\ref{tab1:2l2bMET}.}
\label{H2ttbar:2l2bMET}
\end{table}
\end{center}

\subsection{The semi-leptonic Channel}
\label{sec:Htott:1l}

\begin{figure}
\centering
\includegraphics[scale=0.37]{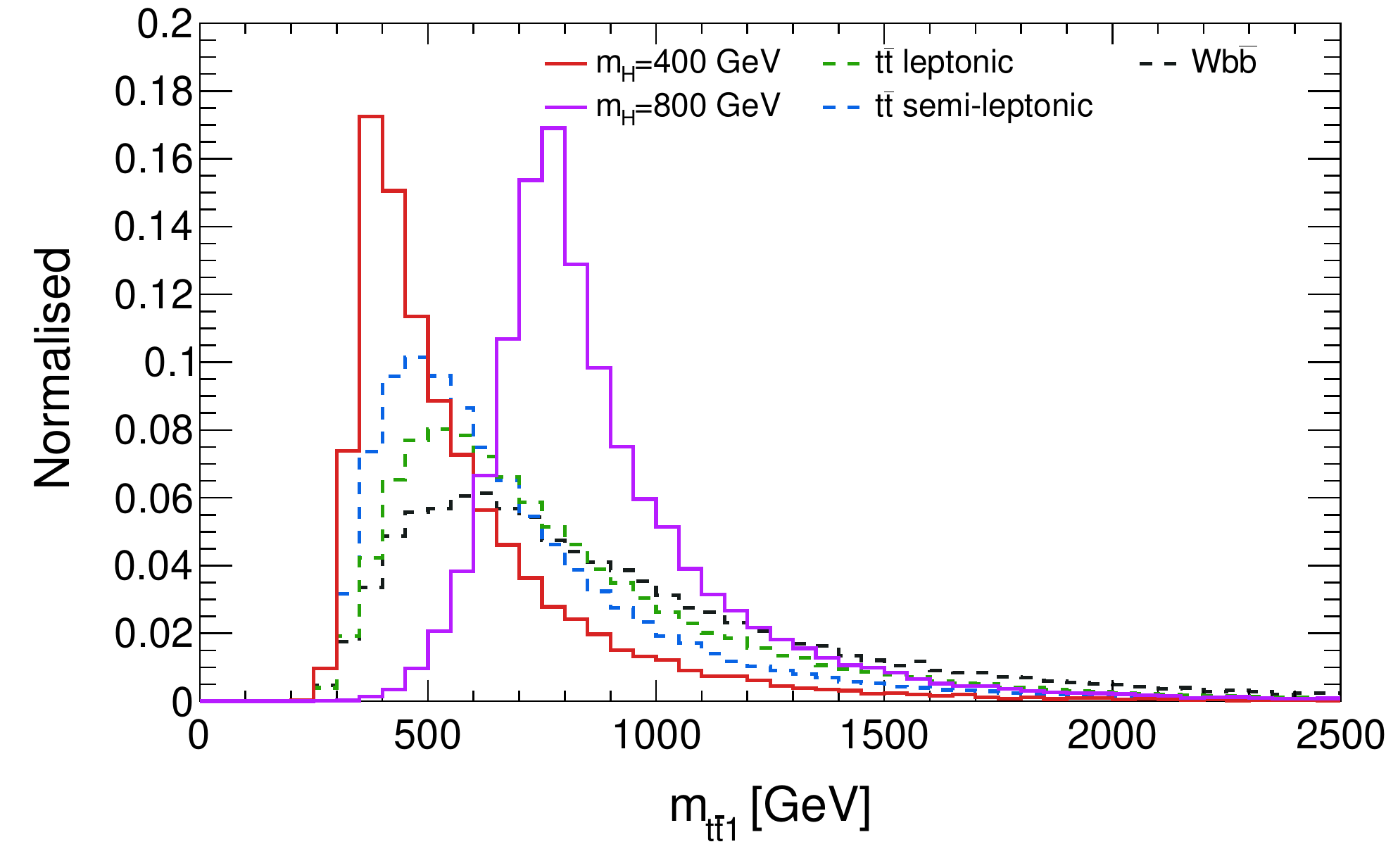}\includegraphics[scale=0.37]{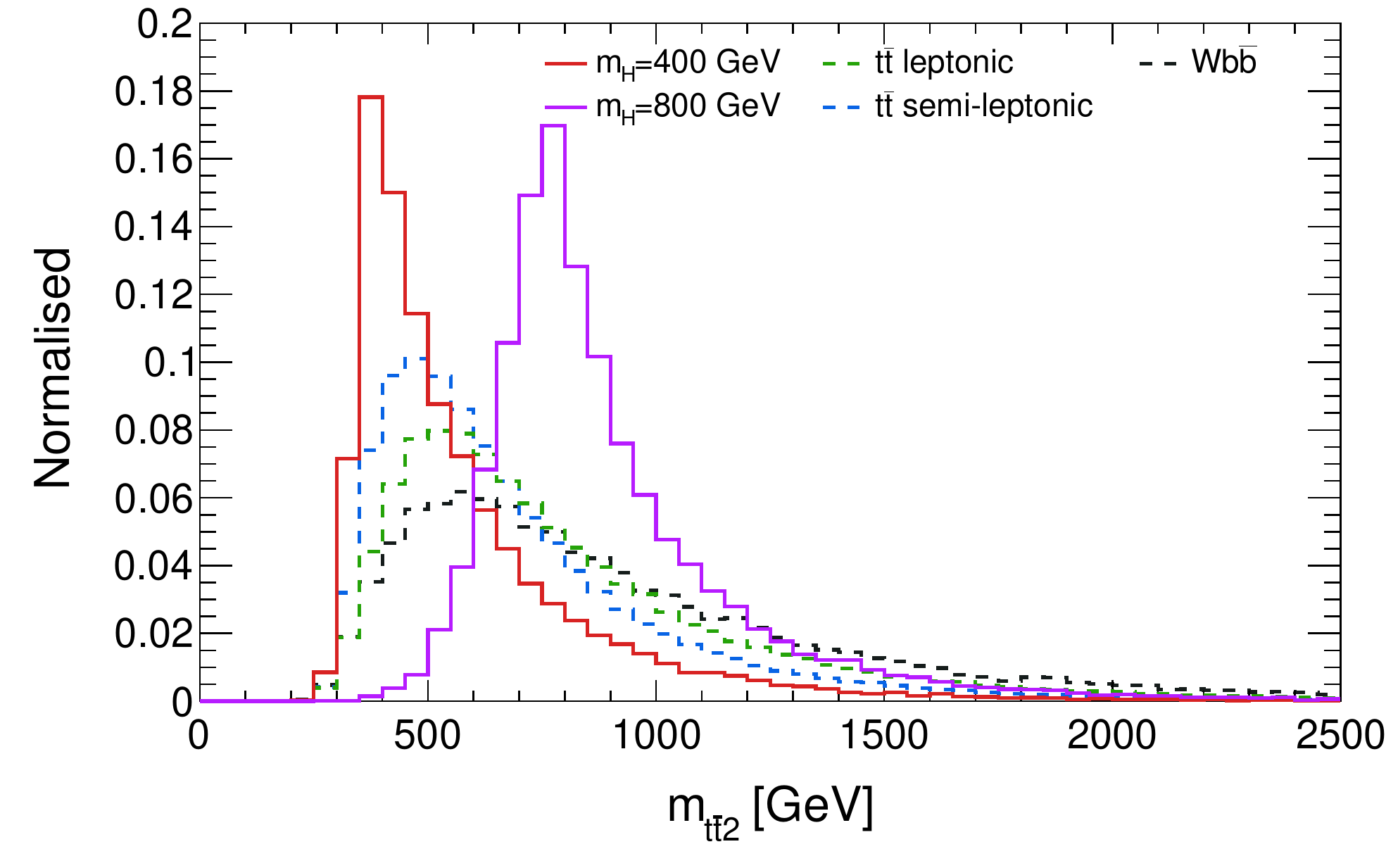}
\caption{The reconstructed invariant mass of the top-quark pair in the semi-leptonic decay of the $H\to t\bar{t}$ channel.}
\label{htt1l:mtt}
\end{figure}

\begin{figure}
\centering
\includegraphics[scale=0.37]{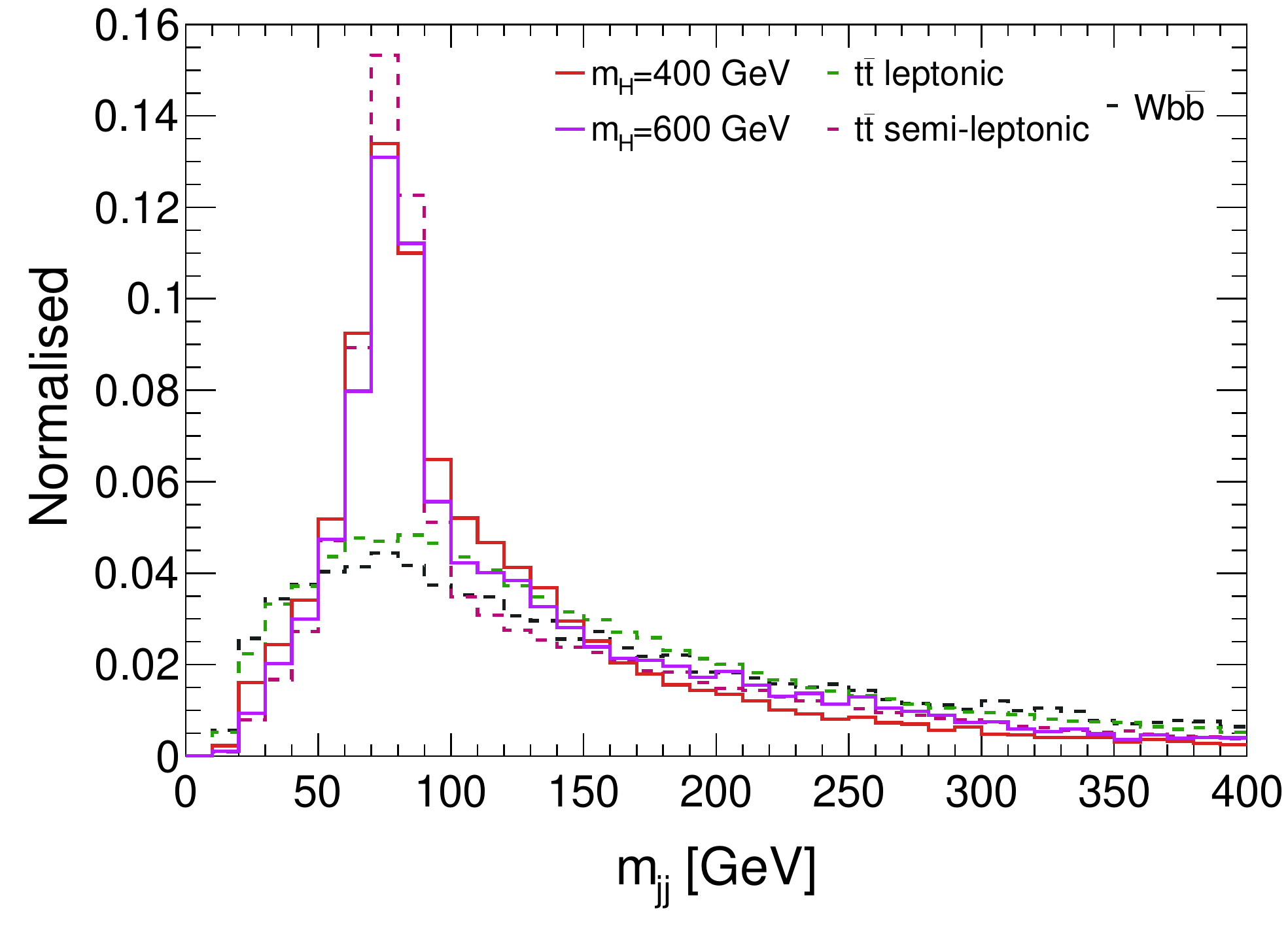}
\includegraphics[scale=0.37]{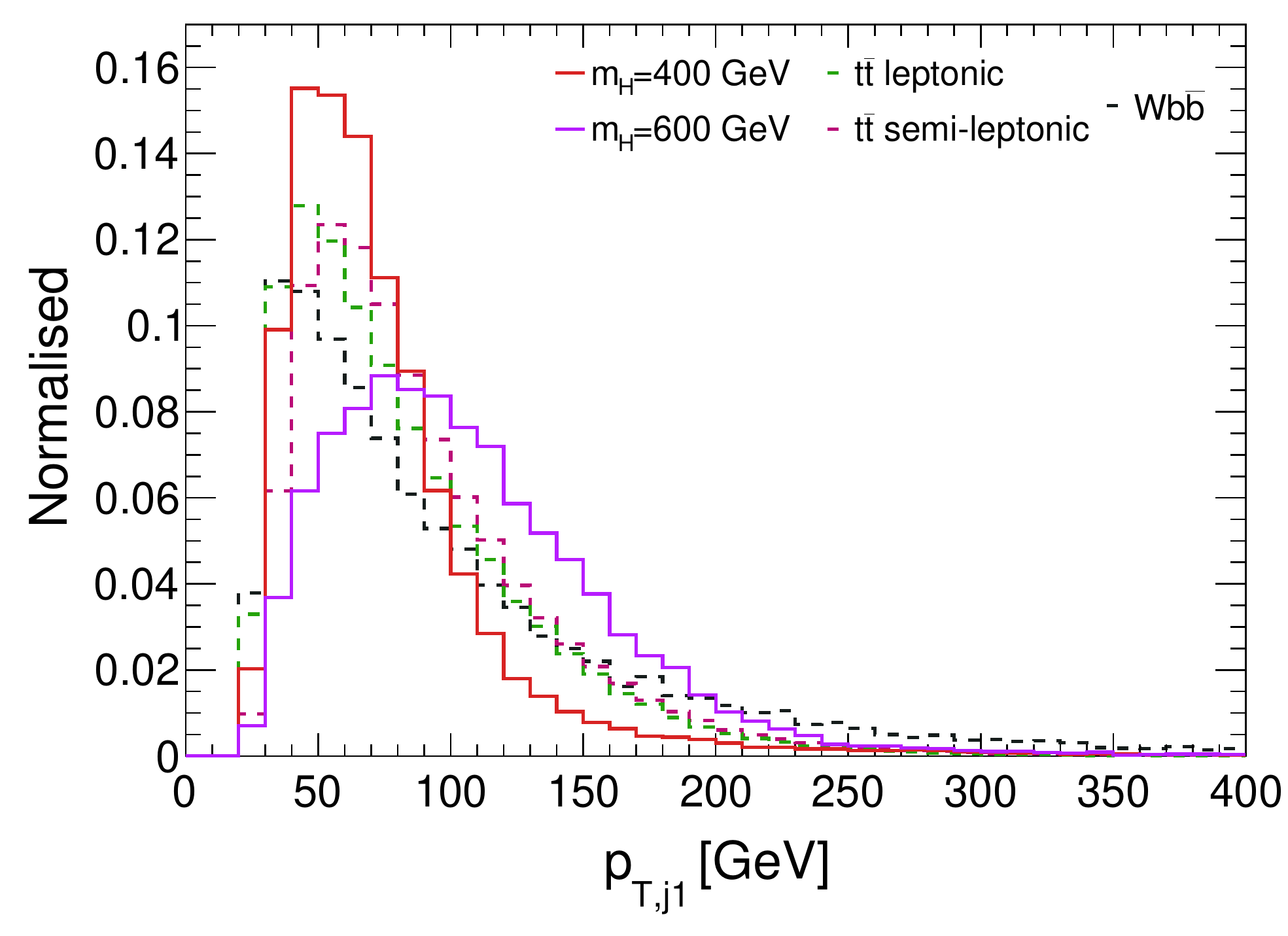} \\
\includegraphics[scale=0.37]{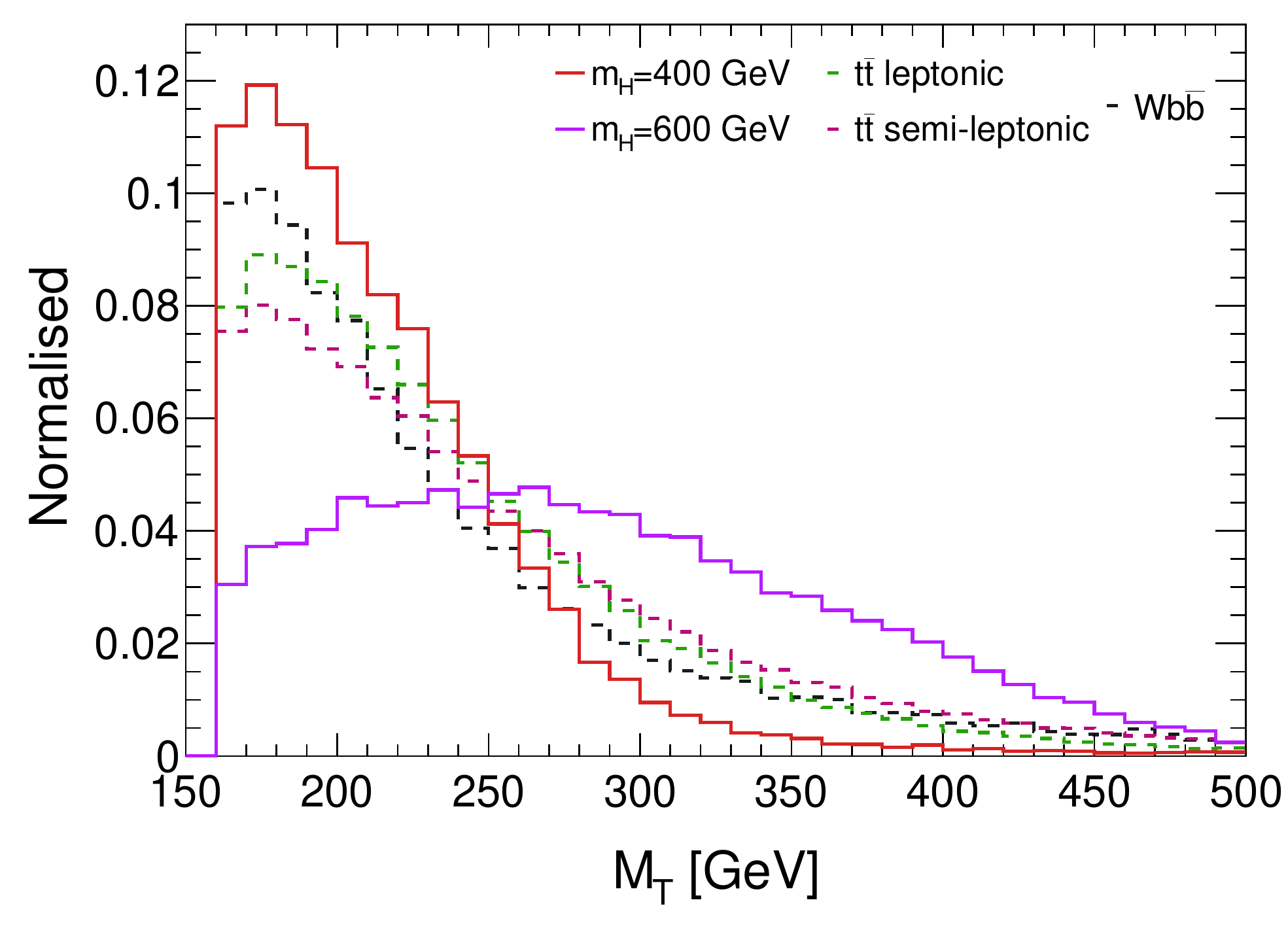}
\includegraphics[scale=0.37]{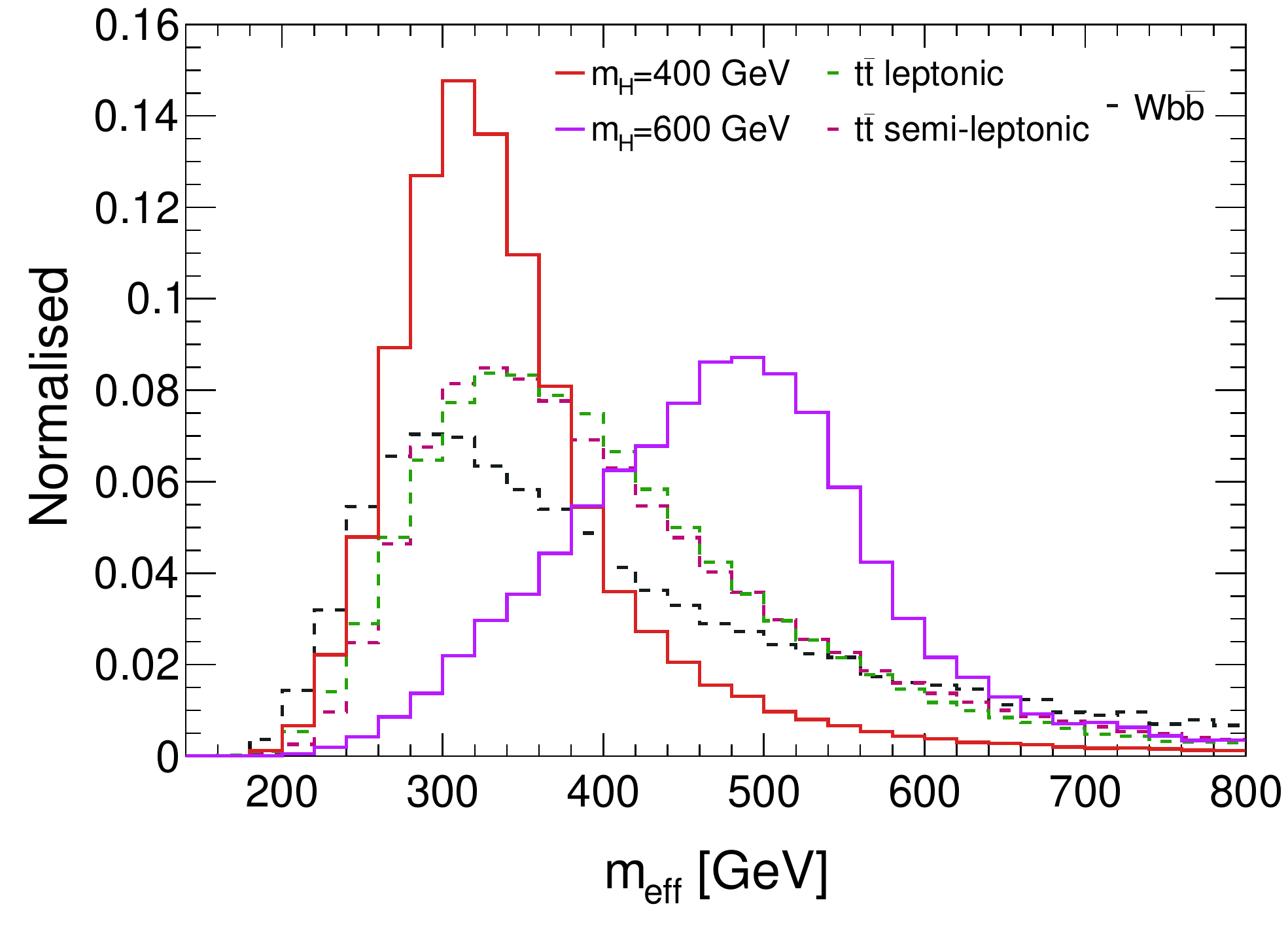}
\caption{The $m_{jj}$, $p_{T,j_1}$, $M_T$ and $m_{\textrm{eff}}$ distributions for the semi-leptonic category for $m_H = 400$ and $600$ GeV with dominant backgrounds. Here the heavy Higgs boson is searched for in the $t\bar{t}$ channel. The distributions are shown before doing the multivariate analysis.}
\label{Htt:1l}
\end{figure}

\begin{figure}
\centering
\includegraphics[scale=0.5]{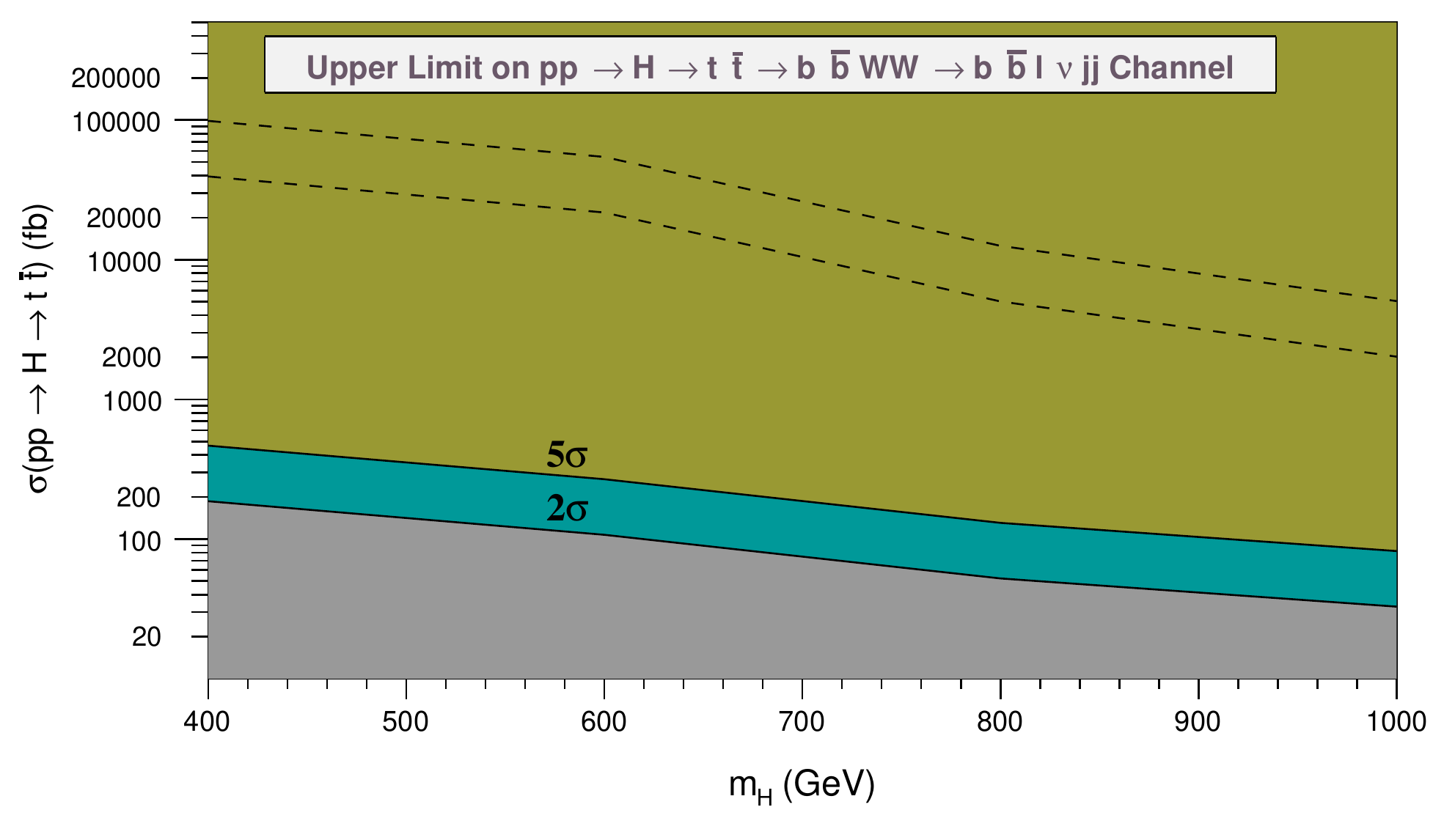}
\caption{Upper limit on $\sigma(pp\to H\to t\bar{t})$ (fb) as a function of $m_{H}$ (GeV) for the $1\ell 2j 2b + \met$ channel. The solid (dashed) lines show the 2$\sigma$-5$\sigma$ band on taking 0\% (5\%) systematic uncertainties.}
\label{h2ttul:1l2jMET}
\end{figure}

We end this section by analysing the semi-leptonic final state ensuing from the semi-leptonic decays of $t\bar{t}$. We select events which contain a single isolated lepton, two $b$-tagged jets and at least two light jets after applying the same set of trigger cuts as discussed in section~\ref{sec:Htohhtobbww}. Finally, we perform a multivariate analysis with the following set of kinematic variables:

\begin{equation}
\begin{split}
p_{T, {bb}},~m_{bb},~\Delta R_{bb},~m_{jj},~m_{\textrm{eff}},~M_T,~m_{t11},\\~m_{t12},
~m_{t\bar{t}1},~m_{t\bar{t}2},~p_{T, \ell\nu},~\Delta R_{bb,jj},~p_{T, \ell_1},~p_{T, j_1}\nonumber,
\end{split}
\end{equation}

\noindent where, $m_{t\bar{t}i}$ are the possible combinations for the invariant mass of the heavy Higgs reconstructed from the top pair. The four most sensitive variables are listed in Fig~\ref{Htt:1l}. The reconstruction procedure is discussed at the beginning of section~\ref{sec:Htohhtobbww}. We show the reconstructed $t\bar{t}$ invariant masses in Fig.~\ref{htt1l:mtt}. Finally, we summarise the boosted decision tree results in Table~\ref{H2ttbar:1l2jMET}. For heavy Higgs mass ranging between 400 GeV and 1 TeV, we show the upper limit on $\sigma(pp\to H\to t\bar{t})$ in Fig.~\ref{h2ttul:1l2jMET}. The $95\%$ CL upper limit varies between $186.57$ fb ($39460.45$ fb) and $32.81$ fb ($2021.51$ fb) for $m_H$ varying between 400 GeV and 1 TeV with zero ($5\%$) systematic uncertainty. The $H \to t\bar{t}$ channel has a small $S/B$ ratio. Hence, adding systematic uncertainty will drastically change the upper limit on the cross-section.

\begin{center}
\begin{table}[htb!]
\centering
\scalebox{0.7}{%
\begin{tabular}{|c|c|c|}\hline
(a) & Process & Events \\ \hline \hline

\multirow{9}{*}{Background}
 & $t\bar{t}$ semi-lep              & $15257053.17$ \\ 
 & $t\bar{t}$ lep                	& $2037363.62$ \\  
 & $Wb\bar{b}+\textrm{jets}$     	& $513737.45$ \\   
 & $t\bar{t}h$                  	& $23963.91$ \\  
 & $t\bar{t}Z$                  	& $20628.78$\\  
 & $t\bar{t}W$                  	& $14852.43$ \\ 
 & $pp\to hh$                   	& $232.54$ \\ 
 & $\ell\ell b\bar{b}$   		    & $25865.11$ \\ \cline{2-3}  
 & \multicolumn{1}{c|}{Total}   	& $17893697.01$ \\ \hline
\end{tabular}}
\bigskip
\scalebox{0.7}{%
\begin{tabular}{|c|c|c|}\hline
(b) & Process & Events \\ \hline \hline
\multirow{9}{*}{Background}
 & $t\bar{t}$ semi-lep            & $14297184.84$ \\ 
 & $t\bar{t}$ lep                 & $1620244.31$ \\  
 & $Wb\bar{b}+\textrm{jets}$      & $435088.64$ \\  
 & $t\bar{t}h$                    & $45147.37$ \\  
 & $t\bar{t}Z$                    & $42620.52$\\  
 & $t\bar{t}W$                    & $29695.00$ \\ 
 & $pp\to hh$                     & $216.70$ \\ 
 & $\ell\ell b\bar{b}$            & $13470.49$ \\ \cline{2-3}   
 & \multicolumn{1}{c|}{Total}     & $16483667.87$ \\ \hline
\end{tabular}}
\quad
\scalebox{0.7}{%
\begin{tabular}{|c|c|c|}\hline
(c) & Process & Events \\ \hline \hline
\multirow{9}{*}{Background}
 & $t\bar{t}$ semi-lep            & $3171586.10$ \\ 
 & $t\bar{t}$ lep                 & $298446.40$ \\  
 & $Wb\bar{b}+\textrm{jets}$      & $185875.57$ \\  
 & $t\bar{t}h$                    & $15846.33$ \\  
 & $t\bar{t}Z$                    & $16523.44$\\  
 & $t\bar{t}W$                    & $12081.76$ \\ 
 & $pp\to hh$                     & $66.60$ \\ 
 & $\ell\ell b\bar{b}$            & $3244.23$ \\ \cline{2-3}   
 & \multicolumn{1}{c|}{Total}     & $3703670.43$ \\ \hline
\end{tabular}}
\quad
\scalebox{0.7}{%
\begin{tabular}{|c|c|c|}\hline
(d) & Process & Events \\ \hline \hline
\multirow{9}{*}{Background}
 & $t\bar{t}$ semi-lep            & $1254581.87$ \\ 
 & $t\bar{t}$ lep                 & $115725.81$ \\  
 & $Wb\bar{b}+\textrm{jets}$      & $123298.89$ \\  
 & $t\bar{t}h$                    & $7762.68$ \\  
 & $t\bar{t}Z$                    & $8767.07$\\  
 & $t\bar{t}W$                    & $6720.41$ \\ 
 & $pp\to hh$                     & $28.36$ \\ 
 & $\ell\ell b\bar{b}$            & $1441.88$ \\ \cline{2-3}   
 & \multicolumn{1}{c|}{Total}     & $1518326.97$ \\ \hline
\end{tabular}}
\caption{Respective background yields for the $1\ell 2j 2b + \met$ channel after the BDT analyses optimised for $m_H = (a)~400$ GeV, $(b)~600$ GeV, $(c)~800$ GeV and $(d)~1$ TeV. The various orders of the signal and backgrounds are same as in Table~\ref{tab1:2l2bMET}.}
\label{H2ttbar:1l2jMET}
\end{table}
\end{center}

\section{The $(H/A)b\bar{b}$ channel}
\label{sec:bbH}

\begin{figure}[htb!]
\centering
\includegraphics[trim=0 130 0 70,clip,width=\textwidth]{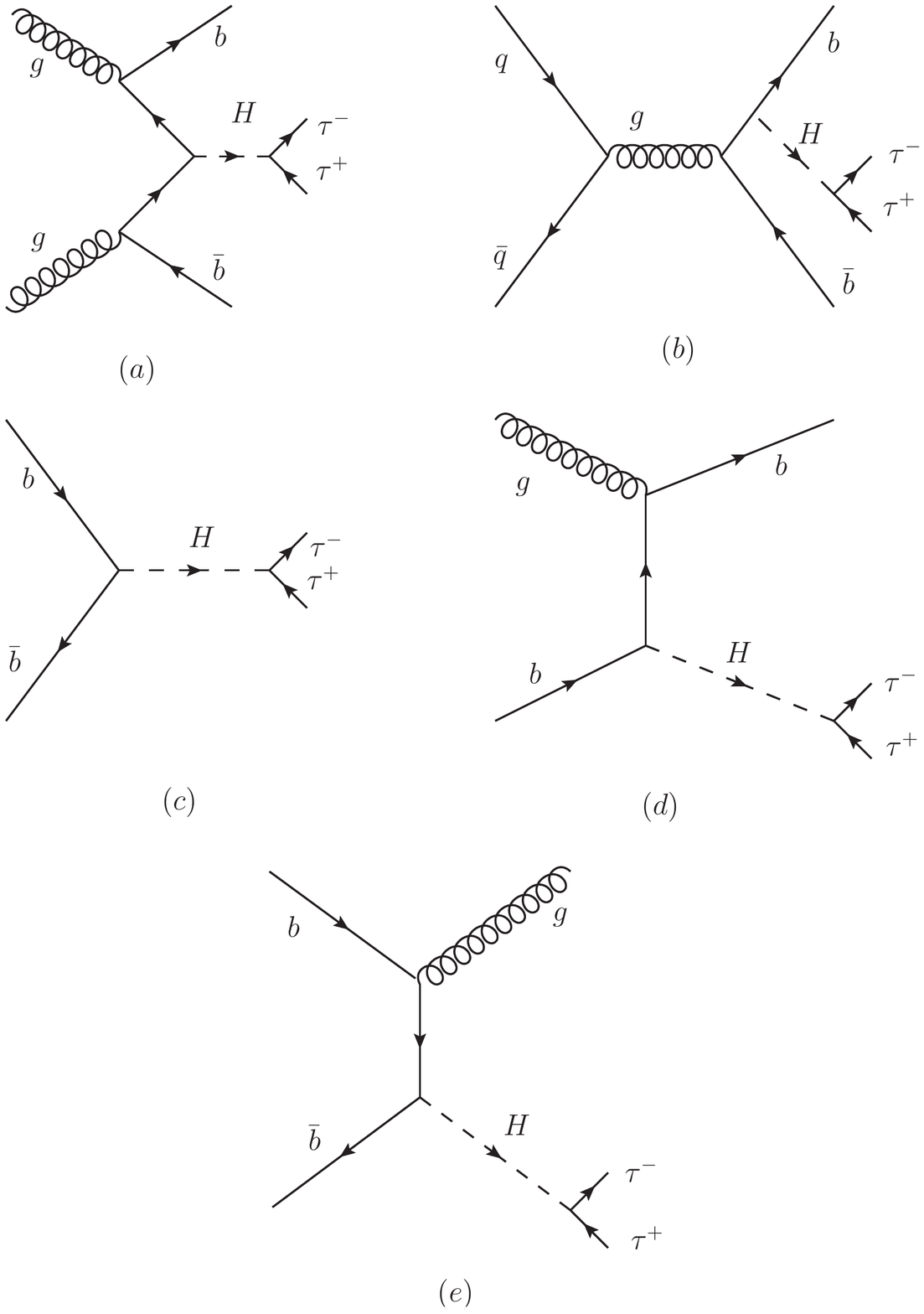}
\caption{The Feynman diagrams for 4F (a) $gg\rightarrow b\bar{b}H$ and (b) $q\bar{q}\rightarrow b\bar{b}H$ process, and 5F (c) $b\bar{b}\rightarrow H$ (LO), (d) $gb\rightarrow bH$ and (e) $b\bar{b}\rightarrow gH$ process.}
\label{FD:bbh}
\end{figure}

Finally, we study the process where the resonant (pseudo)scalar is produced in association with a pair of bottom quarks, \textit{viz.}, $pp \rightarrow (b\bar{b})H/A$. The need to study this process lies in the fact that one can probe and impose strong limits on the lower part in the $m_A-\tan{\beta}$ plane, as will be discussed in section~\ref{sec:pMSSM}. The cross-section of the inclusive $(b\bar{b})H$ process receives contribution from both the 4-flavour (4F) and the 5-flavour (5F) processes. There are two QCD processes (Fig.~\ref{FD:bbh}) contributing to the 4F scheme at LO where the heavy Higgs is produced in association with two $b$-quarks, one is via the gluon fusion process ($gg\rightarrow b\bar{b}H$) and the other is via quarks ($q\bar{q}\rightarrow b\bar{b}H$). The 4F inclusive cross-section suffers from large logarithms due to an almost collinear splitting of a gluon into a pair of bottom quarks. This is of the form $\textrm{ln}(\frac{\mu_F}{m_b})$ ($\mu_F \equiv$ factorisation scale) and may lead to a breakdown of the perturbative theory. However, these logarithms can be absorbed inside the bottom quark parton distribution function (PDF) by re-summing at all orders in the perturbation theory. This forms the basis of the 5F scheme. At leading order (LO), the 5F scheme is dominated by the QCD process $b\bar{b}\rightarrow H$ (Fig.~\ref{FD:bbh}). However, for scenarios involving $b$-jet in the final state, the processes where the resonant scalar is produced in association with a $b$-quark or a gluon, becomes important, \textit{viz.}, $gb~(b\bar{b})\rightarrow bH~(gH)$ (Fig.~\ref{FD:bbh}). Owing to different perturbative expansions, these two schemes give different results when truncated at any finite order. Thus, higher order calculations become important to match these two results. The 4F scheme calculation is available up to NLO in QCD~\cite{Dittmaier:2003ej, Dawson:2005vi, Dawson:2004wq, Dawson:2003kb}, while the 5F scheme is known up to NNLO accuracy in QCD~\cite{Harlander:2003ai}. Here the LO process in the 4F scheme \textit{i.e.} $gg\rightarrow b\bar{b}H$ appears at the NNLO order in 5F scheme. The resonant scalar production in association with a $b$-quark or gluon \textit{i.e.} $gb~(b\bar{b})\rightarrow bH~(gH)$ has been calculated up to NLO in QCD~\cite{Dawson:2004sh} and electroweak (EW)~\cite{Dawson:2010yz}.

It has been argued that with a proper choice of factorisation scale $\sim \frac{m_H}{4}$, the inclusive cross-section in the 4F and 5F schemes agree very well~\cite{Maltoni:2003pn, Boos:2003yi, Plehn:2002vy}. There is a proposed way to combine these two approaches. This is known as the Santander matching~\cite{Harlander:2011} scheme. The total inclusive cross-section is obtained by matching the 4F and 5F scheme numbers in which both these cross-sections are multiplied by their proper weight factors. These weight factors change logarithmically with the heavy scalar mass ($m_H$) because of logarithmic difference between these two scheme approaches. The matched cross-section is computed as follows: $$\sigma^{\text{matched}}=\frac{\sigma^{\text{4FS}}+w\sigma^{\text{5FS}}}{1+w},$$ where $$w=\text{ln}\frac{m_H}{m_b}-2$$ is the weight factor\footnote{Here, $m_b$ is the bottom quark pole mass which enters in the re-summed logarithms.}. The analysis can be subdivided according to the number of $b$-tagged jets. However, we specifically focus on the category with $\geq 1$ b-jets upon following a recent study performed by the ATLAS collaboration~\cite{Aaboud:2017sjh}. Furthermore, we consider the heavy Higgs decaying to a pair of $\tau$-leptons and we specifically focus on the scenario where both the $\tau$s decay hadronically. The $H$ and $A$ masses are varied between $200$ GeV and $1$ TeV.

The various backgrounds at play are $Z/\gamma^*+$ jets, multijets, $W+$ jets, $VV$ ($V = W^{\pm},Z$), $t\bar{t}$ and single top. The $Z/\gamma^*+$ jets with the $Z$-boson decaying to a pair of leptons ($e$, $\mu$ and $\tau$) is the dominant background for the $\tau_h\tau_{\ell}$ category (a category that we will not address in the present work) but also gives significant contribution to the $\tau_h\tau_h$ category. We simulate this background merged with three additional partons and some specific generation level cuts which are described in Appendix-\ref{sec:appendixA}. Similarly, the $W+$ jets is also generated with up to three additional partons and the $W$-boson is then decayed leptonically. In order to include the dominant multijets background in the $\tau_h\tau_h$ category, we generate an exclusive $b\bar{b}jj$ sample where $j$ includes light quarks and gluon. These light jets can fake hadronically decaying $\tau$s. Finally, we include the top-quark related backgrounds \textit{viz.}, $t\bar{t}$ and single top. Next, we describe our analysis for the $\geq 1$ $b$-tagged jets category upon closely following Ref.~\cite{Aaboud:2017sjh}.

\subsection{The $\tau_h\tau_h$ Channel : $b$-tag category}
\label{bbH:hh}

\begin{figure}
\centering
\includegraphics[scale=0.37]{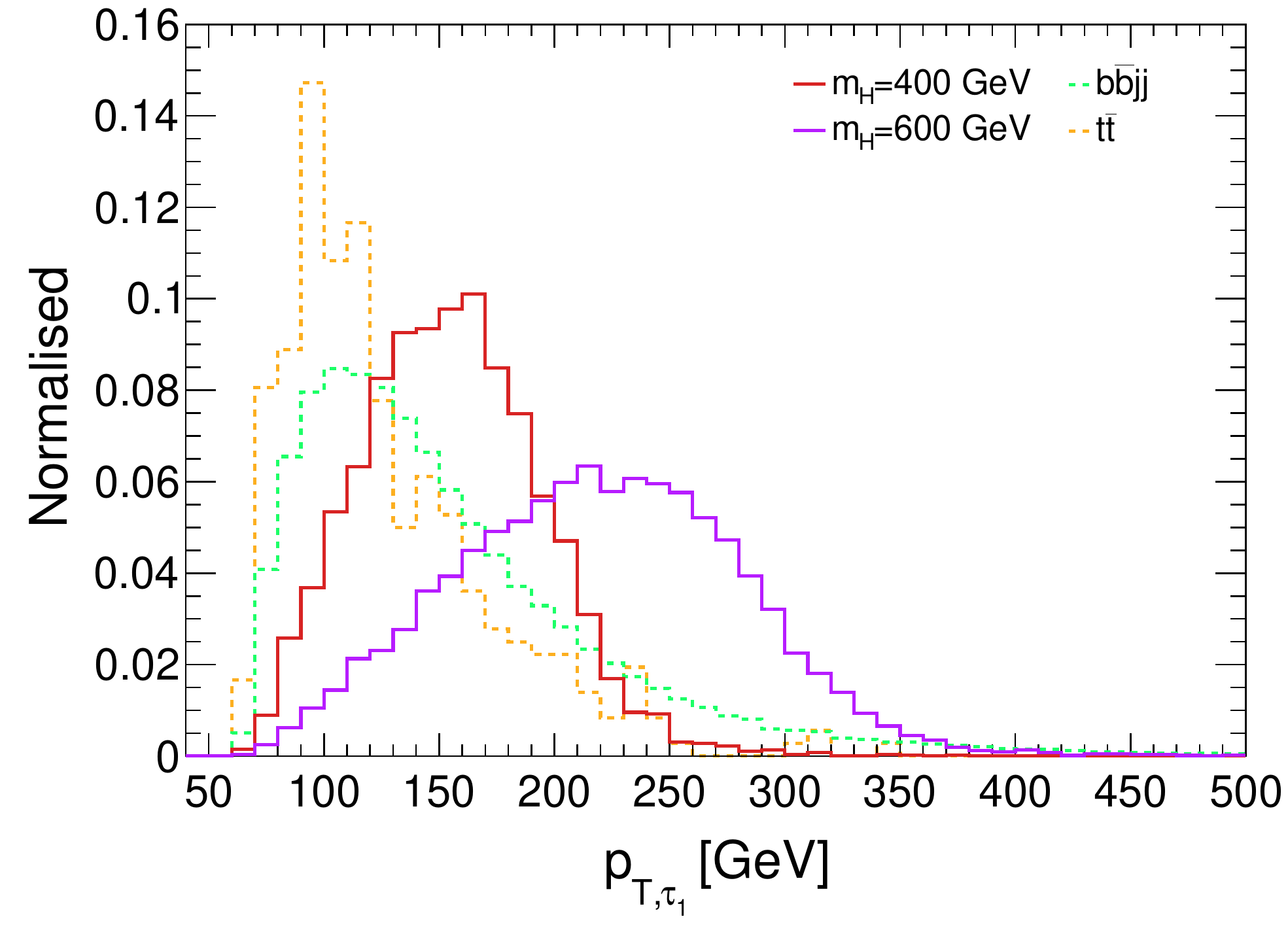}
\includegraphics[scale=0.37]{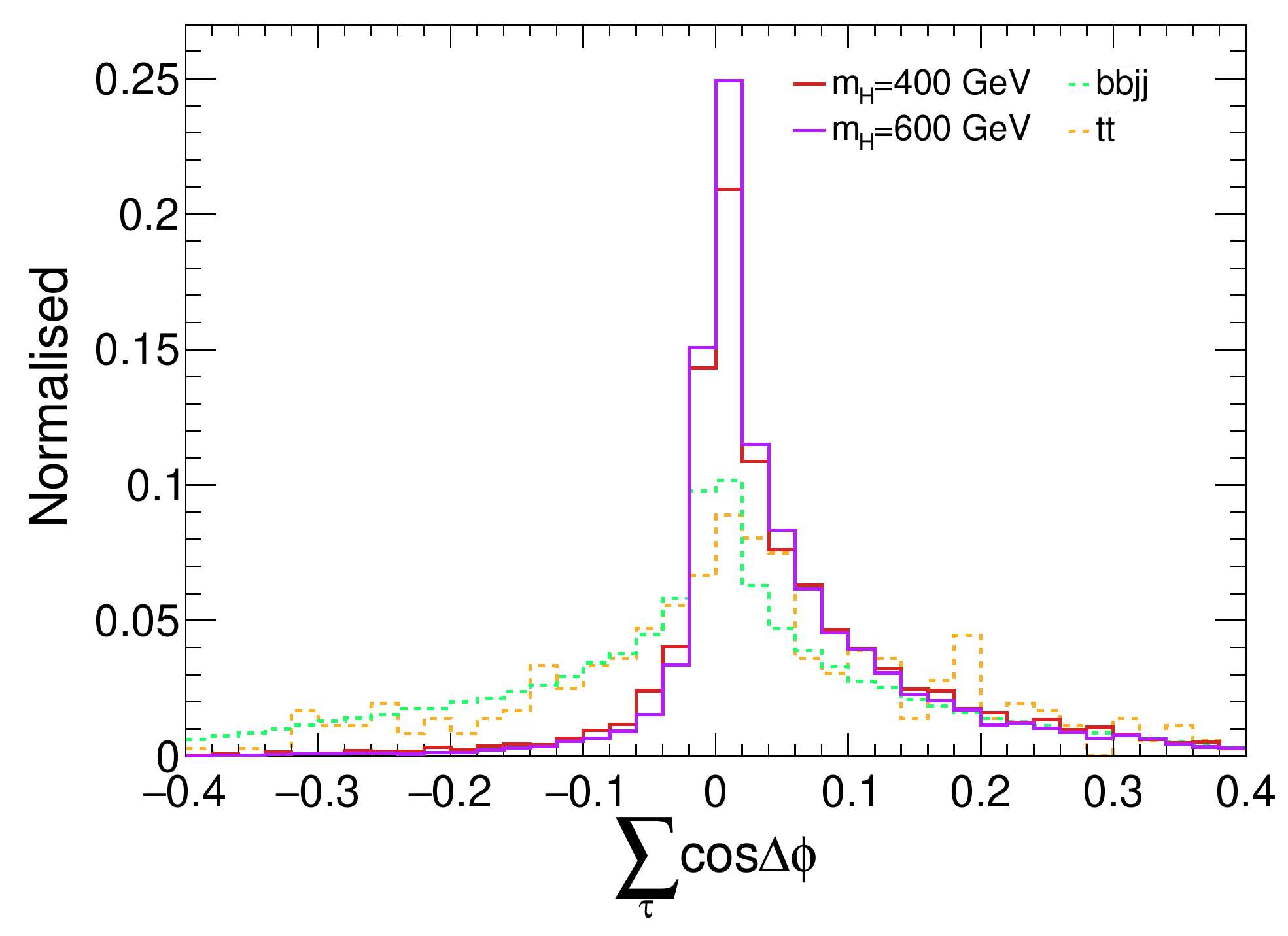} \\
\includegraphics[scale=0.37]{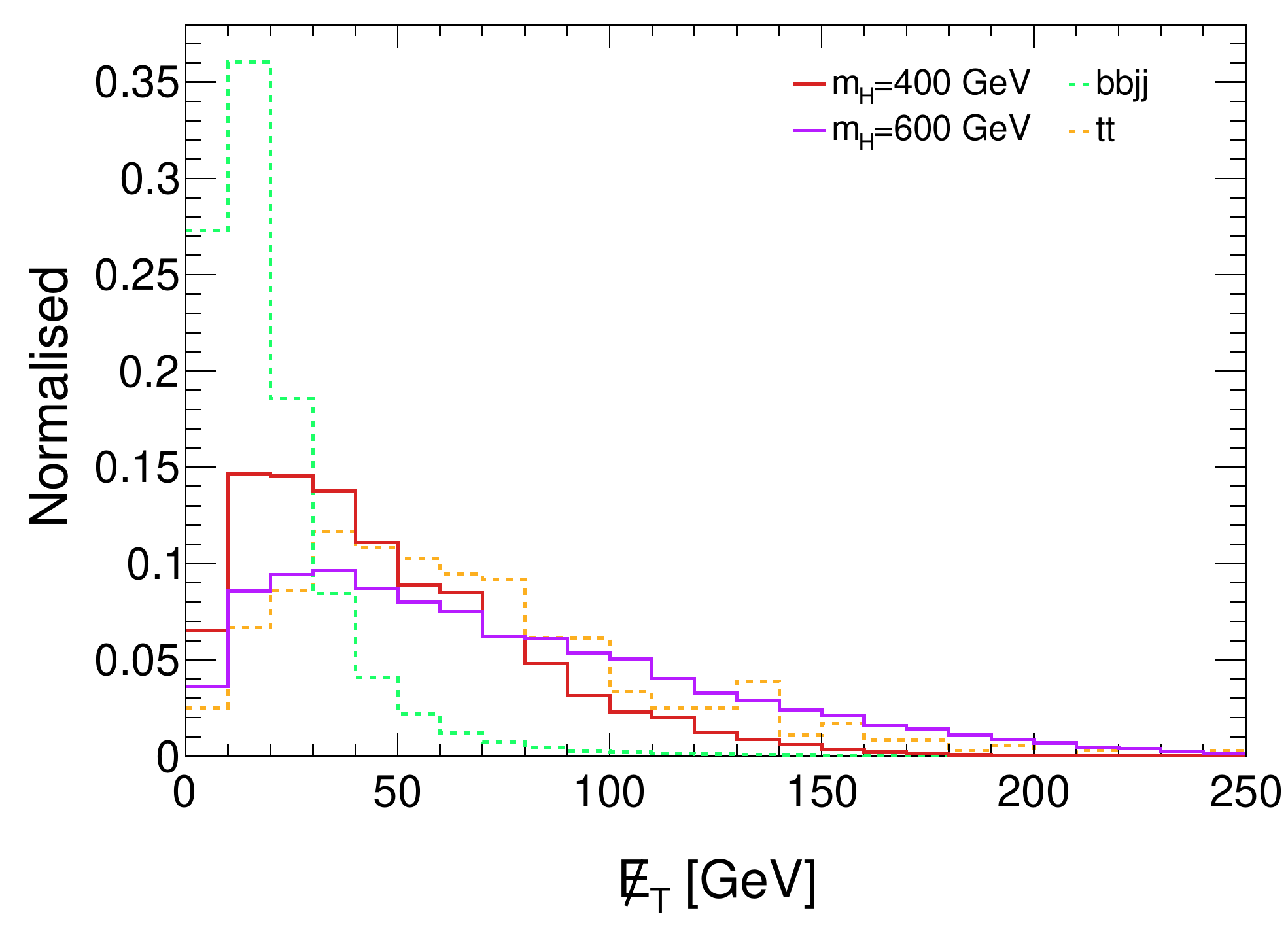}
\includegraphics[scale=0.37]{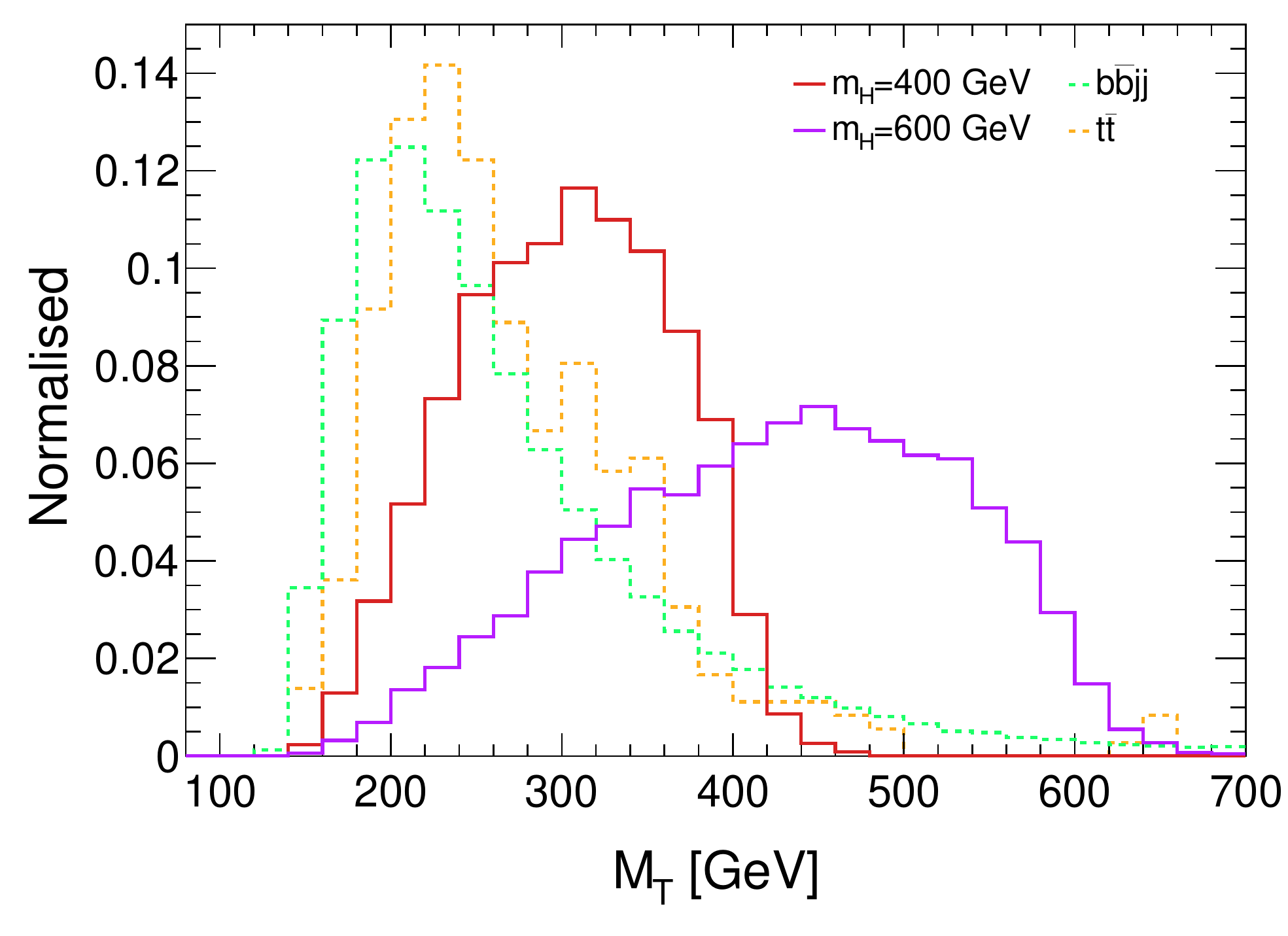}
\caption{The $p_{T,\tau_1}$, $\sum\limits_{\tau_{1,2}}^{} \cos\Delta\phi$, $\met$ and $M_T$ distributions for the $b\bar{b}\tau_h\tau_h$ category for $m_H = 400$ and $600$ GeV with dominant backgrounds. Here the heavy Higgs boson is searched for in the $b\bar{b}H$ channel. The distributions are shown before the optimisation analysis.}
\label{bbH:2tauH1}
\end{figure}

\begin{figure}
\centering
\includegraphics[scale=0.5]{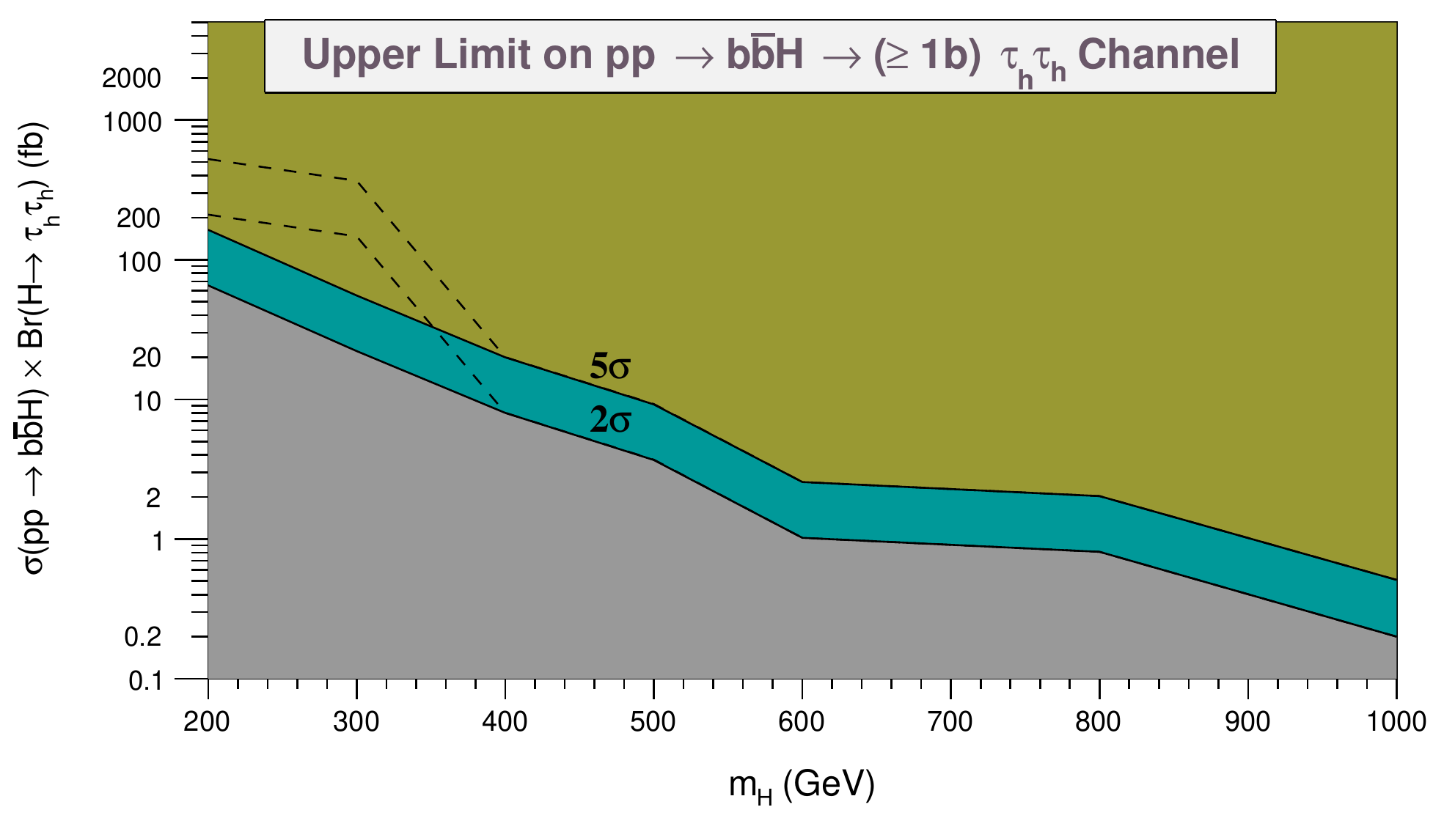}
\caption{Upper limit on $\sigma(pp\to b\bar{b} H \to b\bar{b}\tau^+\tau^-)$ (fb) as a function of $m_{H}$ (GeV) for the $\geq 1b + 2\tau_h$ channel. The solid (dashed) lines show the 2$\sigma$-5$\sigma$ band on taking 0\% (5\%) systematic uncertainties.}
\label{bbH:2tauH2}
\end{figure}

We select events containing at least one $b$-tagged jet with $p_T > 20$ GeV and two $\tau$-tagged jets with $p_T > 65$ GeV~\footnote{Before performing this analysis, we validated our setup with the ATLAS analysis at 13 TeV. The validation is shown in Appendix~\ref{sec:appendixB}.}. These two $\tau$-tagged jets must have opposite electric charge (from their track reconstruction). We also veto events having leptons ($e,\mu$) or $\tau$-tagged jets with $1.37<|\eta_\tau|<1.52$, in the final state. The azimuthal angle separation between the two $\tau$-tagged jets has to fulfil the condition, $|\Delta\phi(\tau,\tau)|>2.7$. The $b$- and the $\tau$-jets must have an angular separation in the $\eta-\phi$ plane, \textit{viz.}, $\Delta R(b,\tau)>0.2$. Besides, we also impose a minimum bound on the visible invariant mass of the two hadronically decaying $\tau$ leptons to be $m_{\tau\tau}^{vis.} >50$ GeV. For the fake $b\bar{b}jj$ background, we demand the two light jets to satisfy the $\tau$ jet configuration during the analysis and we later multiply the event yield with the $j\to\tau$ fake rate. Similarly, for the $W(\to\tau\nu \; \textrm{or} \; \ell\nu)+$ jets background, we demand at least one extra light jet satisfying the $\tau$ jet requirement on top of the $\tau$ jet ensuing from $W$-boson decay. After imposing the aforementioned cuts, we improve our analysis by optimising over some other kinematic variables \textit{viz.}, the transverse momentum of the hardest $\tau$-tagged jet ($p_{T,\tau_1}$), sum of the cosine of the azimuthal angle separation between the $\tau$-jets and $\met$ ($\sum\limits_{\tau_{1,2}}^{} \cos\Delta\phi$) and the transverse mass of the total system which is defined below,

$$M_T = \sqrt{(p_{T,\tau_1} + p_{T,\tau_2} + \met)^2 - (\vec{p}_{T,\tau_1} + \vec{p}_{T,\tau_2} + \vec{\met})^2}~,$$ where the symbols have their usual meanings. These four kinematic variables are shown in Fig.~\ref{bbH:2tauH1}. The optimised cuts along with the signal efficiencies and the background yields for each benchmark point are shown in Table~\ref{tab:bbH_hh}. We perform our analysis upon considering both the 4F and 5F signal samples separately. Finally, we add them by multiplying these cross-sections with the aforementioned weight factor in order to obtain the upper limit on the matched $b\bar{b}H$ production cross section. We show the 95\% and 99.7\% exclusion for $\sigma(p p \to b\bar{b}H)\times \textrm{BR}(H \to \tau_h \tau_h)$ in Fig.~\ref{bbH:2tauH2}. The 95\% upper limit varies between $22.16$ fb and $3.68$ fb (within $[146.91,~3.70]$ fb with $5\%$ systematic uncertainty) for $m_H$ varying between 300 GeV and 500 GeV. This is close to an order of magnitude improvement over the existing bounds at 13 TeV~\cite{Aaboud:2017sjh}. The effects of systematic uncertainties for this $b\bar{b}H$ channel become negligible for $m_H > 400$ GeV. However, for lower masses, the backgrounds are larger and thus the inclusion of uncertainties weaken the limits.

\begin{table}[htb!]
    \begin{bigcenter}
        \scalebox{0.7}{%
            \begin{tabular}{||c||c|c|c|c|c|c||}
                \hline
                Heavy Higgs mass, & \multicolumn{4}{c|}{Optimised cuts (GeV)} & \multicolumn{2}{c||}{After all cuts}   \\\cline{2-7}
                
                $m_H$ (GeV) & $p_{T,\tau_1}>$ & $\sum\limits_{\tau}^{} \cos\Delta\phi>$ & $M_{T}$ & $\met>$ & Signal Efficiency ($\times 10^{-4}$) & Background yield at $3000 \; \textrm{fb}^{-1}$   \\\hline
                
                $200$  & $70$  & $-0.10$ & [$80$ , $200$]   & $0$    & $7.94$  & $3725.90$ \\\hline
                $300$  & $75$  & $-0.06$ & [$160$ , $320$]  & $0$    & $49.40$ & $17172.83$ \\\hline
                $400$  & $180$ & $-0.04$ & [$380$ , $400$]  & $80$   & $1.39$  & $2.22$ \\\hline
                $500$  & $180$ & $-0.02$ & [$380$ , $420$]  & $80$   & $5.14$  & $5.16$ \\\hline
                $600$  & $240$ & $-0.10$ & [$500$ , $580$]  & $140$  & $5.38$  & $0.62$ \\\hline
                $800$  & $260$ & $0.00$  & [$400$ , $840$]  & $220$  & $7.91$  & $0.62$ \\\hline
                $1000$ & $260$ & $-0.02$ & [$780$ , $1020$] & $200$  & $23.03$ & $0.36$  \\\hline
               
        \end{tabular}}
    \end{bigcenter}
    \caption{The details of final optimised cuts with signal efficiency and background yields after all the applied cuts.}
    \label{tab:bbH_hh}
\end{table}

\section{The future of the pMSSM parameter space}
\label{sec:pMSSM}

The Higgs sector in the MSSM comprises two Higgs doublets which give rise to five massive Higgs states after the electroweak symmetry breaking. The Higgs spectrum is thus composed of two $CP$-even scalars, $h$ and $H$, one $CP$-odd scalar, $A$, and two charged scalars, $H^{\pm}$ (detailed studies on the Higgs sector of MSSM can be found in Ref.~\cite{Djouadi:2015jea,Zhao:2017qpe}). In addition to the extended Higgs sector, the SUSY particle spectrum boasts a multitude of particles, \textit{viz.}, the sleptons, squarks, gluinos and electroweakinos. A majority of direct searches at the LHC have excluded stops and gluinos below the TeV scale (Refs.~\cite{Moortgat:2001pp,Belanger:2015vwa,Aaboud:2018kya, Aaboud:2018mna, Aaboud:2018ujj, CMS-PAS-SUS-18-002, CMS-PAS-SUS-16-017, CMS-PAS-SUS-17-012, Sirunyan:2017bsh} show some such limits in various supersymmetric interpretations). This more or less nullifies the prospect of observing these particles unless the luminosity is enhanced significantly.  The electroweakino sector has also been probed in numerous studies~\cite{Zhang:2002fu,Ibrahim:2008rq,ATLAS:2009zmv,Heinemeyer:2015pfa,Medina:2017bke,
Gori:2018pmk} and bounds have been obtained on their masses within simplified scenarios~\cite{Kulkarni:2017xtf,Aaboud:2018htj, Aaboud:2018sua, Aaboud:2018zeb, Aaboud:2018jiw, Aaboud:2017leg, Aaboud:2017mpt, Aaboud:2017nhr, Sirunyan:2018lul, Sirunyan:2018ubx}. In many of these studies, the electroweakino masses are excluded from between a few hundred GeVs to about half a TeV and are comparatively weakly coupled compared to the gluinos and stops. Within a generic SUSY parameter space without any correlation between the choice of the electroweakino mass parameters, these bounds can become considerably weaker. The ATLAS and CMS collaborations have also performed several studies to search for resonant Higgs through their decay into SM final states~\cite{Aaboud:2017rel, Aaboud:2017gsl, Aaboud:2017sjh,CMS:2017epy,
Aaboud:2017yyg,CMS:2017ihs, CMS-PAS-HIG-17-009,Sirunyan:2017djm, 
Aaboud:2018cwk, Aaboud:2018sfw}. However, none of these searches could find any significant excess over the SM expectations and thus only imposed upper limits on the production cross-section of the heavy Higgs bosons times their branching ratio into various SM final states. In this section, we present a brief discussion on the relevant constraints and discuss the parameter scan for the 14 TeV HL-LHC. We follow this up with an analysis to capture the present status of the MSSM parameter space in light of the latest results from the Run-II data ($13~{\rm TeV}, 36~{\rm fb^{-1}}$) of LHC. Finally, we study the implications of the projected heavy Higgs direct search limits derived in sections~\ref{sec:Htohh},~\ref{sec:Htott} and~\ref{sec:bbH}, on the MSSM parameter space. 

The initial constraint on the parameter space ensues from the allowed mass of the 125 GeV SM-like Higgs boson. A combined measurement by the ATLAS and CMS collaborations constrains $m_{h}$ within the range [124.4,125.8] GeV at $3\sigma$. It is to be duly noted that in the context of MSSM, the available calculation of the Higgs mass is not exact. Thus, in order to correctly account for the existing uncertainties, we allow a window of $\pm$3 GeV about 125 GeV and restrict the light Higgs mass in our parameter space to lie within [122, 128] GeV. Furthermore, both collaborations have performed numerous measurements on the coupling strengths of the SM-like Higgs bosons. These results are presented through bounds on the signal strength variable ($\mu^{if}$) which is defined as follows:

\begin{equation}
\mu_{if} = \frac{\sigma_{i} \times \textrm{BR}_{f}}{\sigma_{i}^{SM} \times \textrm{BR}_{f}^{SM}},
\end{equation}

\noindent where, $\sigma_{i}$ represents the MSSM (or any specific model in question) Higgs production cross-section in the $i^{th}$ production mode ($i~=~ggF$, $VBF$, $t\bar{t}h$ or $Vh$) at the LHC and $\sigma_{i}^{SM}$ denotes the corresponding SM cross-section. BR$_{f}$ corresponds to the branching fraction of the SM-like Higgs into a particular SM final state ($f$ = $WW,ZZ,b\bar{b},\gamma\gamma,\tau \tau$) and BR$_{f}^{SM}$ is the corresponding SM value. We apply all these constraints over our parameter space by demanding that all our signal strengths simultaneously lie within 2$\sigma$ of their experimental counterparts. The latest Higgs signal strengths (13 TeV, $36~{\rm fb^{-1}}$) measured by both the CMS and ATLAS collaborations are listed in Table~\ref{table-mu13}.

Additionally, the flavour physics bounds also potentially constrain the MSSM parameter space, as shown in \cite{Barman:2016jov}. In this regard, the bounds on the branching fraction of rare $B$-decay processes: $B \to X_{s}\gamma$, $B_{s} \to \mu^{+}\mu^{-}$, $B^{+} \to \tau^{+}\nu_{\tau}$, are among the most sensitive probes of new physics searches. As shown in \cite{Barman:2016jov}, constraints from $B \to X_{s}\gamma$ disfavours the low $M_{A}$ regime while bounds on $Br(B_{s} \to \mu^{+}\mu^{-})$ constrains the low $M_{A}$ and high $\tan\beta$ regions. The low $M_{A}$ and high $\tan\beta$ region gets further constrained by $B^{+} \to \tau^{+}\nu_{\tau}$. On the other hand, the current limits from direct heavy Higgs searches in $b\bar{b}H/A ,~H/A \to \tau^{+}\tau^{-}$, imposes much stringent constraint in the $M_{A} \gtrsim 300~{\rm GeV}$ region and excludes $\tan\beta$ up to $\sim 18$ for $M_{A} \sim 1~{\rm TeV}$ (a detailed discussion concerning this can be found in \cite{Barman:2016jov}). Consequently, within the scope of this section where our major emphasis is on exploring the future reach of direct heavy Higgs searches at the HL-LHC, we do not consider the implications from flavour physics bounds and impose only the light Higgs mass constraint and Higgs signal strength constraints in order to obtain the allowed parameter space region relevant for studying the current and future reach of direct heavy Higgs searches on the MSSM parameter space.

\begin{table}[!htbp]
\renewcommand{\arraystretch}{1.2}
\begin{center}
\begin{tabular}{| c || c | c || c | c ||}
\hline \hline
\centering 
\multirow{3}{*}{\shortstack{Production \\ mode}}  & \multicolumn{2}{c|}{CMS~\cite{Sirunyan:2018koj}} & \multicolumn{2}{c|}{ATLAS} \\ \cline{2-5}
 & \multirow{2}{*}{\shortstack{Decay \\ channel}}  & \multirow{2}{*}{\shortstack{Best fit \\ value}} & \multirow{2}{*}{\shortstack{Decay \\ channel}} & \multirow{2}{*}{\shortstack{Best fit \\ value}} \\ 
& & & & \\\hline 
\multirow{6}{*}{$ggh$} & $b\bar{b}$ & $2.51^{+2.43}_{-2.01}$  & $WW$ & $1.21^{+0.22}_{-0.21}$~\cite{Aaboud:2018jqu}\\
 & $\tau^+\tau^-$ & $1.05^{+0.53}_{-0.47}$ & $ZZ$ & $1.17^{+0.41}_{-0.50}$~\cite{ATLAS-CONF-2017-047} \\
 & $WW^*$ & $1.35^{+0.21}_{-0.19}$ & $\gamma\gamma$ & $0.81^{+0.19}_{-0.18}$~\cite{Aaboud:2018xdt} \\
 & $ZZ^*$ & $1.22^{+0.23}_{-0.21}$ & & \\
 & $\gamma\gamma$ & $1.16^{+0.21}_{-0.18}$ & & \\ \hline 
 \multirow{5}{*}{$VBF$} & $\tau\tau$ & $1.12^{+0.45}_{-0.43}$ & $b\bar{b}$ & $3.00^{+1.70}_{-1.60}$~\cite{Aaboud:2018gay}  \\
  & $WW^*$ & $0.28^{+0.64}_{-0.60}$ & $WW$ & $0.62^{+0.37}_{-0.36}$~\cite{Aaboud:2018jqu} \\
 & $ZZ^*$ & $-0.09^{+1.02}_{-0.76}$ & $\gamma\gamma$ & $2.00^{+0.60}_{-0.50}$~\cite{Aaboud:2018xdt} \\
 & $\gamma\gamma$ & $0.67^{+0.59}_{-0.46}$ & & \\ \hline 
\multirow{4}{*}{$Wh$} & $b\bar{b}$ & $1.73^{+0.70}_{-0.68}$  & $b\bar{b}$  & $1.08^{+0.47}_{-0.43}$~\cite{Aaboud:2018zhk} \\
 & $WW^*$ & $3.91^{+2.26}_{-2.01}$ & $b\bar{b}$  & $1.21^{+0.45}_{-0.42}$~\cite{Aaboud:2017xsd} \\
 & $ZZ^*$ & $0.00^{+2.33}_{-0.00}$ & & \\
 & $\gamma\gamma$ & $3.76^{+1.48}_{-1.35}$ & & \\\hline 
 \multirow{4}{*}{$Zh$} & $b\bar{b}$ & $0.99^{+0.47}_{-0.45}$  & $b\bar{b}$  & $1.20^{+0.33}_{-0.31}$~\cite{Aaboud:2018zhk}  \\
 & $WW^*$ & $0.96^{+1.81}_{-1.46}$ & $b\bar{b}$  & $0.69^{+0.35}_{-0.33}$~\cite{Aaboud:2017xsd}\\
 & $ZZ^*$ & $0.00^{+4.26}_{-0.00}$ & & \\
 & $\gamma\gamma$ & $0.00^{+1.14}_{-0.00}$ & & \\\hline 
 \multirow{5}{*}{$t\bar{t}h$} & $b\bar{b}$ & $0.91^{+0.45}_{-0.43}$  & $\frac{\sigma_{t\bar{t}h}}{\sigma_{t\bar{t}h_{SM}}}$ & $0.84^{+0.64}_{-0.61}$~\cite{Aaboud:2017rss} \\
 & $\tau^+\tau^-$ & $0.23^{+1.03}_{-0.88}$ & $b\bar{b}$ &  $0.80^{+0.60}_{-0.60}$~\cite{Aaboud:2017jvq} \\
 & $WW^*$ & $1.60^{+0.65}_{-0.59}$ & $\gamma\gamma$ & $0.60^{+0.70}_{-0.60}$~\cite{Aaboud:2017jvq} \\
 & $ZZ^*$ & $0.00^{+1.50}_{-0.00}$ & & \\
 & $\gamma\gamma$ & $2.18^{+0.88}_{-0.75}$ & & \\\hline 

\end{tabular}
\caption{Best-fit value of signal strength variables, along with the associated errors, derived by ATLAS and CMS using LHC Run-II data, which have been imposed on the parameter space region.}
\label{table-mu13}
\end{center}
\end{table}

In order to evaluate the current allowed region in the parameter space of the phenomenological MSSM (pMSSM), we perform a random scan over a wide range of pMSSM input parameters, as described below. The parameters relevant to our study are the pseudo-scalar mass variable ($m_{A}$), ratio of the vacuum expectation values of the two Higgs doublets ($\tan\beta$), the third generation soft squark mass parameters ($M_{\tilde{Q}_{3}},~M_{\tilde{u}_{3}},~M_{\tilde{d}_{3}}$), the trilinear coupling of the stop ($A_{t}$) and sbottom ($A_{b}$) and the gluino mass parameter ($M_{3}$). These parameters are varied in the following range:

\begin{equation}
\begin{aligned}
 1 < \tan\beta < 60, \; 200~{\rm GeV} <  m_{A} < 1~{\rm TeV}, \; 1~{\rm TeV} < M_{3} < 10~{\rm TeV} \\
1~ \mathrm{TeV}~<~M_{\tilde{Q}_{3},\tilde{u}_{3},\tilde{d}_{3}}~<~ 20~{\rm TeV}, \; -10~{\rm TeV} < A_{t,b} < 10~{\rm TeV} \\
1~ \mathrm{TeV}~<~M_{\tilde{Q}_{1},\tilde{u}_{1},\tilde{d}_{1}}~<~ 20\ \mathrm{TeV}, \; M_{\tilde{Q}_{2}} = M_{\tilde{Q}_{1}},
M_{\tilde{u}_{2}} = M_{\tilde{u}_{1}}, M_{\tilde{d}_{2}} = M_{\tilde{d}_{1}}\\
A_{e,\mu,\tau,u,d,c,s} = 0, \; M_{\tilde{e}_{1_{L}},\tilde{e}_{1_{R}},\tilde{e}_{2_{L}},\tilde{e}_{2_{R}},\tilde{e}_{3_{L}},\tilde{e}_{3_{R}}}~=~3~{\rm TeV}, \;600~ \mathrm{GeV}~<~M_{1,2},\mu~<~ 5~{\rm TeV}
\end{aligned}
\label{Eqn:para_space_1}
\end{equation}

The bino, wino and higgsino mass parameters, \textit{viz.}, $M_{1},~M_{2}$ and $\mu$ respectively, are varied from 600 GeV in order to prevent the heavier Higgs bosons from having any decays to the electroweakinos. This choice ensures only SM final states for the heavy Higgs boson decays. The second generation soft squark mass parameters ($M_{\tilde{Q}_{2},\tilde{u}_{2},\tilde{d}_{2}}$) are taken to be equal to their corresponding first generation counterparts ($M_{\tilde{Q}_{1},\tilde{u}_{1},\tilde{d}_{1}}$). The slepton mass parameters ($\tilde{e}_{1_{L},1_{R},2_{L},2_{R},3_{L},3_{R}}$) are fixed at $3~{\rm TeV}$ while the trilinear couplings of the first and the second generation squarks ($A_{u,d,c,s}$) and all three generations of sleptons ($A_{e,\mu,\tau}$) are taken to be zero.

The particle spectra and the branching fractions of the SM and SUSY particles are obtained using \texttt{FeynHiggs}~\cite{Heinemeyer:1998yj}. We consider only those parameter points which satisfy the light Higgs mass constraint defined above. Furthermore, we allow only those points which lie within 2$\sigma$ uncertainty of each of the signal strength variables listed in Table~\ref{table-mu13}. The parameter space points which are allowed by the aforementioned light Higgs mass constraint and the Higgs signal strength constraints are referred to as the allowed parameter space points in the remainder of this section and are shown in grey in Fig.~\ref{fig:bbh_tt_curr}.

\begin{figure}
\begin{center}

\includegraphics[scale=0.15]{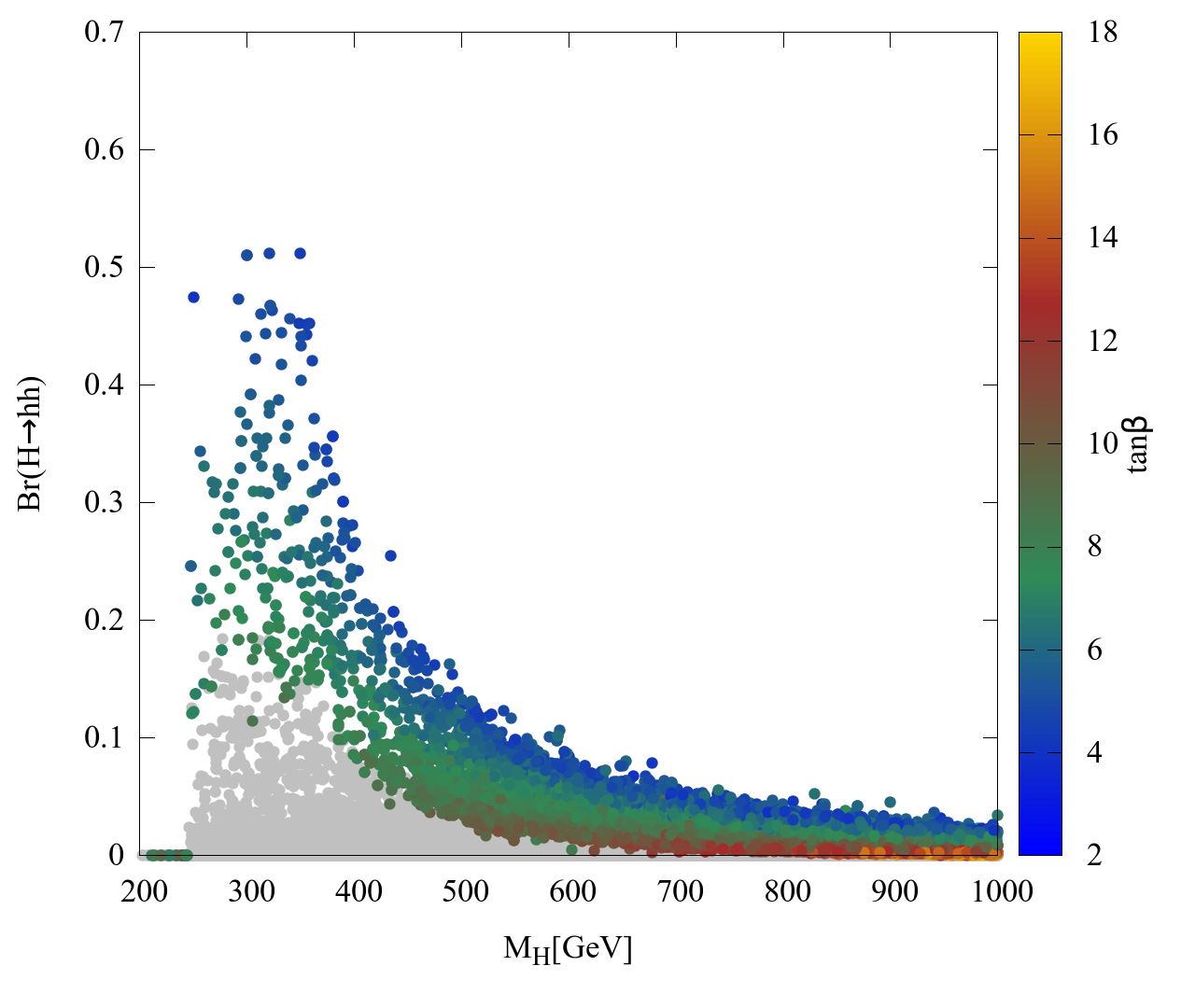}
\includegraphics[scale=0.15]{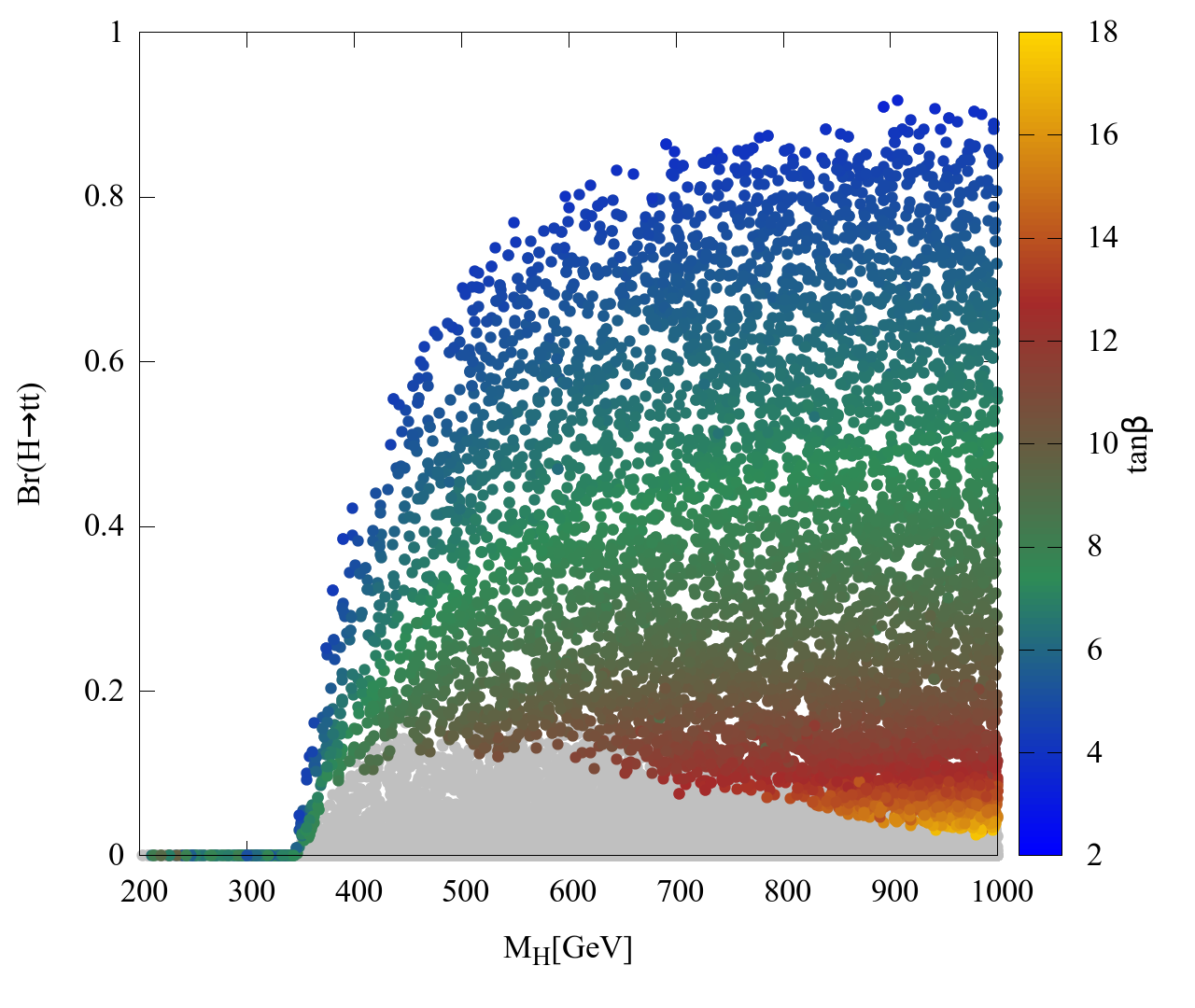}
\includegraphics[scale=0.15]{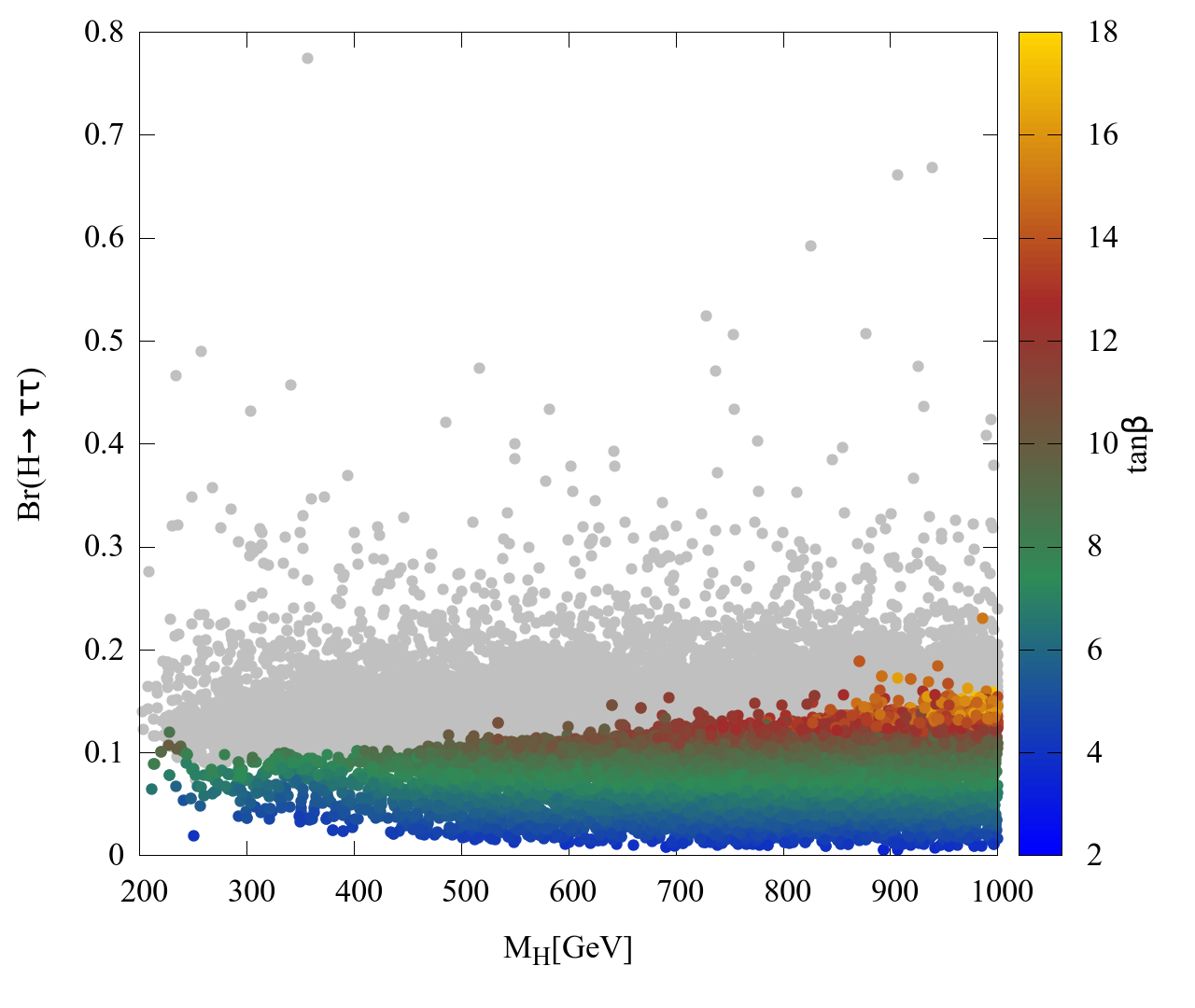}
\caption{Branching ratios of $H \to hh, H \to t\bar{t}$ and $H \to \tau^+\tau^-$ as a function of $m_H$. All the points are allowed by the SM-like Higgs mass and Higgs signal strength constraints. The grey points are excluded by the present direct searches for the heavy Higgs boson.}
\label{fig:SUSY_BR}
\end{center}

\end{figure}

\begin{figure}
\begin{center}

\includegraphics[scale=0.15]{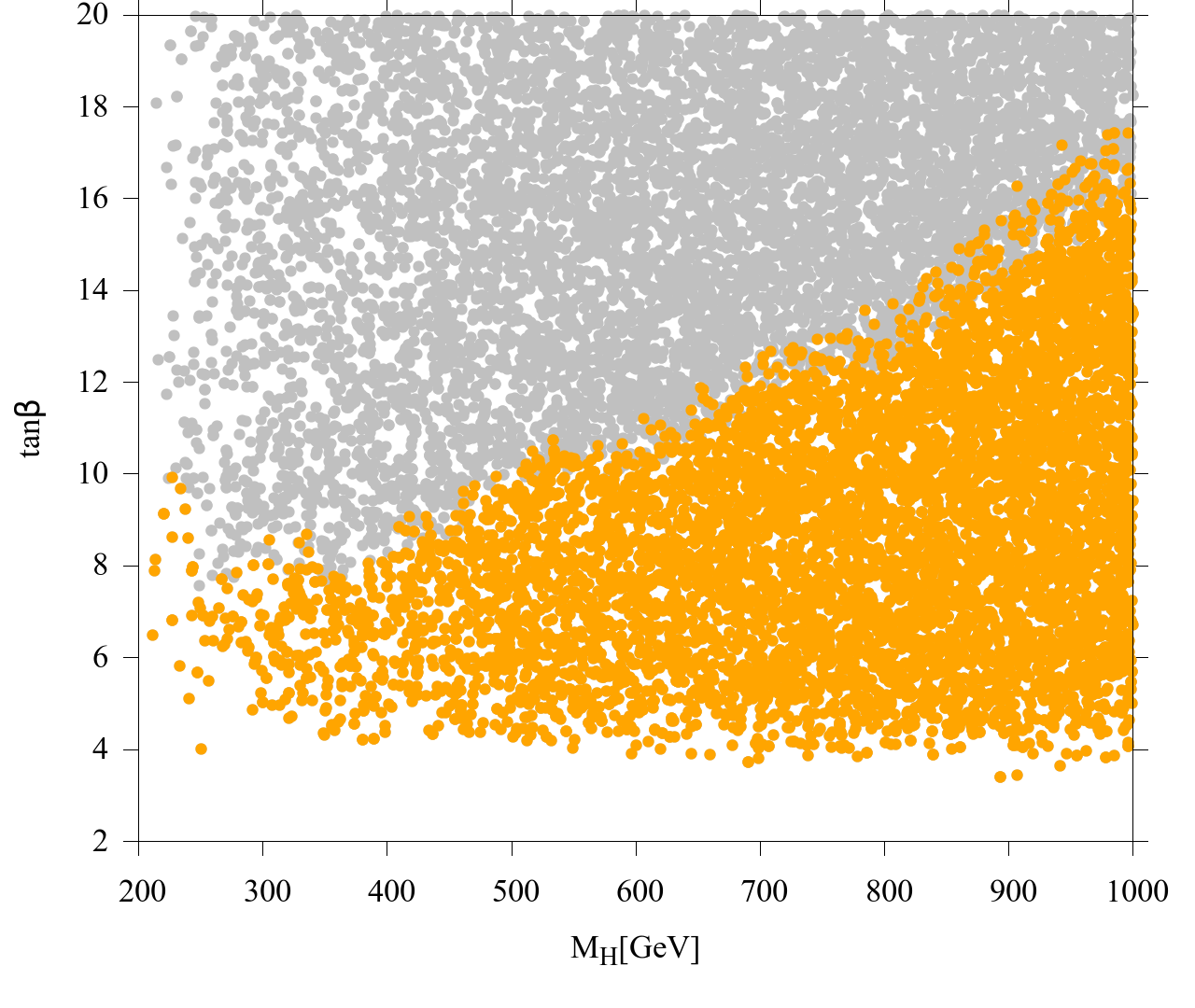}
\caption{Scatter plot in the $m_{A}-\tan\beta$ plane showing the current status of the pMSSM parameter space. All parameter space points satisfy constraints from the Higgs mass measurement and the Higgs signal strengths. The grey coloured points are excluded by the latest direct search limits from $\sigma_{b\bar{b}H/A} \times Br(H/A \to \tau\tau)$ derived by CMS and ATLAS using the Run-II dataset with an integrated luminosity of $\sim 36~ \ifb$.}
\label{fig:bbh_tt_curr}
\end{center}

\end{figure}

The ATLAS and CMS collaborations have also performed numerous searches for the heavy Higgs bosons through their decay into the SM final states, however, none of these searches have been able to observe any significant excess over the SM expectation. Consequently, upper limits have been set on the production cross-section of the heavy Higgs boson ($\sigma_{H/A}$) times its branching ratio into SM states. In this analysis, we consider the latest search limits on $\sigma_{ggH} \times Br(H \to ZZ,WW,\tau\tau)$~\cite{Aaboud:2017rel, Aaboud:2017gsl, Aaboud:2017sjh, CMS:2017epy}, $\sigma_{bbH/A} \times Br(H/A \to \tau \tau)$~\cite{Aaboud:2017sjh, CMS:2017epy}, $pp \to H \to \gamma\gamma$~\cite{Aaboud:2017yyg} and $pp \to H \to hh \to 4b,2b2\gamma,2b2\tau$~\cite{CMS:2017ihs, CMS-PAS-HIG-17-009, Sirunyan:2017djm} derived by the CMS and ATLAS collaborations upon using the Run-II dataset with an integrated luminosity of $\sim 36~\ifb$. The gluon fusion channel is undoubtedly the dominant Higgs production mode at the LHC for low values of $\tan\beta$. However, it gets overrun by the $b\bar{b}H/A$ production channel at high $\tan\beta$ values. In the current analysis, while evaluating the impact of the existing upper limits on $pp \to H \to hh \to 4b,2b2\gamma,2b2\tau$, only the contributions from the gluon fusion production are taken into account. This choice is motivated by the fact that the $H \to hh$ decay modes gain dominance only in the low and intermediate $\tan\beta$ values where the gluon fusion mode overshadows the $b\bar{b}H/A$ channel. Although the current search limits on $H \to hh$ do not impose any constraints on our parameter space, the future runs have the potential to probe the low $m_{A}$ and low $\tan\beta$ regime. The impact of these future limits are discussed in the later part of this section. The $H \to ZZ/WW$ limits also turn out to be ineffective in constraining our parameter space and will require improvements of about three orders of magnitude for making any impact. We would like to mention that the upper limits derived by ATLAS in the $H \to \gamma\gamma$ search channel is on the fiducial cross-section times BR$(H \to \gamma\gamma)$. We compare these upper limits against a combination of the $ggF+b\bar{b}H/A$ production cross-sections and observe that an improvement of around two orders of magnitude will be required in order to affect our parameter space. Limits from searches in the $H/A \to \tau\tau$ channel impose the strongest constraints on the parameter space. Constraints from $\sigma_{b\bar{b}H/A} \times \textrm{BR}(H/A \to \tau\tau)$ yield stronger limits compared to their gluon fusion counterparts and exclude the low $m_{A}$ and high $\tan\beta$ region. The current search limits from ATLAS and CMS furnish roughly equivalent impact and rule out $\tan\beta \gtrsim 16$ for $m_{A} \sim 1~{\rm TeV}$. Before presenting the results in the $m_A-\tan{\beta}$ plane, we show the current allowed branching fractions, \textit{viz.}, $H \to hh, H \to t\bar{t}$ and $H \to \tau^+\tau^-$ in Fig.~\ref{fig:SUSY_BR}. The $H \to hh$ branching ratio dominates for $\tan{\beta} \lesssim 8$ and for $m_H \le 2 m_t$. All points are allowed by the Higgs mass and Higgs signal strength constraints. However, the grey regions are excluded by the present direct searches of the heavy Higgs. In Fig.~\ref{fig:bbh_tt_curr}, we show the impact of the latest direct search limits from $\sigma_{b\bar{b}H/A} \times Br(H/A \to \tau\tau)$ in the $m_{A}-\tan\beta$ plane. The parameter space points shown in Fig.~\ref{fig:bbh_tt_curr} (grey and orange) are obtained after implementing the light Higgs mass constraints and the Higgs signal strength measurements. The grey points are excluded upon imposing the aforementioned direct search limits.

\begin{figure}
\begin{center}
\includegraphics[scale=0.15]{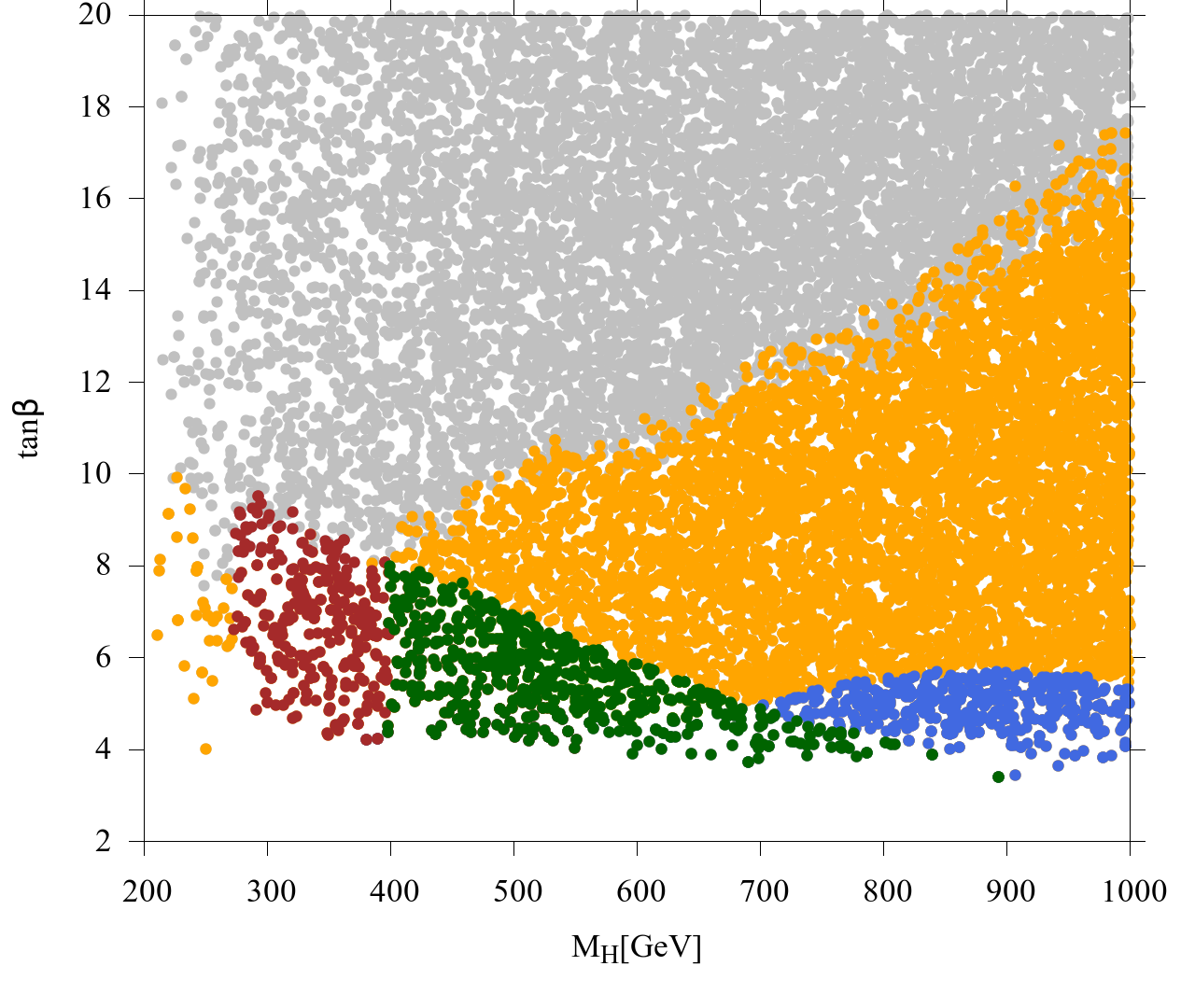}
\includegraphics[scale=0.15]{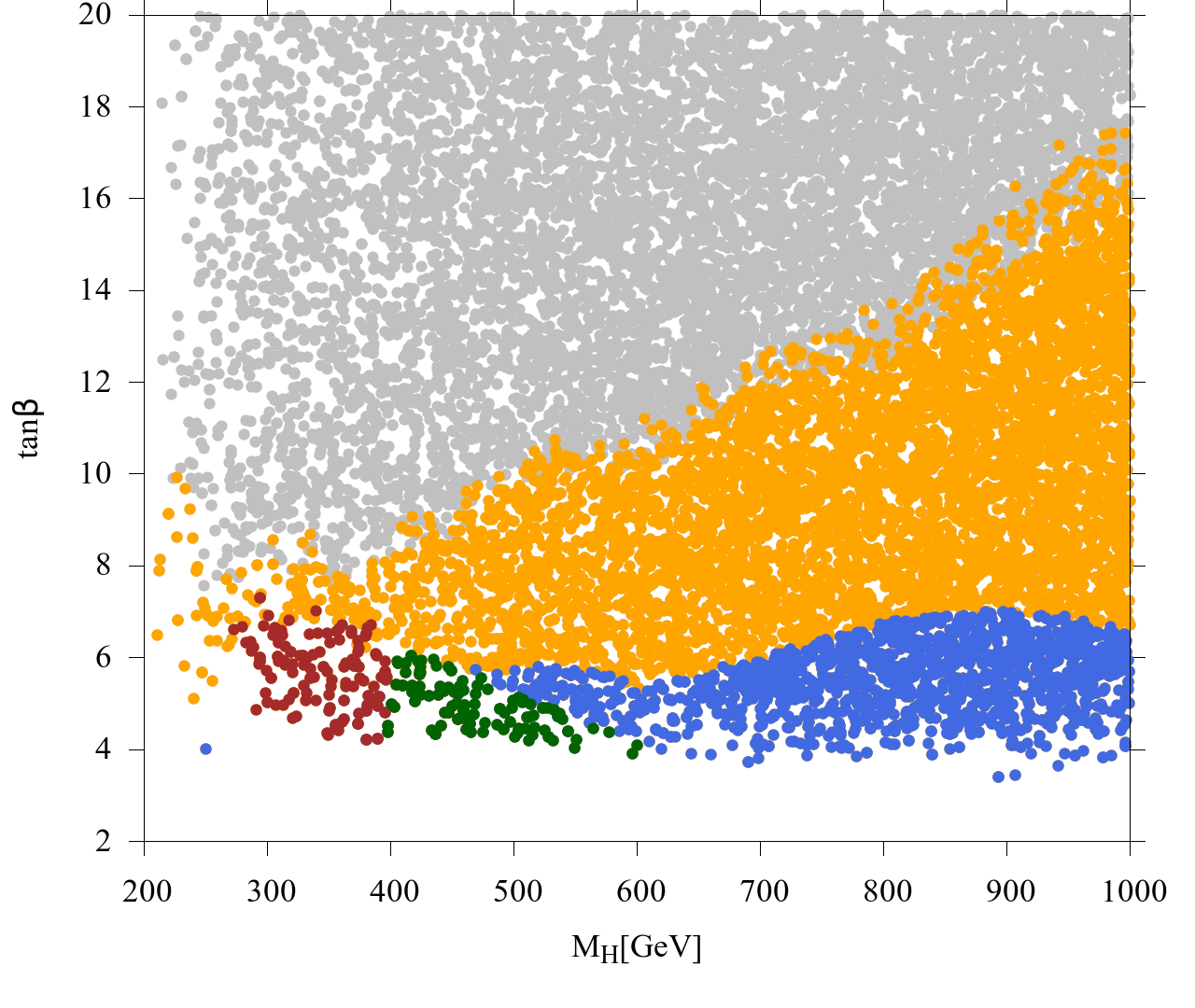}
\caption{Scatter plot in the $m_{A}-\tan\beta$ plane showing the impact of the projected search limits derived in this study, for the case of HL-LHC. The orange and grey colored points represent the same color code of Fig.~\ref{fig:bbh_tt_curr}. The brown colored points are excluded by the $2\sigma$ upper limits on $\sigma_{ggH} \times Br(H \to hh) \times Br(h \to b\bar{b}) \times Br(h \to \gamma\gamma)$, derived in Sec.~\ref{sec:Htohhtobbgaga}, while the green colored points are excluded by the $2\sigma$ upper limits on $\sigma_{ggH} \times Br(H \to t\bar{t})$ derived in Sec.~\ref{sec:Htott}. Upper limits derived for the case of $\sigma_{b\bar{b}H/A} \times Br(H/A \to \tau\tau)$ (Sec.~\ref{sec:bbH}) at $2\sigma$ rule out the orange colored points. The blue colored points represent the parameter space which would remain allowed after the HL-LHC run. The left and the right plots respectively show the exclusion at 2$\sigma$ and discovery reach at 5$\sigma$.}
\label{fig:bbh_tt_fut}
\end{center}
\end{figure}

Our main concern in this section is to quantify the impact of the projected direct search limits for the HL-LHC which were derived in the previous sections. In this regard, we consider the projected direct search limits for the HL-LHC in the $H \to hh$ (Sec.~\ref{sec:Htohh}), $H \to t\bar{t}$ (Sec.~\ref{sec:Htott}) and $b\bar{b}H/A \to b\bar{b}\tau_{h}\tau_{h}$ (Sec.~\ref{sec:bbH}) channels. Among the various final states of the $H \to hh$ channel, the $b\bar{b}\gamma\gamma$ final state furnishes the strongest limit in the $m_{A} \lesssim 600~{\rm GeV}$ regime, while the $4b$ final state imposes the strongest upper limits in the $m_{A} \gtrsim 600~{\rm GeV}$ region. The $H \to hh$ decay mode gains dominance in the low $\tan\beta$ region and especially before the $t\bar{t}$ mass threshold is attained. The same is reflected in the left panel of Fig.~\ref{fig:bbh_tt_fut} where the brown points represent the region excluded at 95\% CL by the projected $2\sigma$ reach from the $H \to hh \to b\bar{b}\gamma\gamma$ channel. The $4b$ final state, on the other hand, is rendered ineffective on account of reduced production cross-section at high values of $m_{A}$. The upper limits derived from searches in the remaining $H \to hh$ channels furnish much weaker bounds and will not be able to probe the pMSSM parameter even at the HL-LHC. The couplings of the heavy Higgs bosons with the up-type quarks have an inverse dependence on $\tan\beta$ and thus consequently the $H \to t\bar{t}$ channel has the potential to play an important role in the low $\tan\beta$ regime. The parameter space points excluded at 95\% CL by the $H \to t\bar{t}$ HL-LHC search limits derived in Sec.~\ref{sec:Htott} are shown in green in Fig.~\ref{fig:bbh_tt_fut}. The strongest future limits are obtained by the $b\bar{b}H \to b\bar{b}\tau\tau$ channel (derived in Sec.~\ref{sec:bbH}). This will be able to exclude (at 2$\sigma$) until $\sim \tan\beta \sim 5.5$ at $m_{A} \sim 1~{\rm TeV}$ as shown in Fig.~\ref{fig:bbh_tt_fut}, where the orange points are excluded by the same. The blue points in Fig.~\ref{fig:bbh_tt_fut} denotes the parameter space which will evade the direct searches at the HL-LHC as well. The right panel in Fig.~\ref{fig:bbh_tt_fut}, however, shows the discovery potential at 5$\sigma$.

At this point, we would like to briefly discuss the implications from direct charged Higgs search limits on the parameter space of our interest. A detailed analysis of the exclusion reach of current limits from direct charged Higgs searches in the $pp \to H^{\pm} \to \tau^{\pm}\nu_{\tau}$ and $pp \to H^{+} \to t\bar{b}$ can be found in \cite{Barman:2016jov,Djouadi2015}. Fig.~11 and Fig.~12 of \cite{Barman:2016jov} shows that the allowed MSSM parameter space points (obtained by imposing the light Higgs mass constraint, Higgs signal strength limits and flavour physics constraints) are outside the current reach of charged Higgs searches in the $\tau\nu_{\tau}$ and $t\bar{b}$ channels, respectively, and, the direct charged Higgs search limits would require an improvement of roughly an order of magnitude in order to be capable of probing some of the MSSM parameter space points considered in \cite{Barman:2016jov}. The implications for future direct charged Higgs search limits for HL-LHC, obtained by scaling the current limits, has been analysed in \cite{Djouadi2015}, where the projected reach of direct searches in the $H^{\pm} \to \tau^{\pm} \nu_{\tau}$ and $H^{\pm} \to t\bar{b}$ channels has been translated to the MSSM parameter space and presented in the $M_{A}-\tan\beta$ plane (see Fig.~20 of \cite{Djouadi2015}). A comparison with the analysis in \cite{Djouadi2015} indicates that the future reach of direct charged Higgs searches at the HL-LHC is weaker than the future reach of direct heavy Higgs searches in the $b\bar{b}H/A,~ H/A \to \tau^{+}\tau^{-}$ channel derived in this work.

\begin{figure}
\begin{center}
\includegraphics[scale=0.15]{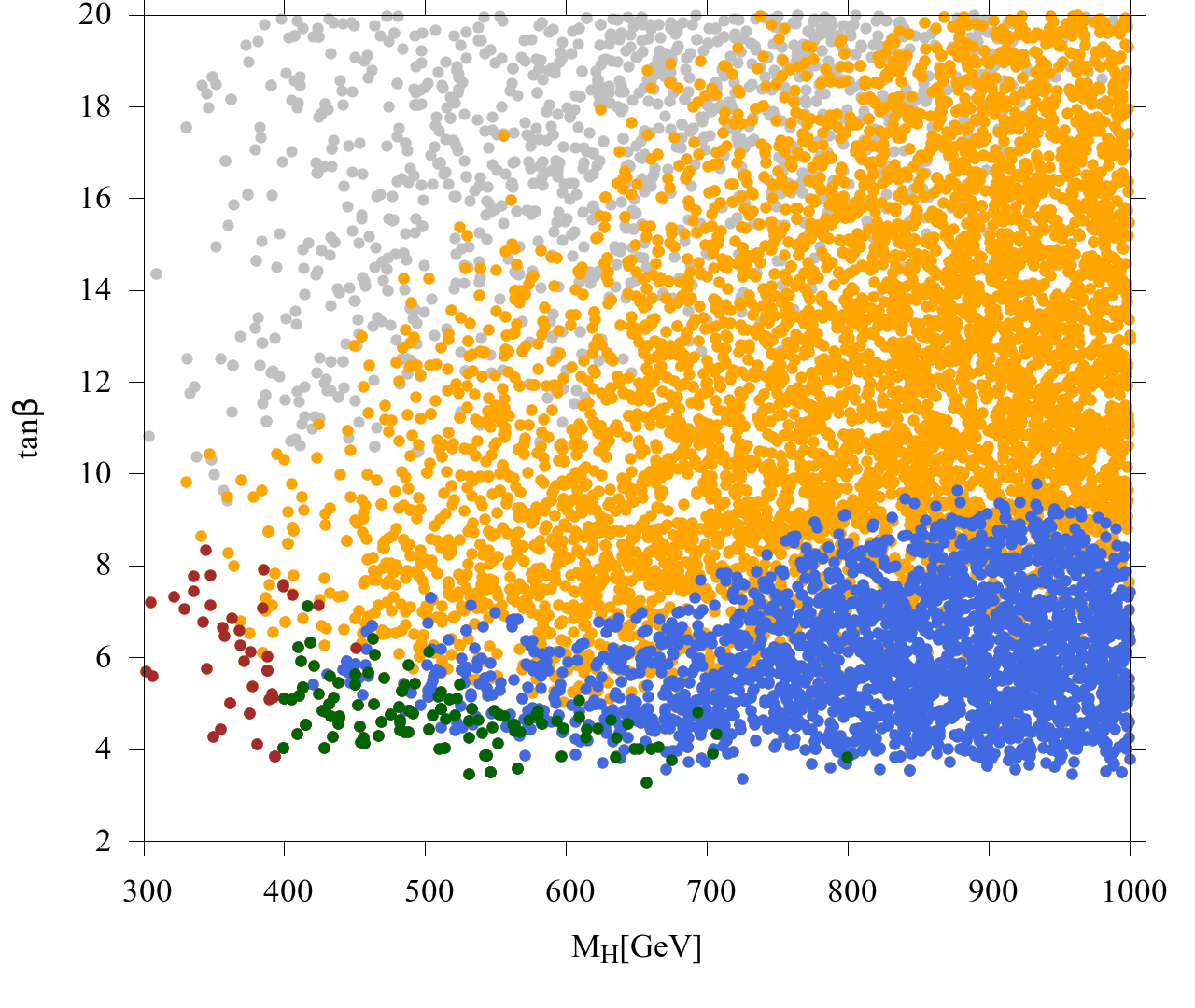}
\caption{Same as the left panel of Fig.~\ref{fig:bbh_tt_fut} but for $|M_{2} -\mu| < 10~{\rm GeV}$, $M_{H/A} > (M_{\chi_{1}^{0}} + M_{\chi_{2}^{0}})$ and $M_{2},\mu > 200~{\rm GeV}$.}
\label{fig:bbh_tt_fut_M2_mu}
\end{center}
\end{figure}

The results discussed till now assume that the heavy Higgs bosons underwent decays only into SM final states. The branching fractions of the heavy Higgs bosons into SM final states can, in principle, undergo significant modifications in the presence of light SUSY particles\footnote{SUSY particles with their masses less than $M_{H/A}/2$, such that it is kinematically possible for the heavy Higgs bosons to decay into them, are referred to as the light SUSY particles}. For example, for intermediate values of $\tan\beta \sim 7-10$, the branching fraction of the heavy Higgs bosons into charginos and neutralinos may attain significantly large values ($\gtrsim 50\%$)~\cite{Ananthanarayan:2015fwa, Barman:2016kgt}. In the remainder of this section we study the impact on the pMSSM parameter space in the presence of non-SM decay modes of the heavy Higgs bosons at the HL-LHC. Here we will restrict ourselves to the case of light electroweakinos. These electroweakinos are required to be an admixture of gauginos (bino and wino) and higgsinos in order to have couplings with the Higgs bosons. The pure gauginos and higgsinos do not couple with the Higgs states. LHC searches in the chargino-neutralino pair production mode furnishes the most stringent constraints on the electroweakino sector and excludes degenerate wino-like $\chi_{2}^{0}$ and $\chi_{1}^{\pm}$ of mass $\lesssim 450~{\rm GeV}$~\cite{Sirunyan:2018ubx}, for an LSP neutralino of mass $\sim 100~{\rm GeV}$. However, such constraints do not apply to scenarios where the LSP and NLSP are almost degenerate in mass. We explore this fact and vary $M_{2}$ and $\mu$ in such a way that $|M_{2} -\mu| < 10~{\rm GeV}$, $M_{H/A} > (\chi_{1}^{0} + \chi_{2}^{0})$ and $M_{2},\mu > 200~{\rm GeV}$~\cite{Sirunyan:2018ubx}. Closeness between $M_{2}$ and $\mu$ ensures that the $\chi_{1}^{0},~\chi_{2}^{0},~\chi_{3}^{0}$ and $\chi_{1}^{\pm}$ have significant admixtures from both winos and higgsinos. The remaining input parameters are randomly varied within the range specified in Eqn.~\ref{Eqn:para_space_1} except for $M_{1}$ which we fix at $1~{\rm TeV}$. In presence of these $H/A \to ino$ decay modes, the branching fraction of the heavy Higgs bosons to SM final states undergoes modifications and manifests in weaker limits on the parameter space, as shown in Fig.~\ref{fig:bbh_tt_fut_M2_mu}. Correspondingly, the orange and blue regions shift upward. The brown and green regions shrink further down. For $m_A$ varying between 400 GeV and 700 GeV, $\tan{\beta}$ as low as 3 is excluded at 95\% CL. In presence of these non-SM decay modes, the current ($13~{\rm TeV}$, $36~\ifb$) limits on $\sigma_{b\bar{b}H/A}\times Br(H/A \to \tau \tau)$ exclude $\tan\beta \gtrsim 22$ for $m_{A} \sim 1~{\rm TeV}$. The HL-LHC reach weakens out till $\tan\beta \sim 10$ at $m_{A} \sim 1~{\rm TeV}$. In the current scenario, the projected limits from $H \to t\bar{t}$ lose sensitivity on the pMSSM parameter space under study. The HL-LHC projections from $H \to hh \to b\bar{b}\gamma\gamma$ also imposes weaker constraints and excludes $\tan\beta < 8$ at $m_{A} \sim 400~{\rm GeV}$.

\section{Summary}
\label{sec:summary}

In this work, we have studied the prospects of observing or excluding a resonant heavy Higgs or pseudoscalar in the purview of the HL-LHC at 14 TeV. Various experimental observations and theoretical motivations necessitate physics beyond the Standard Model (BSM). Several searches are performed, either in the context of specific models or in a generic model-independent fashion, to gauge the type of new physics. In this work, we specifically focus on neutral heavy Higgs bosons (both $CP$ odd and even). Run-II data at the LHC has already constrained strongly interacting BSM particles like gluinos and stops to $\mathcal{O}(\ge 1)$ TeV. However, the LHC still hasn't imposed such strong constraints on extended Higgs sectors. The Standard Model Higgs self-coupling being still unknown, we are yet to fully understand the scalar sector of new physics. Here, we studied three major search channels for such a heavy Higgs (or pseudoscalar). Specific to the $CP$-even heavy Higgs, we studied the prospects of constraining $\sigma(pp \to H \to hh)$ in multifarious channels, \textit{viz.}, $b\bar{b}\gamma\gamma$, $b\bar{b}b\bar{b}$, $b\bar{b}\tau^+\tau^-$, $b\bar{b}WW^*$ and $\gamma\gamma WW^*$. We took guidance from the present searches and optimised each channel carefully to obtain upper limits on $\sigma(pp \to H \to hh)$ from each channel. Corroborating the present searches, we find that the $b\bar{b}\gamma\gamma$ and $b\bar{b}b\bar{b}$ final states serve as the golden channels for $m_H$ in the range $[300,600]$ GeV and $[600,1000]$ GeV, respectively. The $b\bar{b}\gamma\gamma$ sets a 95\% CL upper limit on $\sigma(pp \to H \to hh)$ between $[79.03,14.10]$ fb in the aforementioned mass range. The $4b$ channel on the other hand sets a corresponding cross-section limit between $[5.36,2.51]$ fb for $m_H \in [800,1000]$ GeV. The limits from the remaining three channels are not so promising. On the other hand, if the mass of the scalar or the pseudoscalar Higgs is above the $t\bar{t}$ threshold and the Higgs is dominantly produced via gluon fusion, then it can also have a dominant decay into $t\bar{t}$. Upon studying the fully leptonic as well as the semi-leptonic final states, we find the strongest limits on $\sigma(pp \to H \to t\bar{t})$ from the semi-leptonic category. The 95\% CL upper limits lie between $186.57$ fb and $32.81$ fb for $m_H \in [400,1000]$ GeV. Finally, we studied the $b\bar{b}H/A$ production in the $b\bar{b}\tau\tau$ final state upon demanding at least one $b$-tagged jet in the final state and demanding two hadronic $\tau$s. The 95\% CL upper limit on $\sigma(pp \to b\bar{b}H)$ varies between $22.16$ fb and $3.68$ fb for $m_H$ lying between 300 GeV and 500 GeV. All our searches for the scalars are mostly model independent and can be translated to models with multiple Higgs bosons with narrow widths.

In this work, we considered the specific example of supersymmetry, more specifically the pMSSM. We apply present constraints from the SM-like Higgs boson mass measurement and all its signal strengths into multiple final states. The future limits obtained in this work constrain different regimes of the parameter space. The $H \to hh$ search, mostly in the $b\bar{b}\gamma\gamma$ channel excludes $\tan{\beta}$ to as low as $4$, at $95\%$ CL, for $m_A \sim 2 m_t$ GeV. The $H \to t\bar{t}$ has a similar exclusion on $\tan{\beta}$ for $m_A$ varying between $[400,800]$ GeV. The $b\bar{b}H$ channel in the di-$\tau + \ge 1 b$-tagged jet final state excludes $\tan{\beta}$ as low as $5.5$ for $m_A=1$ TeV. The blue region in Fig.~\ref{fig:bbh_tt_fut} will not be probed even by direct searches if the heavy Higgs bosons decay only to SM particles. We will require higher energy colliders in order to be able to probe this region. This scenario might change if there are light electroweakinos, below $m_{H/A}/2$. In that situation, the $m_A-\tan{\beta}$ parameter region changes. Upon considering a scenario where $|M_2 - \mu| < 10$ GeV, $M_{H/A} > (\chi_1^0 + \chi_2^0)$ and $M_2/\mu > 200$ GeV, one finds that the $H \to hh$ and $H \to t\bar{t}$ excludes $\tan{\beta}$ down to 3 for $m_A \in [400,700]$ GeV. The exclusion bound on $\tan{\beta}$ from the $b\bar{b}H$ search decreases to $10$ for $m_A=1$ TeV at $95\%$ CL.

\newpage

\appendix

\section{Detailing the cross section with generation cuts for the signal and backgrounds}
\label{sec:appendixA}

\begin{table}[htb!]
\centering
\begin{bigcenter}
\scalebox{0.6}{%
\begin{tabular}{|c|c|c|c|c|}

\hline
Process & Backgrounds & \makecell{Generation-level cuts ($\ell=e^\pm,\mu^\pm$)\\ (NA : Not Applied)} & Cross section (fb)  \\ \hline\hline
\multicolumn{4}{|c|}{$pp\to H\to hh$, $pp\to A\to Zh$ and $pp\to H\to t\bar{t}$ final states} \\ \hline\hline
\multirow{14}{*}{$b \bar{b} \gamma \gamma$}
                                           & {$hh\to b \bar{b} \gamma \gamma$} & {NA} & {$0.10$}   \\\cline{2-4}                                                                                      

                                           & {$b\bar{b}\gamma\gamma+$ jets} 
                                           & { \makecell{$p_{T,j/b/\gamma}>20~\text{GeV}$, $|\eta_j|<5.0$, $|\eta_{b,\gamma}|<2.5$, \\ $\Delta R_{b,j,\gamma}$\footnote{$\Delta R_{b,j,\gamma}$ means $\Delta R$ between all possible combination of $b,j$ and $\gamma$.}$>0.2$, $m_{bb}>50$ GeV, $110<m_{\gamma\gamma}<140$ GeV}} 
                                           & {$18.78$}\\\cline{2-4}  

                                           & {$c\bar{c}\gamma\gamma$} 
                                           & {\makecell{$p_{T,j/b/\gamma}>20~\text{GeV}$, $|\eta_j|<5.0$, $|\eta_\gamma|<2.5$ \\$\Delta R_{b,j,\gamma}$$>0.2$, $110<m_{\gamma\gamma}<140$ GeV}} 
                                           & {$162.22$}\\\cline{2-4}                                                                                                                             
                                           
                                           & {$jj\gamma\gamma$} 
                                           & {same as $c\bar{c}\gamma\gamma$} 
                                           & {$2770.67^{*}$}\\\cline{2-4}                                                                                                                                                                      

                                           & {$t\bar{t}h$, $h\to\gamma\gamma$} & {NA}  & {$1.39$}  \\\cline{2-4}

                                           & {$hb\bar{b}$, $h\to\gamma\gamma$} & {NA}  & {$1.32$}   \\\cline{2-4}                                      
                                                                                     
                                           & {$Zh$, $h\to\gamma\gamma$, $Z\to b\bar{b}$} & {NA}  & { $0.33$}   \\\cline{2-4}
                                           
                                           & {$b\bar{b}jj$} 
                                           & {\makecell{$p_{T,j}>10~\text{GeV}$, $p_{T,b}>20~\text{GeV}$, $|\eta_{j/b}|<5.0$, \\  $m_{jj}>50$ GeV, $m_{bb}>50$ GeV}} & $549583730.00^{*}$ \\\cline{2-4}                                                                                 
                                                                                     
                                           & {$b\bar{b}j\gamma$} 
                                           & {\makecell{$p_{T,j/b/\gamma}>20~\text{GeV}$, $|\eta_j|<5.0$, $|\eta_{b/\gamma}|<2.5$, \\ $\Delta R_{b/b/\gamma/\gamma,b/j/j/b}$\footnote{$\Delta R_{a/b,c/d}$ signifies $\Delta R_{ac}$ and $\Delta R_{bd}.$}$>0.2$, $m_{bb}>50$ GeV}}                                 & {$201800^{*}$}                                 \\\cline{2-4}                                                                      
                                           
                                           & {$c\bar{c}j\gamma$} 
                                           & {$p_{T,j/\gamma}>20~\text{GeV}$, $|\eta_j|<5.0$, $|\eta_{\gamma}|<2.5$, $\Delta R_{j,\gamma}$$>0.2$}                                  
                                           & {$1132709.63^{*}$} \\\cline{2-4}                                                     
                                                       
                                           & {$Z\gamma\gamma+$ jet, $Z\to b\bar{b}$} & {\makecell{$p_{T,j/b/\gamma}>20~\text{GeV}$, $|\eta_j|<5.0$, $|\eta_{b/\gamma}|<2.5$, \\  $\Delta R_{b,j,\gamma}$$>0.2$, $m_{bb}>50$ GeV, $110<m_{\gamma\gamma}<140$ GeV}} & {$0.87$}   \\\cline{2-4}                                                                                                                                                                  
                                           
                                           & {$gg\to h~+~c\bar{c}$, $h\to\gamma\gamma$ } & {\makecell{$p_{T,j}>20~\text{GeV}$, $|\eta_j|<5.0$, $\Delta R_{j,j}$$>0.2$ }} & $0.31$   \\\cline{2-4}                                                                                                                                                                  
    
                                           & {$gg\to h~+~jj$, $h\to\gamma\gamma$ } & {same as $gg\to h~+~c\bar{c}$} & $27.89^{*}$   \\\hline                                   
        

\multirow{7}{*}{$b\bar{b}b\bar{b}$}          
                                           & {$hh\to b \bar{b} b \bar{b}$} & {NA}  & {$13.42$}  \\\cline{2-4}
                                           
                                           & {multijet $b\bar{b}b\bar{b}$} & {\makecell{$p_{T,j/b}>50~\text{GeV}$, $|\eta_{j/b}|<3.0$, $\Delta R_{b,j}>0.3$,\\ $H_{T}>250$ GeV }}  & {$14541.30$}  \\\cline{2-4}
                                                                                      
                                           & {multijet $b\bar{b}c\bar{c}$} & {$same$ as multijet $b\bar{b}b\bar{b}$}  & {$28633.60$}  \\\cline{2-4}
                                           
                                           & {multijet $b\bar{b}jj$} & {$same$ as multijet $b\bar{b}b\bar{b}$}  & {$3602560.00^{*}$}  \\\cline{2-4}                                                                                    
                                           
                                           & {$t\bar{t}$, $W^\pm\to c(\bar{c}) \bar{s}(s)$} & {$same$ as multijet $b\bar{b}b\bar{b}$}  & {$860.17$}  \\\cline{2-4}                                                                                                                          
                                                                                                                              
                                           & {$t\bar{t}b\bar{b}$} & {$p_{T,b}>50~\text{GeV}$, $|\eta_{b}|<3.0$, $\Delta R_{b,b}>0.3$, $H_{T}>250$ GeV}  & {$170.58$} \\\hline


\multirow{13}{*}{$b \bar{b} \tau^+ \tau^-$} 
                                           & {$hh\to b \bar{b} \tau^+ \tau^-$} & {NA} & {$2.89$}   \\\cline{2-4}

                                           & {$t\bar{t}$ hadronic} & \makecell{$p_{T,j/b}>20~\text{GeV}$, $p_{T,l}>8~\text{GeV}$, $|\eta_j|<5.0$, \\ $|\eta_{b/\ell}|<3.0$, $\Delta R_{b,j,\ell}>0.2$, $m_{bb}>50$ GeV }  & {$135623.50$}  \\\cline{2-4}
                                                                                      
                                           & {$t\bar{t}$ semi-leptonic} & {$same$ as $t\bar{t}$ hadronic}  & {$173409.88$}  \\\cline{2-4}

                                           & {$\tau\tau b\bar{b}$} & {\makecell{$p_{T,b}>20~\text{GeV}$, $p_{T,\tau}>8~\text{GeV}$, $|\eta_{b/\tau}|<3.0$, \\ $\Delta R_{b,\tau}>0.2$, $m_{bb}>50$ GeV, $m_{\tau\tau}>30$ GeV }}  & {$2128.56$}   \\\cline{2-4}
                                                                                      
                                            & {$b\bar{b}h$, $h\to\tau\tau$} & {\makecell{$p_{T,b}>20~\text{GeV}$, $p_{T,\tau}>10~\text{GeV}$, $|\eta_j|<5.0$,\\ $|\eta_{b/\tau}|<3.0$, $\Delta R_{b,\tau}>0.2$, $m_{bb}>50$ GeV }}  & {$1.23$}  \\\cline{2-4}

                                           & {Zh, $h\to (b\bar{b}+\tau\tau)$, $Z\to (\tau\tau+b\bar{b})$} & {NA}  & {$28.21$}   \\\cline{2-4}
                                           & {$t\bar{t}h$} & {NA}  & {$611.30$}   \\\cline{2-4}
                                           & {$t\bar{t}Z$} & {NA}  & {$731.54$}  \\\cline{2-4}
                                           & {$t\bar{t}W$} & {NA}  & {$437.87$}   \\\cline{2-4}
                                           & {$b\bar{b}jj$} & {\makecell{$p_{T,j}>10~\text{GeV}$, $p_{T,b}>20~\text{GeV}$, $|\eta_{j/b}|<5.0$, \\  $m_{jj}>50$ GeV, $m_{bb}>50$ GeV}}  & {$549583730.00^{*}$}   \\\hline

      \end{tabular}}
       \end{bigcenter}
           \caption{Generation level cuts and cross-sections for the various backgrounds used in the analyses. The backgrounds labelled with $*$ are multiplied by the fake rates before doing the analysis. The fake rates used are $0.05\%$~\cite{ATL-PHYS-PUB-2017-001} for $j\to\gamma$, $\sim 1.75\%$(average from the fake rate function) for $j\to b$ and $0.35\%$~\cite{CMS-PAS-TAU-16-002} for $j\to\tau$.}

\label{app1:1}
          \end{table}

\begin{table}[htb!]
\centering
\begin{bigcenter}
\scalebox{0.6}{%
\begin{tabular}{|c|c|c|c|c|}
\hline
Process & Backgrounds & \makecell{Generation-level cuts ($\ell=e^\pm,\mu^\pm$)\\ (NA : Not Applied)} & Cross section (fb)  \\ \hline\hline
\multicolumn{4}{|c|}{$pp\to H\to hh$, $pp\to A\to Zh$ and $pp\to H\to t\bar{t}$ final states} \\ \hline\hline
\multirow{7}{*}{$b \bar{b} WW^*$}          
                                           & {$hh\to b \bar{b} W^+ W^-$} & {NA}  & {$9.85$}  \\\cline{2-4}
                                           & {$t\bar{t}$ semi-leptonic} & {\makecell{$p_{T,j/b}>20~\text{GeV}$, $p_{T,l}>8~\text{GeV}$, $|\eta_j|<5.0$, \\ $|\eta_{b/\ell}|<3.0$, $\Delta R_{b,j,\ell}>0.2$, $m_{bb}>50$ GeV }}  & {$173409.88$}  \\\cline{2-4}
                                                                                      
                                           & {$t\bar{t}$ leptonic} & {$same$ as $t\bar{t}~$semileptonic}  & {$55319.44$}  \\\cline{2-4}
                                           & {$\ell\ell b\bar{b}$} & {\makecell{$p_{T,b}>20~\text{GeV}$, $p_{T,l}>8~\text{GeV}$, $|\eta_{b/\ell}|<3.0$, \\ $\Delta R_{b,\ell}>0.2$, $m_{bb}>50$ GeV}}  & {$7393.72$}   \\\cline{2-4}       
                                                                                 
                                           & {$Wbb+jets$, $W\to \ell \nu$, $\ell$ also includes $\tau$} & {\makecell{$p_{T,j/b}>20~\text{GeV}$, $p_{T,l}>8~\text{GeV}$, $|\eta_j|<5.0$,\\ $|\eta_{b/\ell}|<3.0$, $\Delta R_{j,b,l}>0.2$}}  & {$32576.60$}  \\\cline{2-4}
                                                                                      
                                           & {$t\bar{t}h$} & {NA}  & {$611.30$}   \\\cline{2-4}
                                           & {$t\bar{t}Z$} & {NA}  & {$731.54$}  \\\cline{2-4}
                                           & {$t\bar{t}W$} & {NA}  & {$437.87$}   \\\hline  

\multirow{6}{*}{$\gamma\gamma WW^*$ }      & {$hh\to \gamma\gamma W^+ W^-$} & {NA} & {$0.04$}  \\\cline{2-4}                                                                   
                                                                                                                                 
                                           & {$t\bar{t}h$, $h\to \gamma\gamma$} & {NA}  & {$1.39$}   \\\cline{2-4}
                                           
                                           & {$Zh$ + jets, $h\to\gamma\gamma$, $Z\to \ell\ell$($\ell$ includes $\tau$ also)} & {\makecell{$p_{T,\gamma/\ell}>10~\text{GeV}$, $|\eta_j|<5.0$, $|\eta_{\gamma/\ell}|<2.5$,\\ $\Delta R_{\gamma,\ell,j}>0.2$, 120 GeV  $<m_{\gamma\gamma}<$ 130 GeV}}  & {$0.12$}  \\\cline{2-4}
                                           
                                           & {$Wh$ + jets, $h\to\gamma\gamma$, $W\to \ell\nu$($\ell$ includes $\tau$ also)} & {same as $Zh$ + jets }  & {$0.70$}   \\\cline{2-4}
                                           
                                           & {$\ell\nu\gamma\gamma$ + jets, $\ell$ also includes $\tau$} & {\makecell{$p_{T,\gamma/\ell}>10~\text{GeV}$, $|\eta_{j}|<5.0$, $|\eta_{\gamma/\ell}|<2.5$, $\Delta R_{\gamma \gamma}>0.2$, \\ $\Delta R_{\gamma \ell}>0.2$, $\Delta R_{\gamma j}>0.4$, 120 GeV  $<m_{\gamma\gamma}<$ 130 GeV }}  & {$3.17$}  \\\cline{2-4}   
                                                 
                                           & {$\ell\ell\gamma\gamma$ + jets, $\ell$ also includes $\tau$} & {same as $\ell\nu\gamma\gamma$ + jets, with $m_{\ell\ell}>$ 20 GeV }  & {$1.00$}  \\\cline{2-4}                       
\hline \hline

\multicolumn{4}{|c|}{$pp\to b\bar{b} H$ final state} \\ \hline\hline

\multirow{13}{*}{$b \bar{b} \tau^+ \tau^-$} 
                                           & {$t \bar{t}$} & {$p_{T,b}>20~\text{GeV}$, $|\eta_b|<3.0$, $\Delta R_{bb}>0.2$} & {$633946.81$}   \\\cline{2-4}

                                           & {single top s-channel} & {NA} & {$11390.00$}  \\\cline{2-4}
                                                                                      
                                           & {single top t-channel} & {NA} & {$248090.00$}  \\\cline{2-4}                                                                                      
                                           
                                           & {single top Wt-channel} & {NA} & {$84400.00$}   \\\cline{2-4}
                                                                                      
                                           & {$\tau\tau$ + jets, via $Z/\gamma^*$} & {\makecell{$p_{T,j/b}>20~\text{GeV}$, $p_{T,\ell}>60~\text{GeV}$, $|\eta_{j/b/\ell}|<3.0$,\\ $\Delta R_{j,b,\ell}>0.2$, $m_{\ell\ell}>50$ GeV }}  & {$884370.24$}  \\\cline{2-4}                                        
                                                                                                                                                                             
                                           & {$W$ + jets, $W\to \ell\nu$($\ell$ includes $\tau$ also)} & {$p_{T,j/b/\ell}>20~\text{GeV}$, $|\eta_{j/b/\ell}|<3.0$, $\Delta R_{j,b,\ell}>0.2$}  & {$112358.64$}   \\\cline{2-4}
                                           
                                           & {$VV$ ($V$ includes $W^\pm$ and $Z$)} & {NA}  & {$106510.72$}   \\\cline{2-4}
                                           
                                           & {$b\bar{b}jj$} & {\makecell{$p_{T,j}>65~\text{GeV}$, $p_{T,b}>20~\text{GeV}$, $|\eta_{j/b}|<3.0$, \\ $\Delta R_{j,b}>0.2$, $m_{jj}>50$ GeV}}  & $12091572.60^{*}$   \\\hline

\end{tabular}}
\end{bigcenter}
\caption{Generation level cuts and cross-sections for the signals and various backgrounds used in the analyses. The backgrounds labelled with $*$ are multiplied by the fake rates before doing the analysis. The fake rates used are $0.35\%$~\cite{CMS-PAS-TAU-16-002} for $j\to\tau$.}

\label{app1:2}
\end{table}

\section{Validation of the $b\bar{b}H$ analysis}
\label{sec:appendixB}

\begin{table}[htb!]
    \begin{bigcenter}
        \scalebox{0.7}{%
            \begin{tabular}{||c|c||c|c|}
                \hline
                \multicolumn{2}{||c||}{Process} & Event rates at $13$ TeV with $36.1 \; \textrm{fb}^{-1}$ of integrated luminosity & Total \\ \hline
                \multicolumn{2}{||c||}{}                         & b-tag category &  \\ \hline 
                \multicolumn{2}{||c||}{multijet}                 & $97.06$ & \multirow{6}{*}{$215.26$}\\ \cline{2-3}     
                
                \multicolumn{2}{||c||}{$Z/\gamma^*+$ jets}       & $11.03$ &  \\   \cline{2-3}
                \multicolumn{2}{||c||}{$W+$ jets}                & $2.82$ &  \\   \cline{2-3}
                \multicolumn{2}{||c||}{$t\bar{t}$}               & $83.66$ &  \\   \cline{2-3}
                \multicolumn{2}{||c||}{$VV+$ jets}               & $1.87$ &  \\ \hline
                \multicolumn{2}{||c||}{$500$ GeV Signal}         & $18.82$ &  \\ \hline\hline

                \multicolumn{4}{||c||}{ATLAS numbers~\cite{Aaboud:2017sjh}} \\\hline\hline 
                \multicolumn{2}{||c||}{multijet}                   & $106\pm 32$  & \multirow{6}{*}{$180\pm 60$}\\ \cline{2-3}
                \multicolumn{2}{||c||}{$Z/\gamma^*\to \tau\tau$}   & $7.5\pm 2.9$ &  \\\cline{2-3}  
                \multicolumn{2}{||c||}{$W(\to\tau\nu)+$ jets}      & $4.0\pm 1.0$ &  \\\cline{2-3}
                \multicolumn{2}{||c||}{$t\bar{t}+$ single top}     & $60\pm 50$ &  \\\cline{2-3}
                \multicolumn{2}{||c||}{Others}                     & $1.0\pm 0.5$ &  \\ \hline 
                \multicolumn{2}{||c||}{$500$ GeV Signal}           & $28\pm 12$ &  \\ \hline\hline 
                
        \end{tabular}}
    \end{bigcenter}
    \caption{Comparison table for the $\tau_h \tau_h$ channel in the $b$-tag category.}
    \label{tab:13_TeV}
\end{table}

Before performing our analysis for the $b\bar{b}H$ category, we validate our setup with the existing analysis in this channel~\cite{Aaboud:2017sjh,Pickering:2258362,Wahrmund:2279938}.
We generate the signal events (at LO in SM) with \texttt{MG5\_aMC@NLO} and shower them via \texttt{Pythia-8}~\cite{Sjostrand:2014zea}. We use different parton distribution functions (PDFs) for the sample generations. Specifically, we use the \texttt{CT10nlo\_nf4}~\cite{Dulat:2015mca} for the 4F $b\bar{b}H$ process, \texttt{MSTW2008nnlo68cl}~\cite{Martin:2009iq} for the 5F $b\bar{b}H$ process and \texttt{CT10}~\cite{Lai:2010vv} for the ggF process. Next, we impose the following cuts in sequence. For the $b$-tag category, we demand at least one $b$-tagged jet in the final state. The events are required to have at least two $\tau$ jets with opposite charge (from their reconstructed charged tracks). The leading and sub-leading ($p_T$ ordered) $\tau$-tagged jets are required to have $p_T > 65$ GeV. The $\tau$ jets lying inside the transition region \textit{viz.}, $1.37 < |\eta| < 1.52$, are removed. Furthermore, the azimuthal angle separation between the leading and the sub-leading $\tau$ jets is required to be $\Delta\phi(\tau,\tau)>2.7$. Furthermore, we require $m_{\tau\tau}^{vis.}>50$ GeV, $\met >20$ GeV and $\sum\limits_{\tau}^{}\cos(\Delta\phi)>0$. We show the validation in Table~\ref{tab:13_TeV}.

\acknowledgments
We thank Mikael Chala, Arghya Choudhury and Marius Wiesemann for helpful discussions at various stages of the work. The work of BB is supported by the Department of Science and Technology, Government of India, under the Grant Agreement number IFA13-PH-75 (INSPIRE Faculty Award). SB is supported by a Durham Junior Research Fellowship COFUNDed by Durham University and the European Union, under grant agreement number 609412.

\providecommand{\href}[2]{#2}
\addcontentsline{toc}{section}{References}
\bibliographystyle{JHEP}
\bibliography{refs}

\end{document}